\documentclass[5p,twocolumn,10pt,times]{elsarticle}
\usepackage{amsmath}
\usepackage{hyperref}
\usepackage{cleveref}
\usepackage{booktabs} 
\usepackage{pgf,tikz,pgfplots}
\usetikzlibrary{arrows.meta}
\pgfplotsset{compat=1.14}
\usepackage{mathrsfs}
\usetikzlibrary{arrows}
\usepackage{scalefnt}
\usepackage[ruled]{algorithm2e} 

\SetAlFnt{\small}
\SetAlCapFnt{\small}
\SetAlCapNameFnt{\small}
\SetAlCapHSkip{0pt}
\IncMargin{-\parindent}
\usetikzlibrary{arrows}

\usepackage{verbatim}

\newcommand{\pw}[1]{{\textcolor{blue}{#1}}}

\usepackage{graphicx}
\usepackage{subfigure}
\usepackage{tikzscale}
\usepackage{tikz}
\usepackage{enumerate}
\newcommand{\tabincell}[2]{\begin{tabular}{@{}#1@{}}#2\end{tabular}}

\usepackage{soul}
\usetikzlibrary{arrows.meta}
\hypersetup{draft}

\newcommand*{\affaddr}[1]{#1} 
\newcommand*{\affmark}[1][*]{\textsuperscript{#1}}

\begin{document}
\baselineskip11pt

\begin{frontmatter}

\author{%
	Wei Pan\affmark[1], Xuequan Lu\affmark[2], Yuanhao Gong\affmark[1], Wenming Tang\affmark[1], Jun Liu\affmark[1], Ying He\affmark[3] and Guoping Qiu\affmark[1]\\
	\affaddr{\affmark[1]College of Information Engineering, Shenzhen University, Guangdong Key Laboratory of Intelligent Information Processing}\\
	\affaddr{\affmark[2]School of Information Technology, Deakin University}\\
	\affaddr{\affmark[3]School of Computer Science and Engineering, Nanyang Technological University}%
}



\title{HLO: Half-kernel Laplacian Operator for Surface Smoothing}


\begin{abstract}
This paper presents a simple yet effective and efficientw method for feature-preserving surface smoothing. Through analyzing the differential property of surfaces, we show that the conventional discrete Laplacian operator with uniform weights is not applicable to feature points at which the surface is non-differentiable and the second order derivatives do not exist. To overcome this difficulty, we propose a Half-kernel Laplacian Operator (HLO) as an alternative to the conventional Laplacian. Given a vertex $v$, HLO first finds all pairs of its neighboring vertices and divides each pair into two subsets (called half windows); then  computes the uniform Laplacians of all such subsets and subsequently projects the computed Laplacians to the full-window uniform Laplacian to alleviate flipping and degeneration. The half window with least regularization energy is then chosen for $v$. We develop an iterative approach to apply HLO for surface denoising. Our method is conceptually simple and easy to use because it has a single parameter, i.e., the number of iterations for updating vertices. We show that our method can preserve features better than the popular uniform Laplacian-based denoising and it significantly alleviates the shrinkage artifact. Extensive experimental results demonstrate that HLO is better than or comparable to state-of-the-art techniques both qualitatively and quantitatively and that it is particularly good at handling meshes with high noise. Also, it outperforms all other compared methods in terms of computational time. We make our executable program publicly available\footnotemark.

\end{abstract}

\end{frontmatter}


\section{Introduction}
\label{sec:introduction}
Surface meshes acquired through scanning or sensing equipment are inevitably contaminated with noise. The user has to process them with smoothing techniques before applying to down stream applications, such as shape analysis, animation and rendering. The main technical challenge of surface smoothing is preserving features while removing noise and alleviating shrinkage. The design of robust surface smoothing methods is therefore of particular need in nowadays.

The discrete Laplacian operator on surface meshes has proven highly useful in various tasks of digital geometry processing,  
for example, mesh fairing, parameterization, reconstruction, editing and compressing \cite{taubiny2000geometric,zhang2004discrete,Floater2005,sorkine2006differential,Vasa2014,HU2016}, just name a few.  The differential surface representation encodes information about the local shape of a surface such as the curvature and the orientation. Despite that the uniform Laplacian operator can effectively smooth surfaces, it fails to preserve features and leads to shrinkage.

Among the feature-preserving methods, the majority of them typically use the normal information of surfaces, such as face or vertex normals. In contrast, very few works are based on vertex position \cite{He2013,Lu2016}. Some recent works extended these two types by adding more delicate steps \cite{Wei2015,Fan2010,Bian2011,Wang2012,Zhu2013,Lu2016,Lu2017-1,Lu2017-2}. However, these methods generally consist of multiple steps and numerous parameters, which are difficult to use/tune especially for complex models. Moreover, the users especially those out of the field may find it difficult and tedious to tune the involved parameters. 

To overcome the above issues, we propose a novel, robust approach for feature-preserving mesh smoothing in this paper. Our key idea is to construct a ``half-kernel'' uniform Laplacian operator (HLO) that can approximate the Laplacians at feature and non-feature points using half windows. Specifically, we first analyze the differential property of feature points and found it conflicts with the existence assumption of the uniform Laplacian. We then naturally propose a half-window algorithm to generate different pairs of subsets (half windows) for each vertex by pairing one immediate neighbor to the other unique neighbor. We compute the Laplacians of all subsets which are further projected to the full-window Laplacians to alleviate flipping and degeneration. The final half-kernel Laplacian is automatically determined as the one that incurs the smallest regularization energy. The surface is finally updated with the determined half-kernel Laplacians in an iterative way.

Taking a noisy surface mesh as input, 
our approach can automatically output a quality version with preserving features and resisting shrinkage. The \textbf{main contributions} of this work are:
\begin{itemize}
    \item mathematical analysis of the differential property at feature points;
    \item a half-window algorithm to generate multiple half windows (subsets) for each vertex;
    \item a half-kernel uniform Laplacian operator (HLO) for feature-preserving and shrinkage-resisting surface smoothing.
\end{itemize}

Our method\footnotemark[\value{footnote}] is conceptually simple and easy to use since it involves \textit{only a single parameter} (i.e., the number of vertex update iterations). We demonstrate that the proposed HLO substantially outperforms the uniform Laplacian in preserving features and resisting shrinkage. We evaluate our method on both synthetic and real-world models, and observe that our method can produce results with comparable or higher quality than the state-of-the-art methods. We also show that our method is faster than the state-of-the-art methods.

\footnotetext{Our EXE has been released on https://github.com/WillPanSUTD/hlo. See https://github.com/xuequanlu for more tools and results. }

\section{Related Work}
\label{sec:relatedwork}
There exists a large body of literature of mesh denoising.  Due to space limit, we review only the mostly related works to ours and  refer the readers to the comprehensive surveys~\cite{botsch2008,botsch2010}. 

The classical Laplacian smoothing methods~\cite{Vollmer1999,Field1988} are simple and fast. However, its isotropic property leads to feature-wiping and shrinking artifacts. Taubin~\cite{taubin1995signal} proposed a non-shrinking, two-step smoothing method with positive and negative damping factors. 
Desbrun et al.~\cite{desbrun1999} proposed a fairing method based on diffusion and curvature flow to process irregular meshes. 
Later, various isotropic smoothing methods have been introduced based on volume preservation, pass frequency controlling, differential properties, etc,.~\cite{liu2002,kim2005,Nehab2005,Nealen2006,su2009}.

The above isotropic methods are effective to remove noise, however they also wipe out features. Various anisotropic methods have been proposed to preserve features. Representative methods are diffusion/differential-based methods~\cite{Tasdizen2002,Clarenz2000,Desbrun2000,Bajaj2003,Hildebrandt2004,Ohtake2000,Ouafdi2008:global,Ouafdi2008:smart,He2013}, bilateral filters~\cite{Jones2003,Fleishman2003,Zheng2011,Lee2005,Wang2006}, methods combining normal filtering and vertex update~\cite{Zheng2011,Lee2005,Taubin2001,Ohtake2001,chen2005sharpness,Zhang2015,Sun2008,Sun2007,Shen2004,Jones2003,Yagou2002,Yagou2003,Zhang2015,Wang-2016-SA,Lu2017-1,Lu2017-2}. Bilateral filtering is initially used in image denoising~\cite{Tomasi1998}, and it is successfully extended to mesh denoising. Regarding the feature-preserving techniques, the utilization of filtered normals has seen noticeable progress in recent years~\cite{Zheng2011,Lee2005,Taubin2001,Ohtake2001,chen2005sharpness,Zhang2015,Sun2008,Sun2007,Shen2004,Jones2003,Yagou2002,Yagou2003,Zhang2015,Wang-2016-SA,Lu2017-1,Lu2017-2,Yadav2018-1,Yadav2018-2}. Some recent methods have focused on vertex and face classification before mesh denoising~\cite{Wei2015,Fan2010,Bian2011,Wang2012,Zhu2013,Wang2009,Wei2017}. Nevertheless, the classification results depends largely on the level of noise. Lu et al. \cite{Lu2016,Lu2017-1,Lu2017-2} presented the ideas of pre-filtering before real smoothing and can robustly handle heavy noise with frequent flipped triangle faces. Arvanitis et al. \cite{Arvanitis2019} introduced a novel coarse-to-fine graph spectral processing approach for mesh denoising.

Another line of anisotropic methods focused on the sparse perspective. Compared with non-feature vertices, feature vertices in a mesh are usually sparse, which can be reconstructed or detected by solving a sparse problem~\cite{Avron2010}. Sparsity was introduced into mesh smoothing in some recent works~\cite{He2013,Wang2014,Lu2016,Zhao2018}. For instance, He et al.~\cite{He2013} developed a $ L_{0} $ minimization framework with an area-based edge operator which is generally sparse in a surface mesh. It is, however, non-convex and difficult to solve. Zhao et al. \cite{Zhao2018} designed an improved alternating optimization strategy to solve the $L_0$ minimization which incorporates both vertex positions and face normals. Wang et al.~\cite{Wang2014} proposed a method to decouple noise and features by weighted $ L_{1} $-analysis compressed sensing, and they prove that the pseudo-inverse matrix of the Laplacian of a mesh is a coherent dictionary for sparsely representing sharp feature signals on the shape.  
Recently, Lu et al. \cite{Lu2016} detected features by introducing a novel $L_1$ minimization. Lu and his colleagues \cite{Lu2018} proposed a low-rank matrix approximation approach for geometry filtering and demonstrated various geometry processing applications. More recently, the low-rank optimization was further extended to mesh denoising~\cite{Li2018,Wei2018}. 

\section{Half-kernel Laplacian Operator}
\label{sec:method}
To better understand the proposed half-kernel Laplacian operator (HLO), we briefly introduce the uniform Laplacian on meshes and the Laplacian diffusion flow in the first place. We then analyze the differential property of feature points and finally explain how we construct the HLO.

\subsection{Uniform Laplacian}
Given a surface mesh $M=(V,E,F)$ with $N$ vertices, we have the set of vertices $V$, the set of edges $E$ and the set of faces $F$. The $i$-th vertex $ v_{i} \in V $ is represented by the coordinates $ v_{i}=(x_{i},y_{i},z_{i}) $.
The differential coordinates (i.e., $ \delta $-coordinates) of vertex $v_i$ are defined as
\begin{equation}
\label{diff_form}
\delta_{i} = v_{i} - \frac{1}{|NV(v_i)|}\sum_{v_{k}\in NV(v_i)} v_{k},
\end{equation}
where $\delta_{i}=(\delta_{x},\delta_{y},\delta_{z})$, $NV(v_i)$ is the set of the neighboring vertices of the vertex $v_{i}$, and $|NV(v_i)|$ is the degree (number of neighbors) of $v_i$. It is also called the \textit{uniform Laplacian} due to the equal weights. The direction of the differential coordinates approximates the local normal direction and the magnitude linearly approximates the local mean curvature $H(v_{i})$ of $v_{i}$ \cite{sorkine2006differential}.

\subsection{Laplacian Diffusion Flow}
Fairing a mesh can be viewed as a filtering process on mesh signal. In the discretized setting, it is usually done by solving a heat-diffusion-like partial differential equation as follows.
\begin{equation}
\label{eqn:01}
\frac{\partial V(x,t)}{ \partial t }=\lambda\triangle V(x,t), 
\end{equation}
where $ \lambda $ is the diffusion speed. The Laplacian operator is defined as
\begin{equation}
\label{eqn:02}
\triangle V =\nabla ^{2} V ,
\end{equation}
where $\nabla^{2}V=\nabla\cdot\nabla$ is the divergence of the gradient on the vertices of a given mesh. The above equation (Eq. \eqref{eqn:01}) is usually solved in an iterative way
\begin{equation}
\label{eqn:03}
V^{t+1} = V^{t}+\lambda dt \nabla ^{2} V ^{t},
\end{equation}
where $dt$ is the time step length.

The surface smoothing based on the uniform Laplacian often causes the issues of shrinkage and smoothing out features (Figure~\ref{fig:front}a). Motivated by these issues, we attempt to first analyze the differential property of feature points, which will be described in Sec. \ref{sec:differentialproperty}.

\subsection{Differential Analysis of Feature Points}
\label{sec:differentialproperty}

Intuitively speaking, feature points can be classified as the points with abrupt normals (i.e., with large dihedral angles) \cite{Lu2017-1,He2013}. Denote by $f$ a surface manifold and $a$ a feature point at a sharp edge on the cube model (Figure \ref{fig:HLO}(a)). We simply analyze the situation when $f$ crosses the sharp edge (Figure \ref{fig:HLO}(a)). It is obvious that $f'(a+\epsilon u) \neq f'(a+\epsilon v)$, where $f'$ means the first-order derivative of $f$ and $\epsilon>0$. Thus, $f$ is not differentiable at the feature point $a$, which means that the second-order derivative $f''(a)$ does not exist either. This reveals that it does not make sense to apply the uniform Laplacian (i.e., second-order derivative) operator to a feature vertex with all its neighbors (i.e., full window, see Figure \ref{fig:HLO}(a)). Figure \ref{fig:front}(a) shows the feature-wiping and shrinking results with the uniform Laplacian operator in a full-window sense.

Ideally, the approximations for locations ($a+\epsilon v$) and ($a+\epsilon u$) should come from the upper region and the front region, respectively. The approximation for $a$ should base on its two neighboring vertices on the edge since it is differentiable along the edge \cite{Wang2014}. Corner, the intersection of several edges, is theoretically non-differentiable and usually fixed in mesh denoising \cite{Wang2014}. Sharp edge points and corners are difficult to approximate because they are identified with some means which makes surface smoothing less robust due to sensitivity to the noise level. To overcome this difficulty, we attempt to approximate each (feature or non-feature) vertex with a half-kernel Laplacian operator which targets at computing the Laplacians using half-window neighbors. It will be discussed in Section \ref{sec:hlo}.

%

\subsection{The Half-kernel Laplacian Operator}
\label{sec:hlo}

\addtolength{\subfigcapskip}{5pt}
\begin{figure} [htbp]
	\begin{center}
		\begin{tabular}{c} 
			\subfigure[Feature point]{
			\includegraphics[height=2.2cm]{Fig/tikz0.tikz}		
		
	}
			\subfigure[Uniform Laplacian]{
				\includegraphics[height=2.2cm]{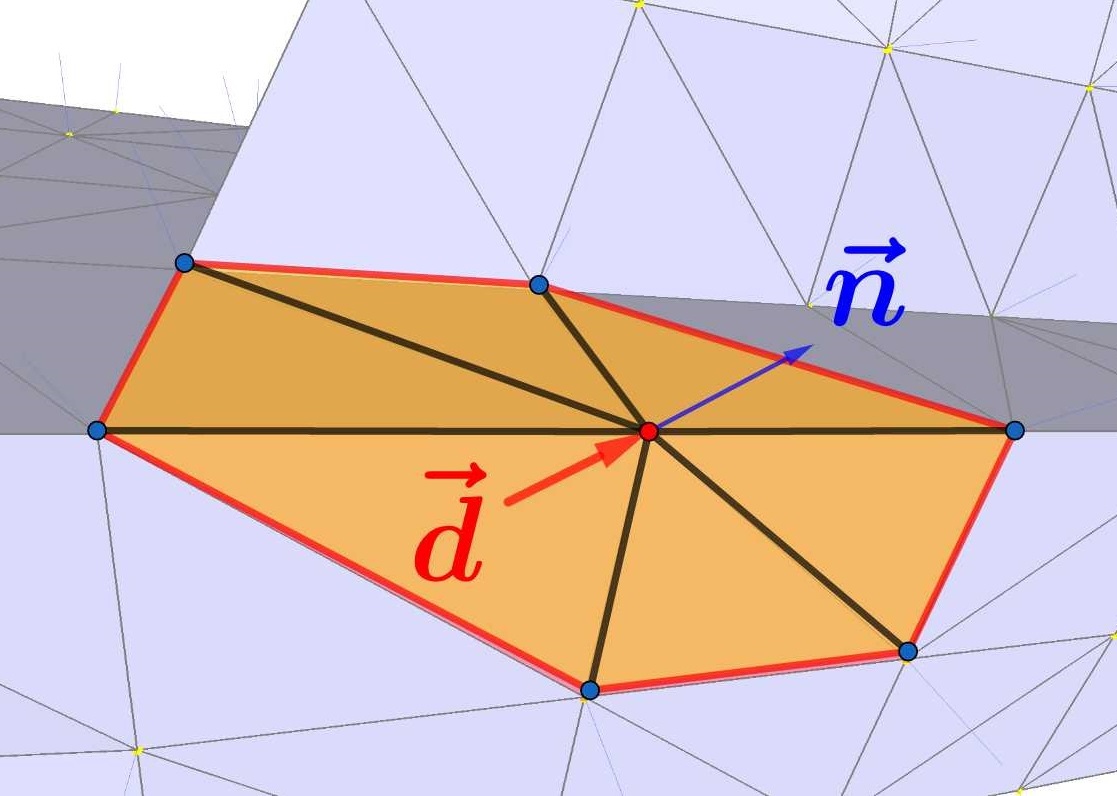}
			}
			\subfigure[Half-window Laplacian]{
				\includegraphics[height=2.2cm]{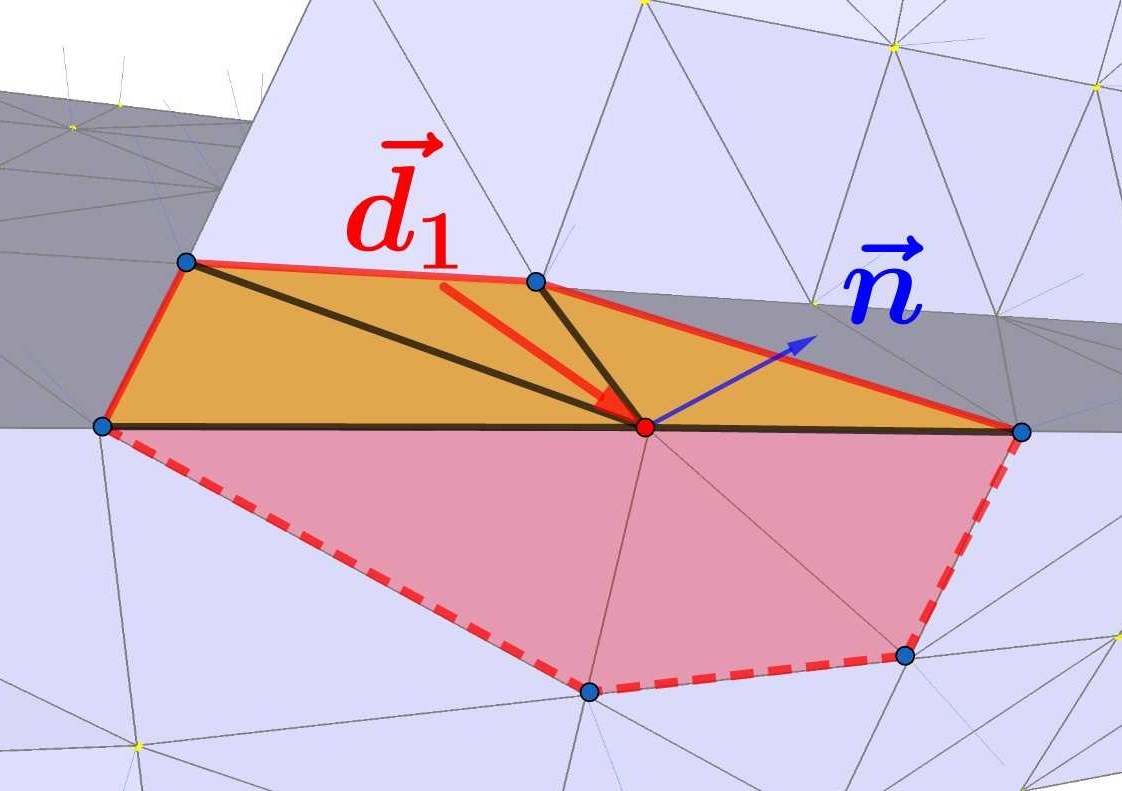}
			}
		\end{tabular}
	\end{center}
	\caption[weights] 
	{ \label{fig:HLO} 
		(a)	Analysis of a feature point $a$ on an edge.  The locations ($a+\epsilon v$) and ($a+\epsilon u$) should be approximated in the half windows which have the same colors with them, rather than the local full windows centered at them. The illustration of the uniform Laplacian operator (b) and the proposed HLO (c). The blue vector $\vec{n}$ represents the local normal of vertex $v_{i}$, and the red vector $\vec{d}$ indicates the orientations of the full-window and half-window Laplacians. }
\end{figure} 
\setlength{\subfigcapskip}{-1pt}

\begin{figure} [htbp]
	\begin{center}
		\begin{tabular}{c} 
			\includegraphics[height=6cm]{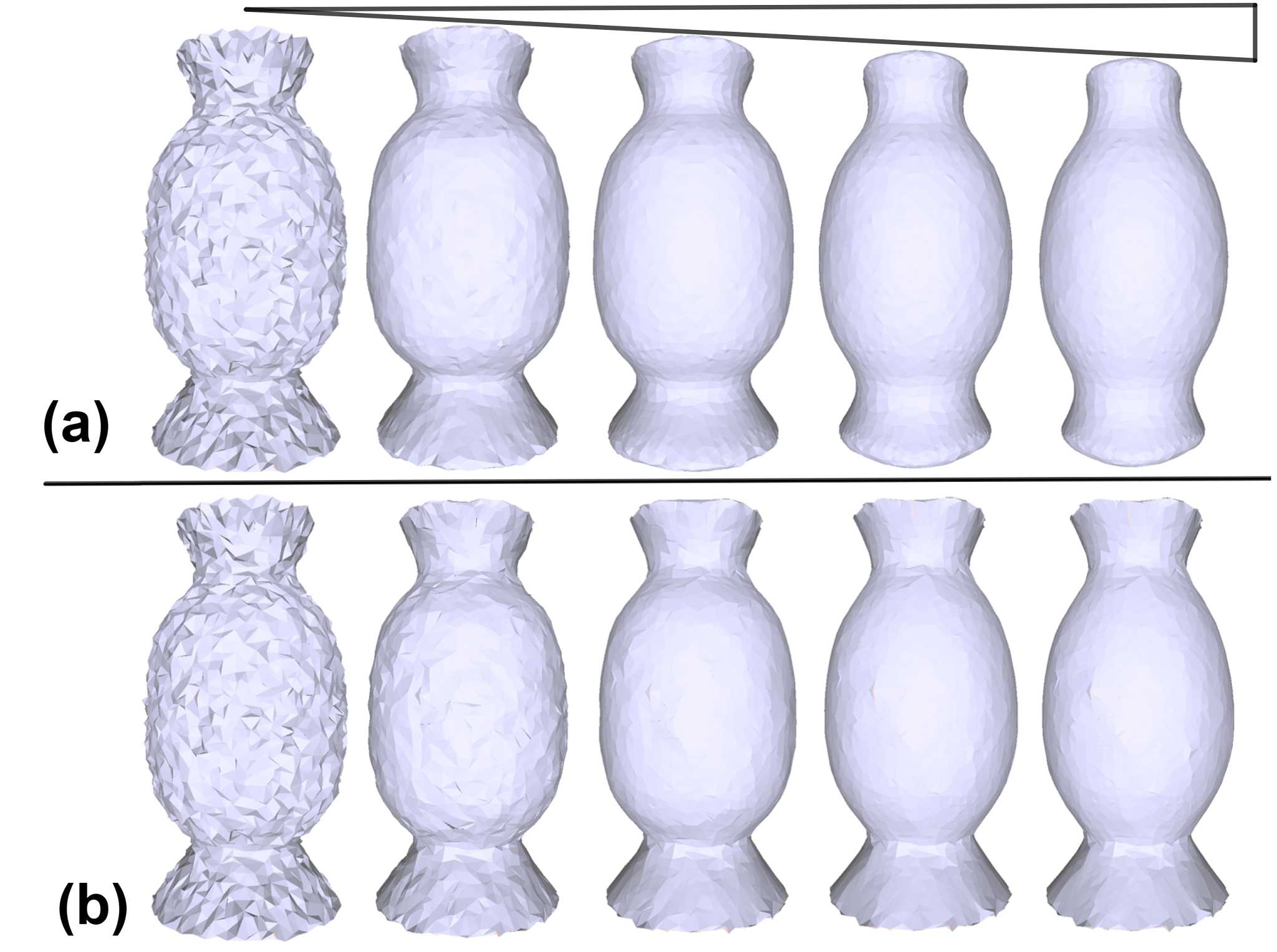}
		\end{tabular}
		\makebox[0.5in]{}
		\makebox[0.5in]{Input}
		\makebox[0.5in]{1st iter}
		\makebox[0.5in]{3rd iter}
		\makebox[0.5in]{5th iter}
		\makebox[0.5in]{15th iter}
	\end{center}
	\caption[front] 
	{ \label{fig:front} 
		Comparing the uniform Laplacian operator and the half-kernel Laplacian operator (HLO) on the noisy Vase. One can clearly see the shrinkage artifact of the uniform Laplacian. In contrast, HLO is geometry-aware and feature-preserving. }
\end{figure}

\begin{algorithm}[t]
	\SetAlgoNoLine
	\KwIn{vertex $v_{i}$}
	\KwOut{all subsets $V_{sub}$}
	$v_{i}' = \frac{1}{|NV(v_i)|} \sum_{v_{k}\in NV(v_i)} v_{k}$ \\
	\For{each $v_{k} \in NV(v_i)$}
	{compute the distance of other neighbors (except $v_k$) to the plane (or line in a degenerated case) defined by $v_i$, $v_i'$ and $v_k$ \\
	select the neighbor with the shortest distance, and pair it with $v_k$ \\
	partition $NV(v_i)$ with the line $v_i,v_k$ into the left and right subsets (i.e., two half windows)
	}
	enumerate all subsets $V_{sub}$
	\caption{Half Windows Generation}
	\label{alg:two}
\end{algorithm}

\begin{algorithm}[t]
	\SetAlgoNoLine
	\KwIn{vertex $v_i$}
	\KwOut{$\delta$--final half-kernel Laplacian for $v_i$}
$E_{init} = 10^6$\\
	compute $V_{sub}$ via Algorithm \ref{alg:two}\\
	compute the centroid $v_i'$ with $NV(v_i)$\\
	compute the local normal: $n_i = normalize(v_i - v_i')$\\
	\For{each subset $v_{sub} \in V_{sub}$}
	{	$d_{i} = \frac{1}{|V_{sub}|}\sum_{v_{m} \in v_{sub}} (v_{i} -v_{m})$\\
		$\delta_i = (d_{i} \cdot n_i) n_i$\\
		\If{$\mathbb{E}(\delta_i)<E_{init}$}
		{
		$E_{init} = \mathbb{E}(\delta_i)$\\
		$\delta = \delta_i$\\
		}
	}
	\caption{Half-kernel Laplacian Operator}
	\label{alg:one}
\end{algorithm}

The above observation and analysis indicate that the uniform Laplacian operator should be performed only on half windows (i.e., subsets of neighbors) of the feature point instead of its whole neighborhood. This key insight encourages us to introduce a half-kernel Laplacian operator (HLO) for surface meshes.


\begin{figure}[ht]
\centering
\subfigure[]{
    \includegraphics[width=2.25cm]{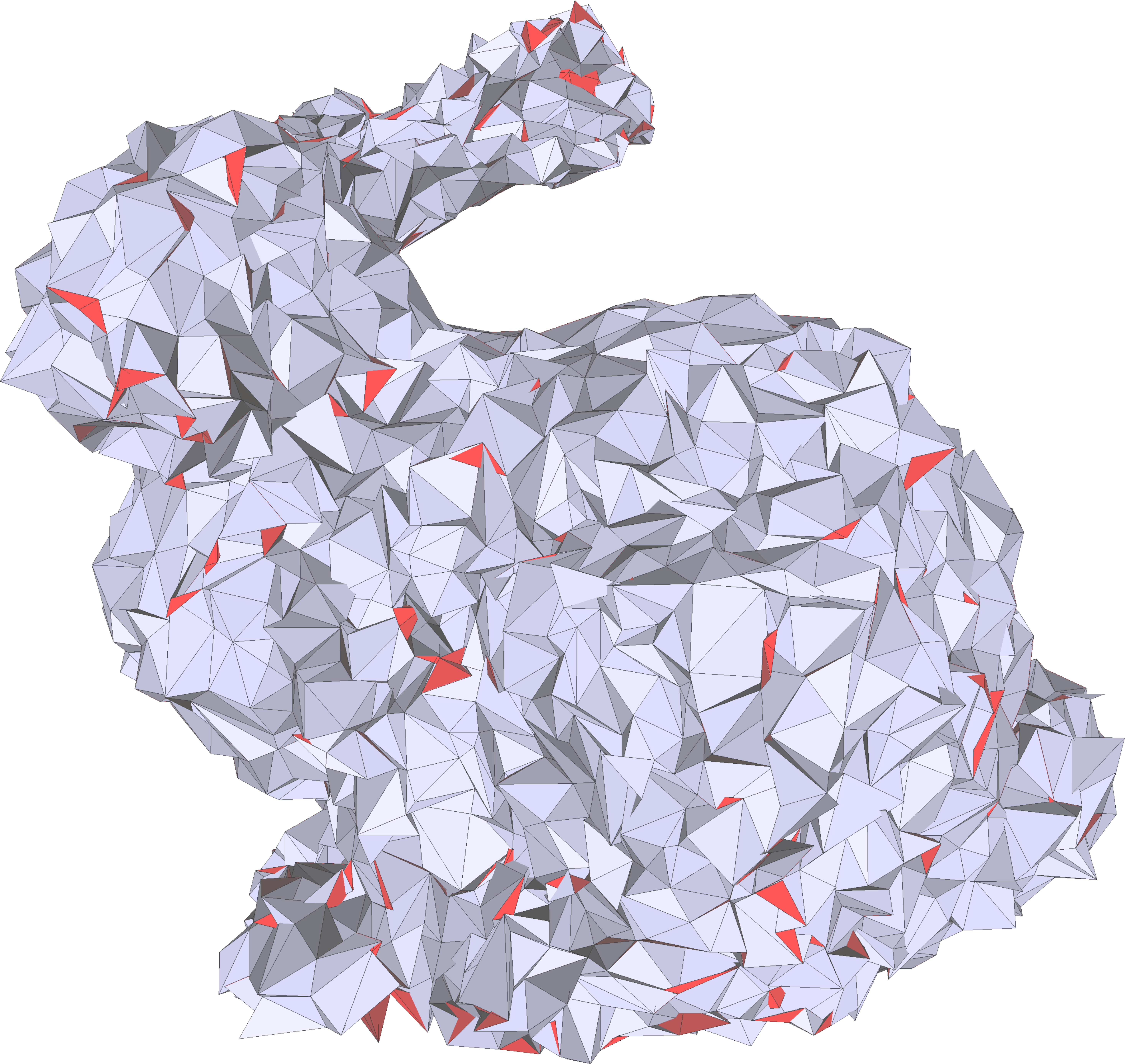}
}
\subfigure[]{
	\includegraphics[width=2.25cm]{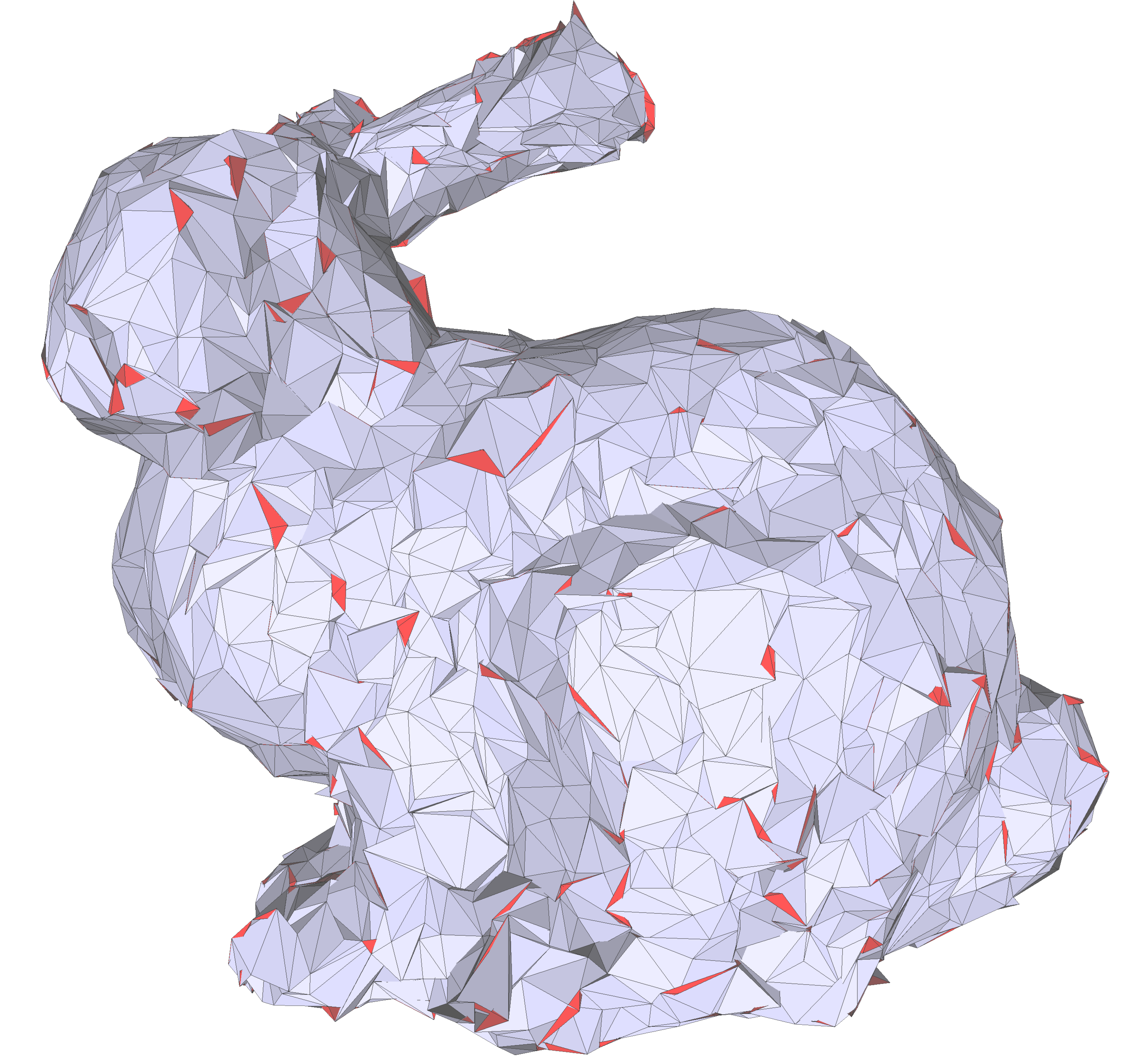}
}\subfigure[]{
 \includegraphics[width=2.25cm]{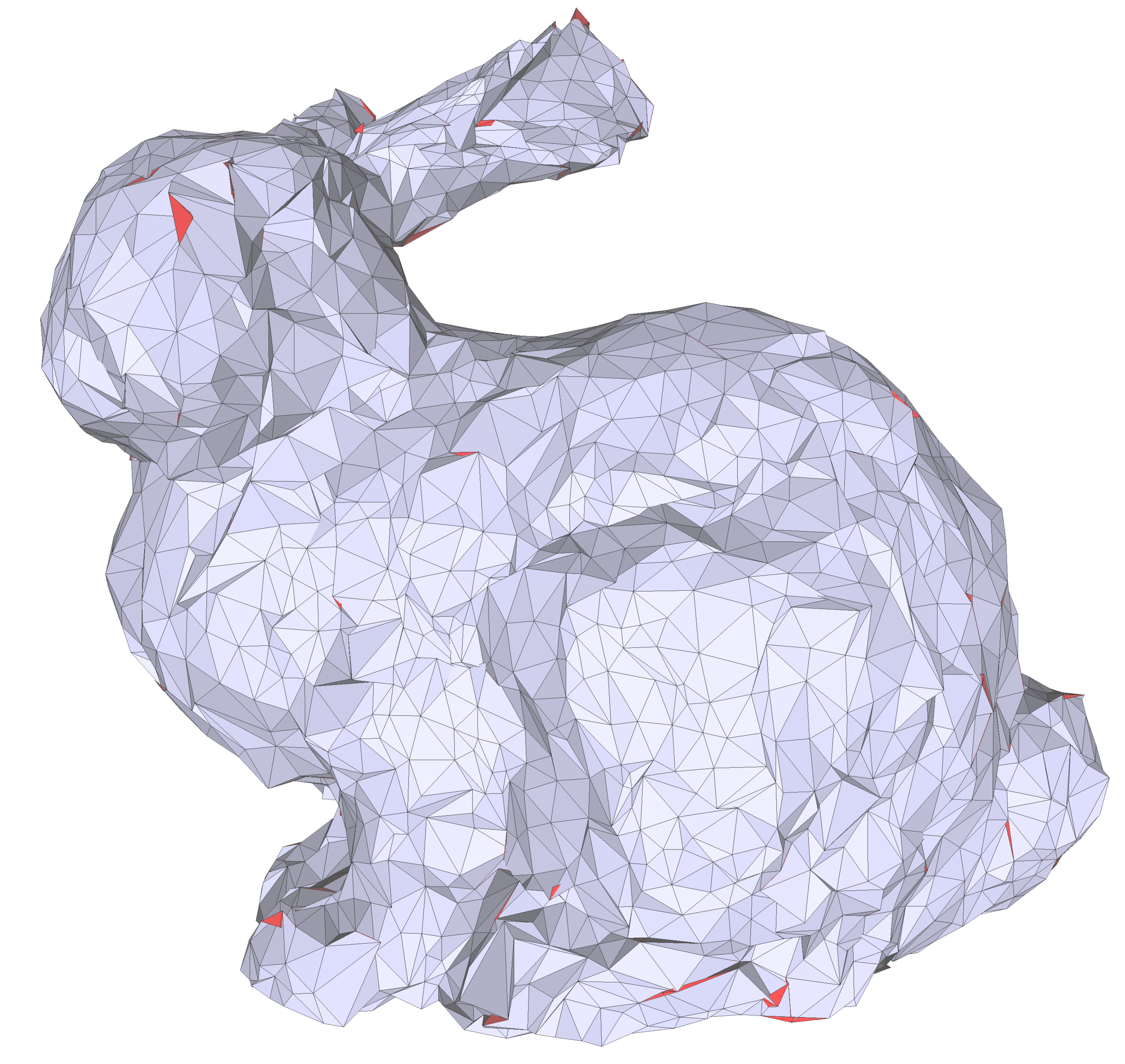}
}
\subfigure[]{\includegraphics[width=1.25cm]{Fig/tikz5.tikz} }

\caption{With and without projection. (a) Noisy bunny. (b) 1 iteration of vertex update of the half-window Laplacian without projecting onto the uniform Laplacian. (c) 1 iteration of vertex update with the half-kernel Laplacian projecting onto the uniform Laplacian. The flipped triangles are rendered in red. (d) A flipped triangle. The clock-wise or counter-clock wise order of vertices have been changed after noise contamination.} 
\label{fig:projection}
\end{figure}

\begin{figure}
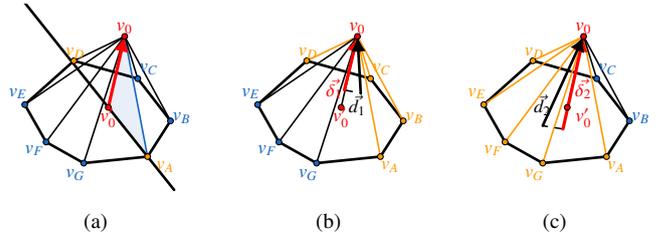
   
	\centering
	\definecolor{ffqqqq}{rgb}{1,0,0}
	\definecolor{rvwvcq}{rgb}{0.08235294117647059,0.396078431372549,0.7529411764705882}
	\definecolor{ffzzqq}{rgb}{1,0.6,0}
	\definecolor{ffqqqq}{rgb}{1,0,0}
	\definecolor{ttttqq}{rgb}{0.2,0.2,0}
	\definecolor{ffxfqq}{rgb}{1,0.4980392156862745,0}
	\definecolor{ttqqqq}{rgb}{0.2,0,0}
\subfigure[]{
	\includegraphics[width=2.8cm]{Fig/tikz1.tikz}
}
\subfigure[]{
	\includegraphics[width=2.8cm]{Fig/tikz2.tikz}
}\subfigure[]{
	\includegraphics[width=2.8cm]{Fig/tikz3.tikz}
}

	\caption[hk]
	{\label{fig:fig4} 
		The half-kernel Laplacian operator on a vertex $v_{0}$. (a) $v_A$, the starting neighbor of $v_0$, is paired with the other neighbor $v_D$ which has the shortest distance to the plane $v_0 v_A v_0'$. $v_0'$ is the centroid of the neighbors. The line $v_A v_D$ partitions the neighborhood into the left and right half windows (subsets).
		(b) and (c):  $d_1$ and $d_2$ are computed by performing the uniform Laplacian operator to the half windows. $\delta_1$ and $\delta_2$ are achieved by projecting $d_1$ and $d_2$ to the full-window uniform Laplacians, respectively. } 
\end{figure}

Our operator is to approximate the Laplacians at vertices with half windows. 
To produce half windows, each of the immediate neighboring vertex of the current vertex is paired with another neighboring vertex to partition the local neighborhood into two half windows (left and right). The other neighbor paired by the starting neighbor is determined as the neighboring vertex which has the shortest distance to the plane defined by the current vertex, the starting neighbor and the centroid of the neighborhood. Refer to  Figure~\ref{fig:fig4}(a). Consider a vertex $v_0$ with a neighbor $v_D$. Denote by $v_A$ the starting neighbor to be paired. $v_0'$ is the centroid of the neighborhood of $v_0$. Figure \ref{fig:HLO}(b) also shows a half window of a feature vertex. As a result, each vertex $v_i$ and its neighborhood has $|NV(v_i)|$ partition choices which further generate $2|NV(v_i)|$ paired subsets (half windows). The half-window generation algorithm is shown in Algorithm \ref{alg:two}. 

\textbf{Remark 1.} Theoretically, the full window neighbors can be split into more than two half windows. We simply select the choice of two half windows, because corners are typically far fewer than the edge features in a shape, and recognizing corners may introduce new parameters and result in less robustness. Also, this half-window scheme is sufficient for non-feature vertices. 

\textbf{Remark 2.} In general, there might exist multiple candidates with the same shortest distances when determining the other neighbor. We randomly select one candidate in this case, which we found no noticeable differences in mesh denoising. Note that the plane defined by the current vertex, the starting neighbor and the centroid of the neighborhood may degenerate into a line. We simply determine the other neighbor using the distance to the line instead. 

We apply the uniform Laplacian operator independently to each of these subsets and compute the half-window Laplacians for each vertex $v_i$. Because half-kernel Laplacians introduce a shift in tangential direction, directly using the half-kernel Lapalcians may result in inferior results like degenerating and flipping (Figure~\ref{fig:projection}). 

We thus project them onto their corresponding full-kernel Laplacians (i.e., using full neighbors) to remove the tangential component and obtain the final half-kernel Laplacians. \textit{This way ensures small Laplacian magnitudes for feature vertices and large Laplacian magnitudes for non-feature vertices, thus smoothing non-features and preserving features.}  Figure \ref{fig:fig4}(b-c) shows two such examples. We define the projection in Eq. (\ref{eq:projection}).
\begin{equation}  
\begin{aligned}
\delta = (d\cdot n)n,
\end{aligned}
\label{eq:projection}
\end{equation}
where $d$ indicates the half-kernel Laplacian and $n$ is the orientation of the full-kernel Laplacian.

We can easily enumerate all the intermediate half-kernel Laplacians for $v_i$: $\dots, \delta_{L}(v_{k}), \delta_{R}(v_{k}), \dots$, where $\delta_{L}(v_{k})$ and $\delta_{R}(v_{k})$ respectively denote the half-kernel Laplacians for the left subset and right subset with the starting neighbor $v_k$ ($v_k \in NV(v_i)$). We next need to determine the optimal Laplacian among the $2|NV(v_i)|$ half-kernel Laplacians. The optimal one will be selected based on the regularization energy which may vary in different applications. 
In this work, we define the regularization energy of each vertex as the sum of the norm of the Laplacian and the the distance to the original vertex position, shown as below.
\begin{equation}
\mathbb{E}(\delta_{i}^t) = \Arrowvert \delta_{i}^t \Arrowvert+\Arrowvert v_{i}^{t}-v_{i}^{0} \Arrowvert  ,
\label{eq:energy}
\end{equation}

where $\delta_{i}^t$ can be any among the calculated $2|NV(v_i)|$ half-kernel Laplacians in the $t$-th iteration. $v_{i}^{t}$ is the position of vertex $v_i$ in the $t$-th iteration and $ v_{i}^{0} $ is the initial position of $v_{i}$. The former term, which naturally uses the computed $\delta_{i}^t$ and is a feature-preserving term, represents the movement between two consecutive iterations, and the latter term, a typical data term, describes the distance between the position at the $t$-th iteration and the initial position of vertex $v_i$. This energy enables a closest walking from the initial positions and positions in the previous iteration, thereby preserving features and the original shape. We compute the energy $\mathbb{E}$ for each half-kernel Laplacian of vertex $v_i$ and select the optimal Laplacian that incurs the smallest energy. Algorithm \ref{alg:one} summarizes the proposed half-kernel Laplacian operator. To smooth surfaces, the final half-kernel Laplacians in the $t$-th iteration are used to update vertex positions in the $(t+1)$-th iteration via Eq. \eqref{eqn:03}. $ \lambda dt $ in Eq. \eqref{eqn:03} and the number of iterations work in a proportional way to each other. In other words, increasing $\lambda dt $ would generally induce a decrease of the number of update iterations, and vice versa. A larger $ \lambda dt $ would also possibly cause instability. To make our method more robust, we empirically set $ \lambda dt $ to 1, which works very well in all our experiments. Figure \ref{fig:front}(b) shows an example for our HLO.



%
%

\section{Experimental Results}
\label{sec:results}
We first describe the parameter setting, and then compare the proposed HLO with the uniform Laplacian operator (ULO). 
Finally, We compare our method with the state-of-the-art denoising methods on both synthetic and scanned data, visually and quantitatively.

\subsection{Parameter Setting}
The compared state-of-the-art techniques are: the bilateral mesh filter (BMF)~\cite{Fleishman2003}, the unilateral normal filter (UNF)~\cite{Sun2007}, the bilateral normal filter (BNF)~\cite{Zheng2011}, the $L_{0}$ minimization method ($L_{0}$)~\cite{He2013}, the guided normal filter (GNF)~\cite{Zhang2015}, the cascaded normal regression (CNR)~\cite{Wang-2016-SA}, and the non-local low-rank normal filtering method (NLLR)~\cite{Li2018}. The source code of these methods are available or the authors provide test results.  
We summarize the parameters of these methods in Table \ref{table:tablepara}. Table \ref{table1} shows the parameter values of all methods on most models. \textit{Interested readers are referred to the papers for more specific details.}

\begin{table}[htbp]\scriptsize
    \centering
    \setlength\tabcolsep{1pt}
    \caption{Parameters of the state-of-the-art mesh smoothing methods.}\label{table:tablepara}
    \begin{tabular}{|l|c|l|}\hline
    Method & \tabincell{l}{Number of \\ Parameters} & Parameters\\
    \hline
    BMF & 1 & \tabincell{l}{$ k_{iter} $: number of iterations.}
  \\ \hline
    UNF & 3 & \tabincell{l}{$ T $: threshold for controlling the averaging weights. \\ $ n_{iter} $: number of iterations for normal update. \\ $ v_{iter} $: number of iterations for vertex update.}
    \\ \hline
    BNF(Local)& 3 &\tabincell{l}{$ v_{iter} $: number of iterations for vertex update.\\ $ \sigma_{s} $: variance parameter for the spatial kernel. \\ $ n_{iter} $: number of iterations for normal update.}
        \\ \hline
  	L0 & 6 & \tabincell{l}{$ \lambda $: weight for the L0 term in the target function..\\ $ \alpha_{0},\beta_{0} $: initial values for $ \alpha $ and $ \beta $ \\ $ \mu_{\alpha},\mu $: update ratios for $ \alpha $ and $ \beta $. \\$ \beta_{max} $: maximum value of $ \beta $.}
  	\\ \hline
  	GNF & 3 &\tabincell{l}{$ v_{iter} $: number of iterations for vertex update.\\ $ \sigma_{r} $: variance parameter for the range kernel. \\ $ n_{iter} $: number of iterations for normal update.}
  	\\ \hline
  	NLLR & 3 &\tabincell{l}{ $ \sigma_{M} $:  the noise variance. \\ $ N_{k} $: number of similar vertices.\\$ v_{iter} $: number of iterations for vertex update.}
       \\ \hline
  	Ours & 1 & \tabincell{l}{$ k_{iter} $: number of iterations.}
        \\ \hline            
    \end{tabular}
\end{table}

\subsection{Comparison with the Uniform Laplacian}
Figure~\ref{fig:front} shows results processed by the uniform Laplacian and our HLO, respectively. We listed the corresponding results after 1, 3, 5, and 15 iterations. 
It is obvious that our HLO significantly alleviates the shrinking effect. The feature places (e.g., upper and lower parts) of the vase are preserved by HLO.

\begin{figure} [htbp]
	\begin{center}
		\begin{tabular}{c} 
		
			\subfigure[Noisy]{
    \includegraphics[width=2.3cm]{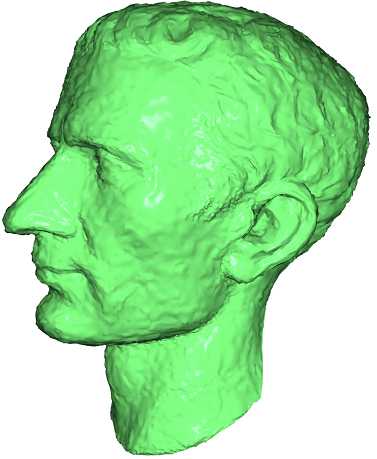}
      \vspace{2\baselineskip}
}
\subfigure[ULO]{
 \includegraphics[width=2.3cm]{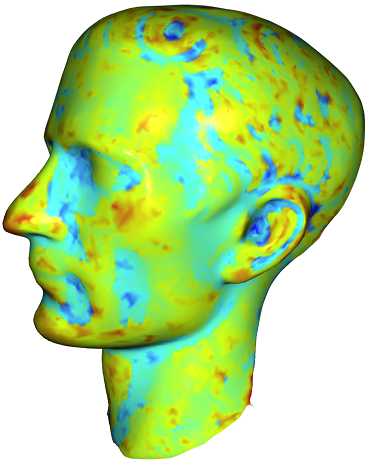}
   \vspace{2\baselineskip}
}
\subfigure[HLO]{
	\includegraphics[width=2.3cm]{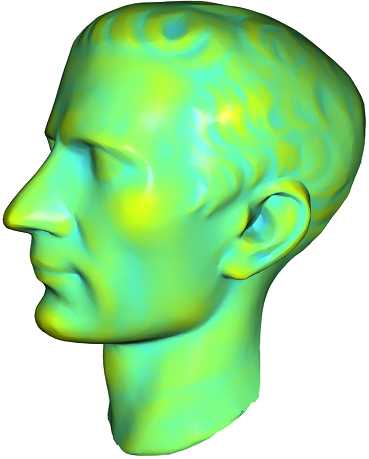}
	  \vspace{2\baselineskip}
}
			\includegraphics[height=2.5cm]{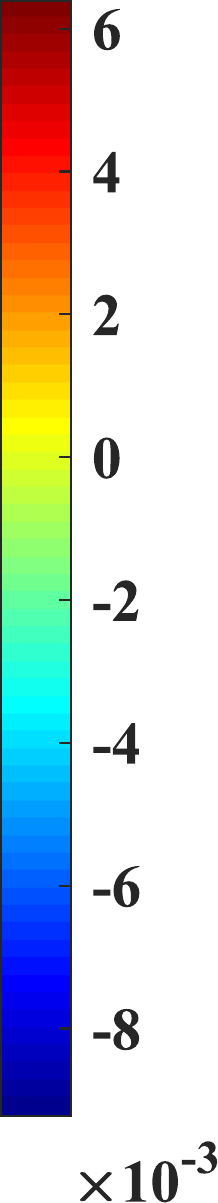}
		\end{tabular}
	\end{center}
	\caption[res3] 
	{ \label{fig:res3} 
		The visualization of one-to-one vertex errors on the Julius model. (a) Noisy shape ($\sigma_{n}=0.2l_{e}$). (b) 10 iterations of the uniform Laplacian operator (ULO). (c) 10 iterations of HLO. The errors are visualized using color maps, where the warm (resp. cold ) colors denote positive (resp. negative) displacements. }
\end{figure}

\begin{figure}
	\centering
	\begin{tikzpicture}[scale=1, line cap=round,line join=round,>=triangle 45]
	\begin{axis}[
	height=4.2cm,
	width=4.2cm,
	xlabel=Iterations,
	ylabel=Avg. Vert. Err.
	]
	
	\addplot[smooth,mark=*,blue,mark size=1pt,line width=1pt] coordinates {
		(0,0.42376)
(1,0.1345)
(2,0.11391)
(3,0.12345)
(4,0.13421)
(5,0.14412)
(6,0.15308)
(7,0.16117)
(8,0.16852)
(9,0.17525)
(10,0.18143)
(11,0.18716)
(12,0.19248)
(13,0.19746)
(14,0.20212)
(15,0.20651)
(16,0.21065)
(17,0.21458)
(18,0.21831)
(19,0.22186)
(20,0.22526)
};
	
	
	\addplot[smooth,color=red,mark=x,mark size=1pt,line width=1pt] coordinates {
(0,0.42376)
(1,0.19066)
(2,0.13221)
(3,0.11574)
(4,0.11226)
(5,0.11367)
(6,0.11703)
(7,0.12129)
(8,0.12592)
(9,0.13064)
(10,0.13536)
(11,0.14003)
(12,0.14462)
(13,0.1491)
(14,0.15347)
(15,0.15773)
(16,0.16187)
(17,0.16589)
(18,0.1698)
(19,0.17359)
(20,0.17725)
};
	\end{axis}
	\end{tikzpicture}	
	\definecolor{ffqqqq}{rgb}{1,0,0}
	\definecolor{rvwvcq}{rgb}{0,0,1}
	\definecolor{zzttqq}{rgb}{0.6,0.2,0}
	\definecolor{ududff}{rgb}{0.30196078431372547,0.30196078431372547,1}
	\begin{tikzpicture}[scale=1, line cap=round,line join=round,>=triangle 45,x=0.4cm,y=0.4cm]
	\begin{axis}[
	height=4.2cm,
	width=4.2cm,
	xlabel=Iterations,
	ylabel=Mean Cur. Ene.,
	legend style={draw=none},
	legend style={fill=none}
	]
	
	\addplot[smooth,mark=*,blue,mark size=1pt,line width=1pt] coordinates {
		(0,197.065)
		(1,47.2996)
		(2,29.3303)
		(3,25.74)
		(4,23.672)
		(5,22.1962)
		(6,21.0616)
		(7,20.1421)
		(8,19.3743)
		(9,18.724)
		(10,18.1593)
		(11,17.6637)
		(12,17.2241)
		(13,16.8309)
		(14,16.4746)
		(15,16.1514)
		(16,15.8572)
		(17,15.5868)
		(18,15.3367)
		(19,15.1048)
		(20,14.8892)
	};
	legend style={
		area legend,
		at={(0.5,-0.15)},
		anchor=north,
		legend columns=2,
		legend style={draw=none}}]
	
	\addlegendentry{Uniform}
	
	\addplot[smooth,color=red,mark=x,mark size=1pt,line width=1pt] coordinates {
		(0,197.065)
		(1,68.8363)
		(2,41.9529)
		(3,34.4461)
		(4,31.536)
		(5,30.0199)
		(6,28.9729)
		(7,28.1354)
		(8,27.5149)
		(9,27.0387)
		(10,26.6523)
		(11,26.3637)
		(12,26.1305)
		(13,25.9749)
		(14,25.8158)
		(15,25.6594)
		(16,25.5234)
		(17,25.4256)
		(18,25.3711)
		(19,25.3136)
		(20,25.2224)
	};
	\addlegendentry{HLO}
	\end{axis}
	\end{tikzpicture}
	\caption[res]
	{\label{fig:res2} 
		 The average vertex errors (Avg. Vert. Err.) and total mean curvature energy (Mean Cur. Ene.) of Figure \ref{fig:res3} with increasing iterations. 
		 }
\end{figure}
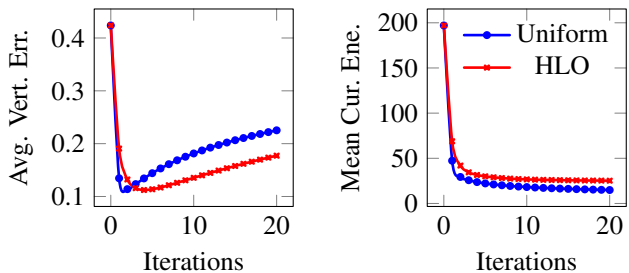

Figure~\ref{fig:res3} visualizes the one-to-one vertex errors of both the uniform Laplacian and our HLO on the Julius model. 
Our method better preserves surface features and produces more accurate smoothing results than the uniform Laplacian operator.
The average vertex errors and mean curvature energy \cite{desbrun1999implicit,sorkine2006differential,gong:cf} of Figure \ref{fig:res3} are displayed in Figure~\ref{fig:res2}. The HLO induces much lower average vertex errors than the uniform Laplacian operator, which means it preserves the shape better. The first iteration sees a remarkable decrease on the mean curvature energy, and the two methods have a similar decreasing trend in mean curvature led by diffusion flow.  

\begin{figure}[ht]
\centering
\includegraphics[width=2cm]{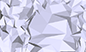}
\includegraphics[width=2cm]{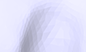}
\includegraphics[width=2cm]{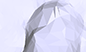}
\includegraphics[width=2cm]{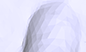}\\
\includegraphics[width=2cm]{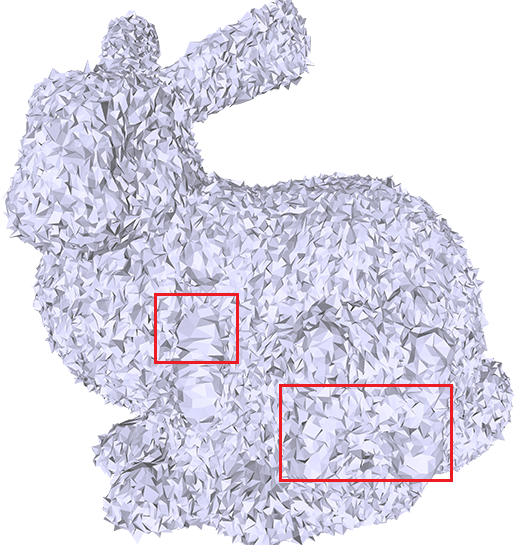}
\includegraphics[width=2cm]{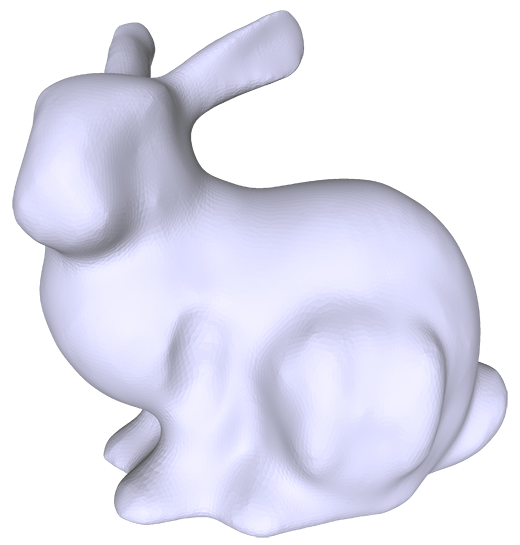}
\includegraphics[width=2cm]{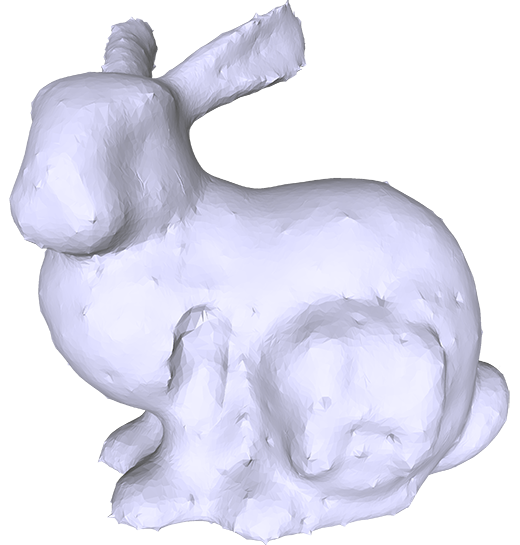}
\includegraphics[width=2cm]{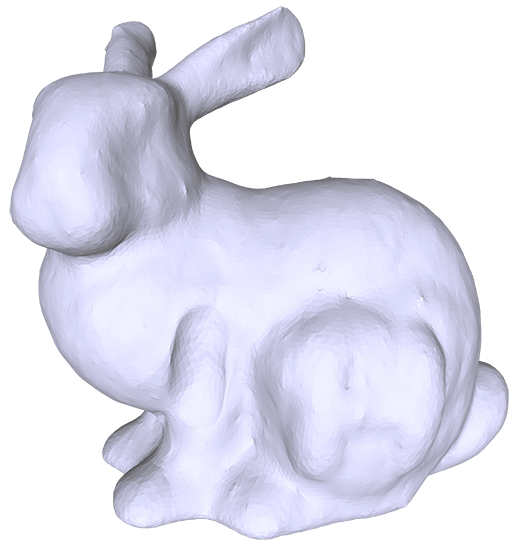}\\

\subfigure[Noisy]{
\includegraphics[width=2cm]{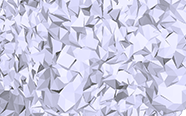}
}
\subfigure[UNI]{
\includegraphics[width=2cm]{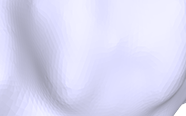}
}
\subfigure[COT]{
\includegraphics[width=2cm]{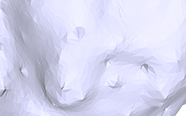}
}
\subfigure[HLO]{
\includegraphics[width=2cm]{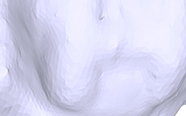}
}
\caption{The comparison of smoothing results after 10 iterations on (a) a noisy bunny model ($\sigma_{n}=0.8l_e$). (b) The uniform Laplacian operator. (c) The cotangent Laplacian operator. (d) The half-kernel Laplacian operator (HLO). }
\label{fig:cotangent}
\end{figure}

We also compared with the cotangent Laplacian. Figure \ref{fig:cotangent} shows the results of the uniform Laplacian, cotangent Laplacian and our HLO. The cotangent Laplacian is more geometry aware than the uniform Laplacian. However, the cotangent Laplacian tends to generate sharp ``thorns'' on the surface (please zoom in to observe).

\addtolength{\subfigcapskip}{5pt}

\setlength{\subfigcapskip}{1pt}

\subsection{Visual Results}
\textit{Synthetic models.} We compare our method with the selected state-of-the-art methods on various models corrupted with synthetic noise. 
Following state-of-the-art mesh smoothing techniques, we generate synthetic models by adding zero-mean Gaussian noise with standard deviation $\sigma_n$ to the corresponding ground truth. $\sigma_n$ is proportional to the mean edge length $l_{e}$ of the input mesh.

\begin{figure*} [htbp]
	\begin{center}
		\begin{tabular}{c} 
		\includegraphics[width=3.5cm]{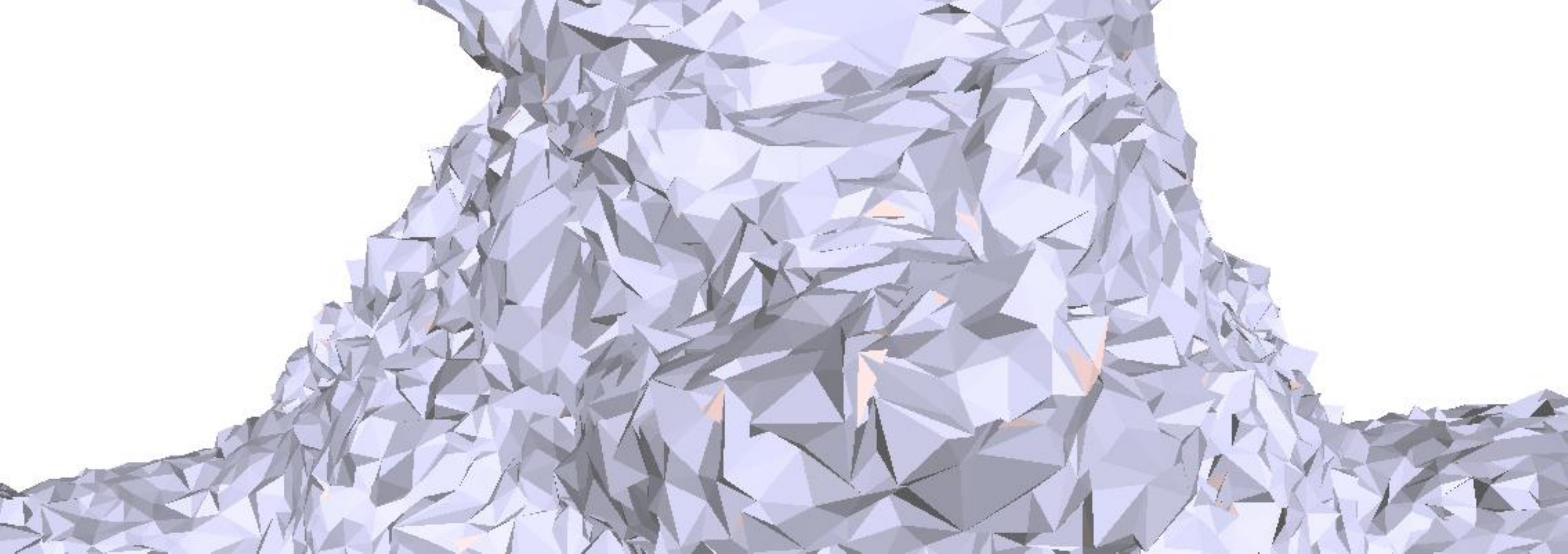}
		\includegraphics[width=3.5cm]{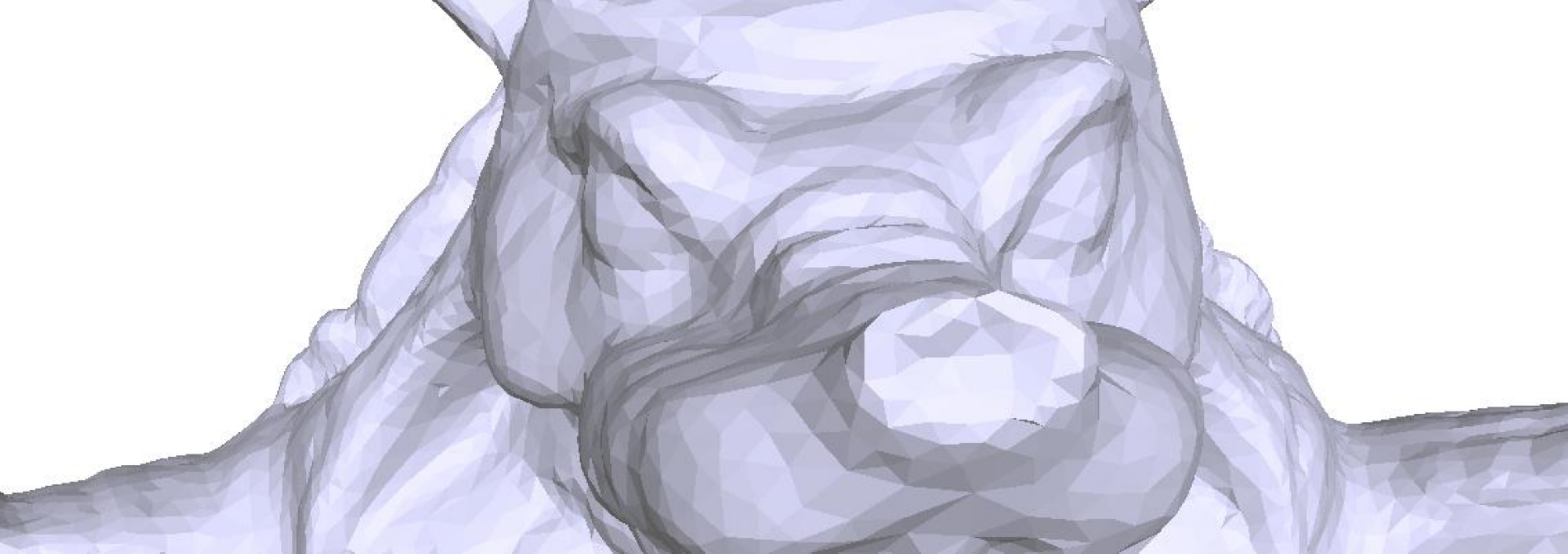}
		\includegraphics[width=3.5cm]{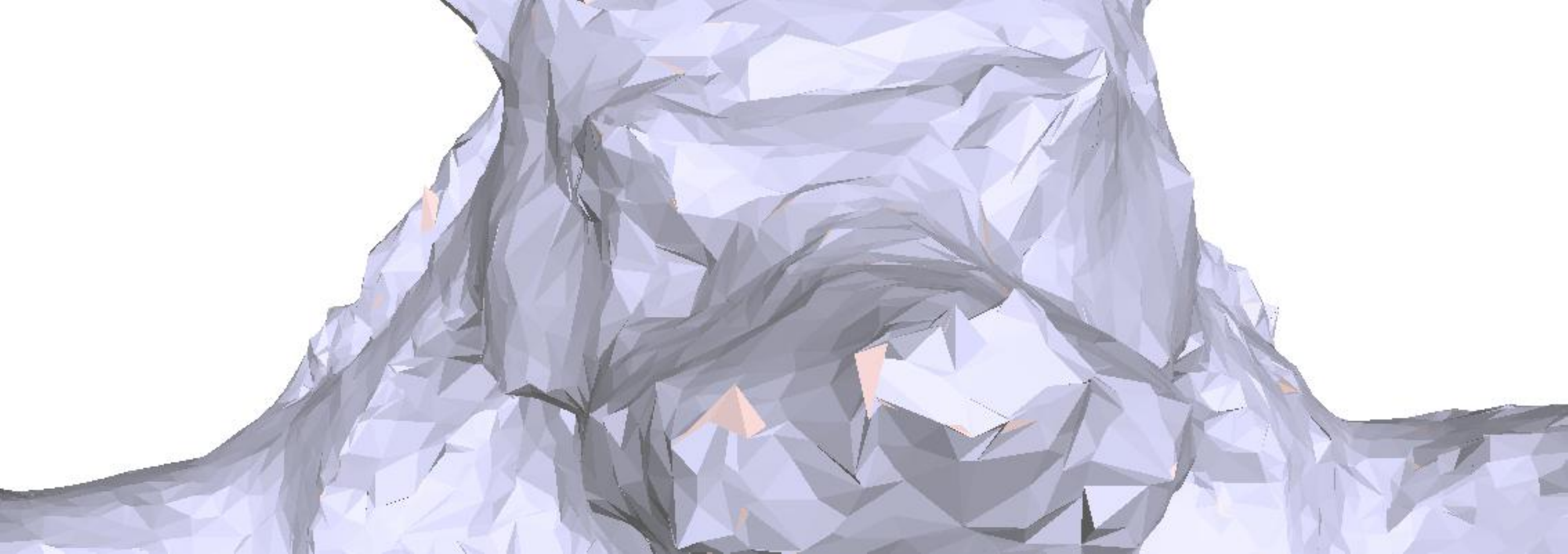}
			\includegraphics[width=3.5cm]{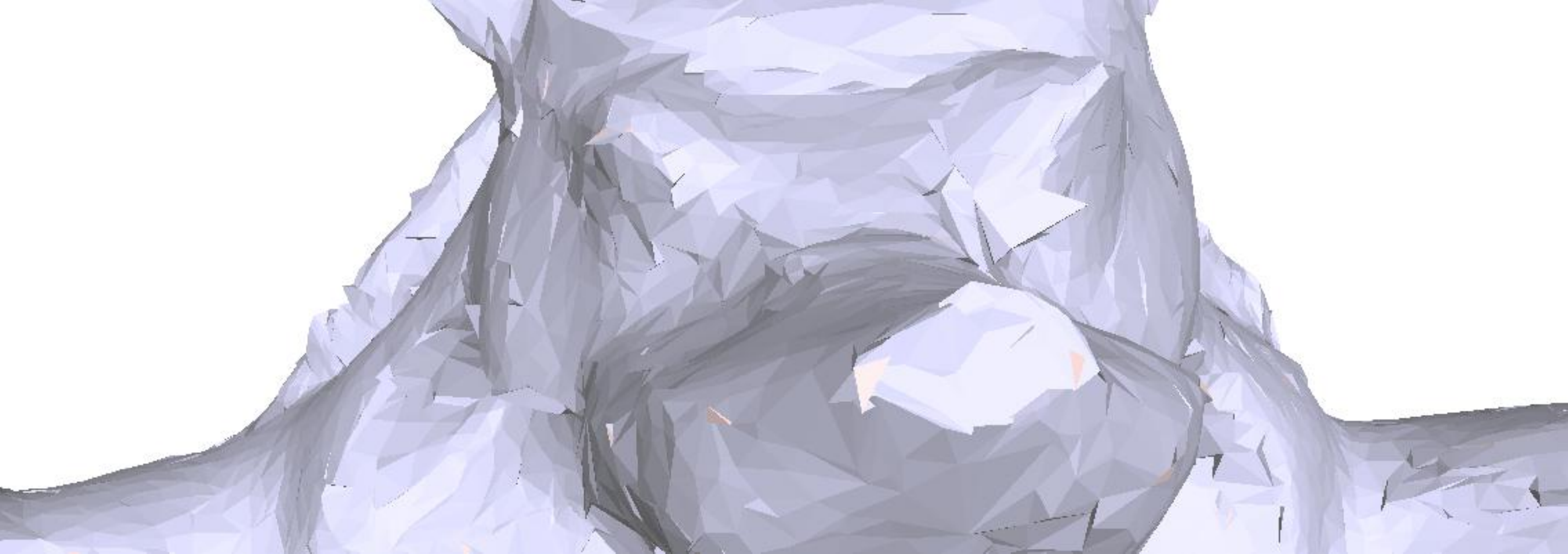}
			\includegraphics[width=3.5cm]{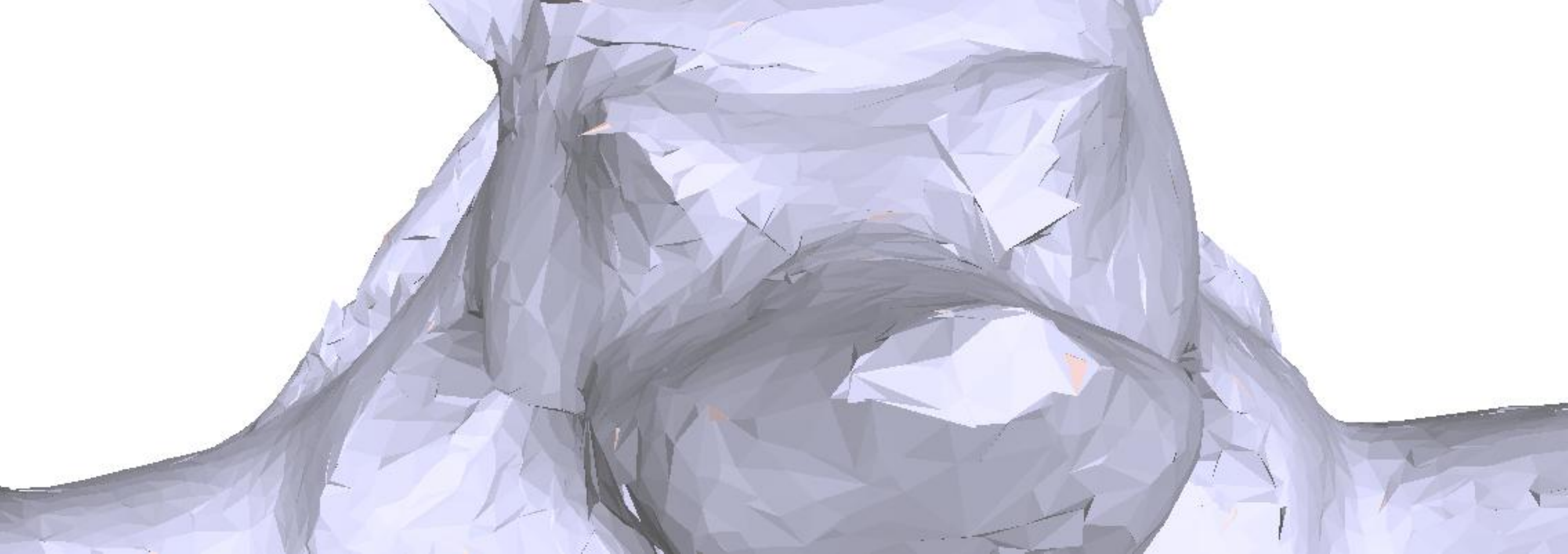}\\
				\subfigure[Noisy]{
						\includegraphics[width=3.5cm]{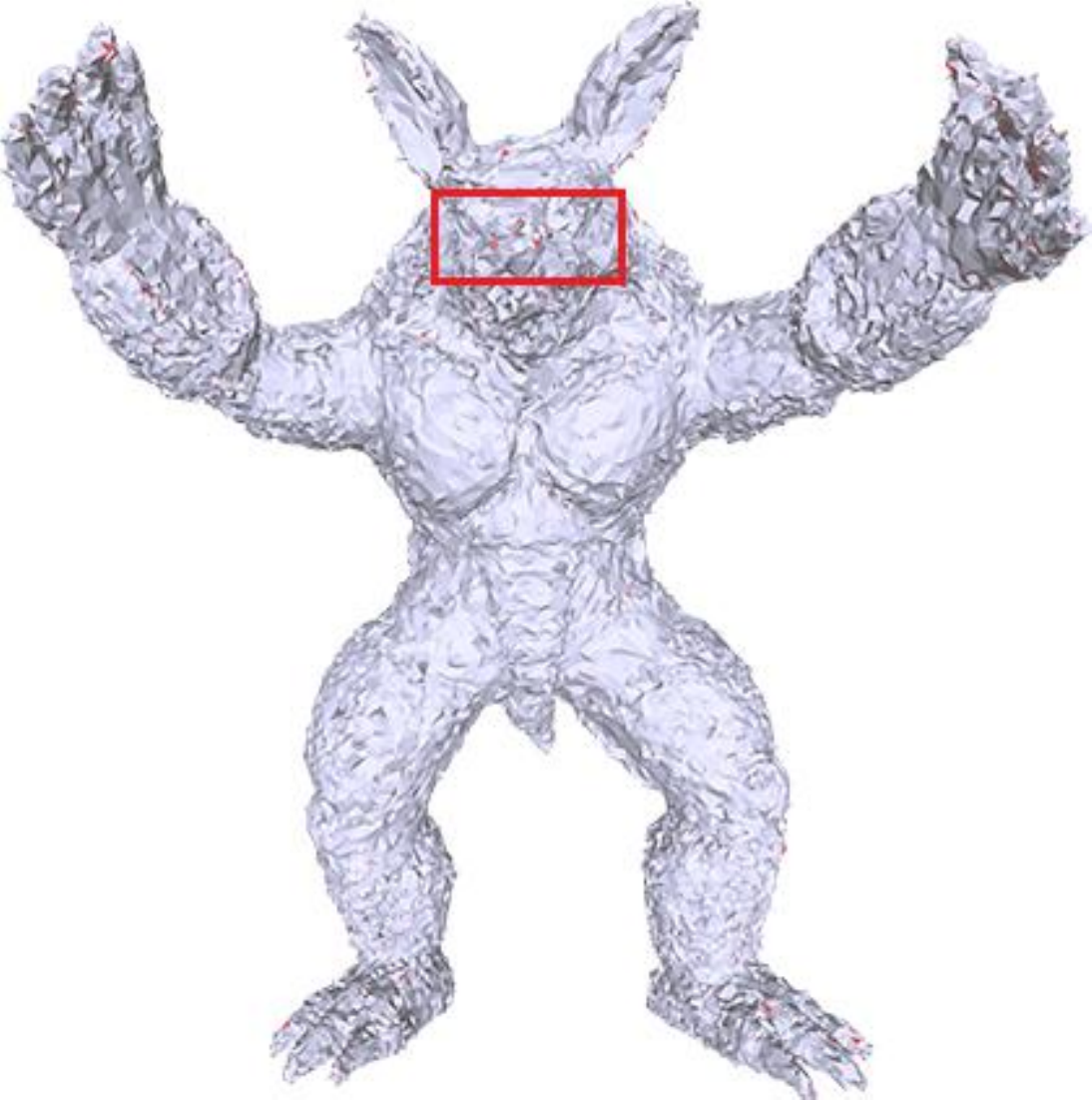}}
				\subfigure[Original]{
				\includegraphics[width=3.5cm]{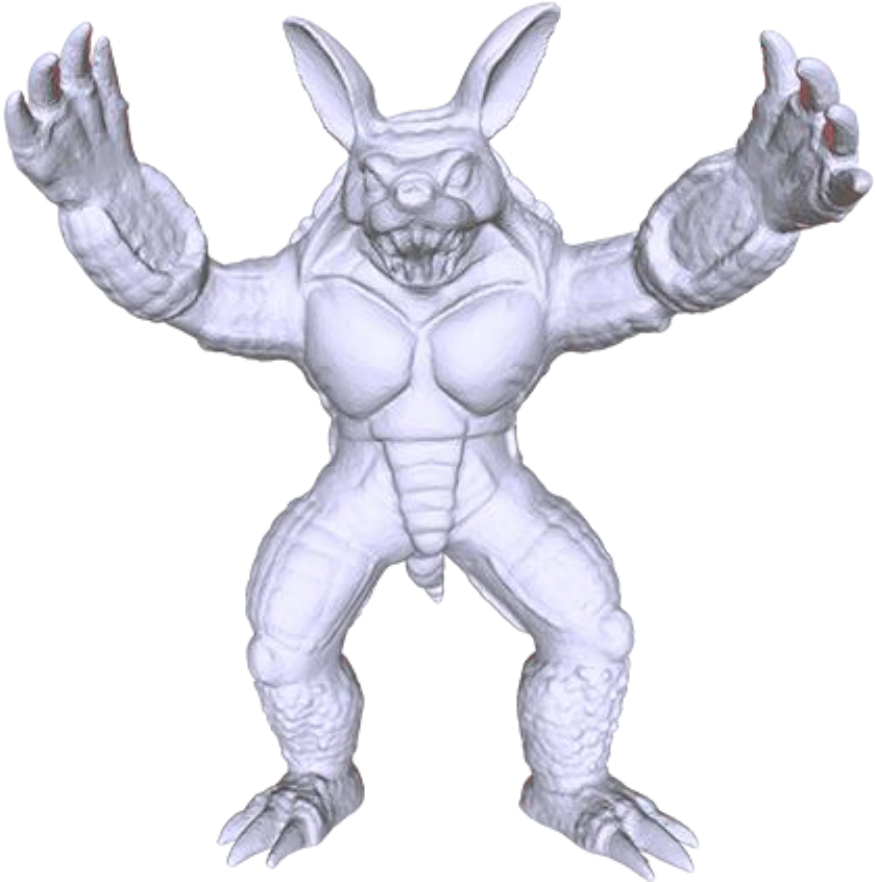}}
					\subfigure[BMF~\cite{Fleishman2003}]{
						\includegraphics[width=3.5cm]{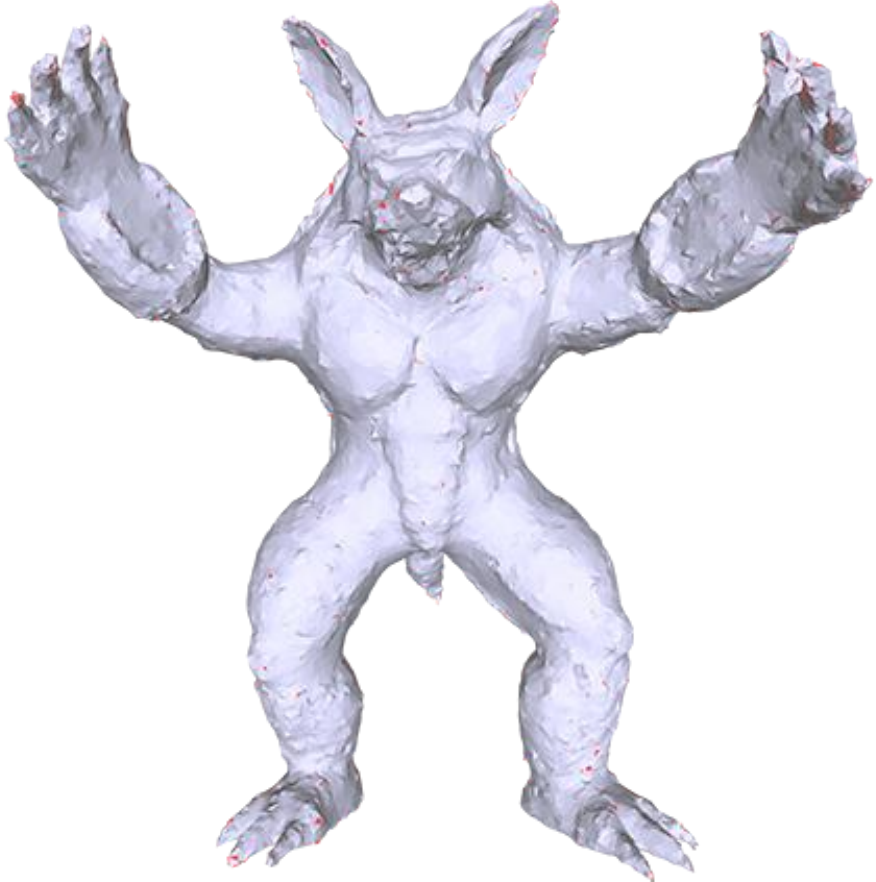}}
						\subfigure[UNF~\cite{Sun2007}]{
						\includegraphics[width=3.5cm]{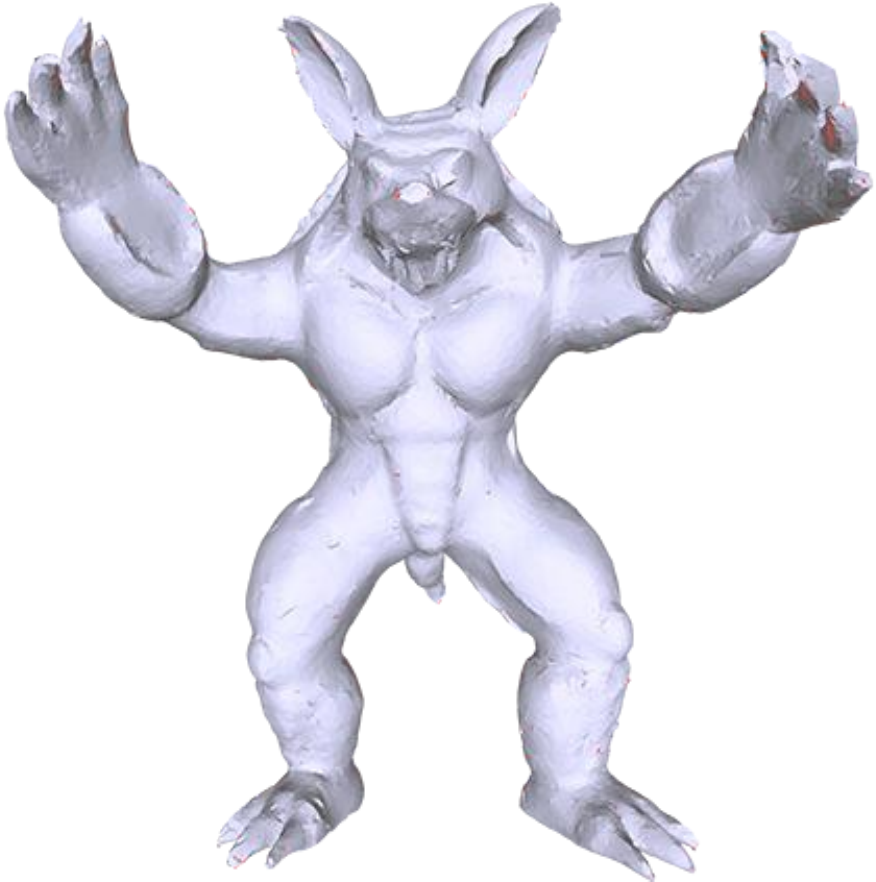}}
				\subfigure[BNF~\cite{Zheng2011}]{			
						\includegraphics[width=3.5cm]{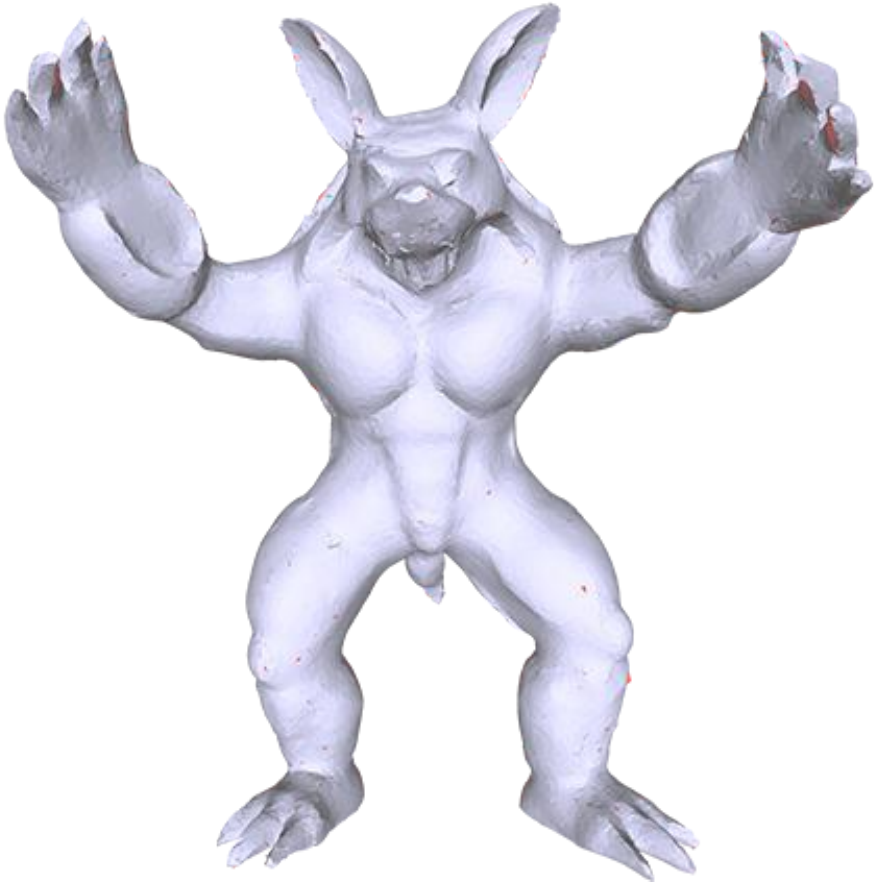}}\\							\includegraphics[width=3.5cm]{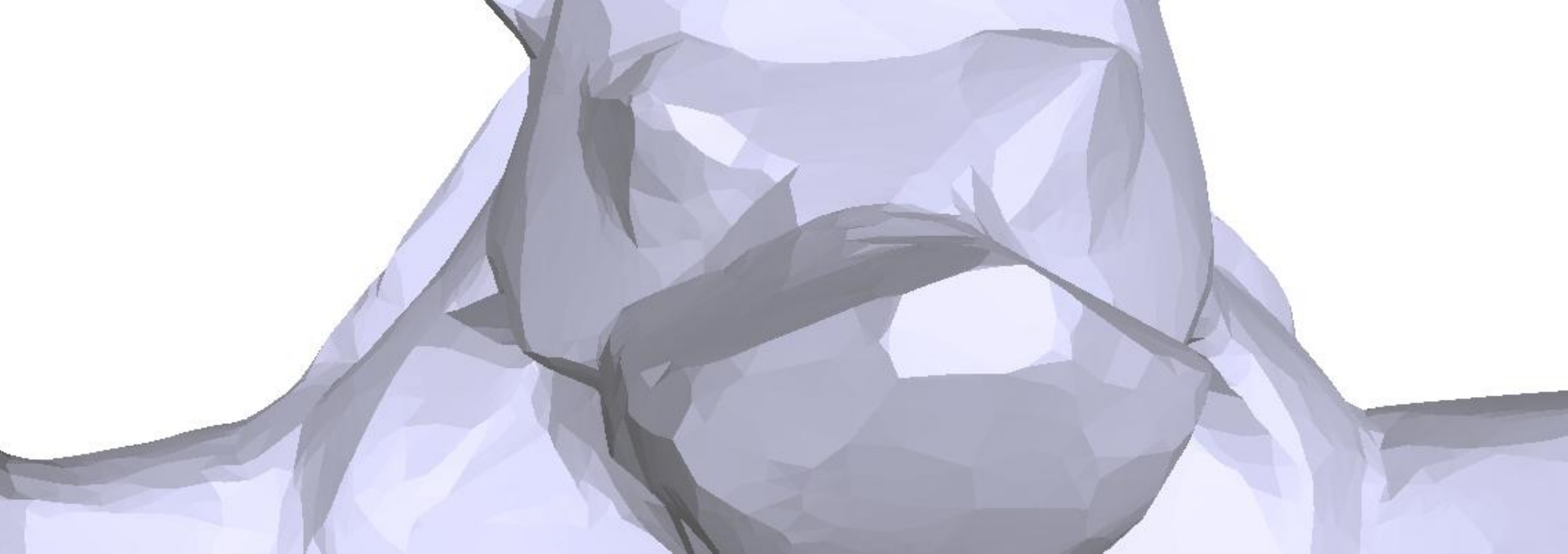}
			\includegraphics[width=3.5cm]{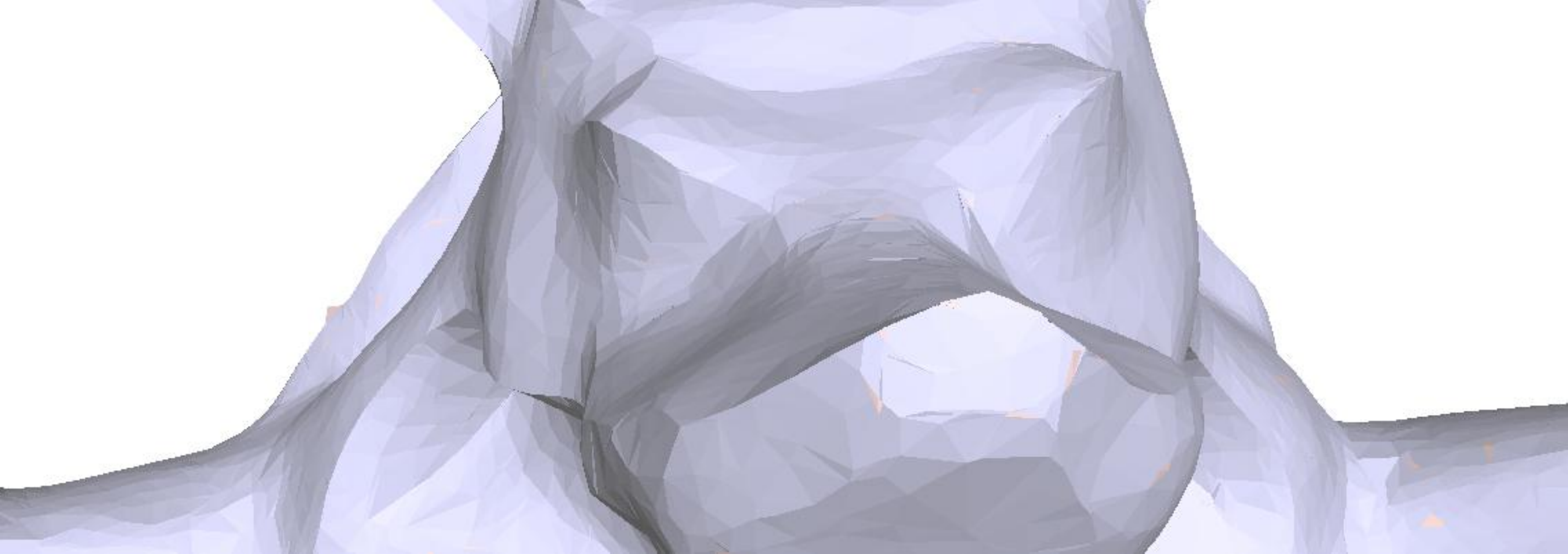}
			\includegraphics[width=3.5cm]{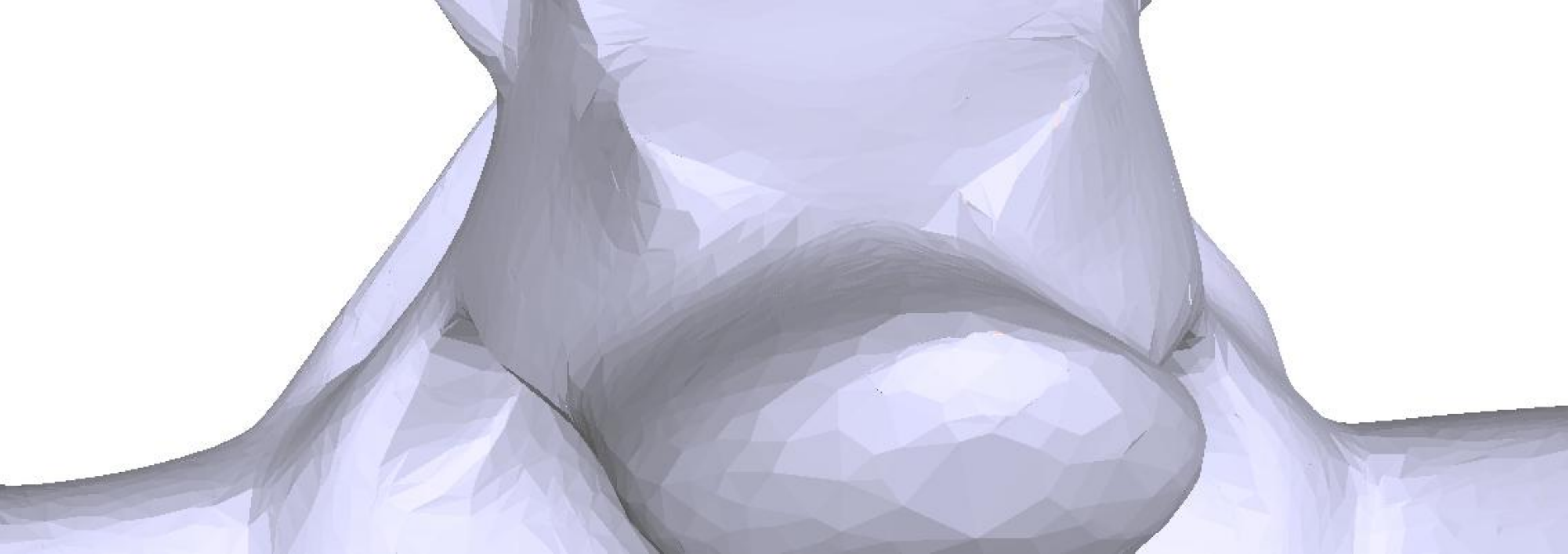}
			\includegraphics[width=3.5cm]{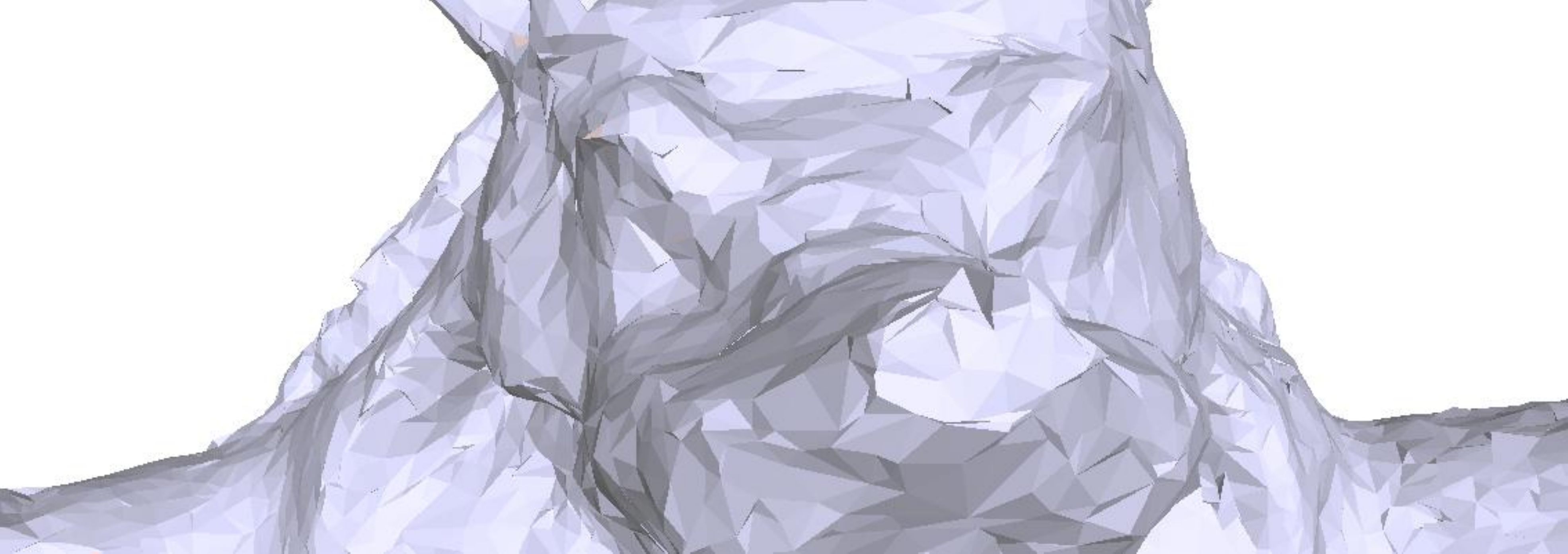}
			\includegraphics[width=3.5cm]{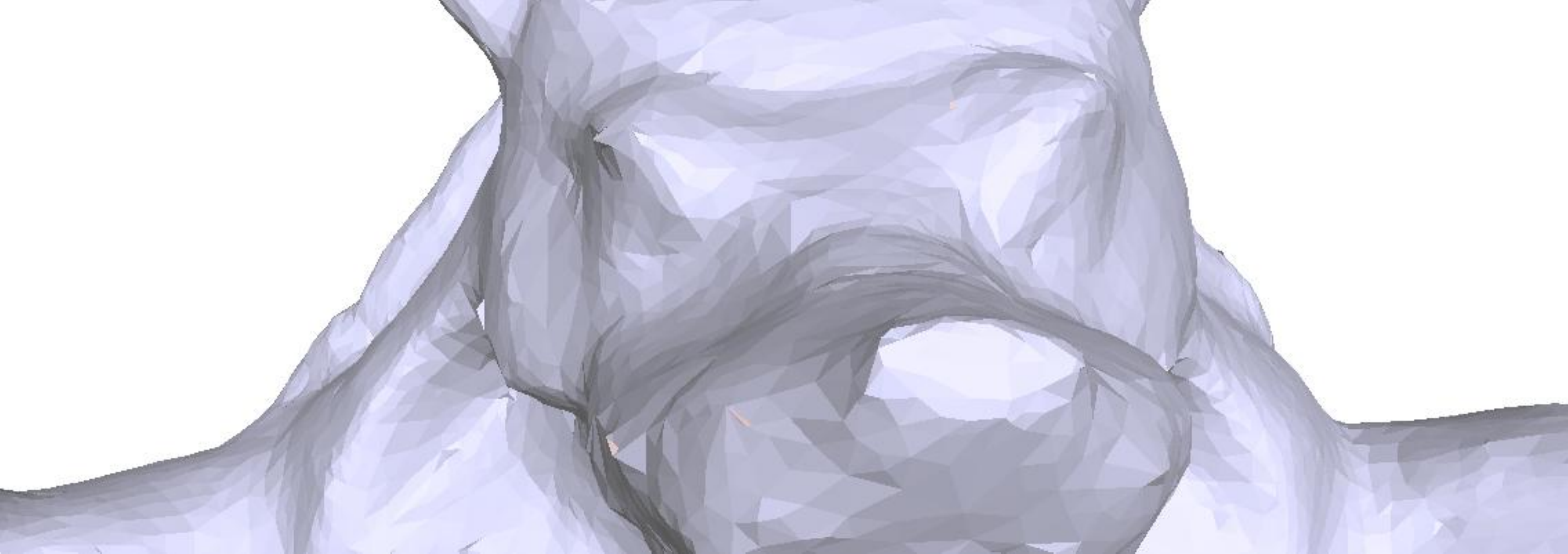}\\
				\subfigure[L0~\cite{He2013}]{
						\includegraphics[width=3.5cm]{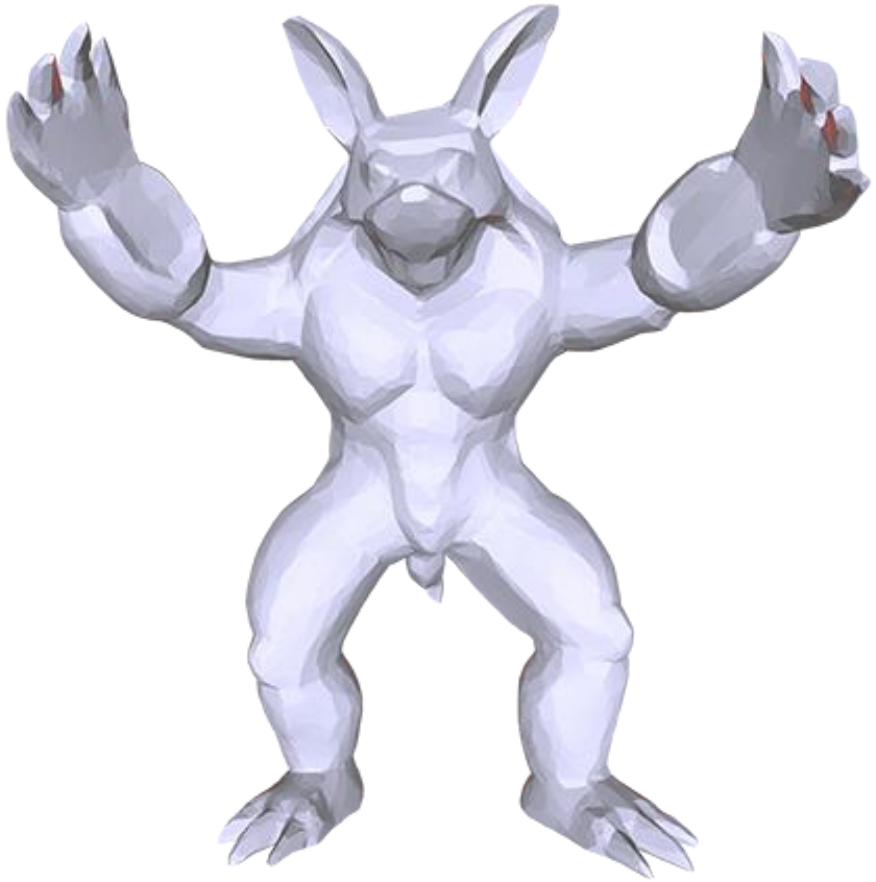}}
				\subfigure[GNF~\cite{Zhang2015}]{			
						\includegraphics[width=3.5cm]{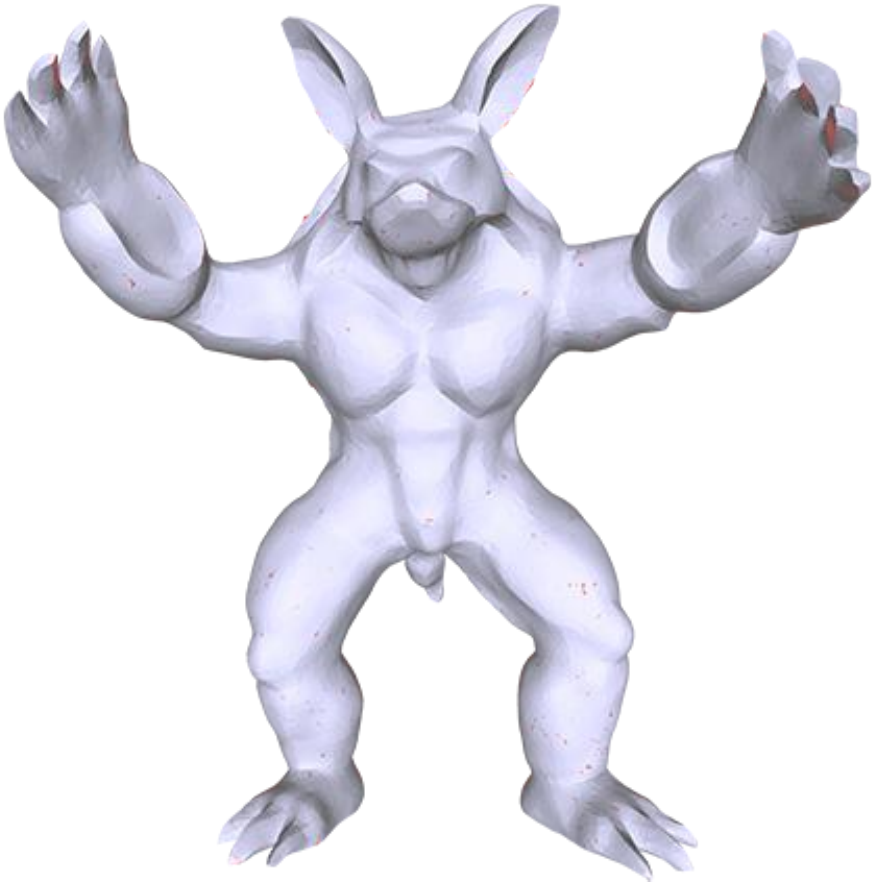}}
				\subfigure[CNR~\cite{Wang-2016-SA}]{			
						\includegraphics[width=3.5cm]{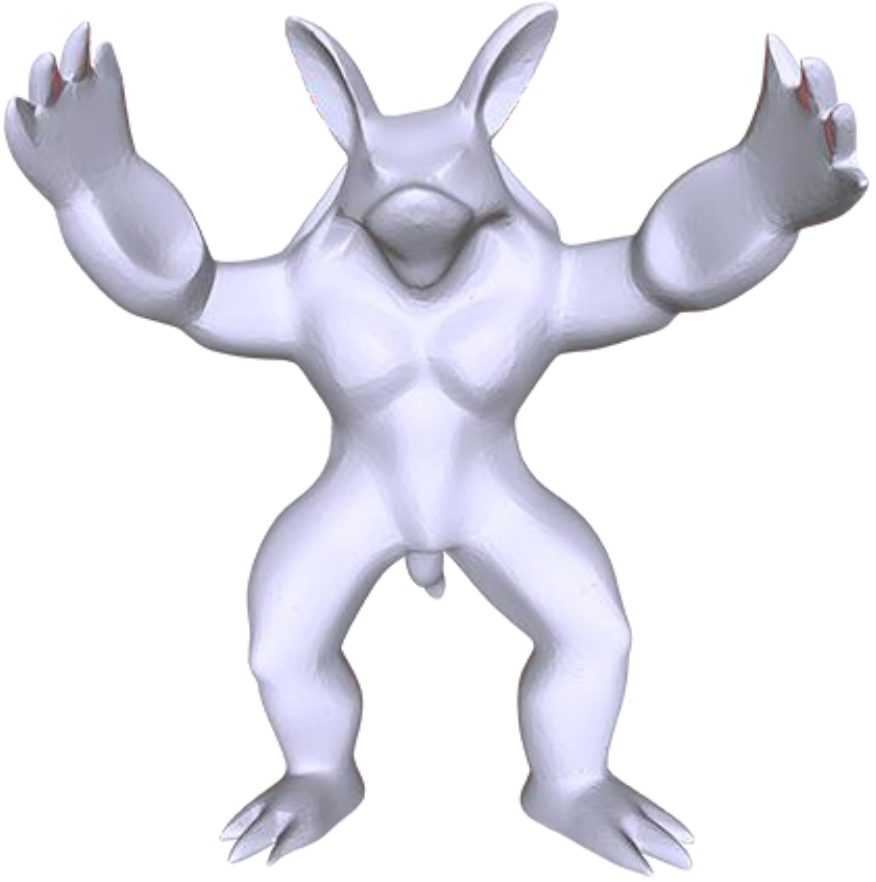}}
				\subfigure[NLLR~\cite{Li2018}]{					
						\includegraphics[width=3.5cm]{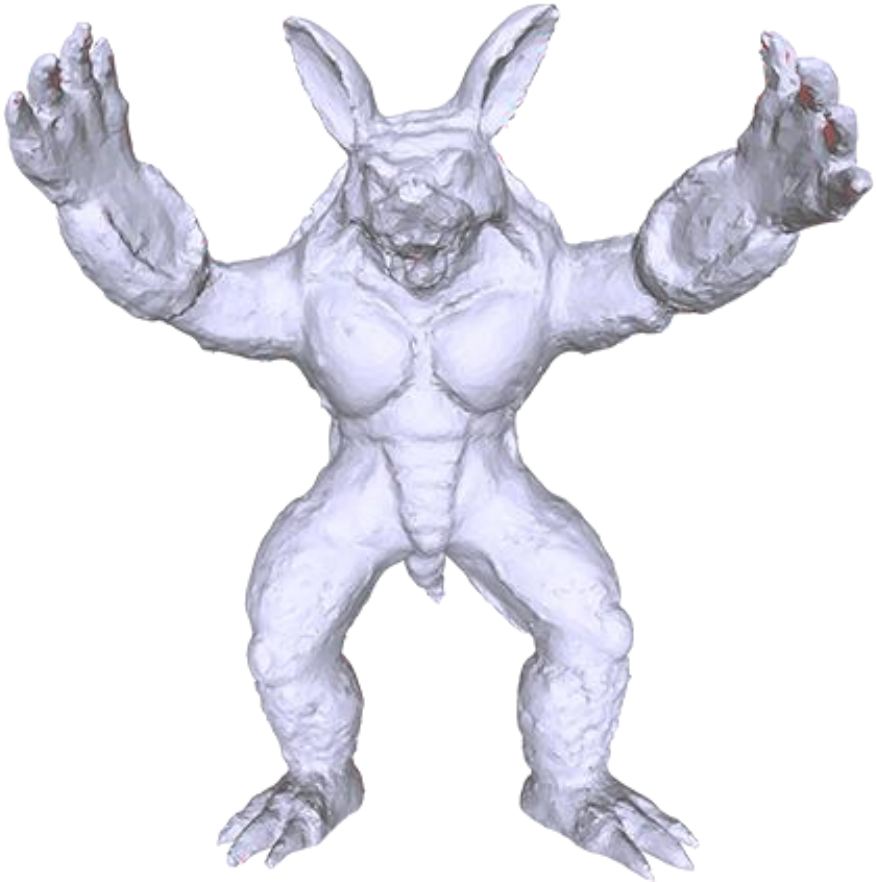}}
				\subfigure[Ours]{				
						\includegraphics[width=3.5cm]{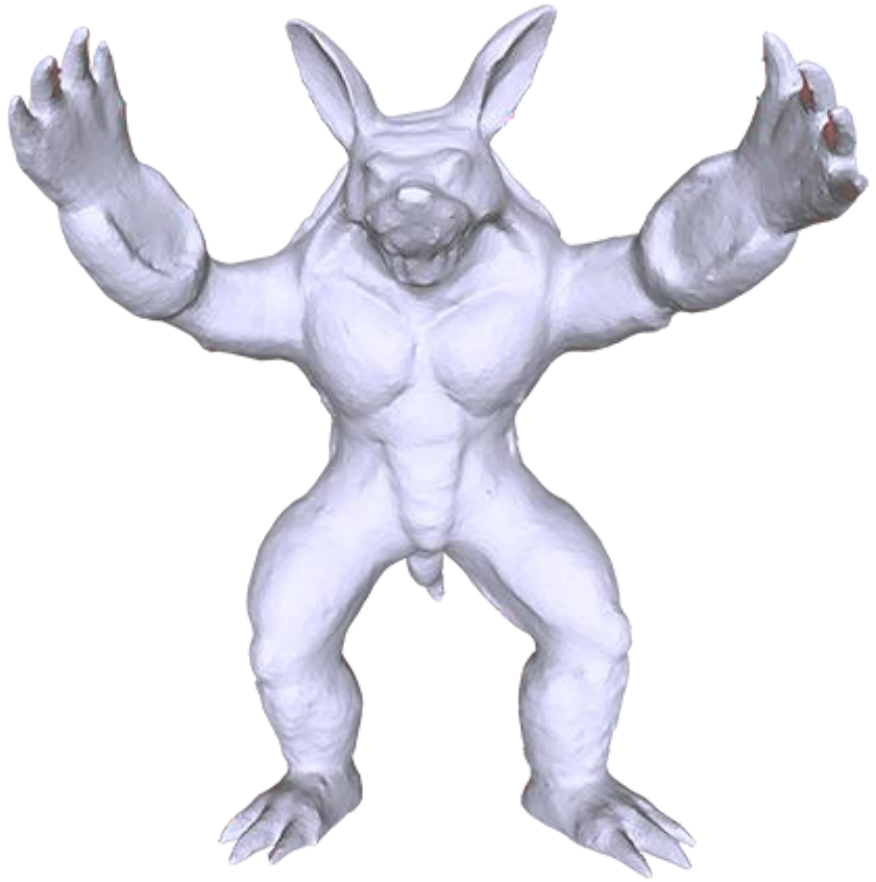}}
		\end{tabular}
	\end{center}
	\caption[res] 
	{ \label{fig:com_amadillo0.5} 
		Amadillo with Gaussian noise $\sigma_{n}=0.5l_{e}$. The flipped triangles are rendered in red.
		}
\end{figure*}

\begin{figure*} [htbp]
\centering
			\includegraphics[width=2.1cm]{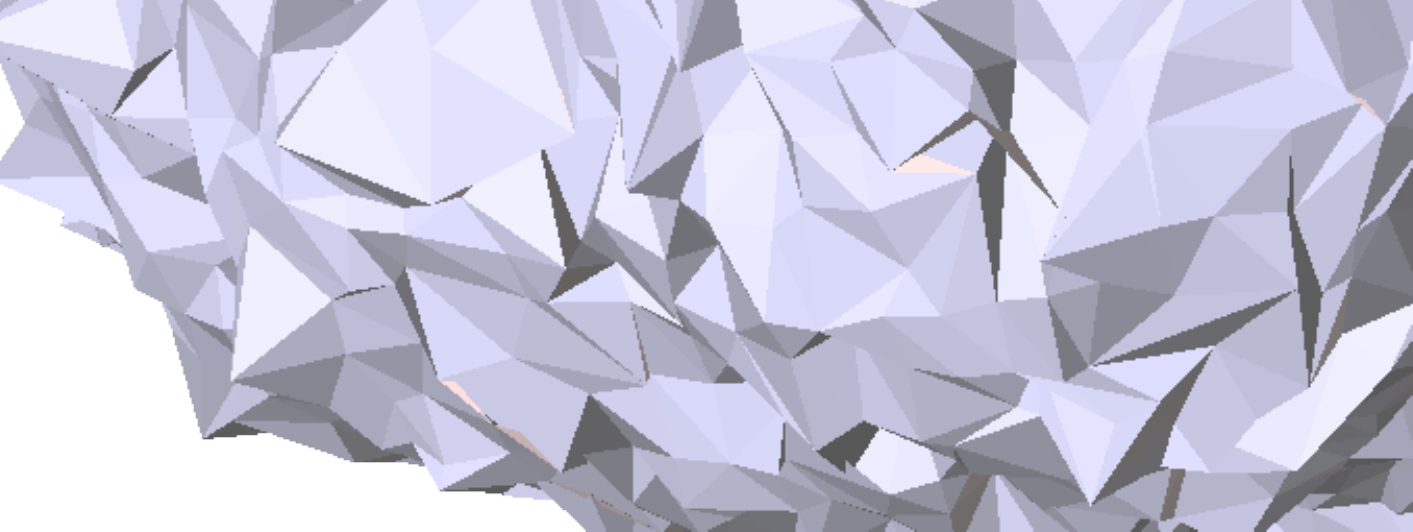}
			\includegraphics[width=2.1cm]{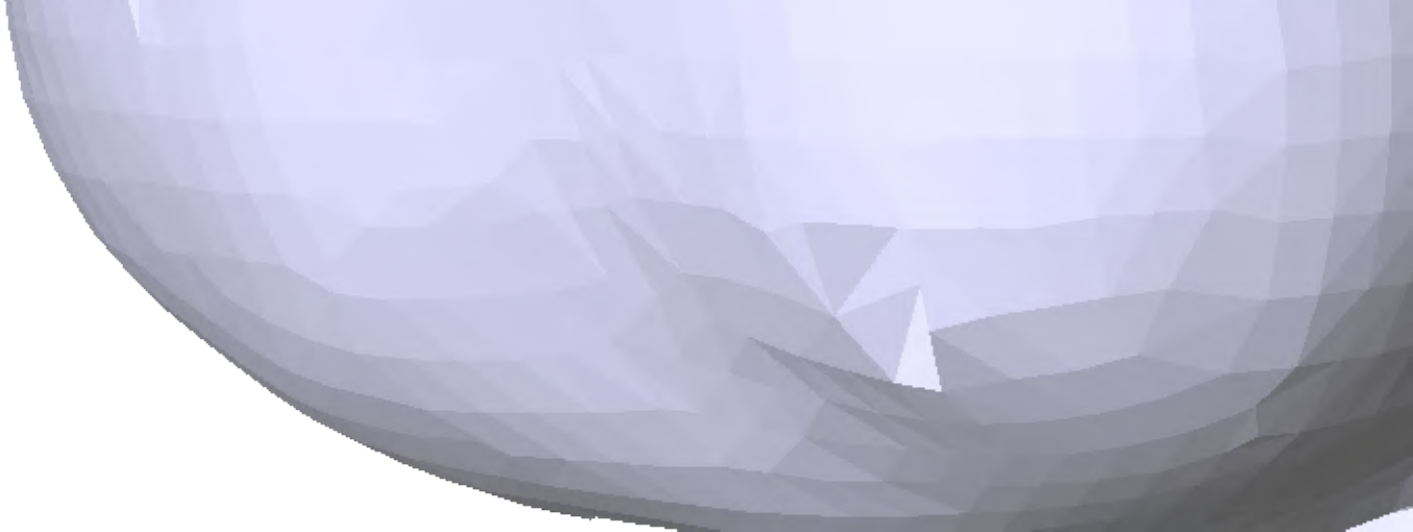}
			\includegraphics[width=2.1cm]{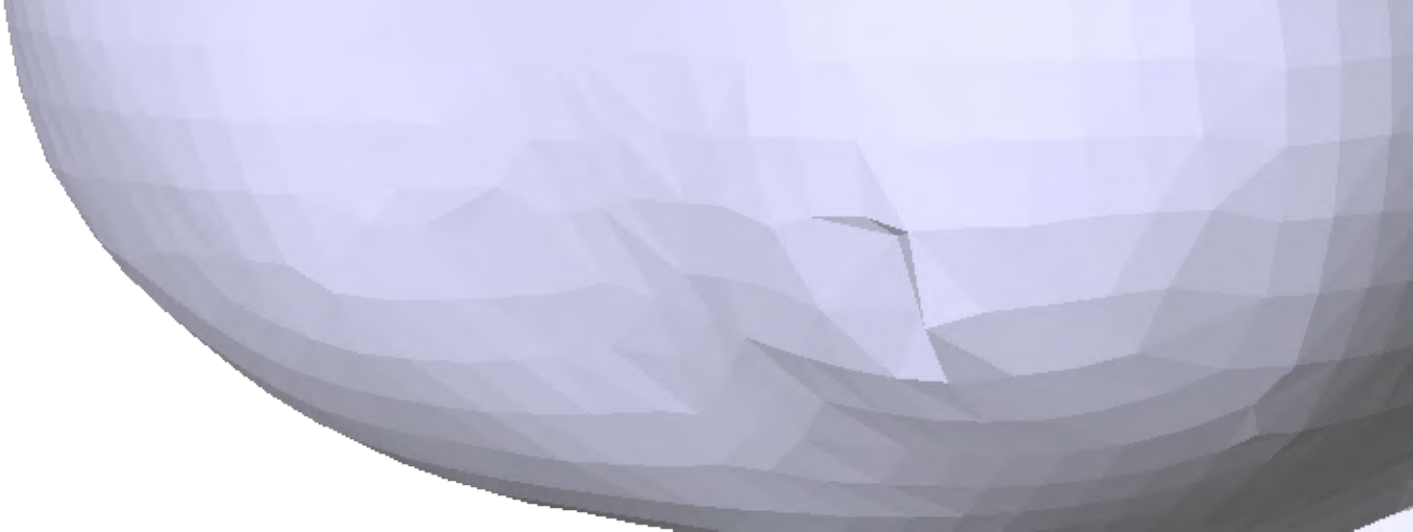}
			\includegraphics[width=2.1cm]{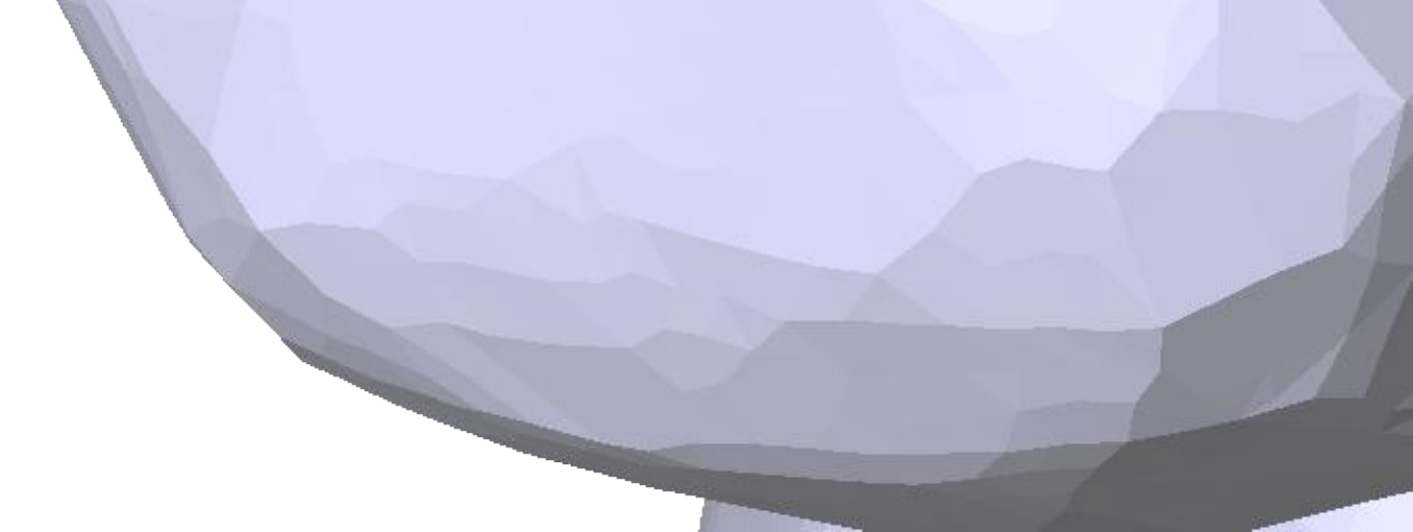}
			\includegraphics[width=2.1cm]{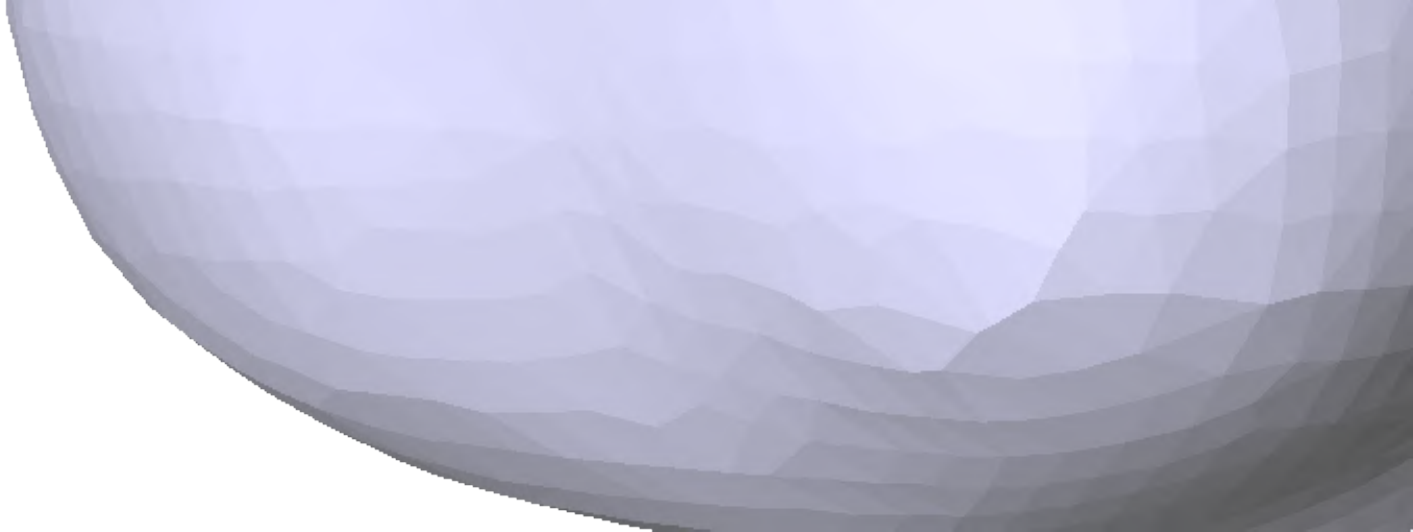}
			\includegraphics[width=2.1cm]{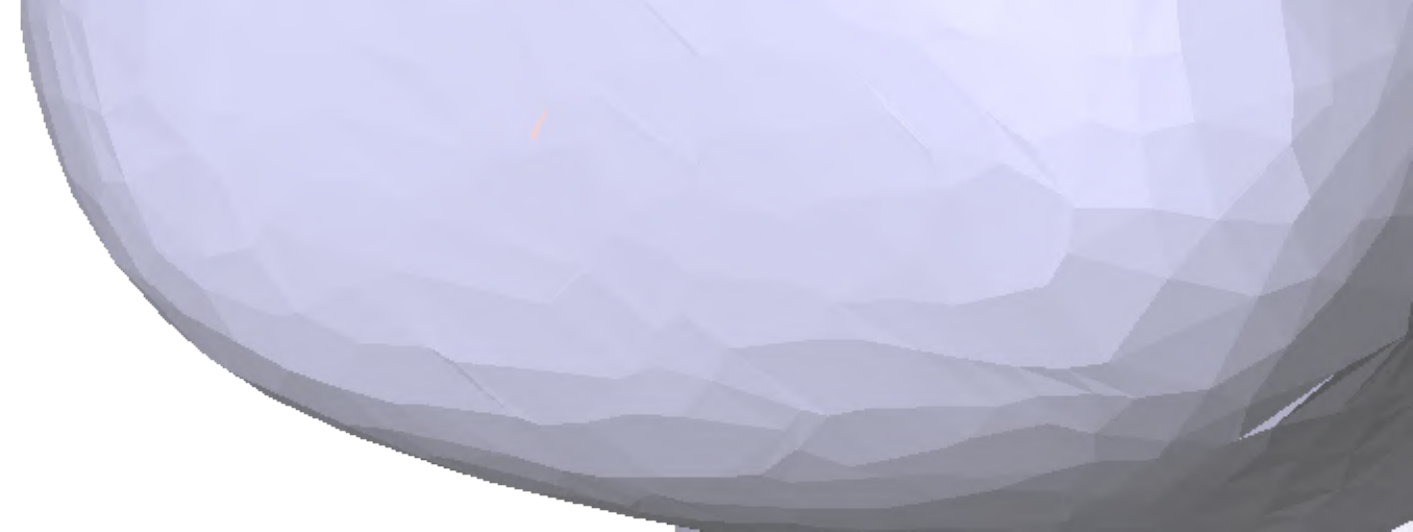}
			\includegraphics[width=2.1cm]{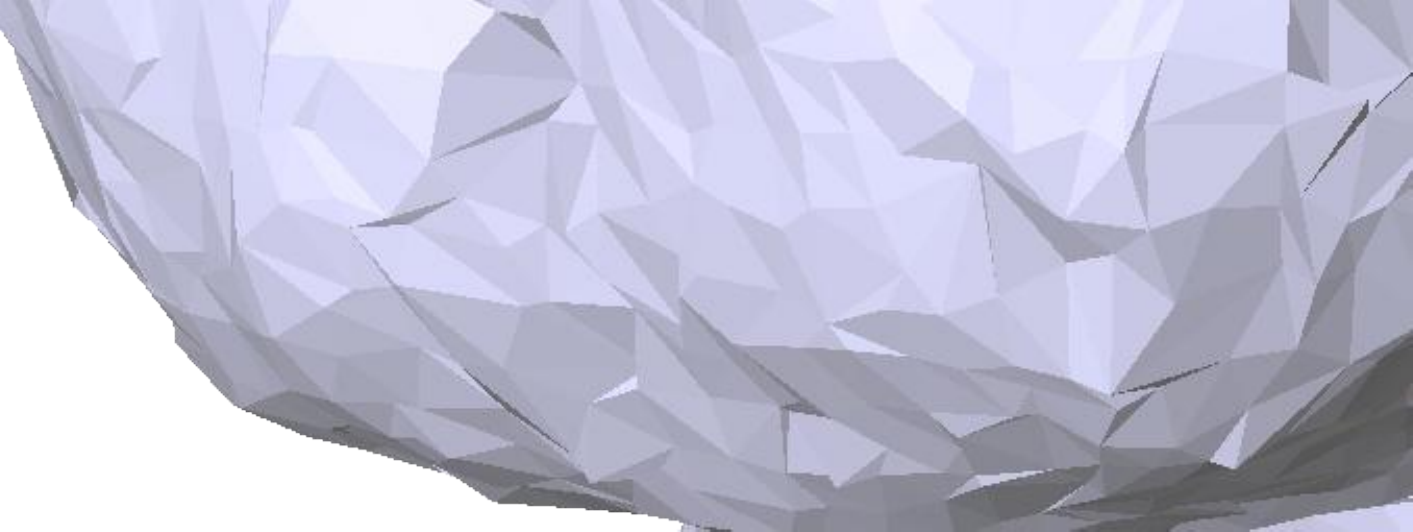}
			\includegraphics[width=2.1cm]{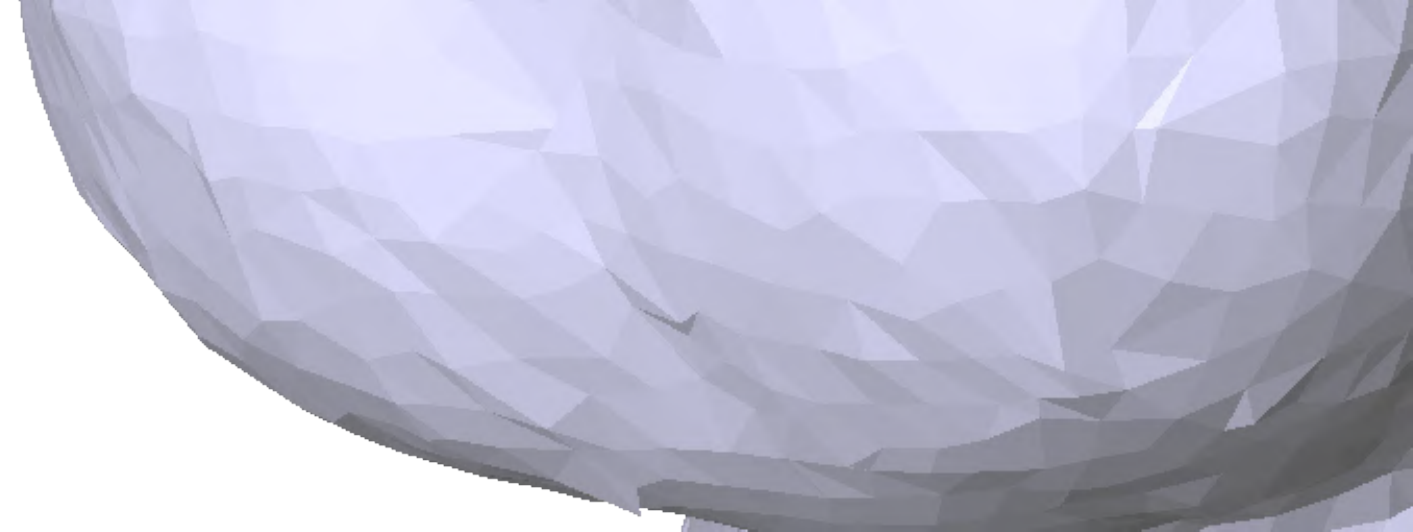}\\
			\includegraphics[width=2.1cm]{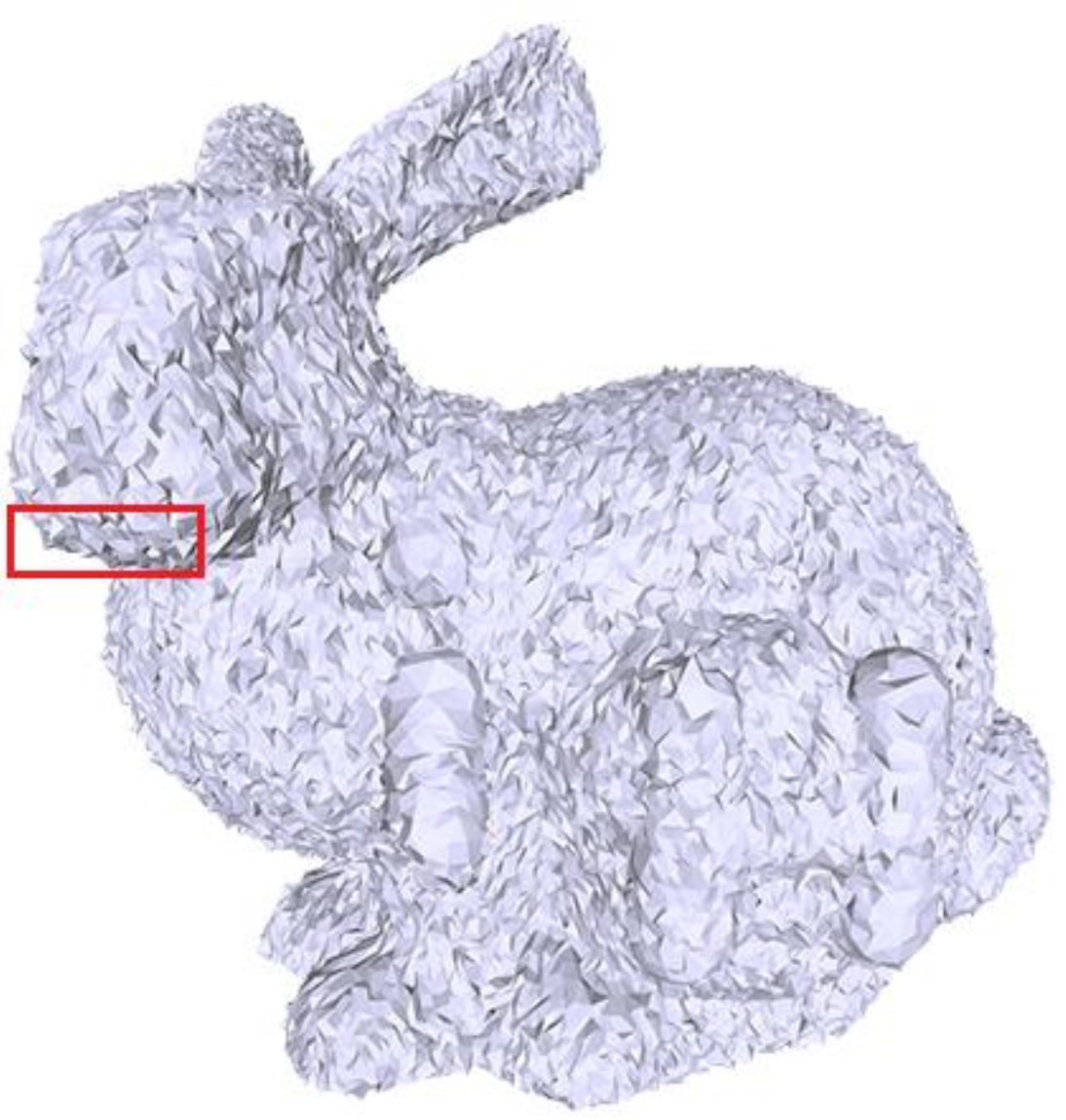}
			\includegraphics[width=2.1cm]{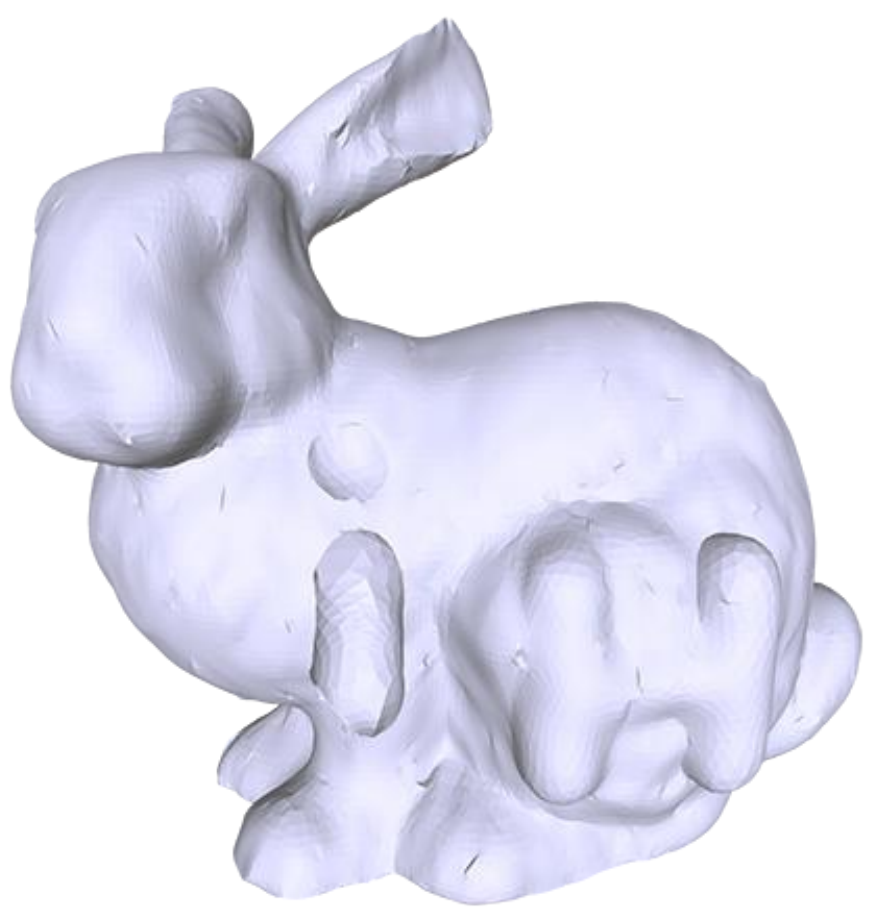}
			\includegraphics[width=2.1cm]{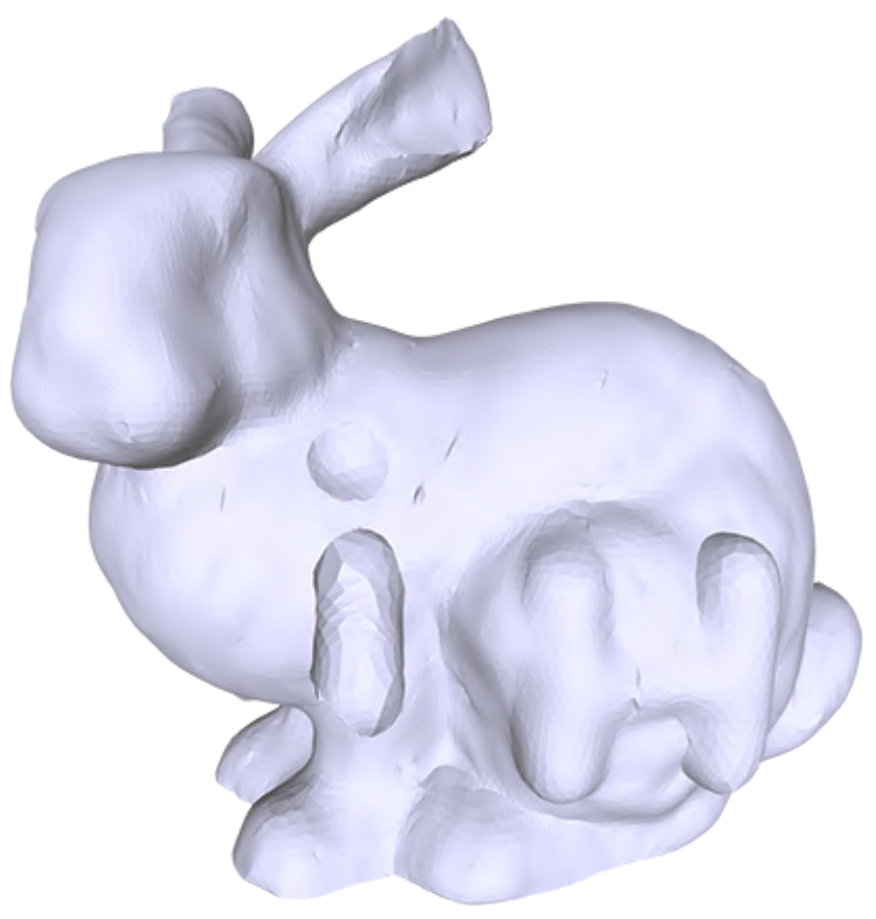}
			\includegraphics[width=2.1cm]{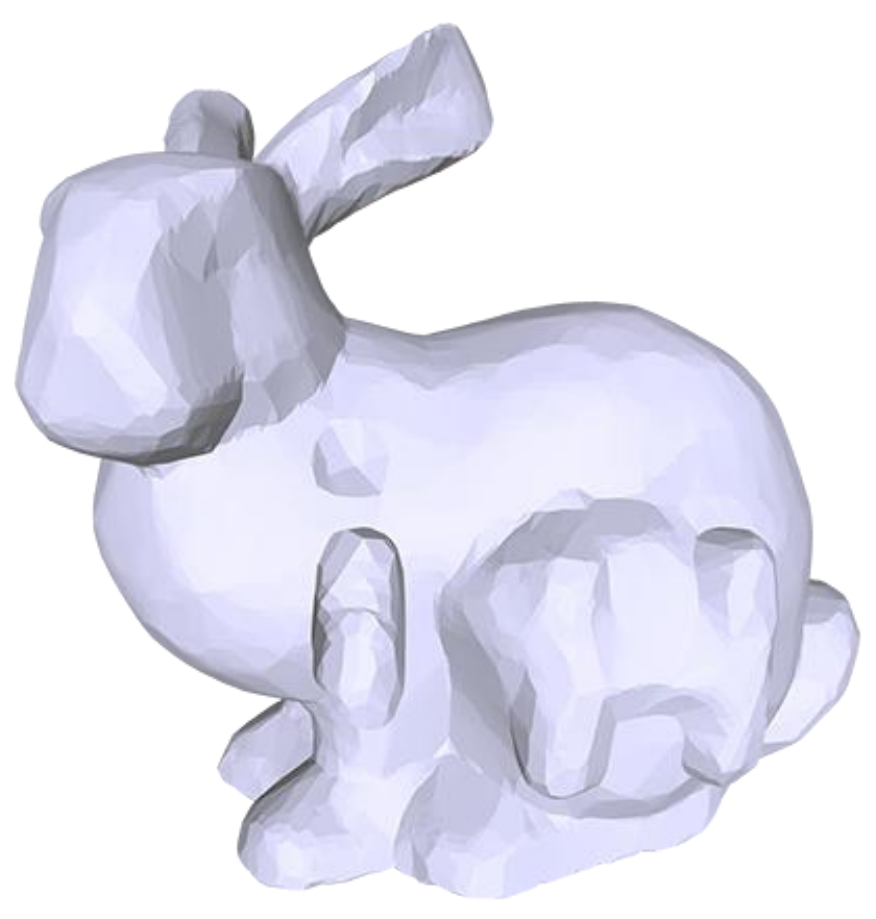}
			\includegraphics[width=2.1cm]{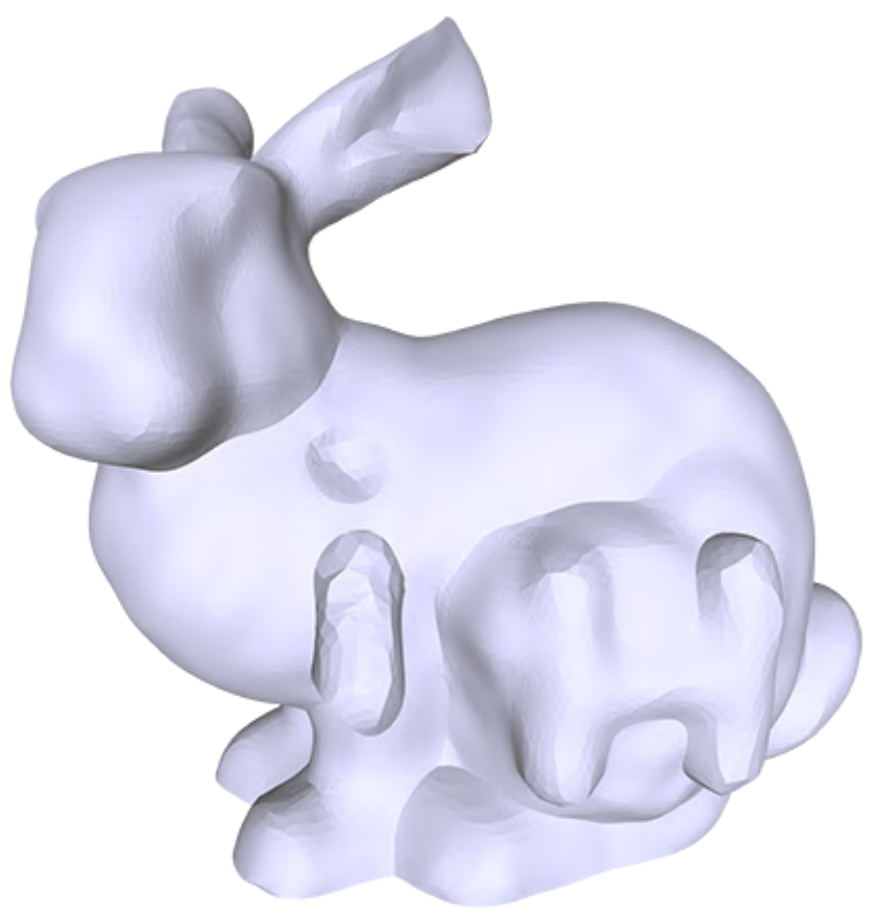}
			\includegraphics[width=2.1cm]{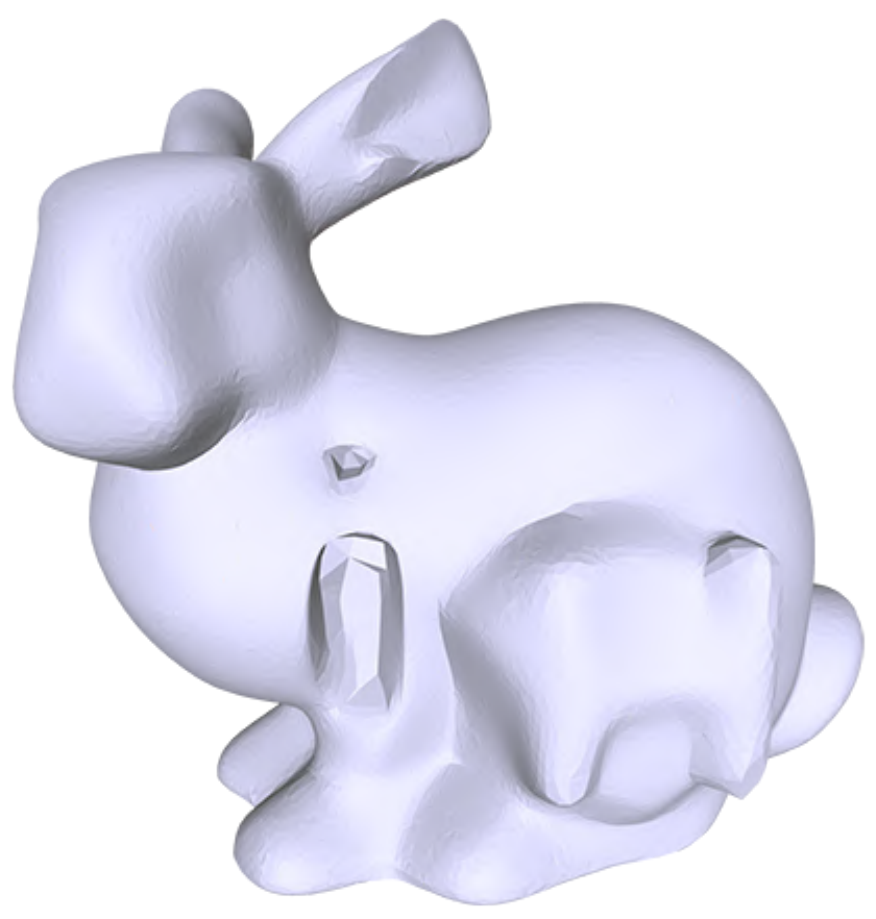}
			\includegraphics[width=2.1cm]{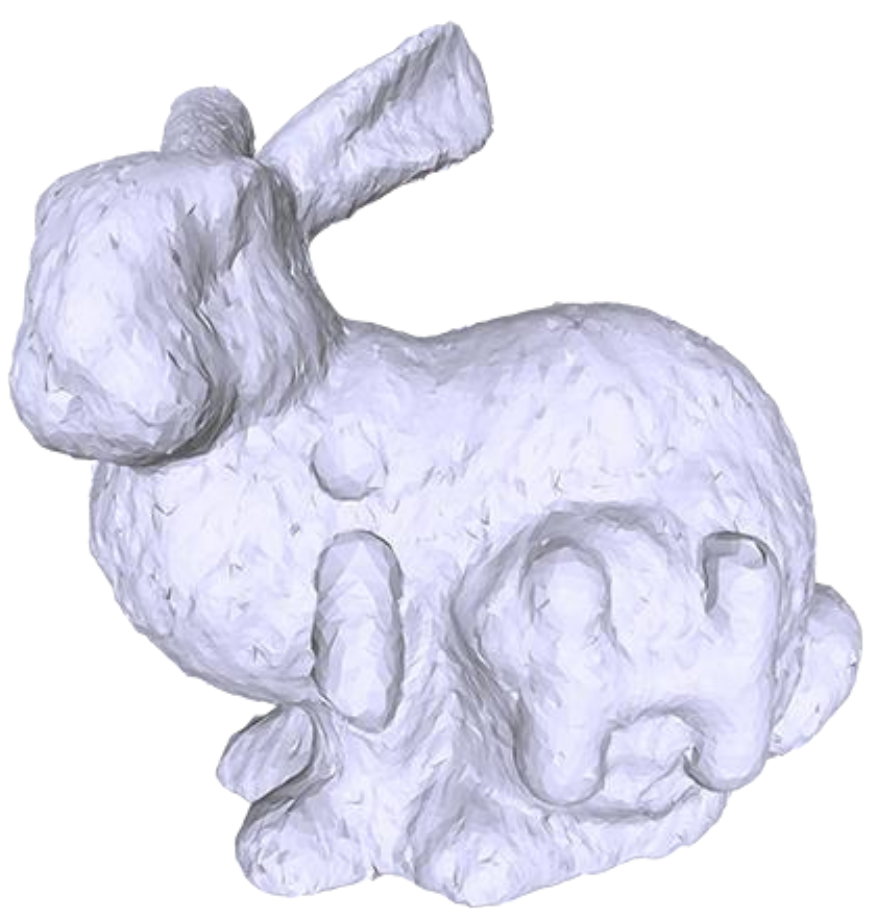}
			\includegraphics[width=2.1cm]{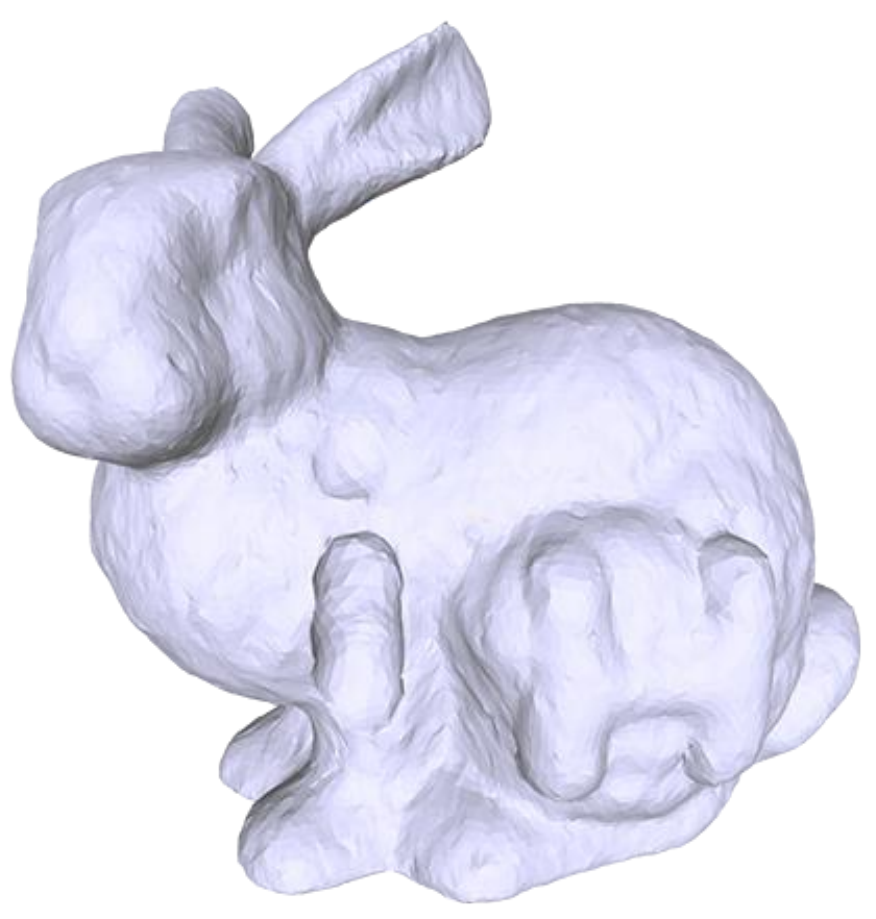}\\
			
			\includegraphics[width=2.1cm]{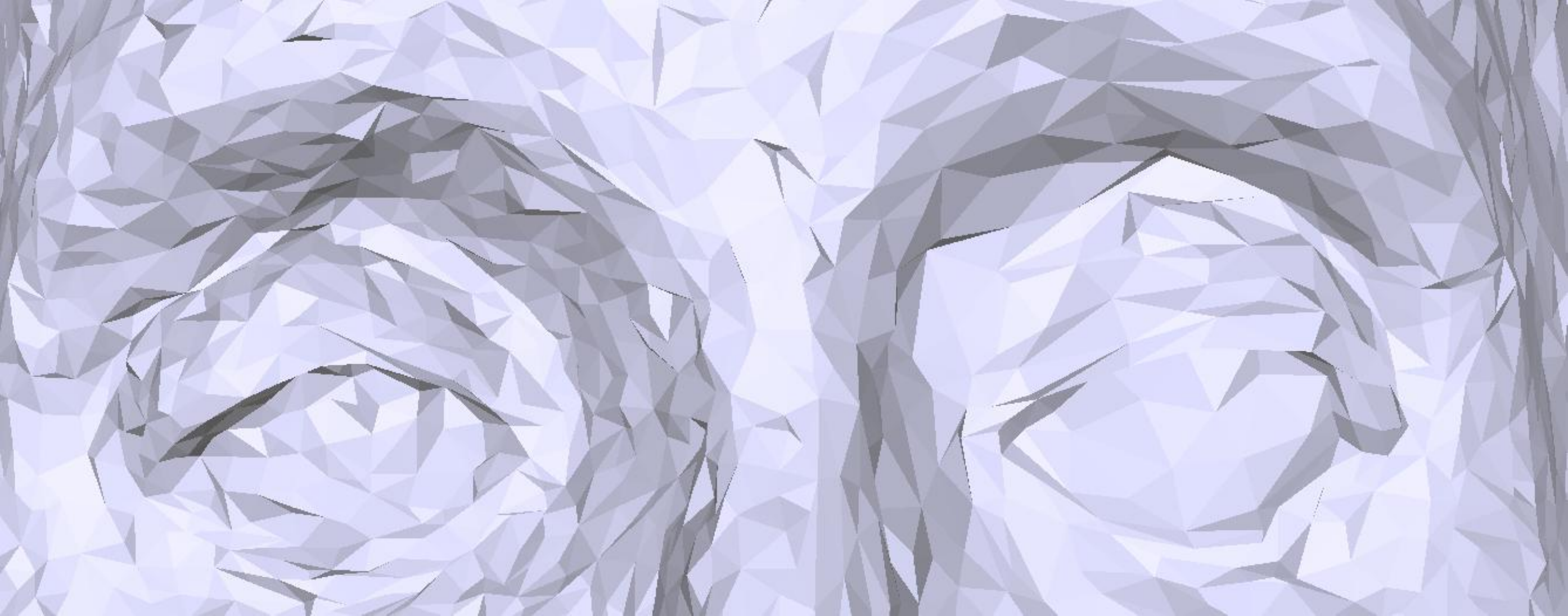}
			\includegraphics[width=2.1cm]{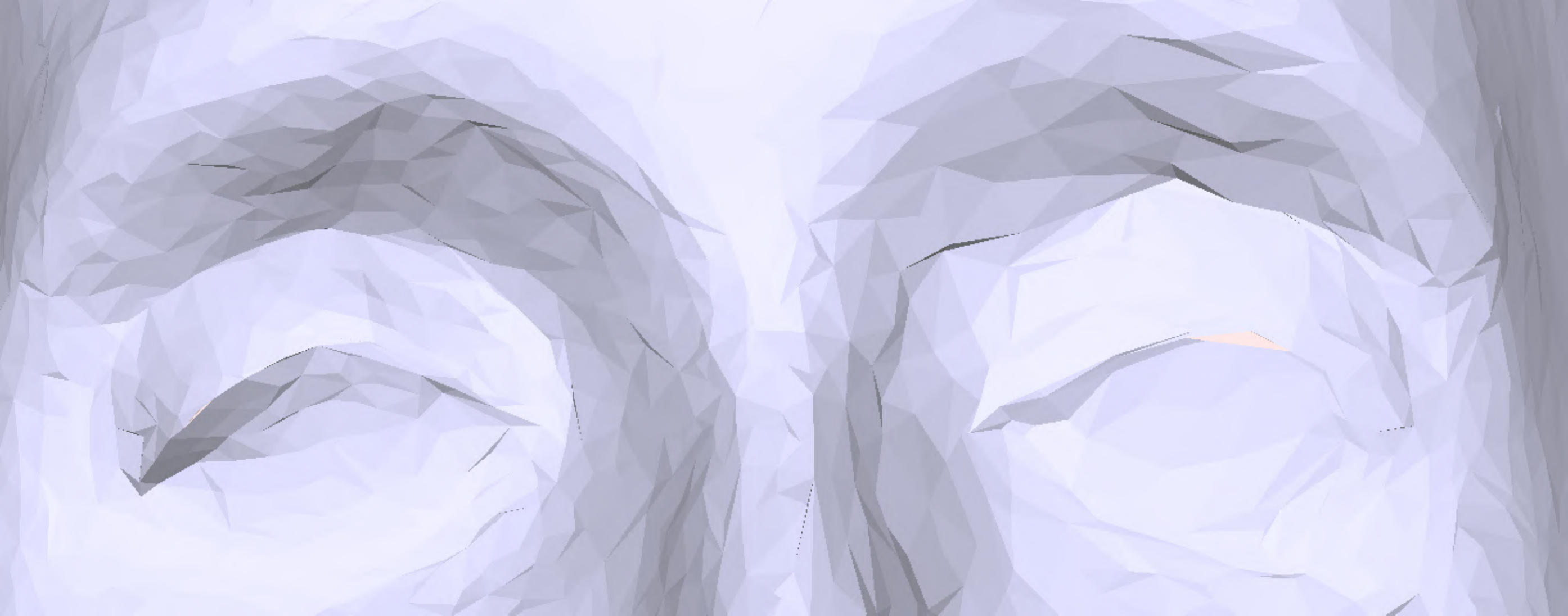}
			\includegraphics[width=2.1cm]{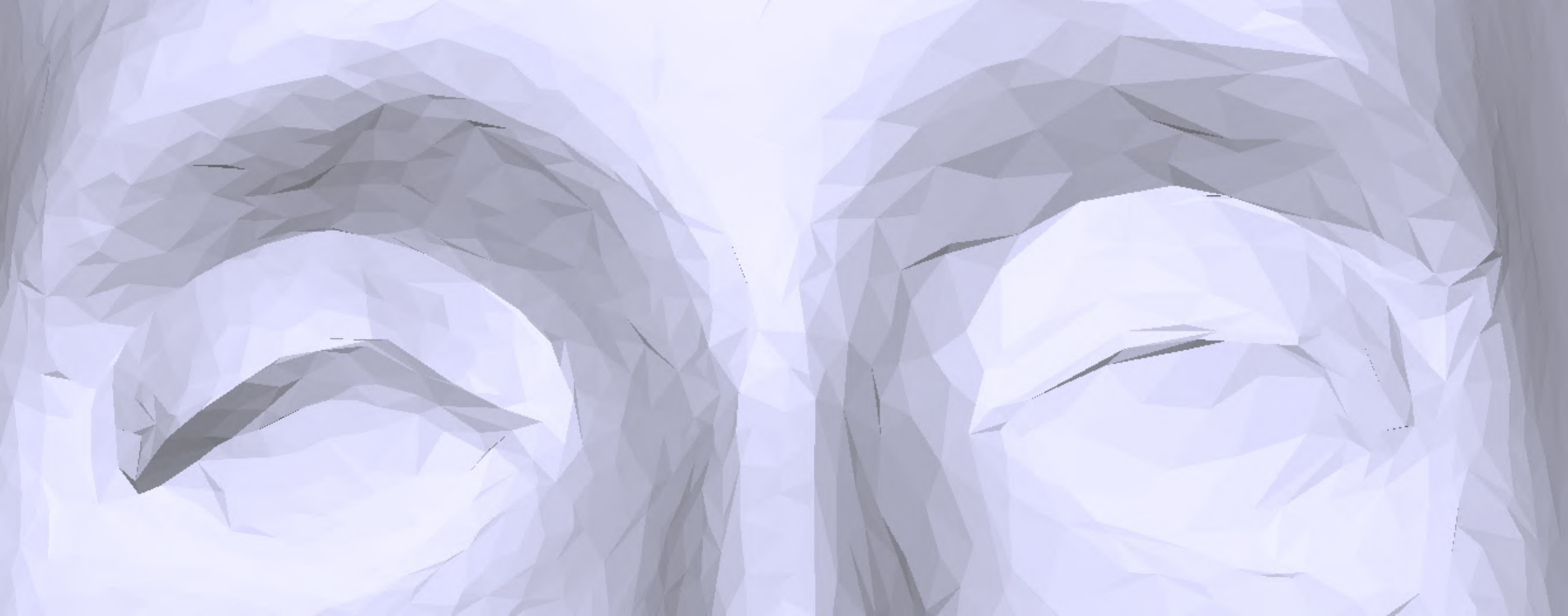}
			\includegraphics[width=2.1cm]{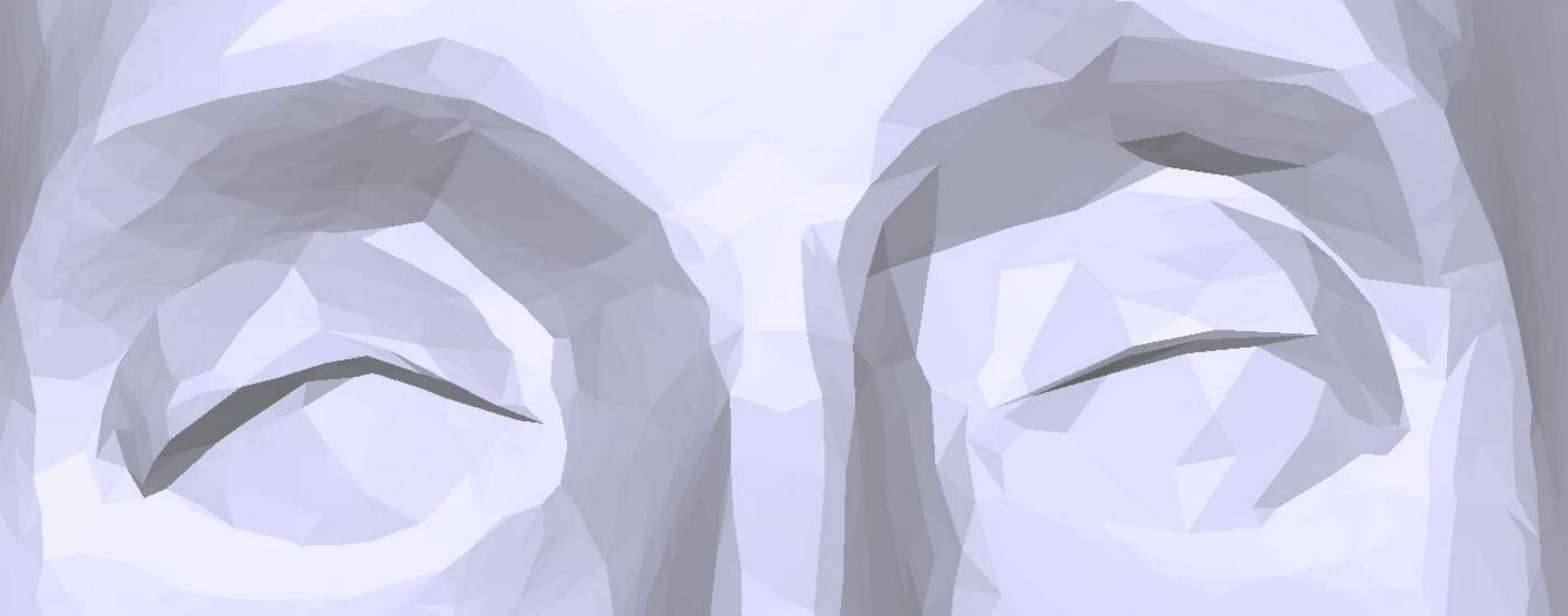}
			\includegraphics[width=2.1cm]{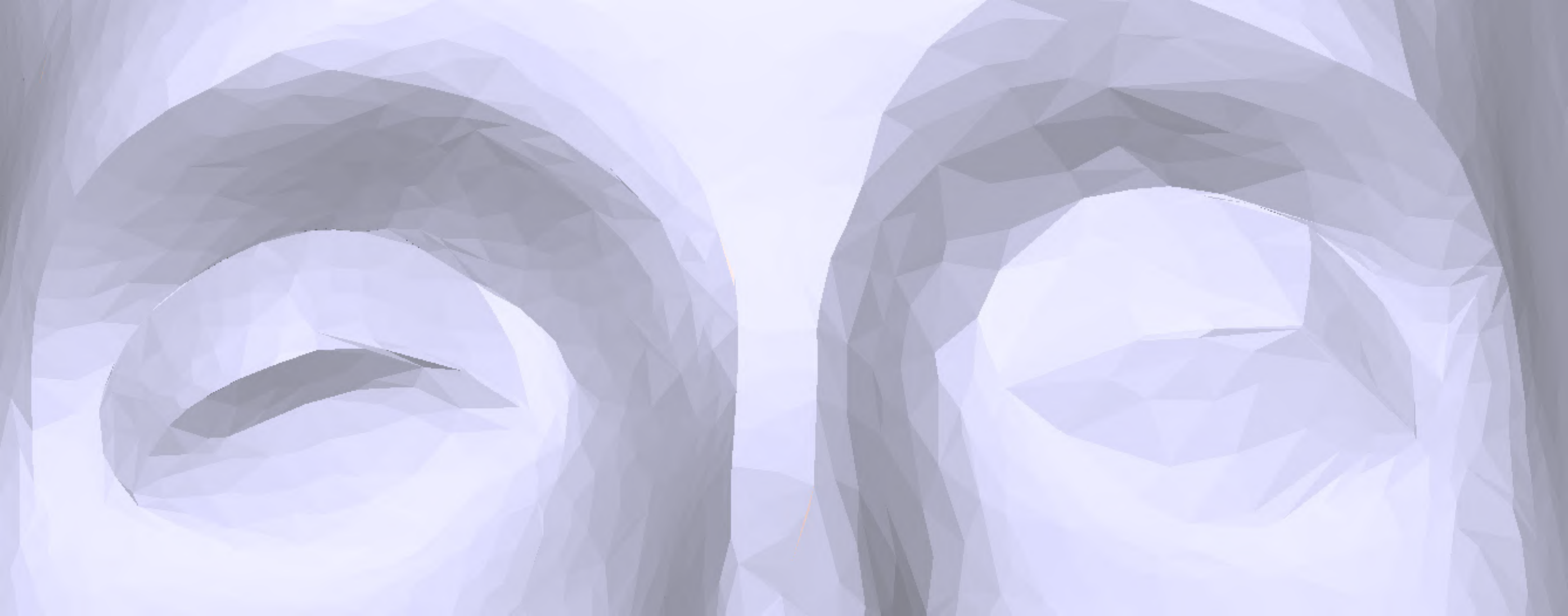}
			\includegraphics[width=2.1cm]{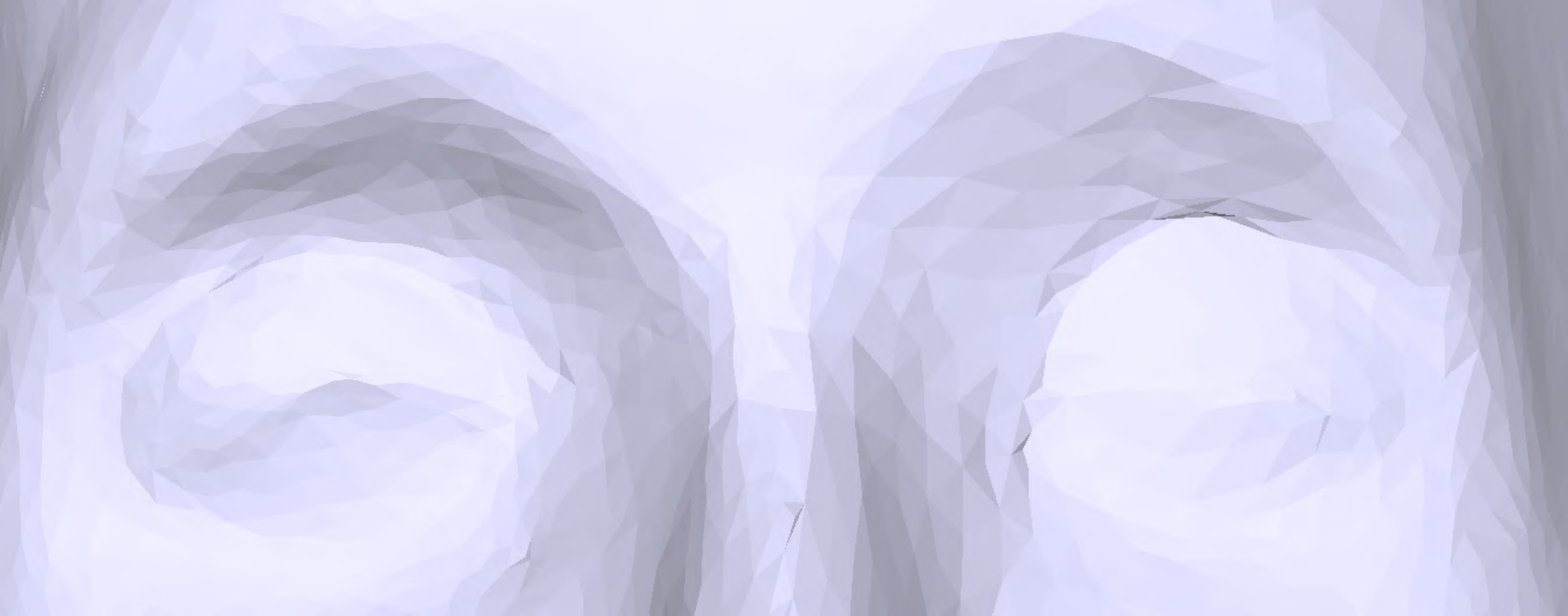}
			\includegraphics[width=2.1cm]{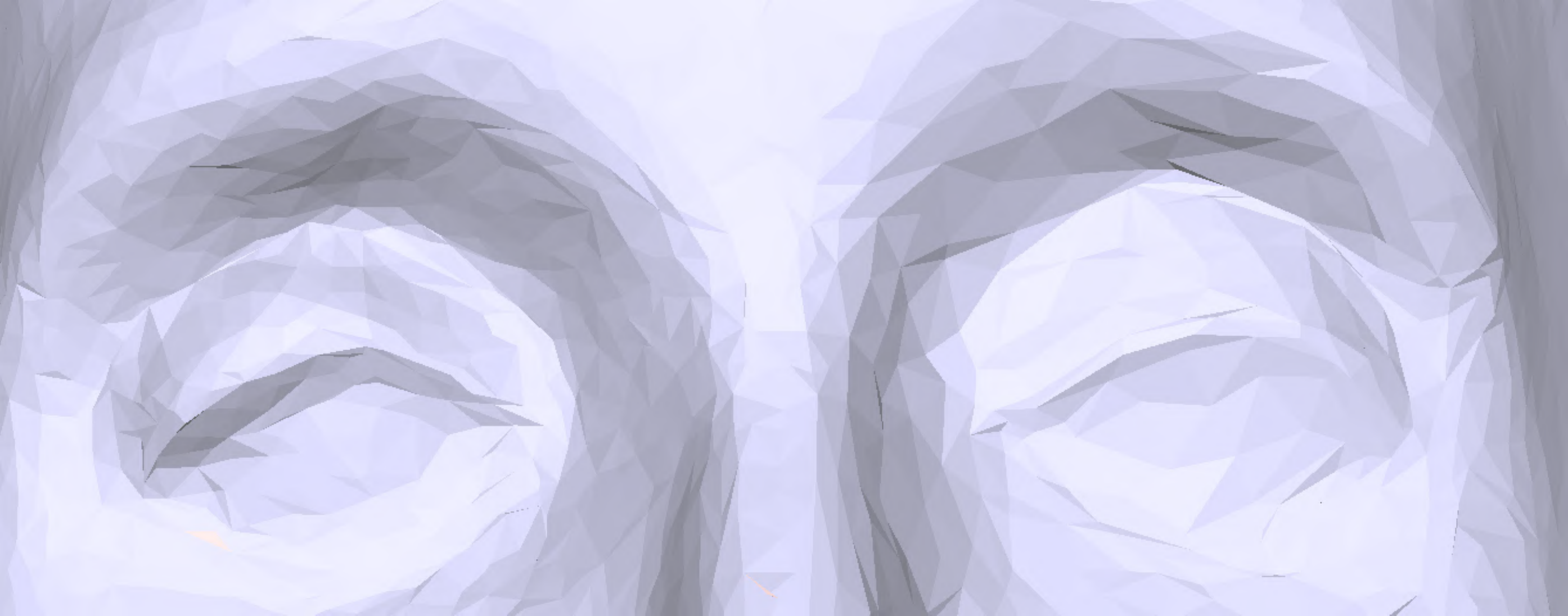}
			\includegraphics[width=2.1cm]{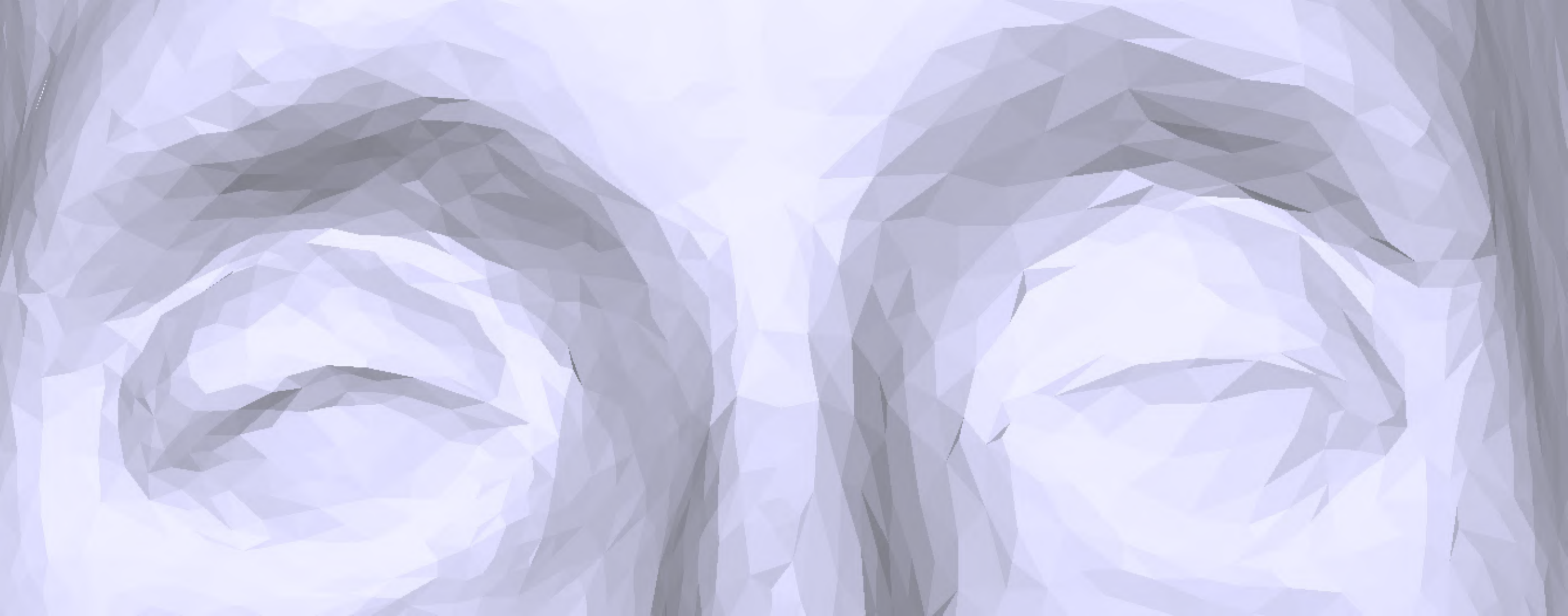}\\
			\includegraphics[width=2.1cm]{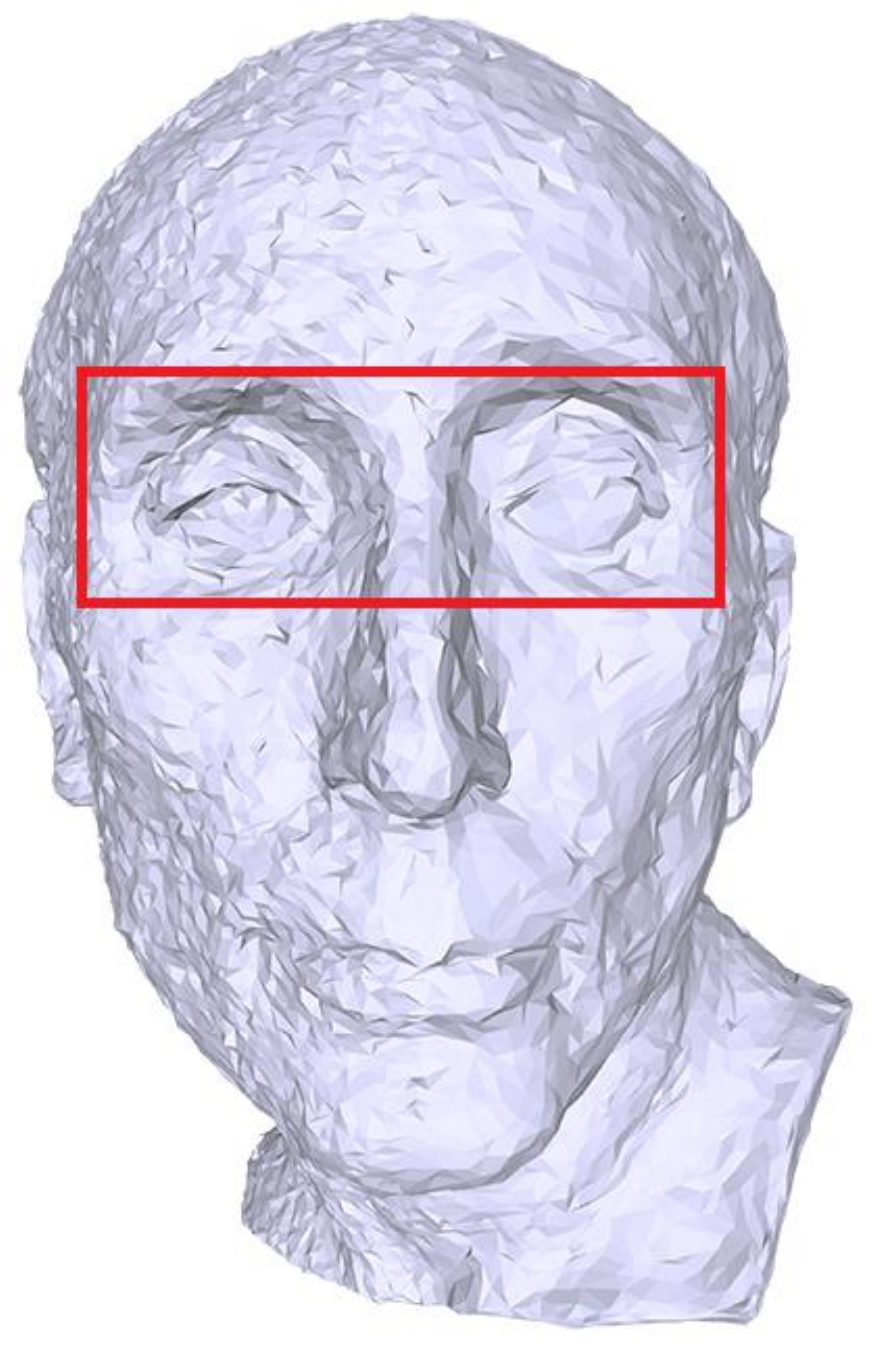}
			\includegraphics[width=2.1cm]{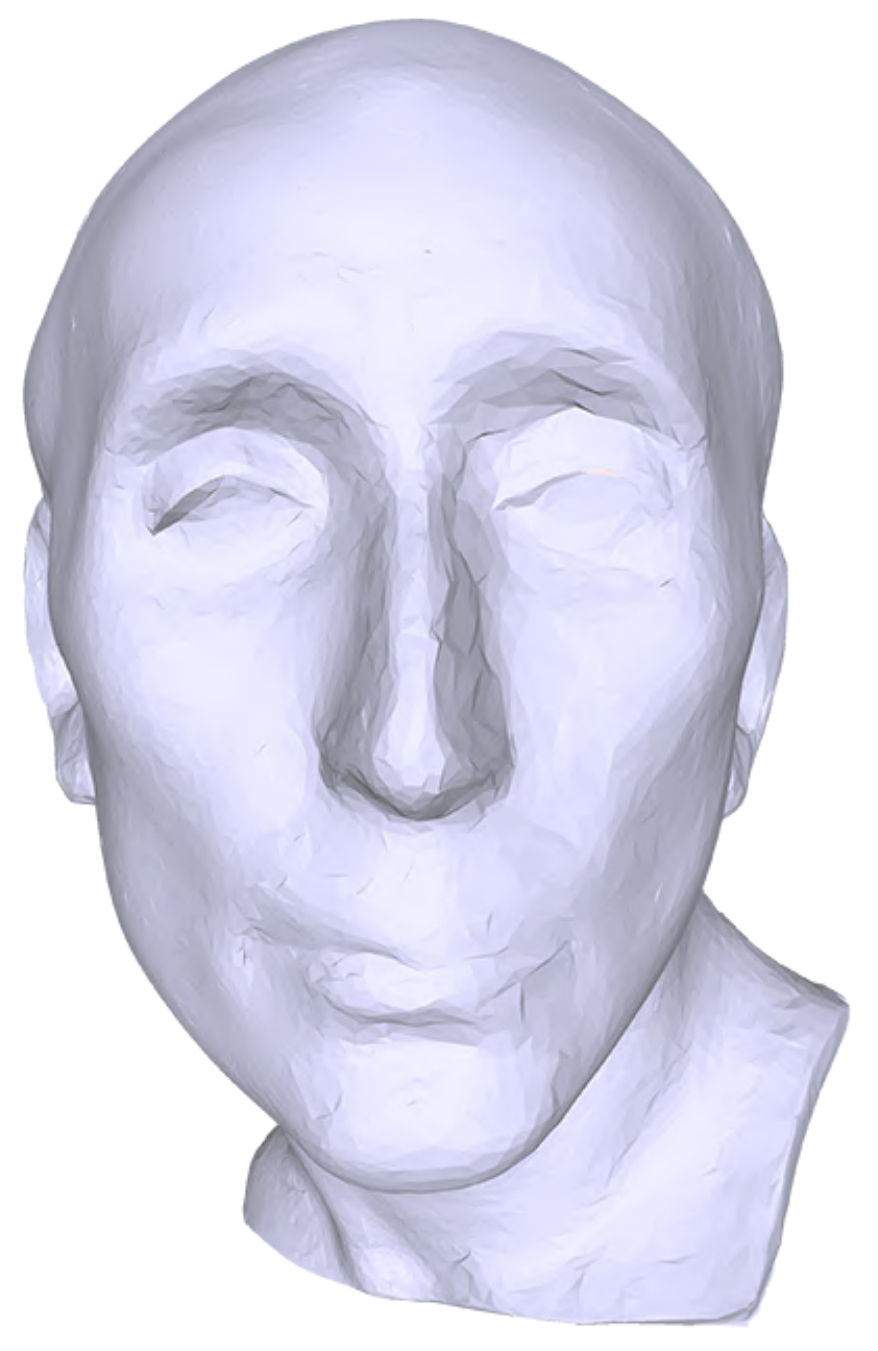}
			\includegraphics[width=2.1cm]{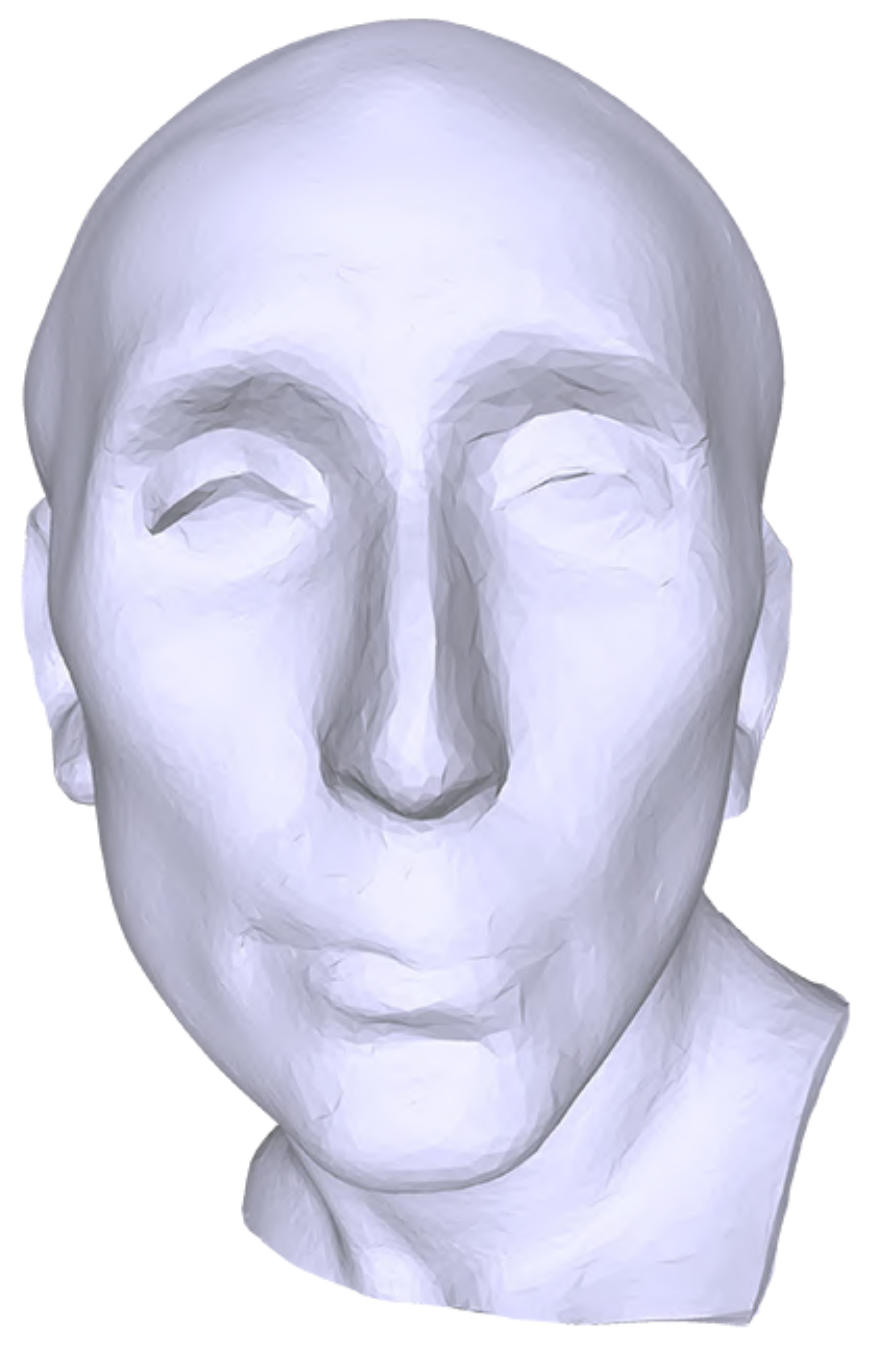}						\includegraphics[width=2.1cm]{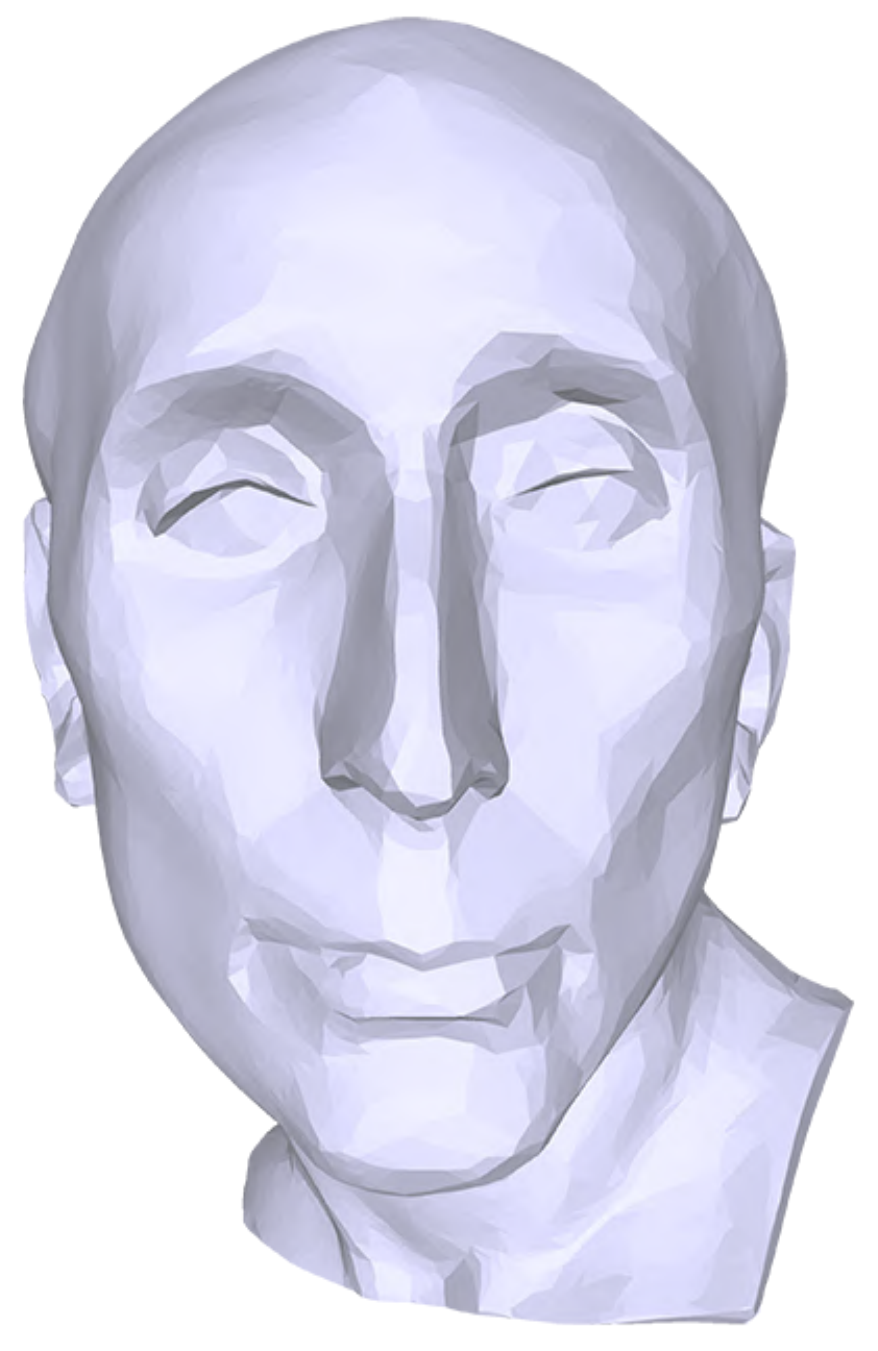}
			\includegraphics[width=2.1cm]{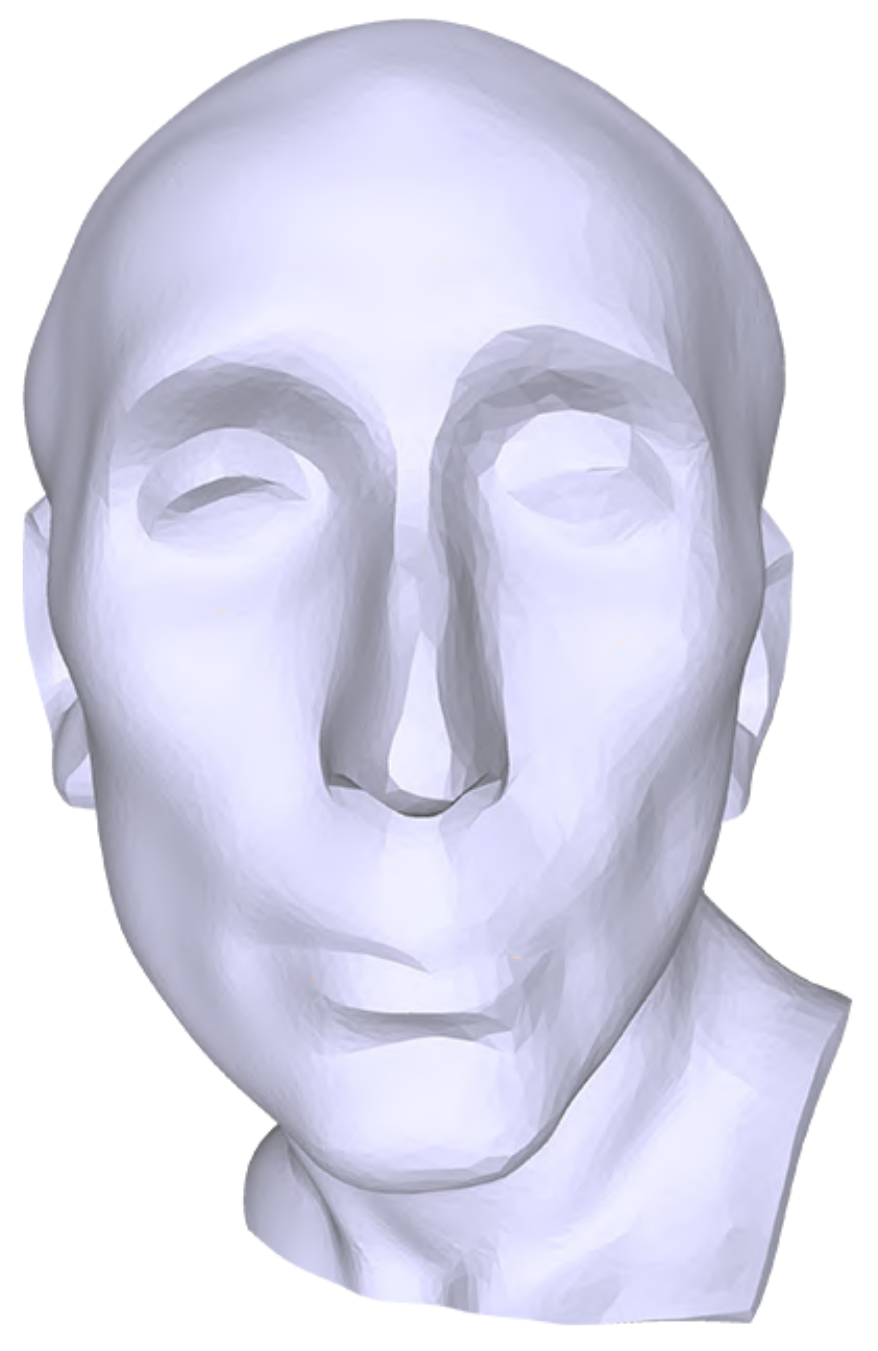}
			\includegraphics[width=2.1cm]{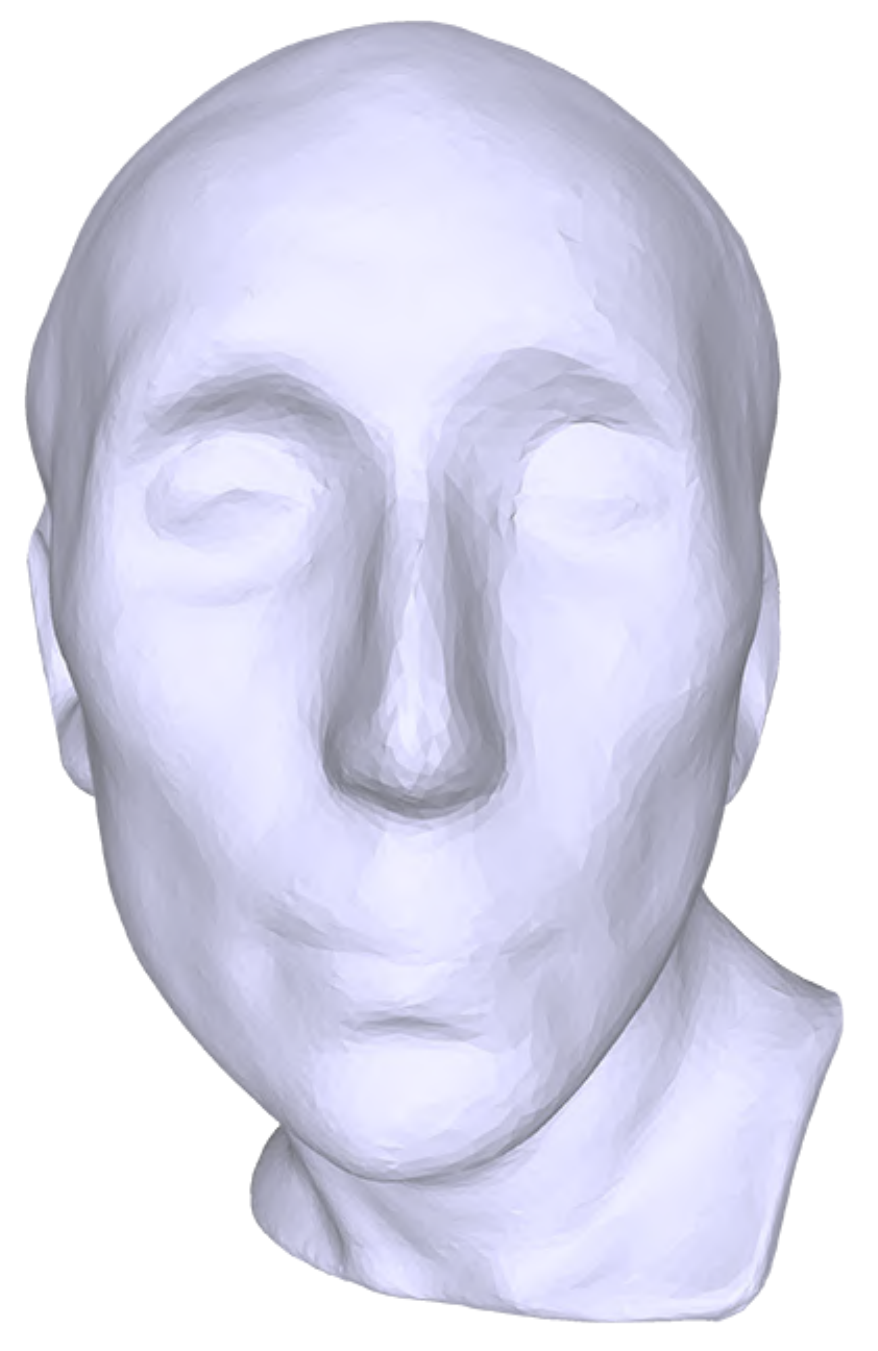}
			\includegraphics[width=2.1cm]{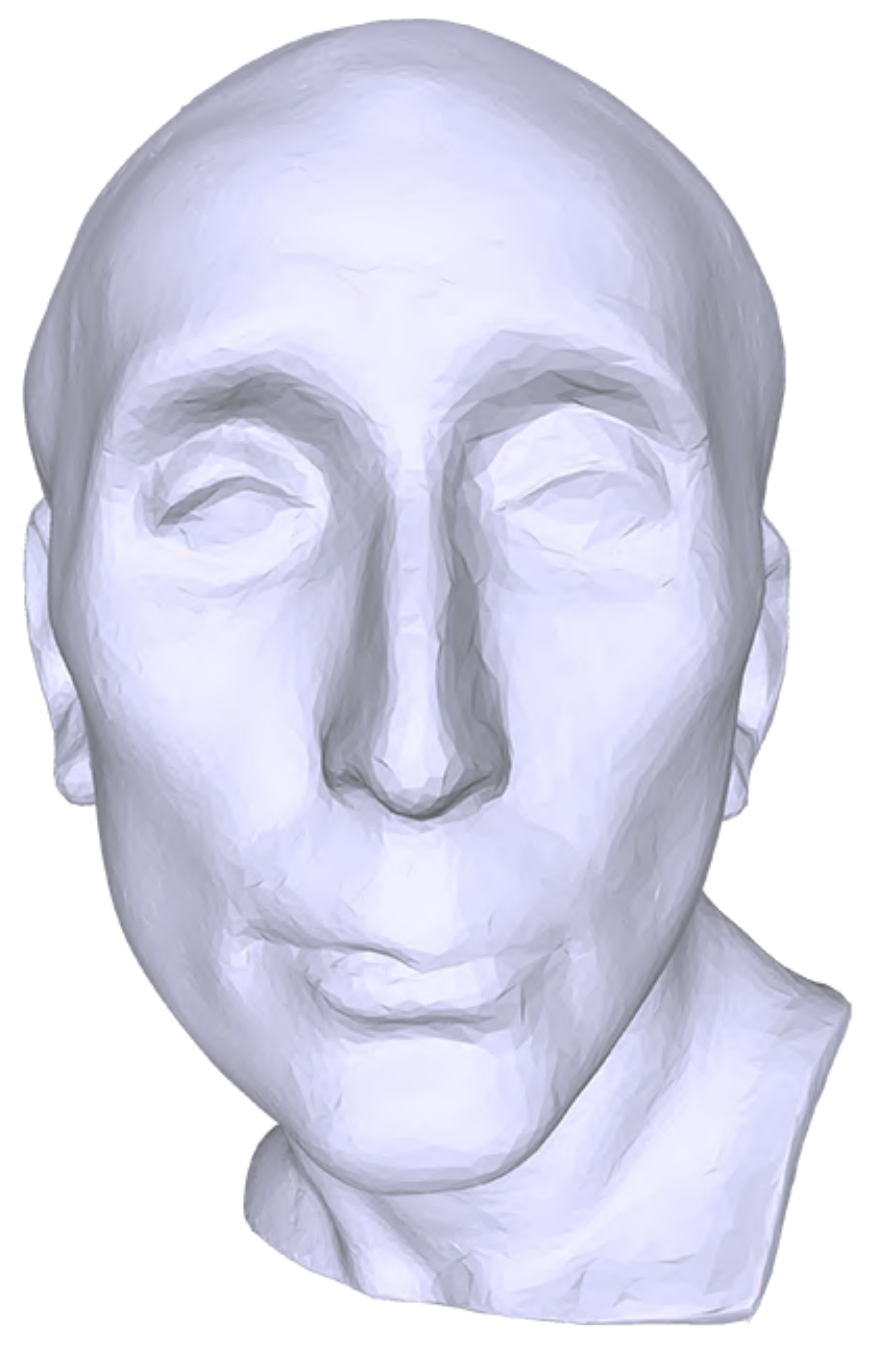}
			\includegraphics[width=2.1cm]{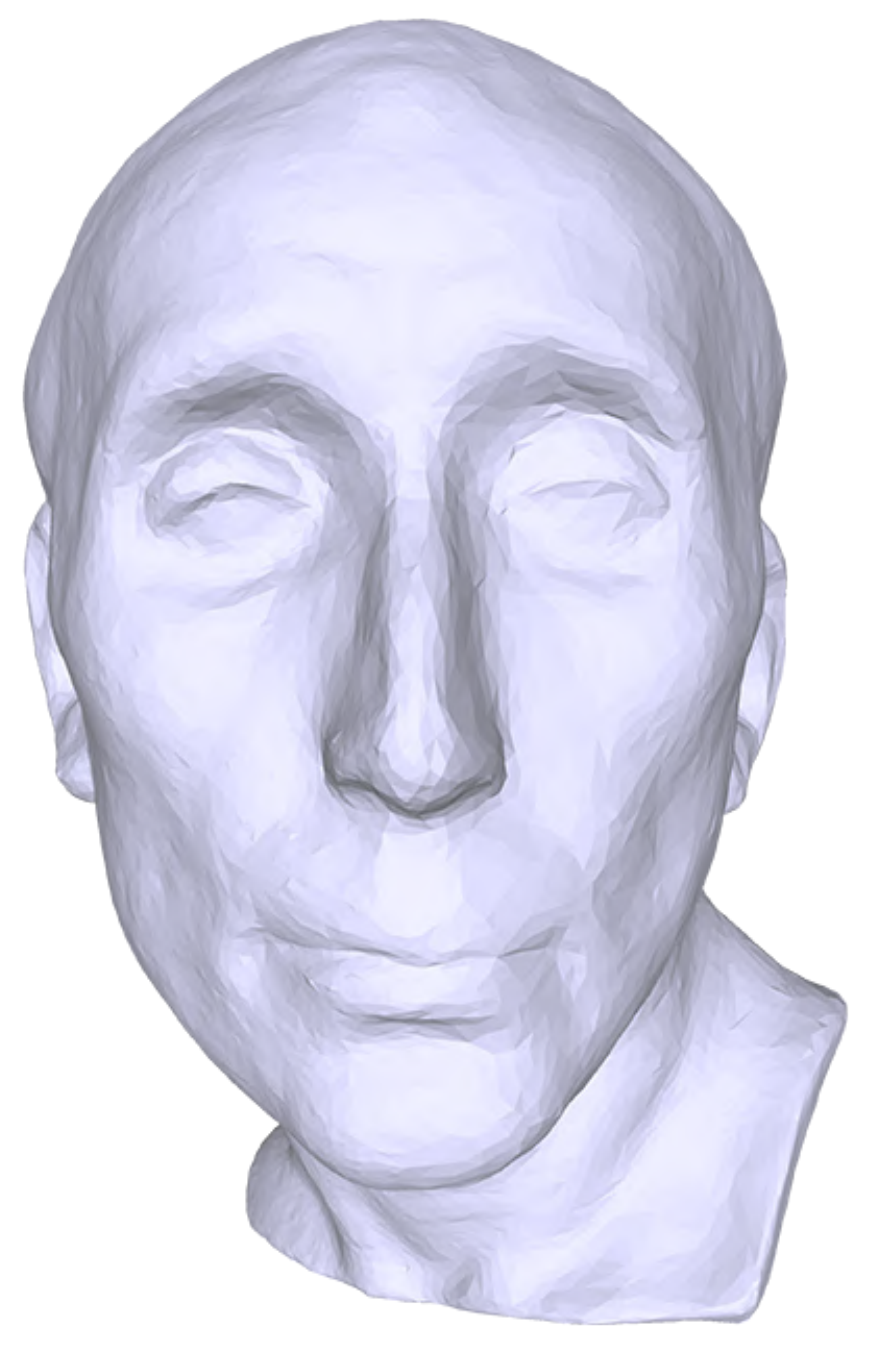}
						
			\includegraphics[width=2.1cm]{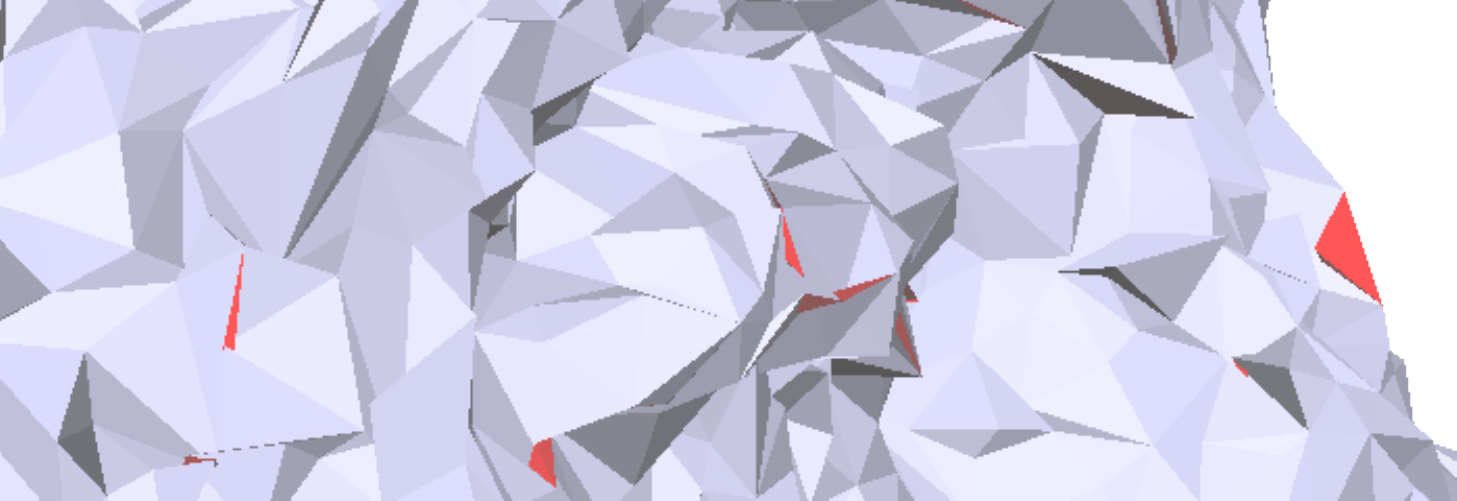}
			\includegraphics[width=2.1cm]{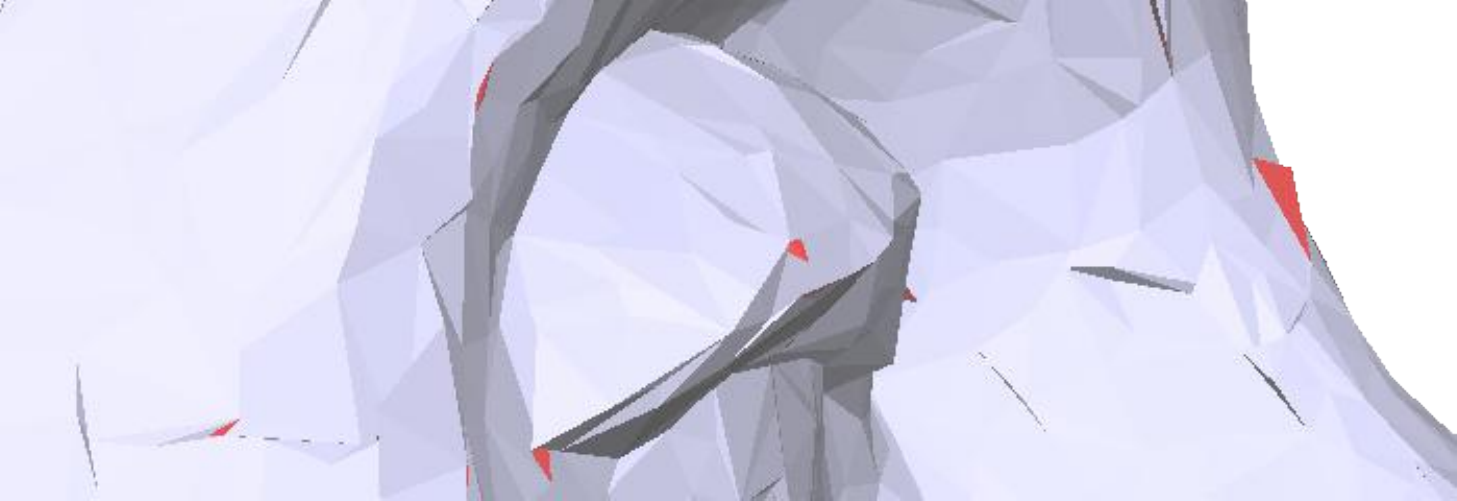}
			\includegraphics[width=2.1cm]{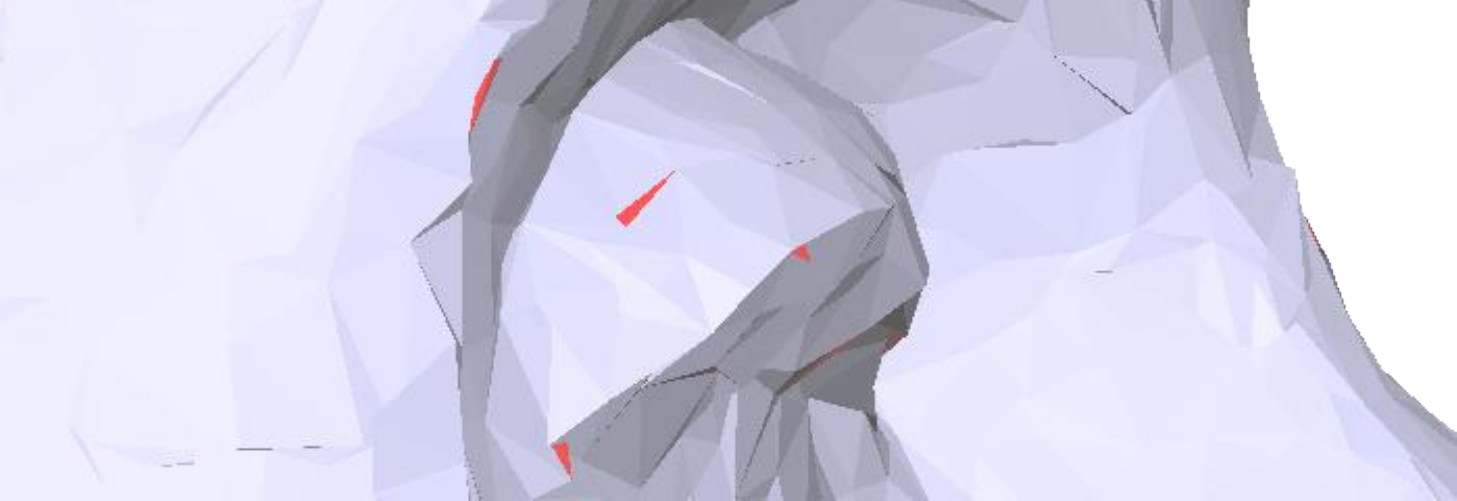}
			\includegraphics[width=2.1cm]{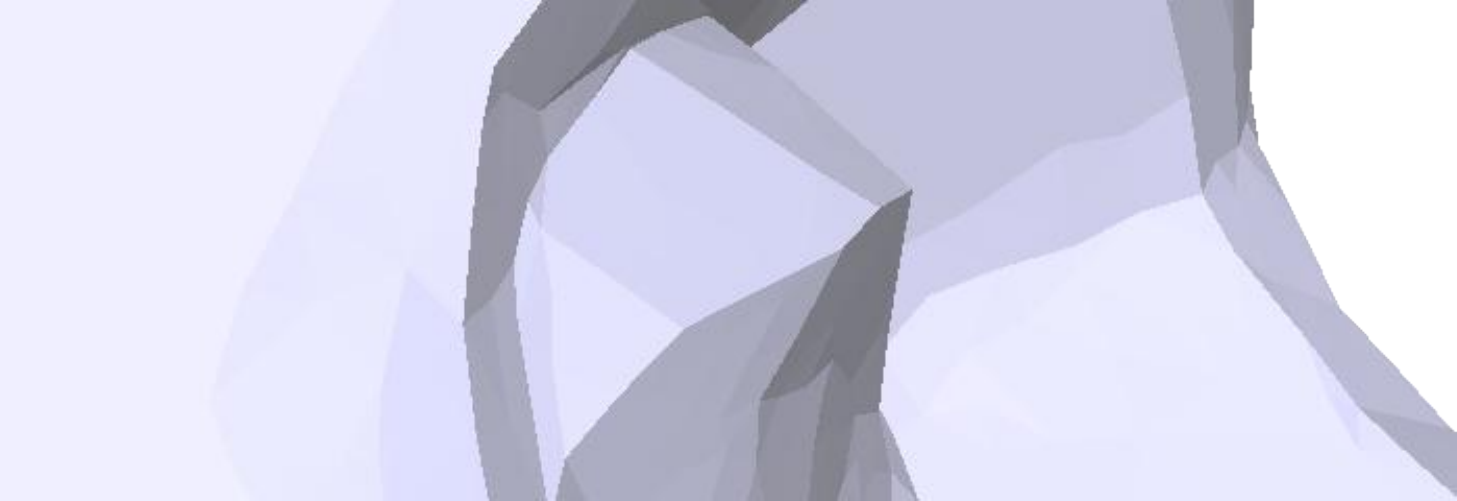}
			\includegraphics[width=2.1cm]{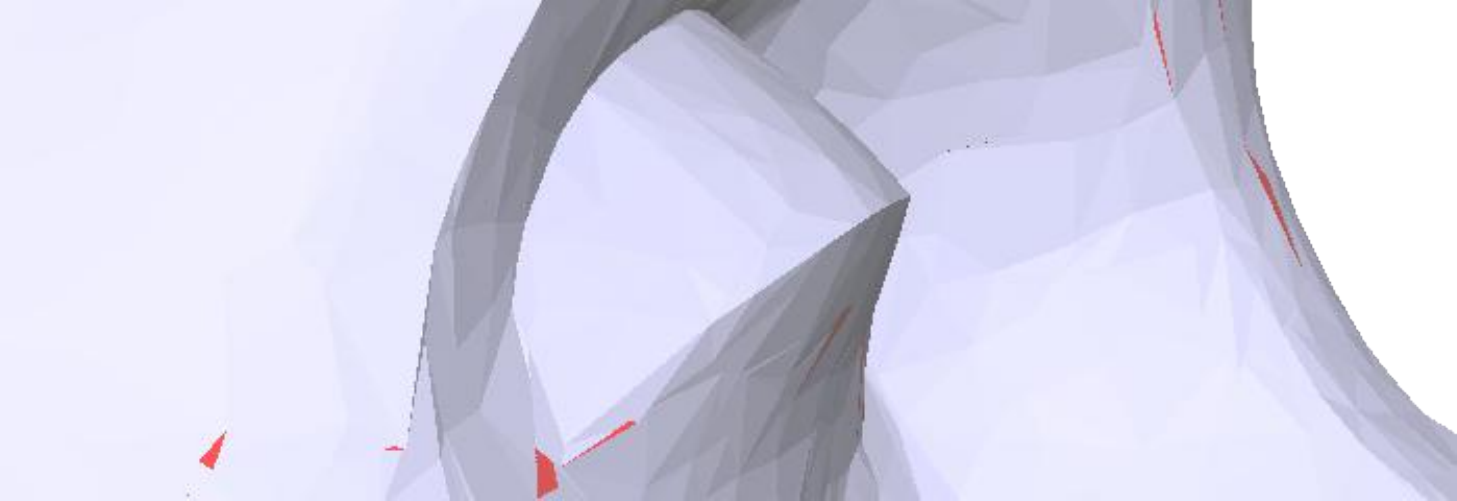}
			\includegraphics[width=2.1cm]{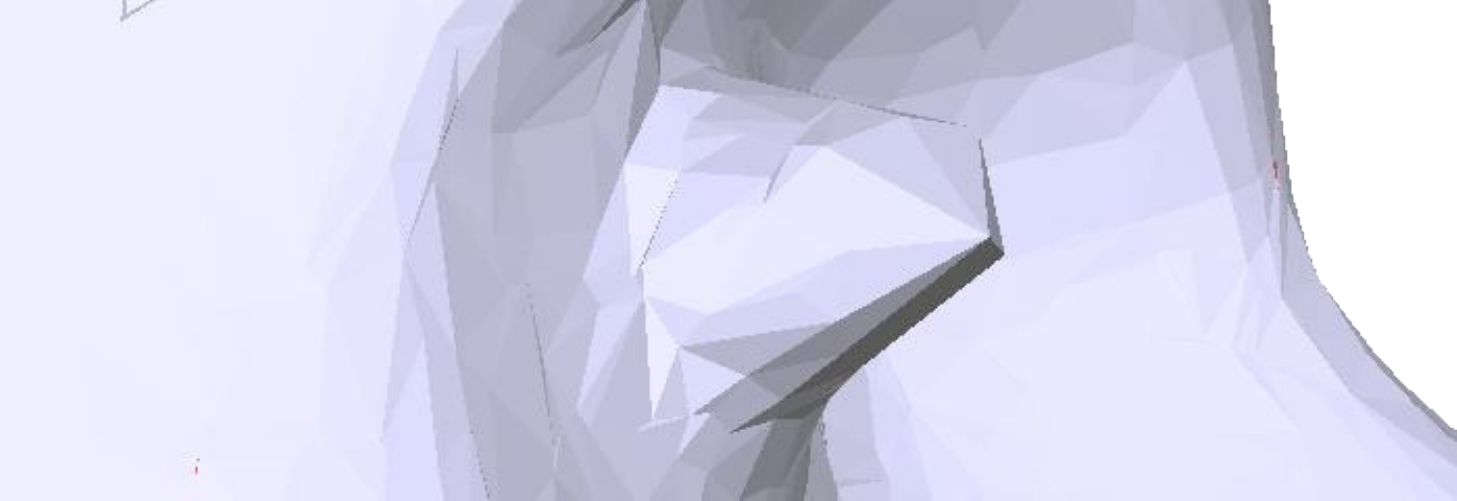}
			\includegraphics[width=2.1cm]{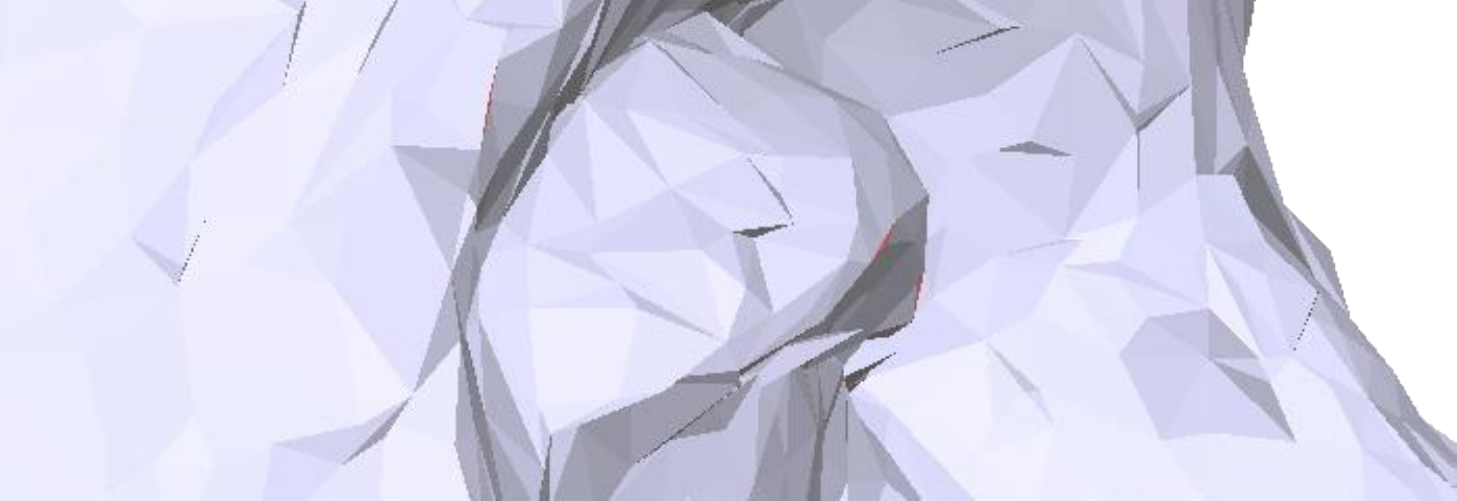}
			\includegraphics[width=2.1cm]{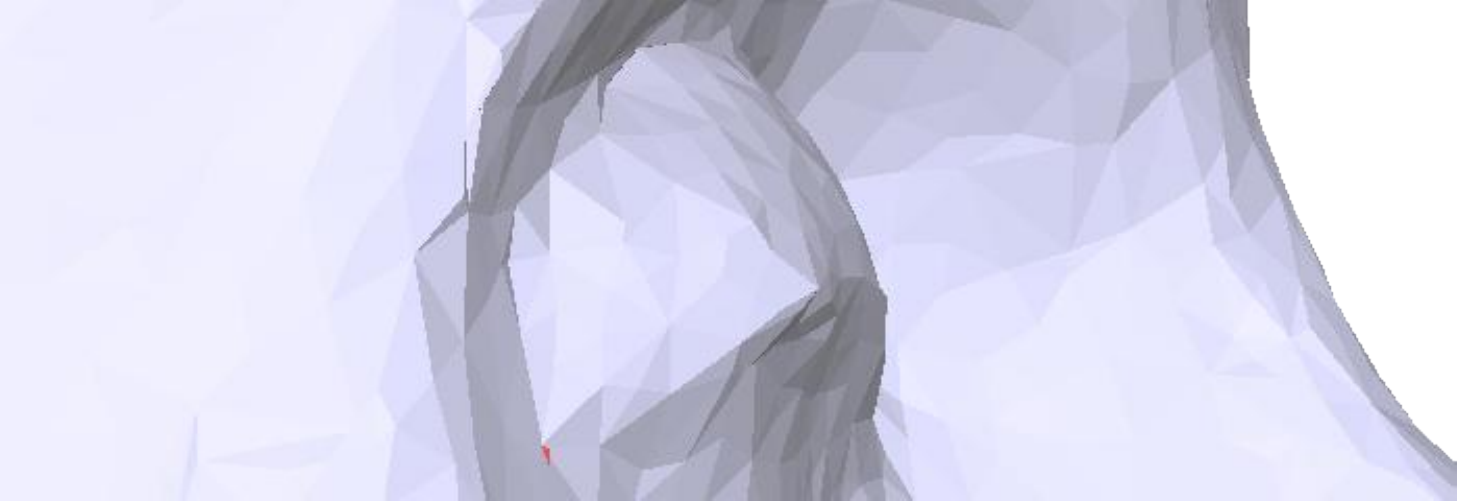}\\		
			\includegraphics[width=2.1cm]{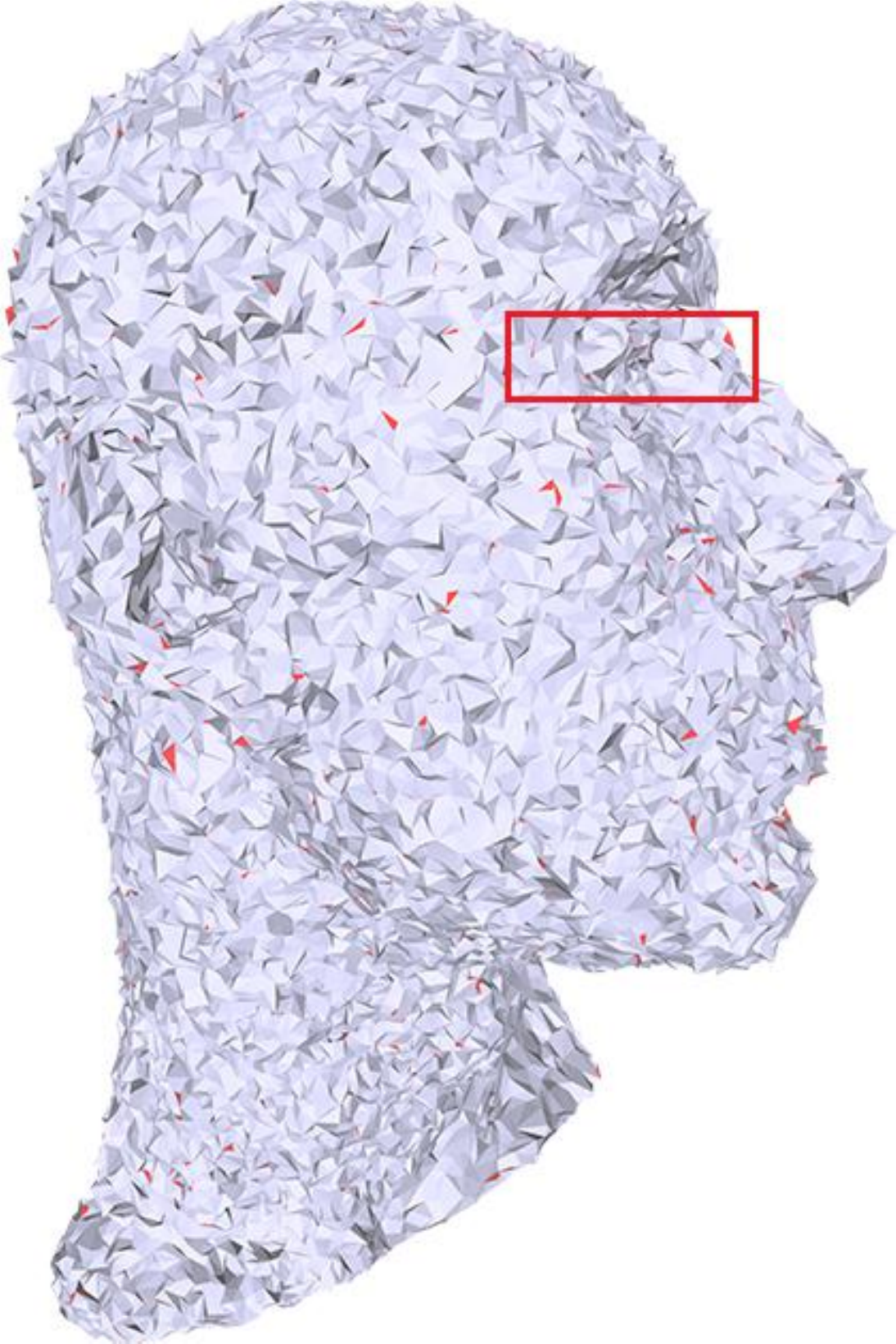}
			\includegraphics[width=2.1cm]{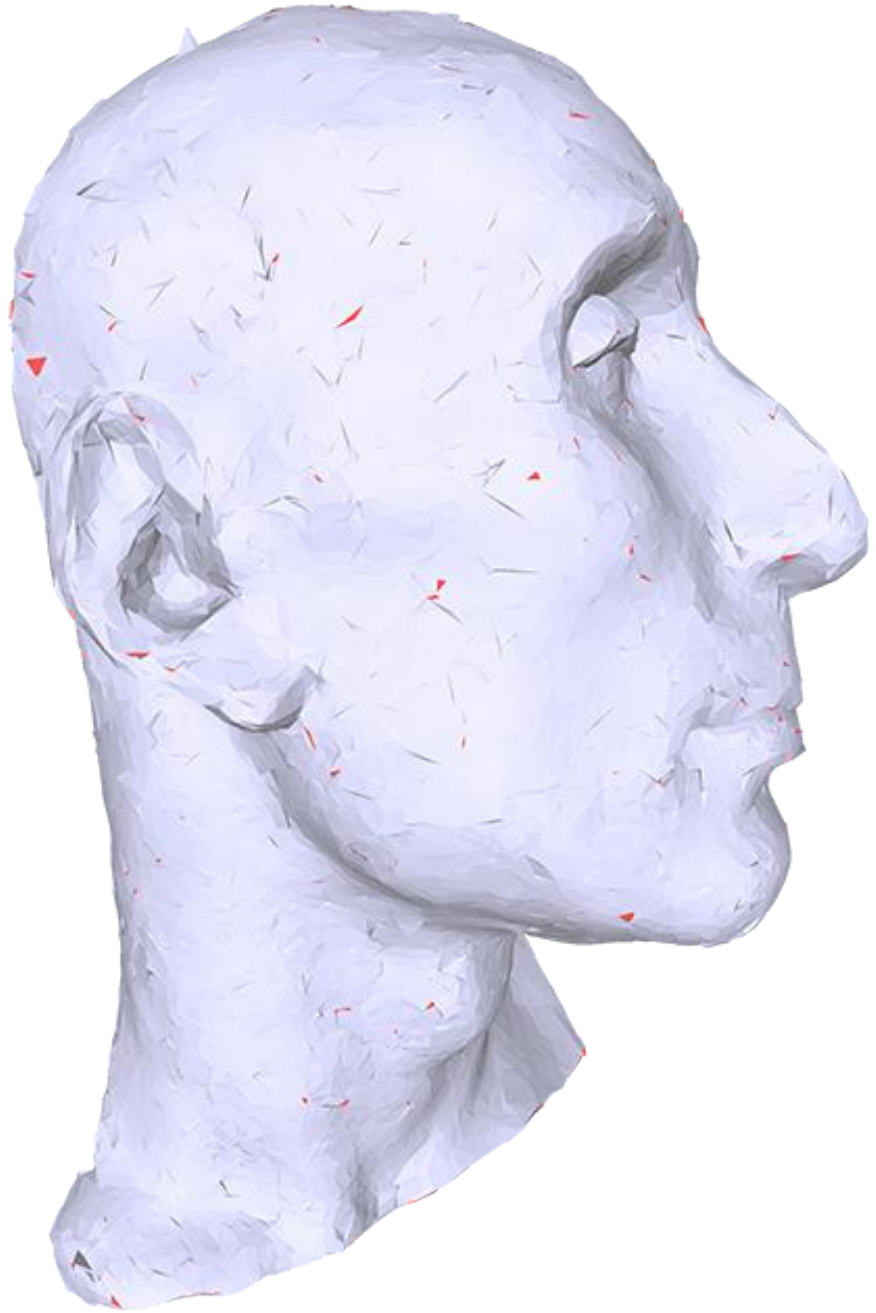}
			\includegraphics[width=2.1cm]{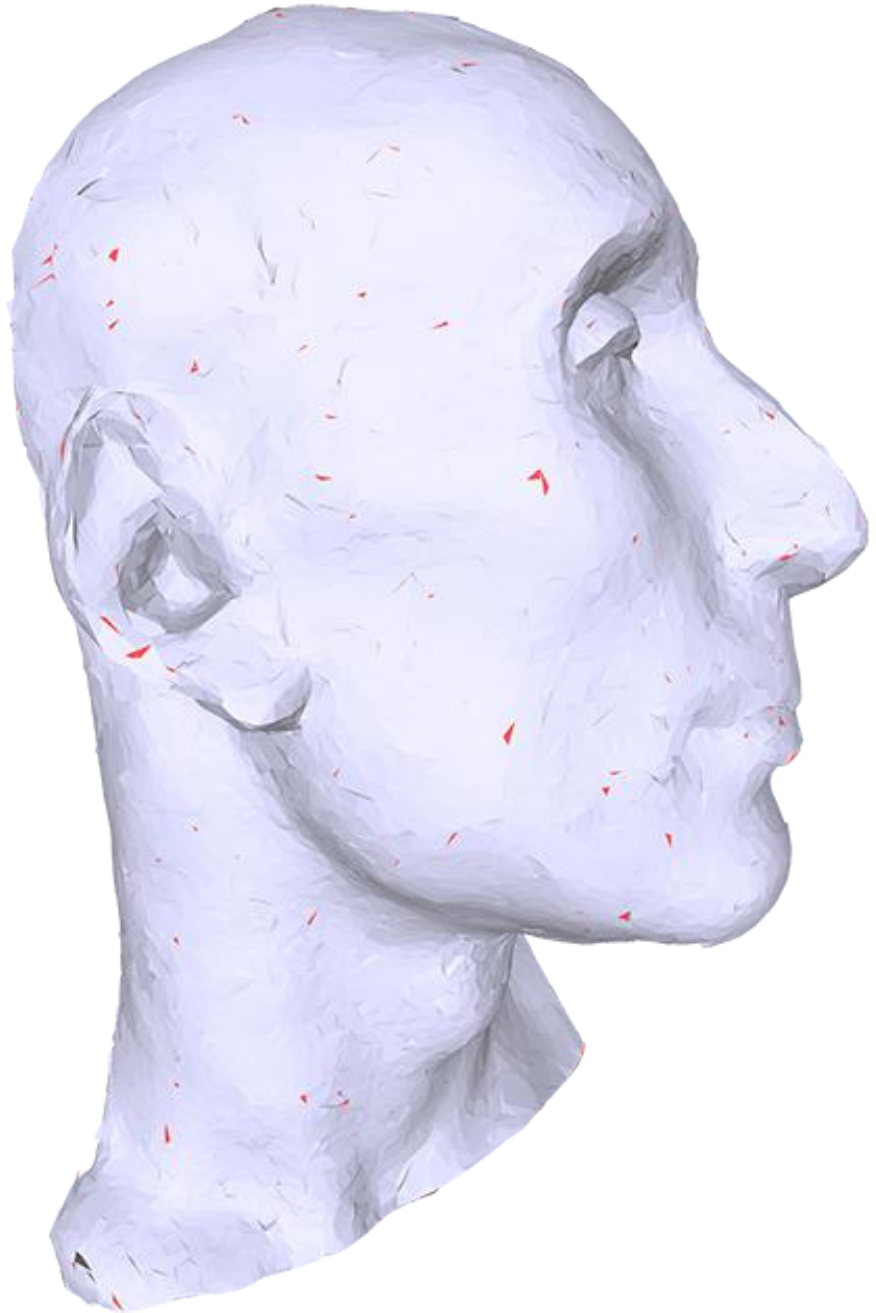}
			\includegraphics[width=2.1cm]{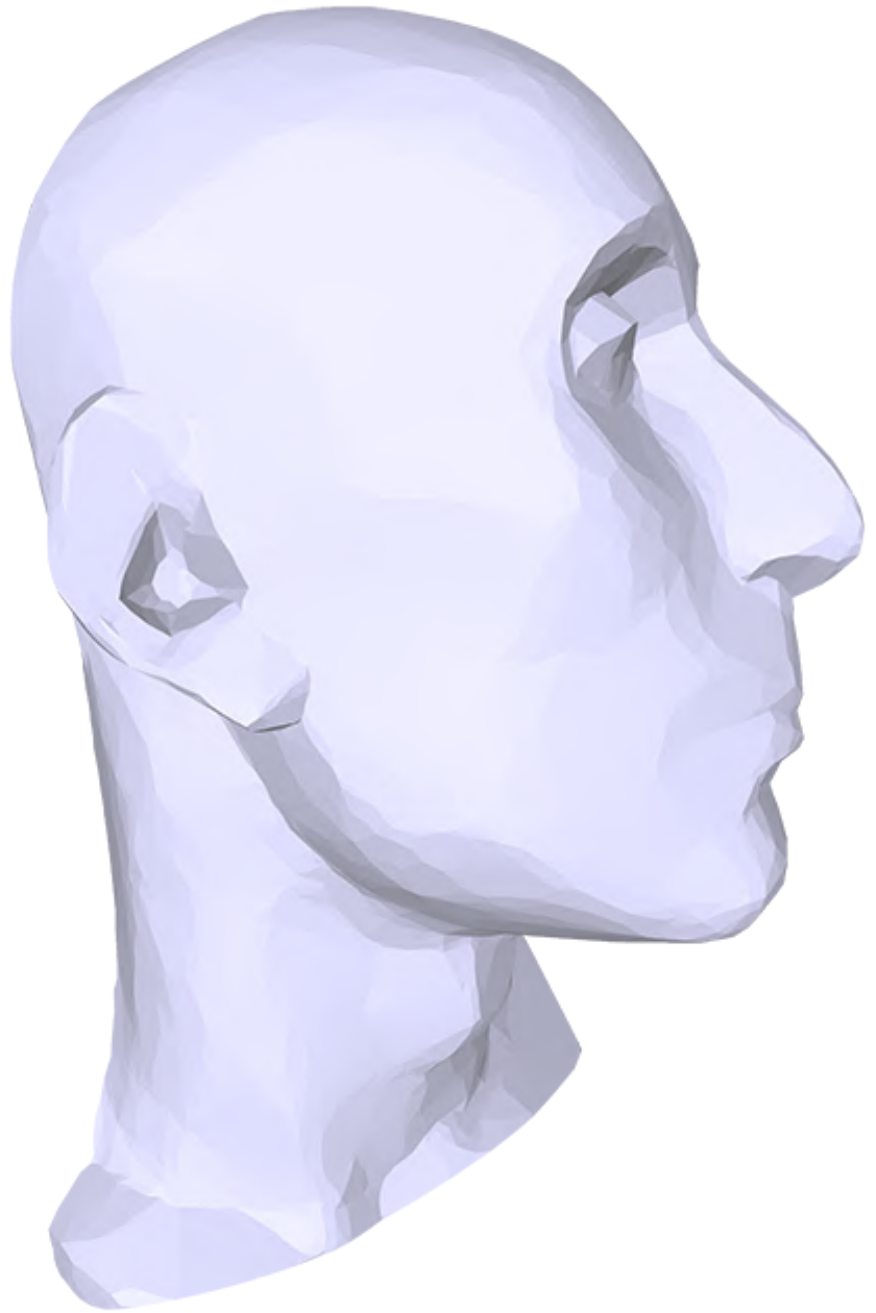}
			\includegraphics[width=2.1cm]{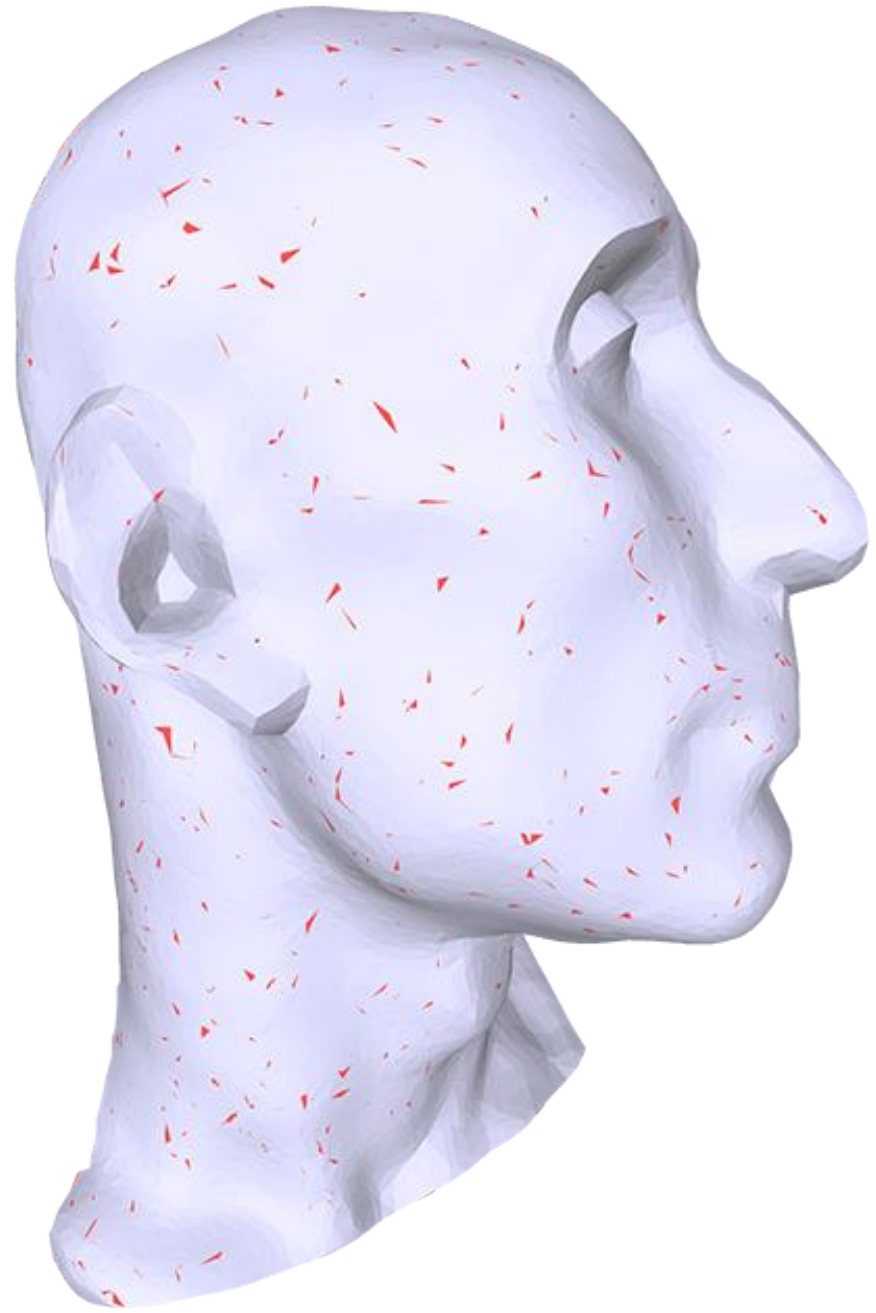}
			\includegraphics[width=2.1cm]{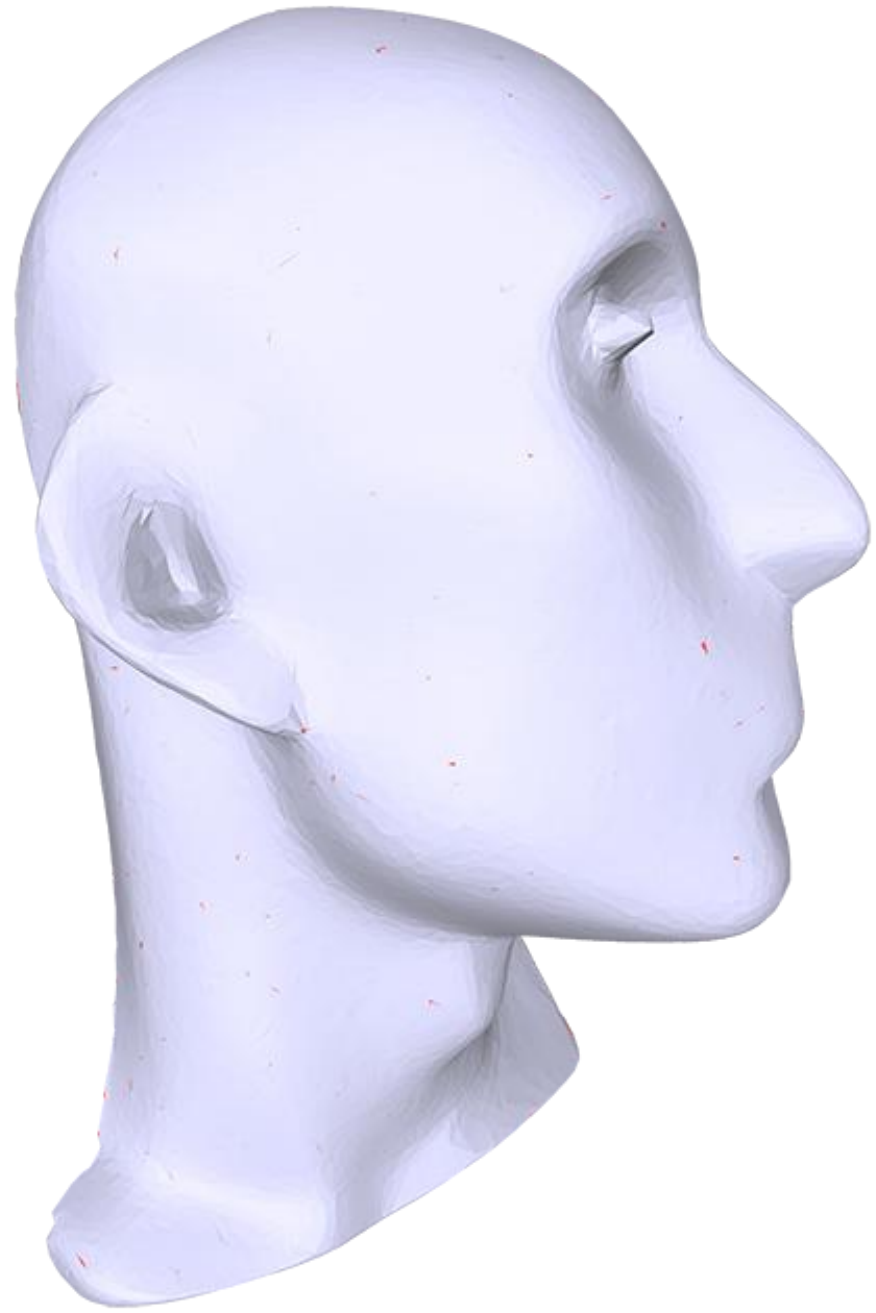}
			\includegraphics[width=2.1cm]{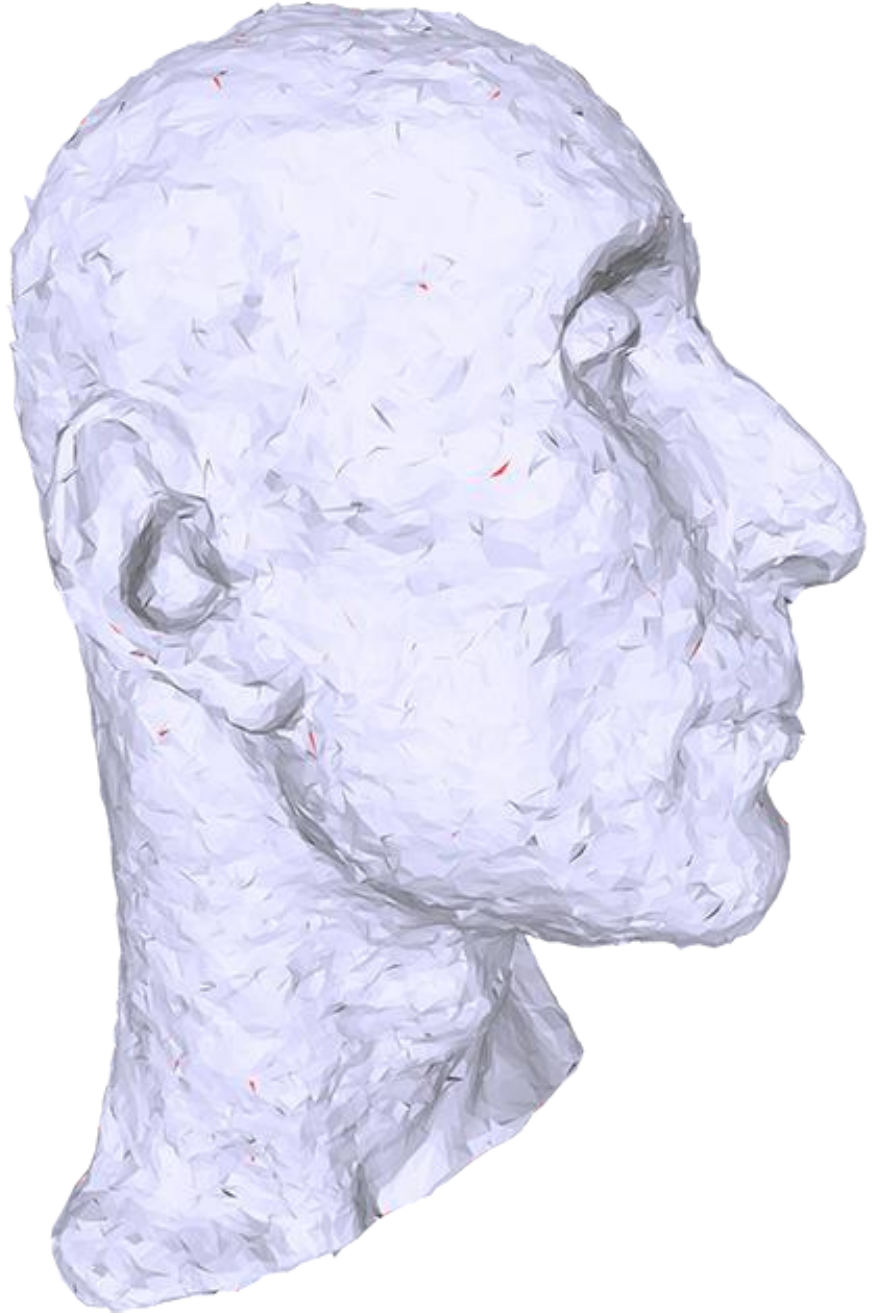}
			\includegraphics[width=2.1cm]{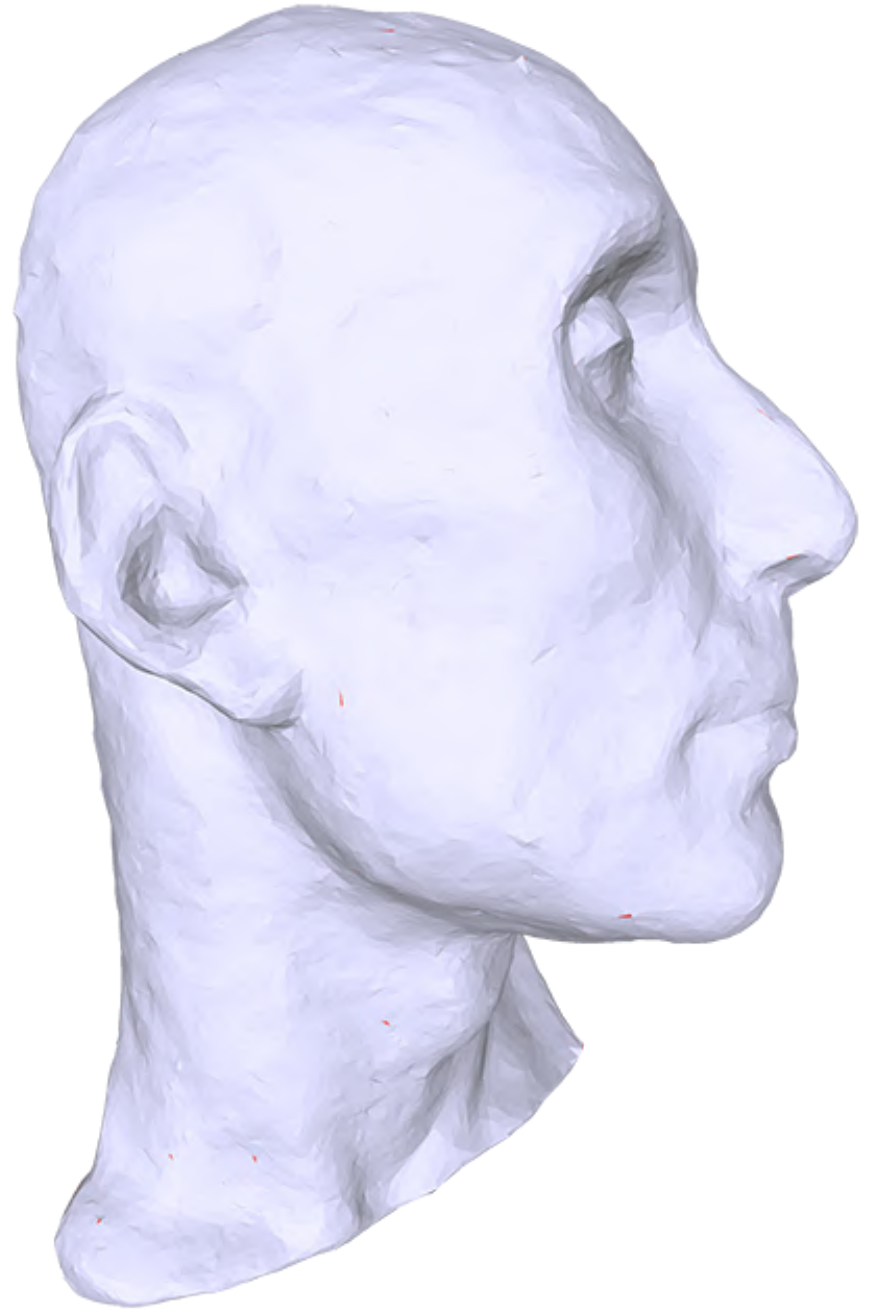}\\
			
			\includegraphics[width=2.1cm]{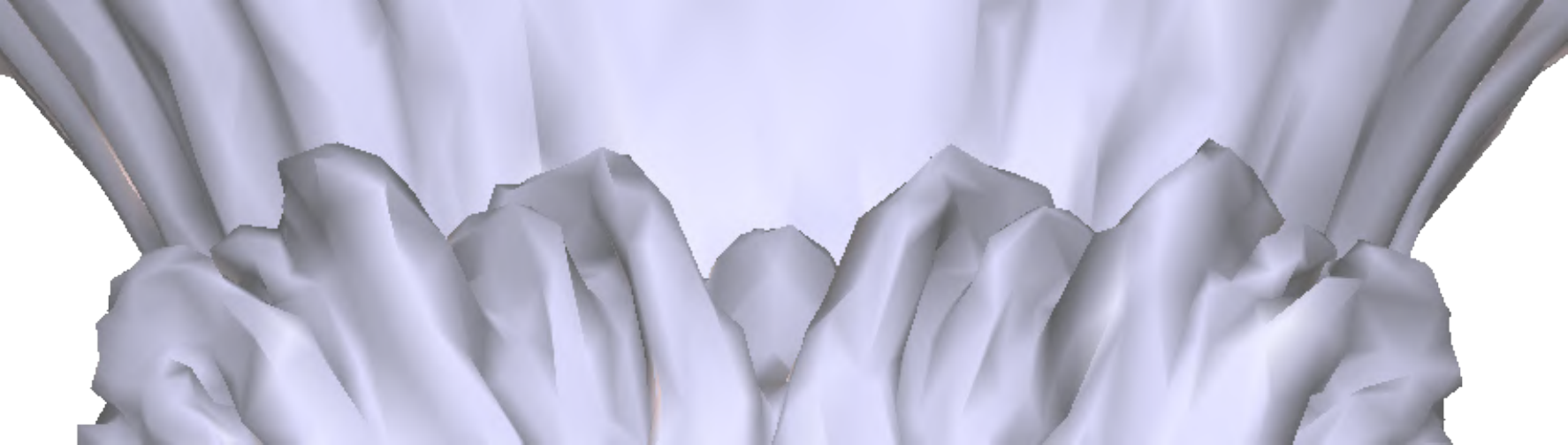}
	\includegraphics[width=2.1cm]{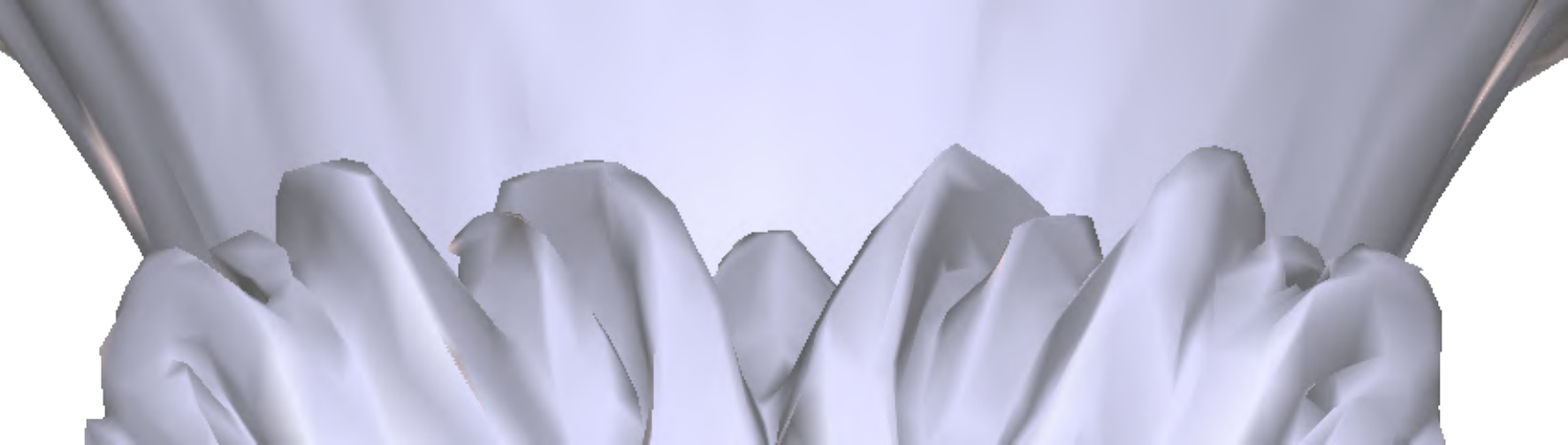}
	\includegraphics[width=2.1cm]{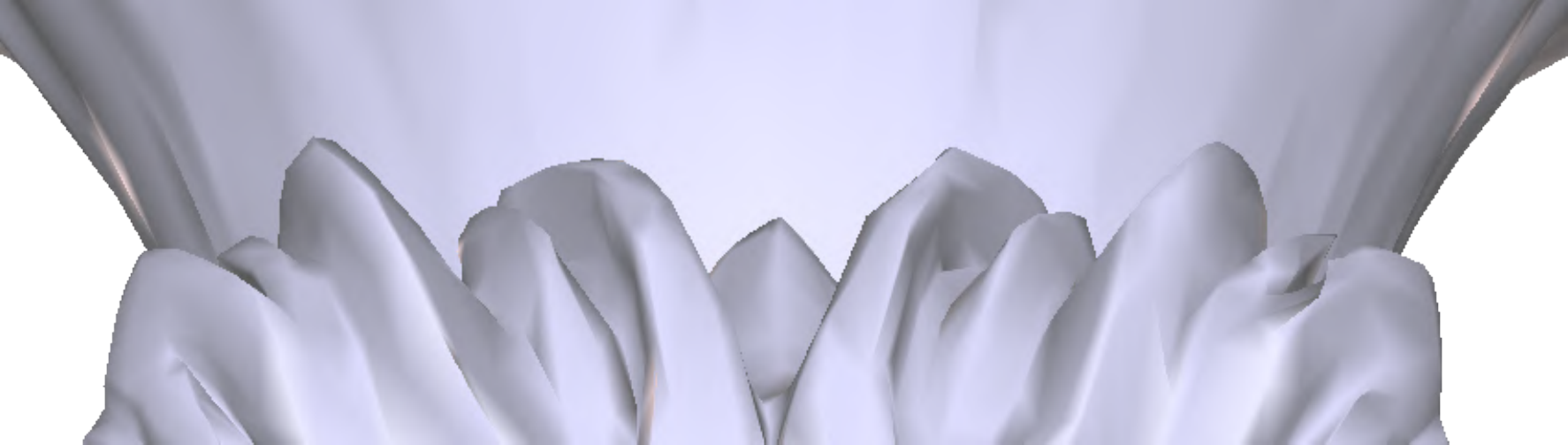}
	\includegraphics[width=2.1cm]{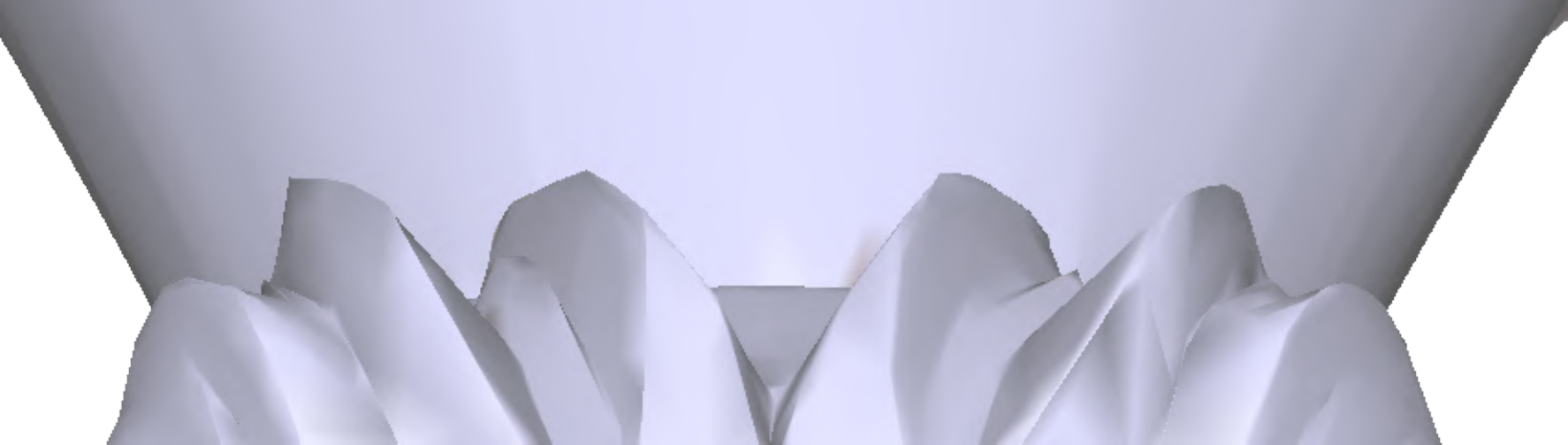}
	\includegraphics[width=2.1cm]{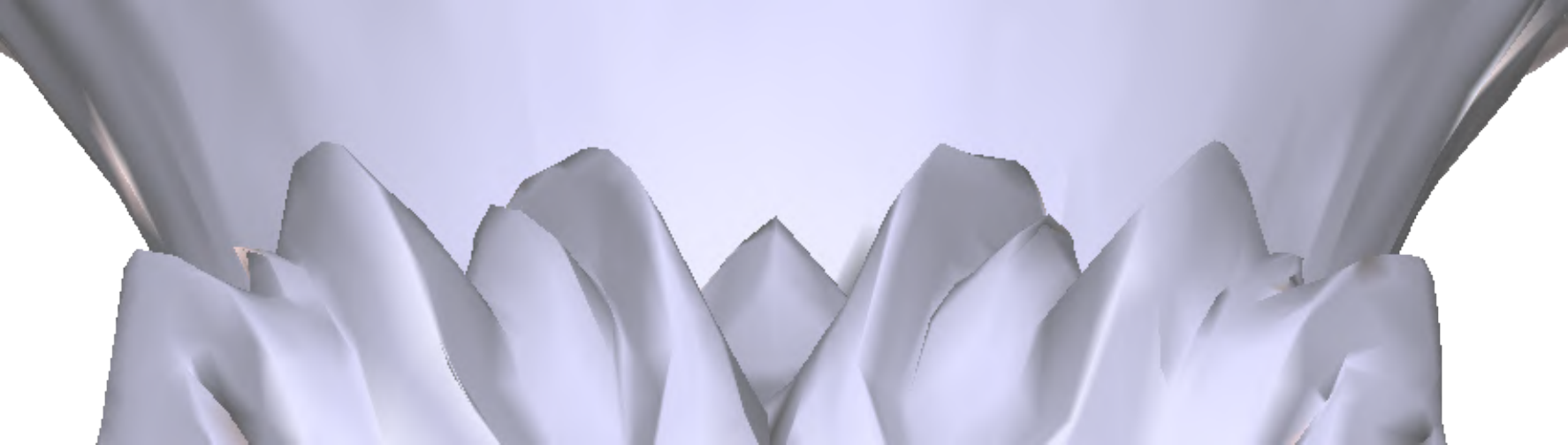}
	\includegraphics[width=2.1cm]{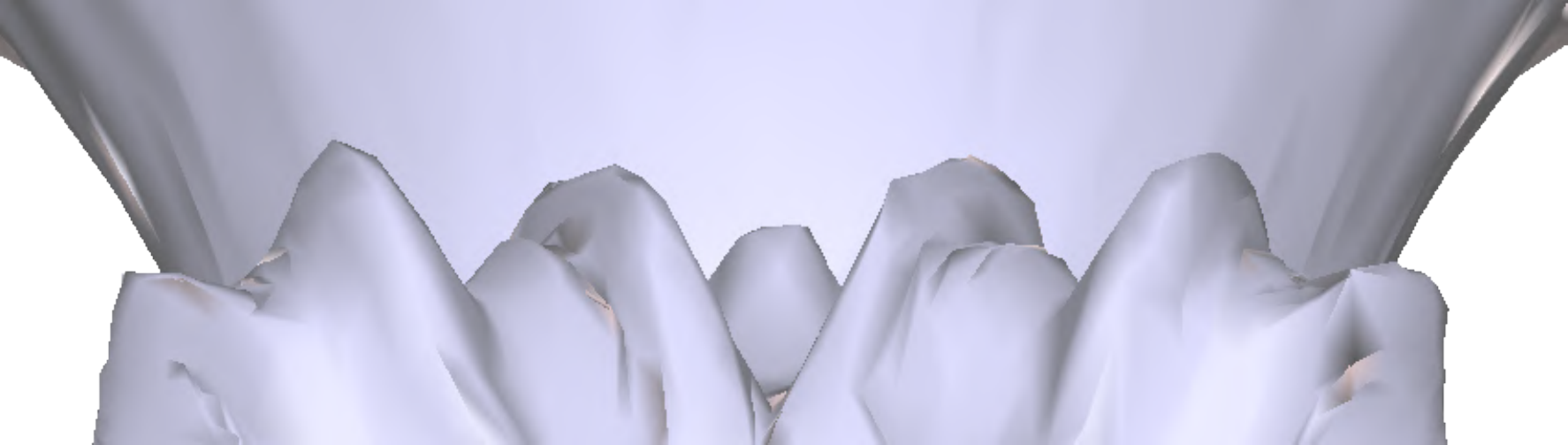}
	\includegraphics[width=2.1cm]{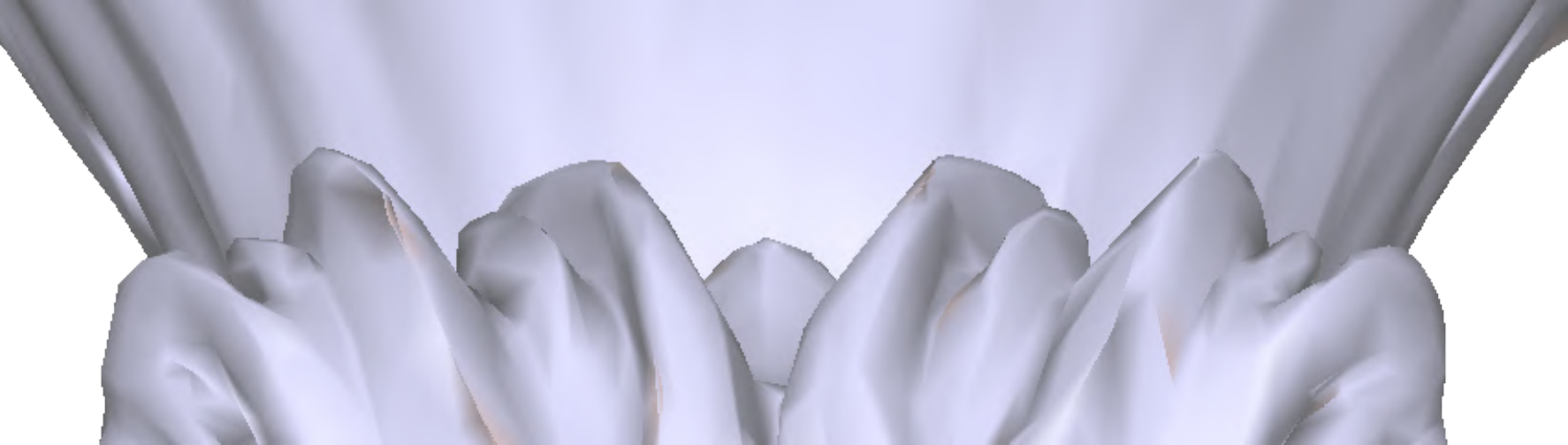}
		\includegraphics[width=2.1cm]{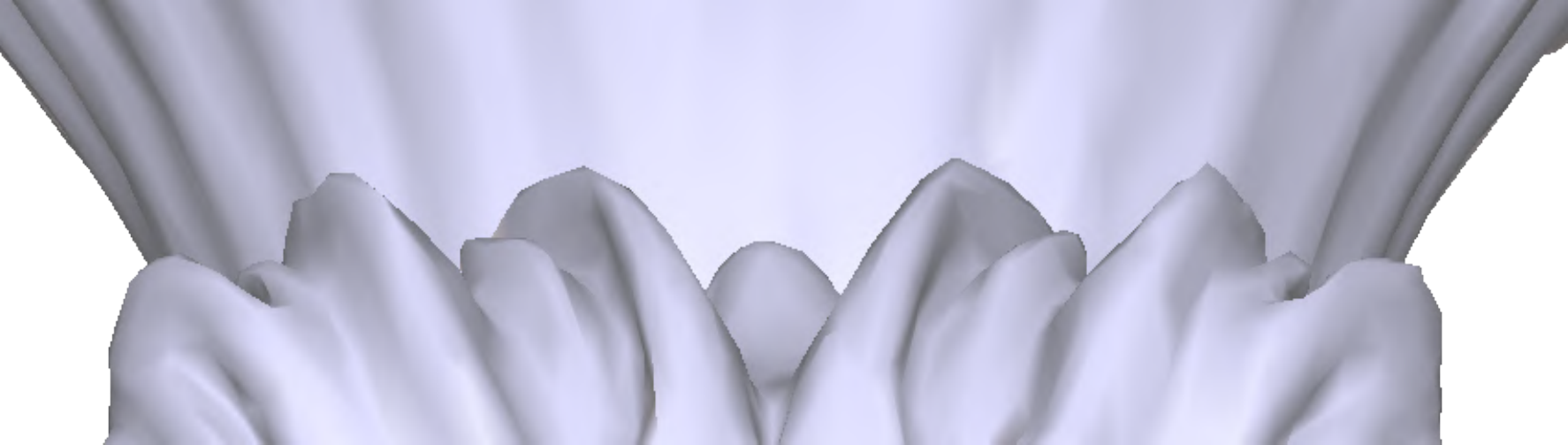}\\	
			\includegraphics[width=2.1cm]{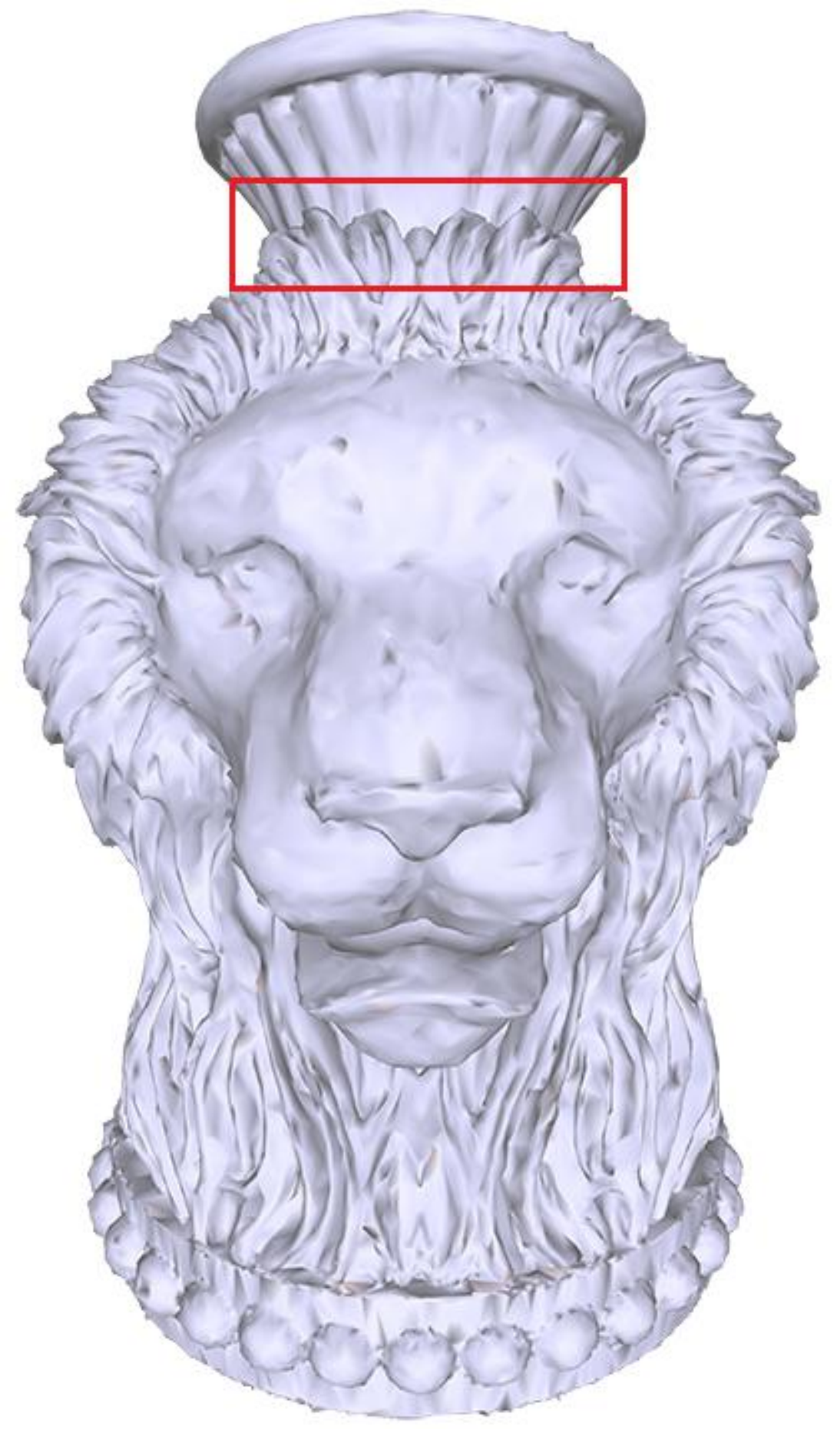}
			\includegraphics[width=2.1cm]{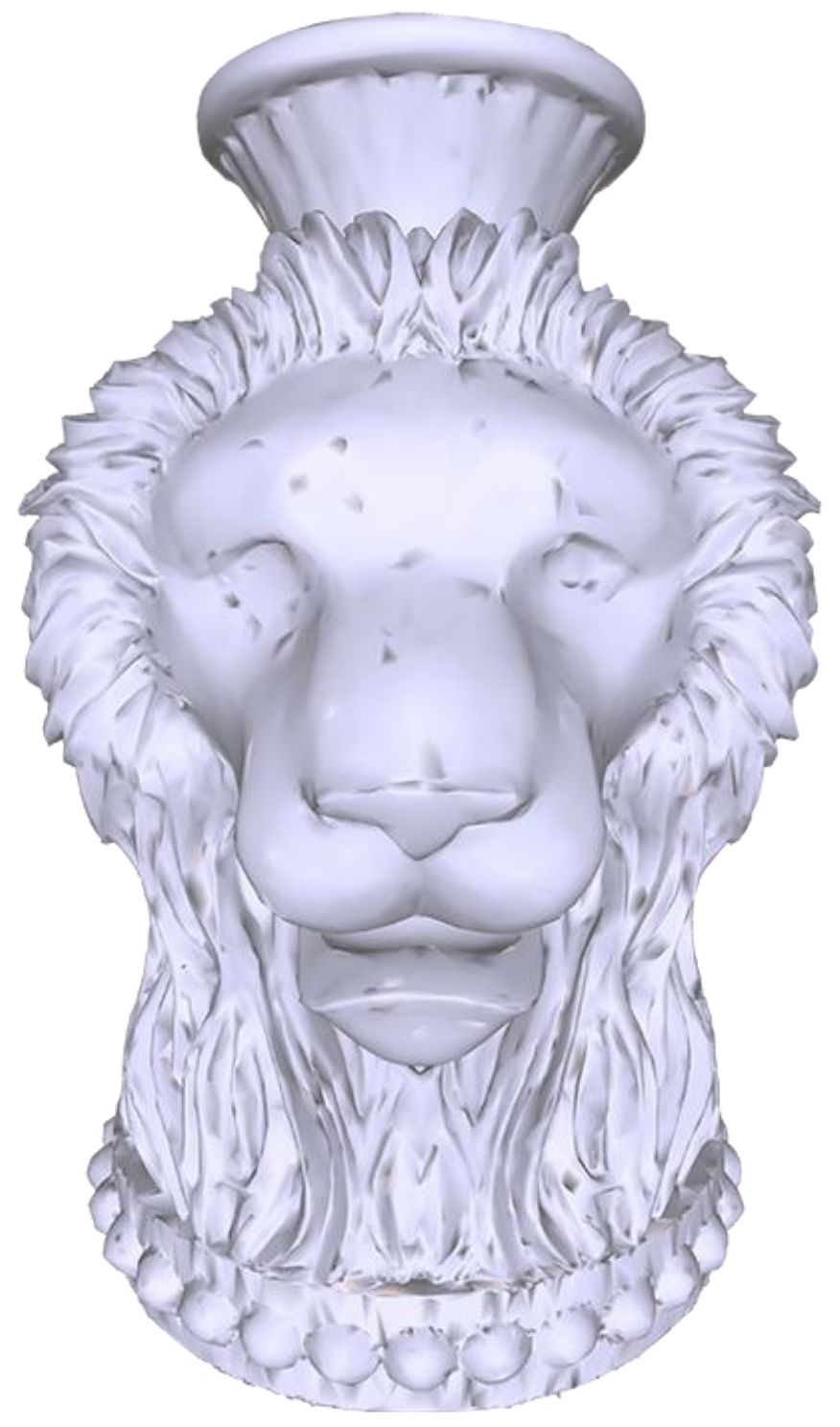}
			\includegraphics[width=2.1cm]{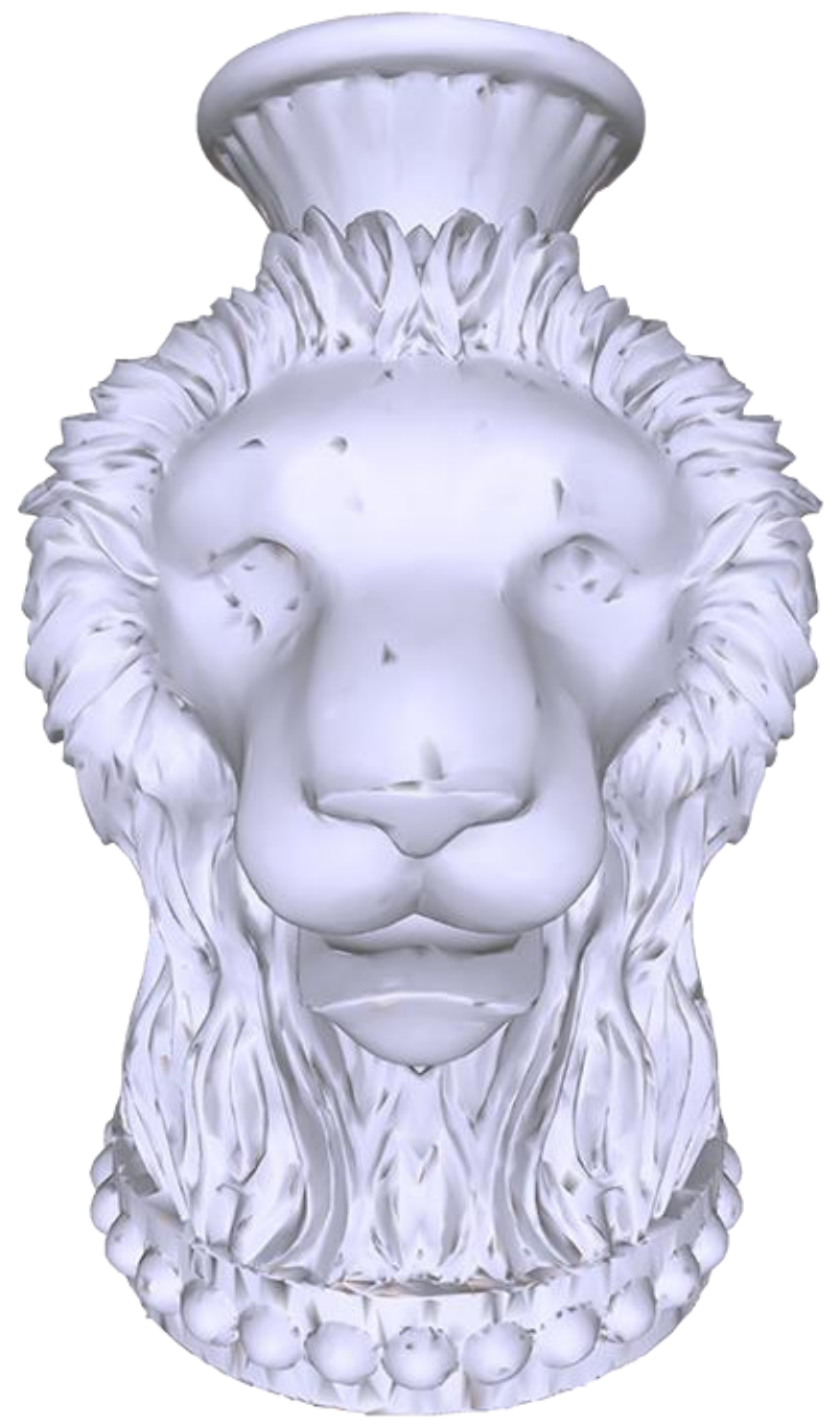}
			\includegraphics[width=2.1cm]{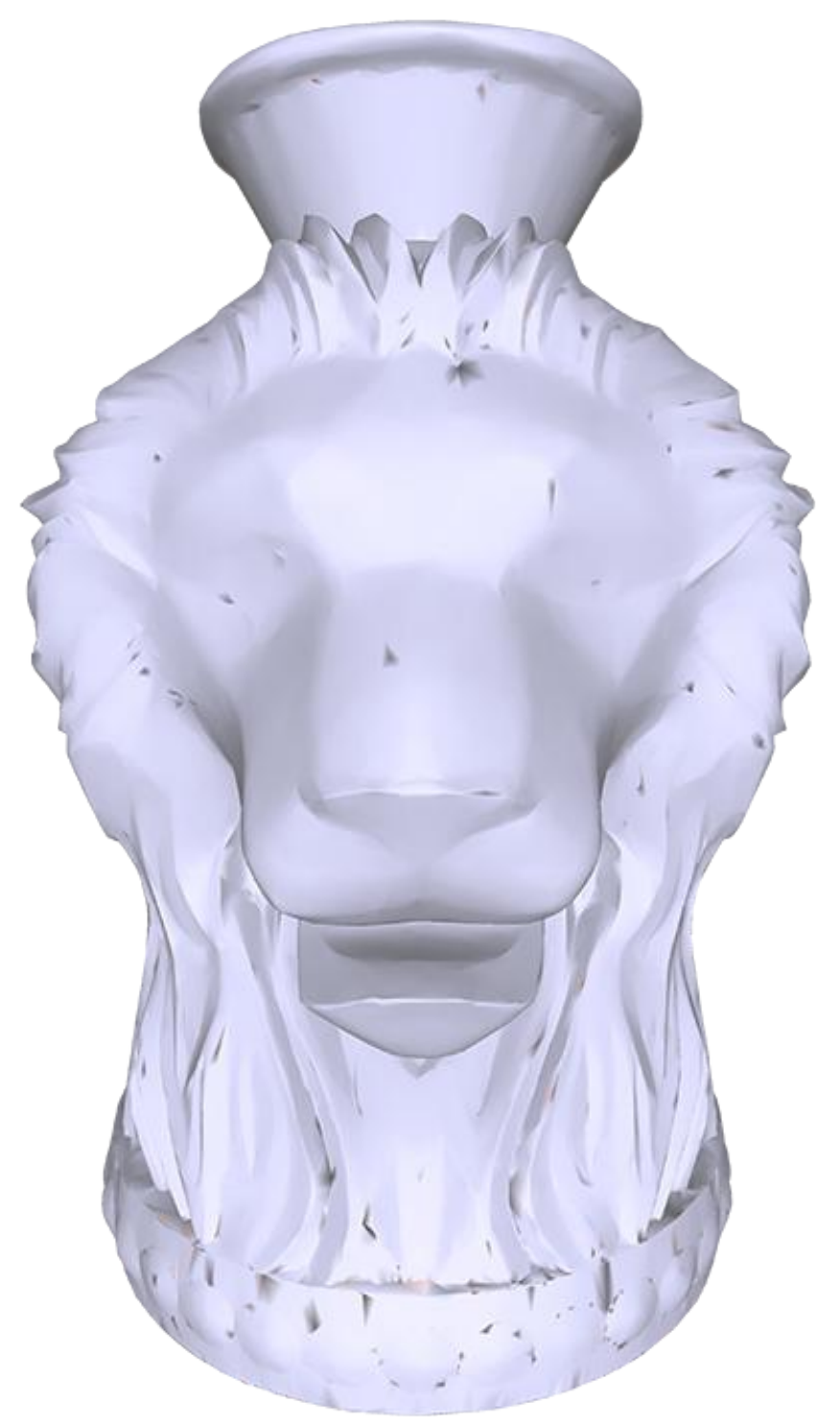}
			\includegraphics[width=2.1cm]{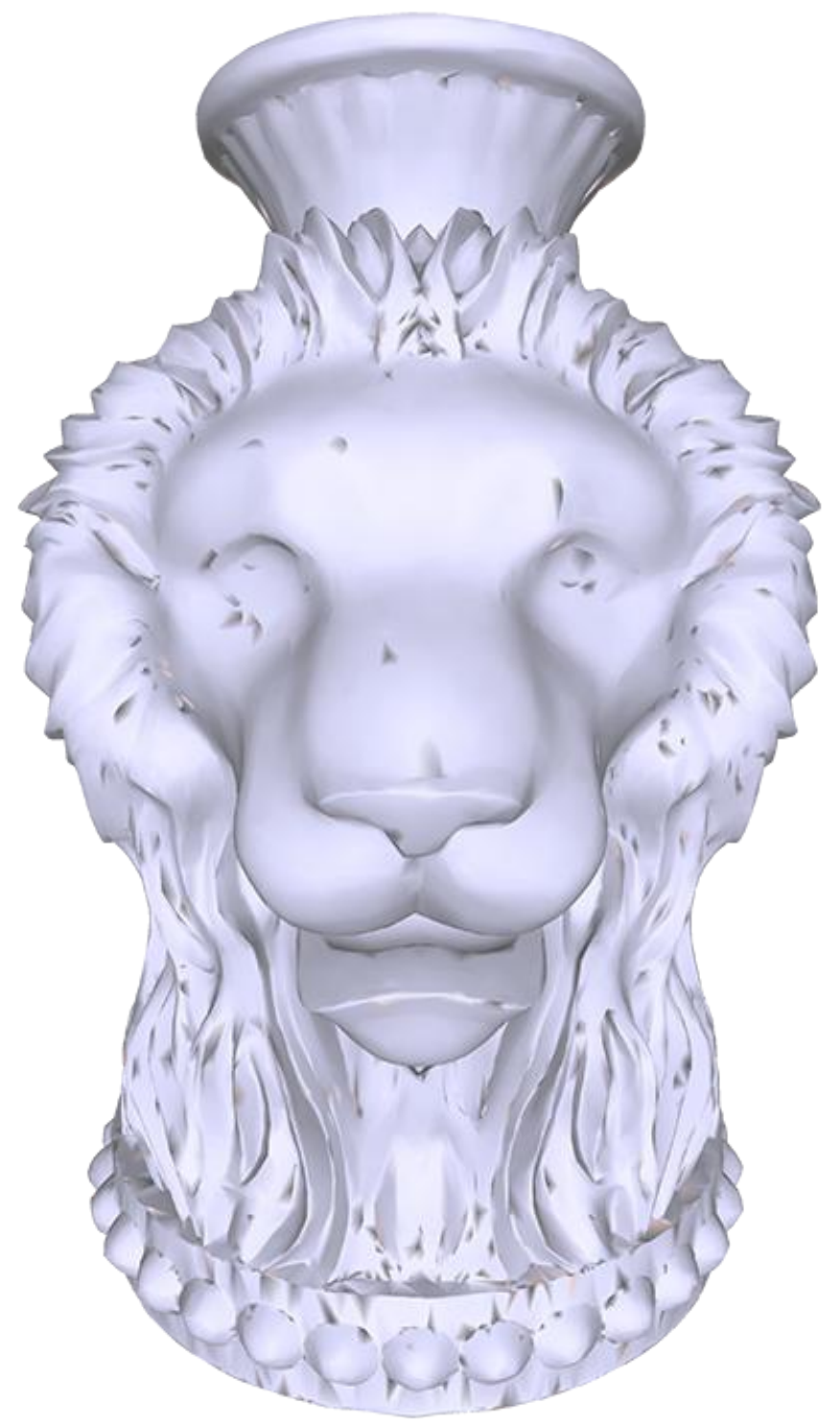}
			\includegraphics[width=2.1cm]{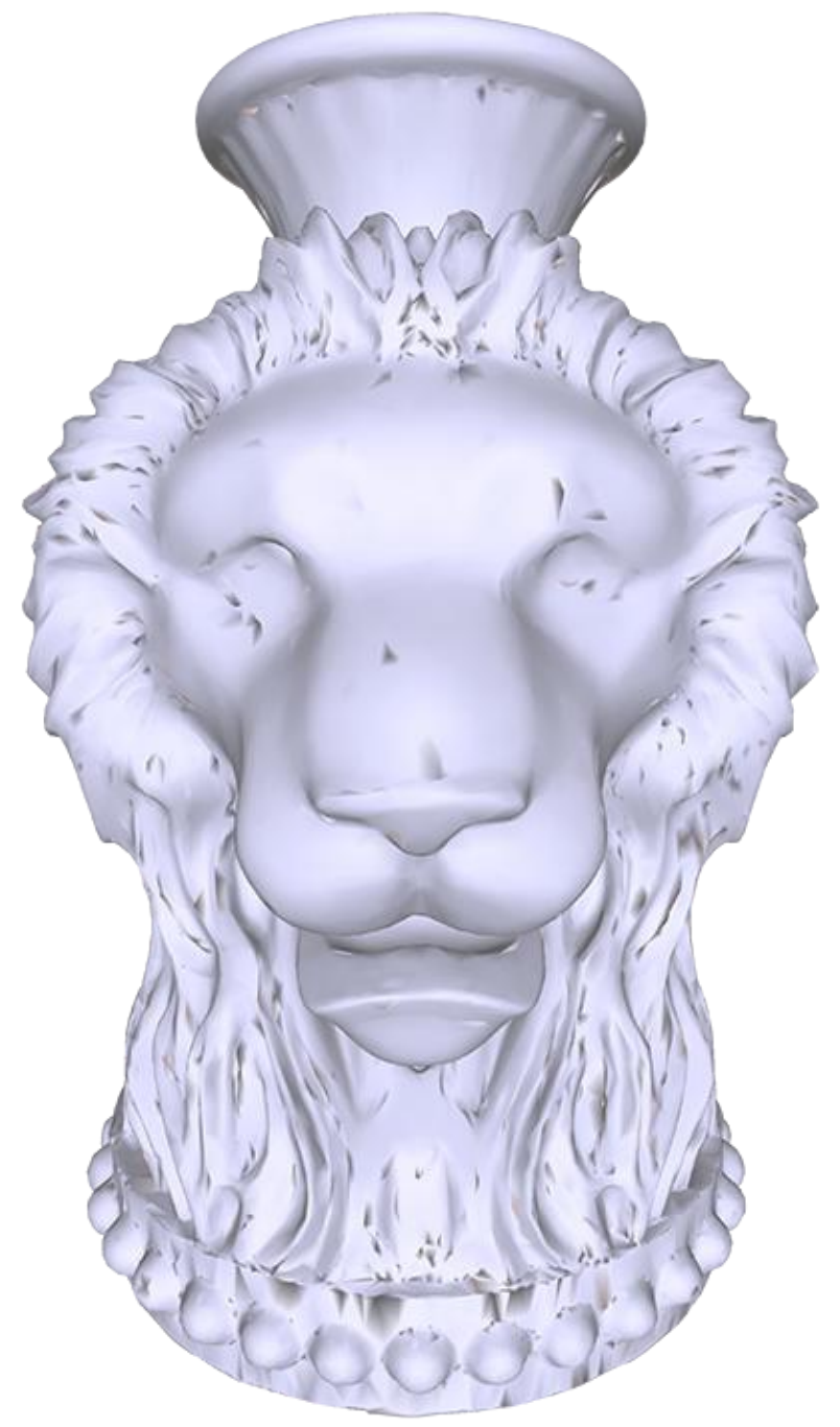}
			\includegraphics[width=2.1cm]{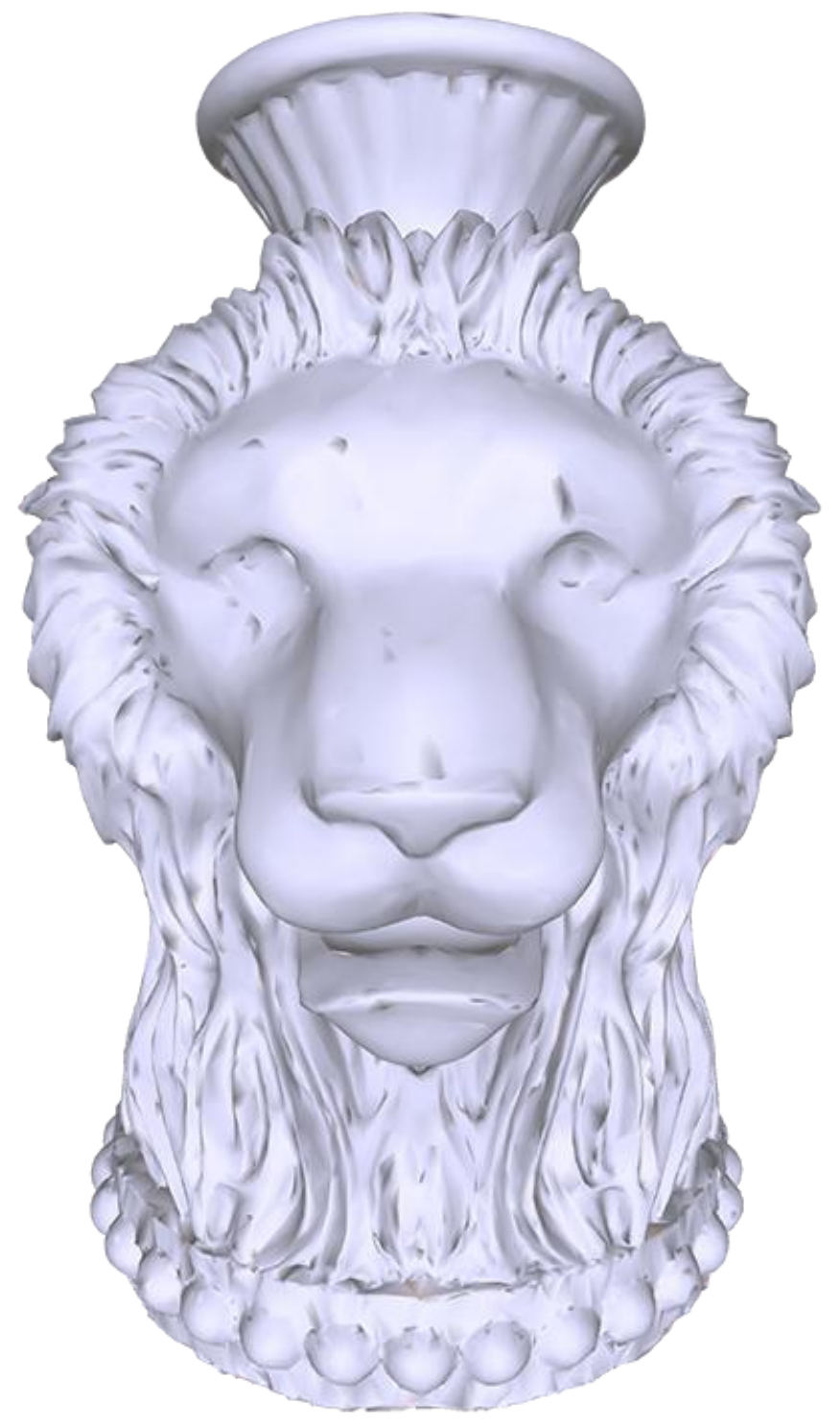}
			\includegraphics[width=2.1cm]{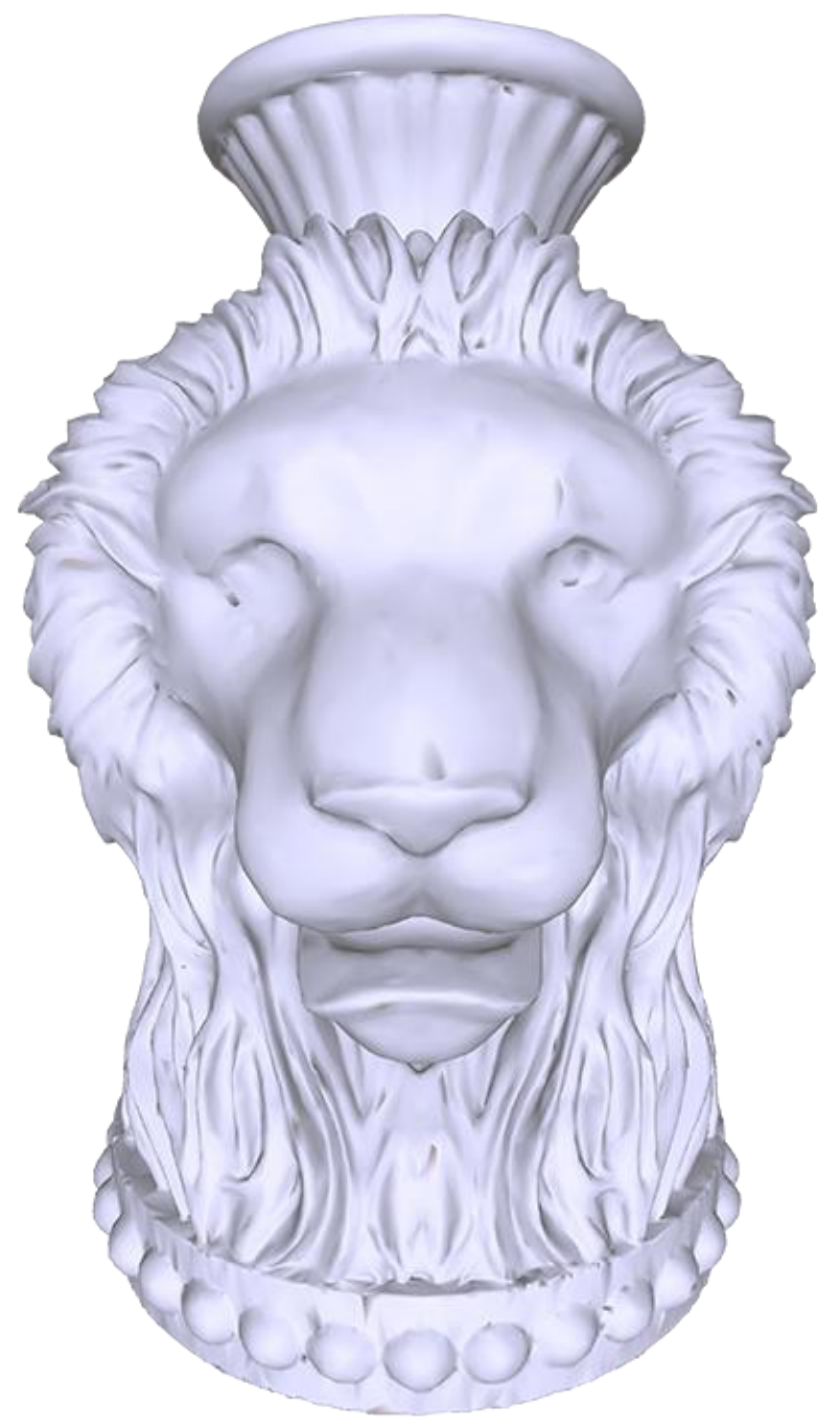}\\
	
					\includegraphics[width=2.1cm]{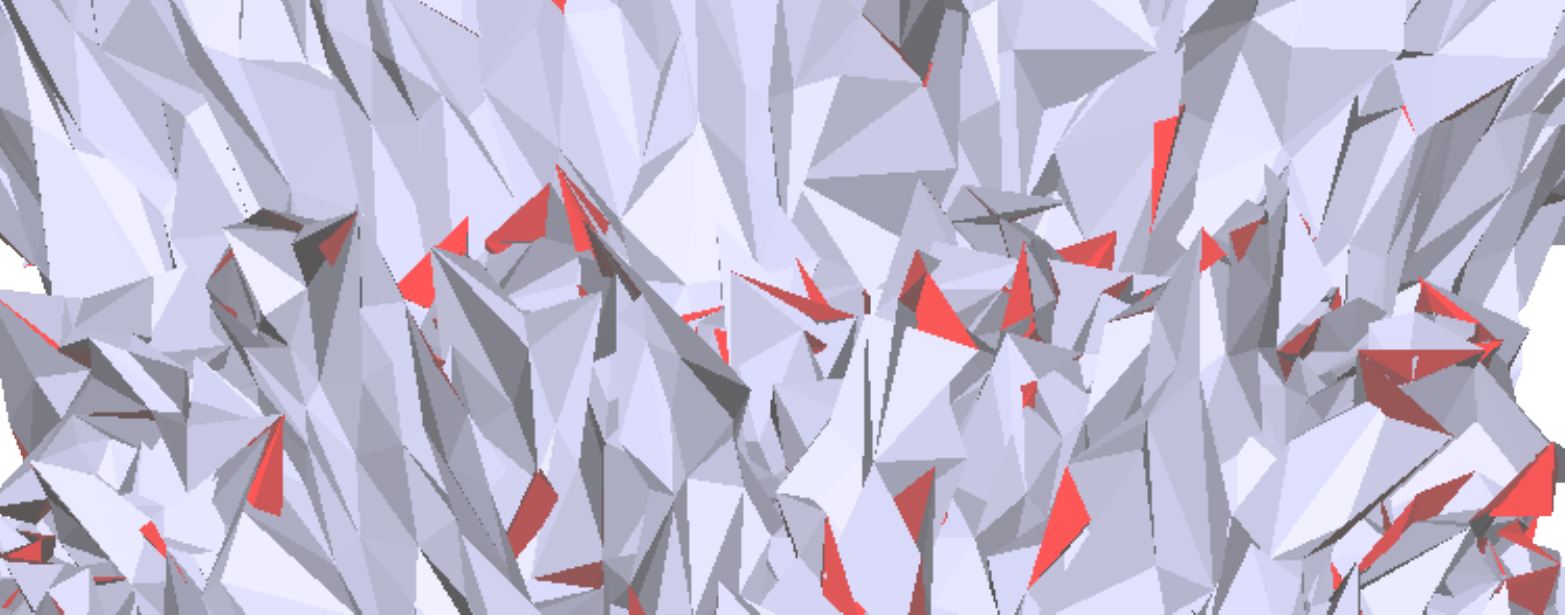}
	\includegraphics[width=2.1cm]{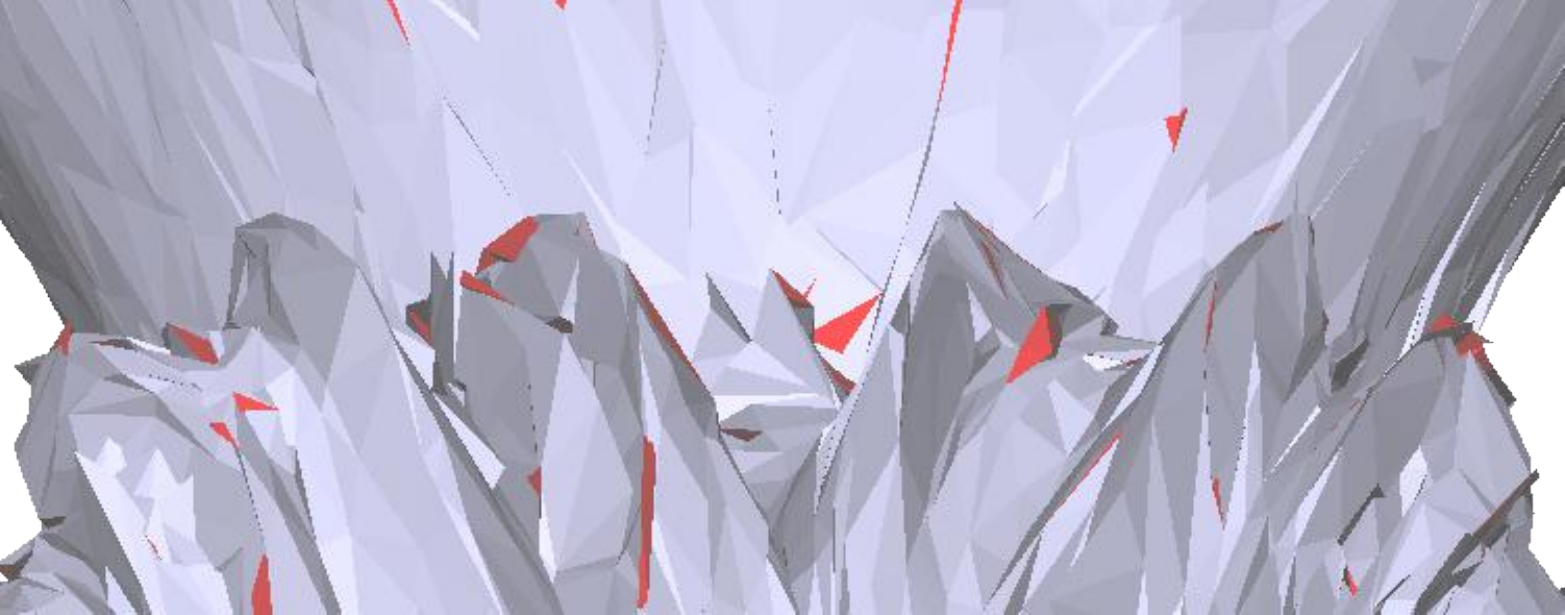}
	\includegraphics[width=2.1cm]{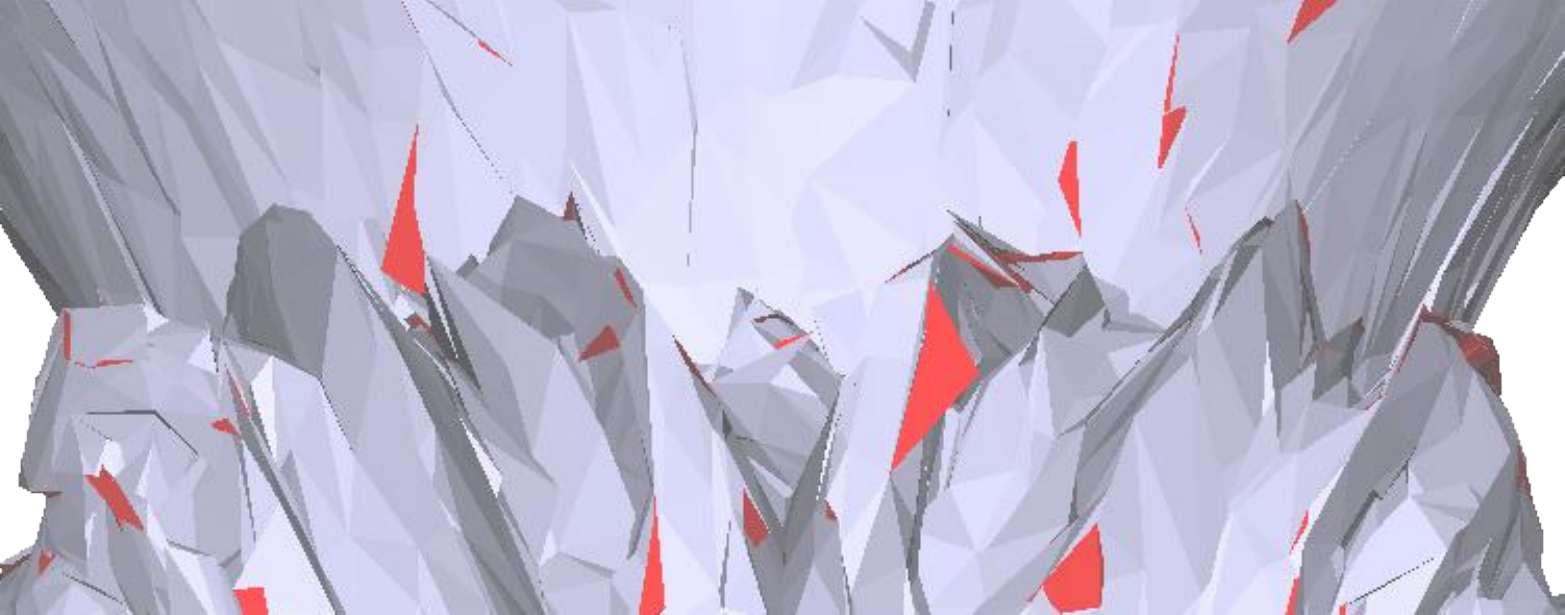}
	\includegraphics[width=2.1cm]{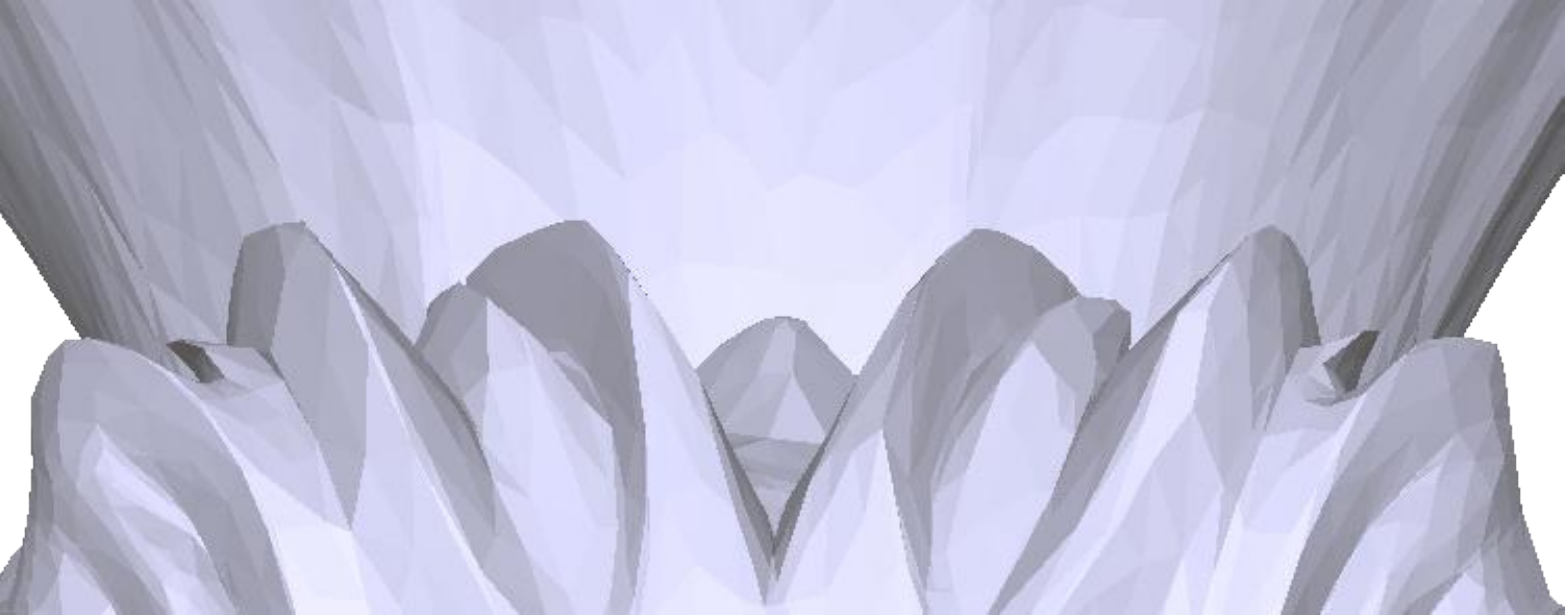}
	\includegraphics[width=2.1cm]{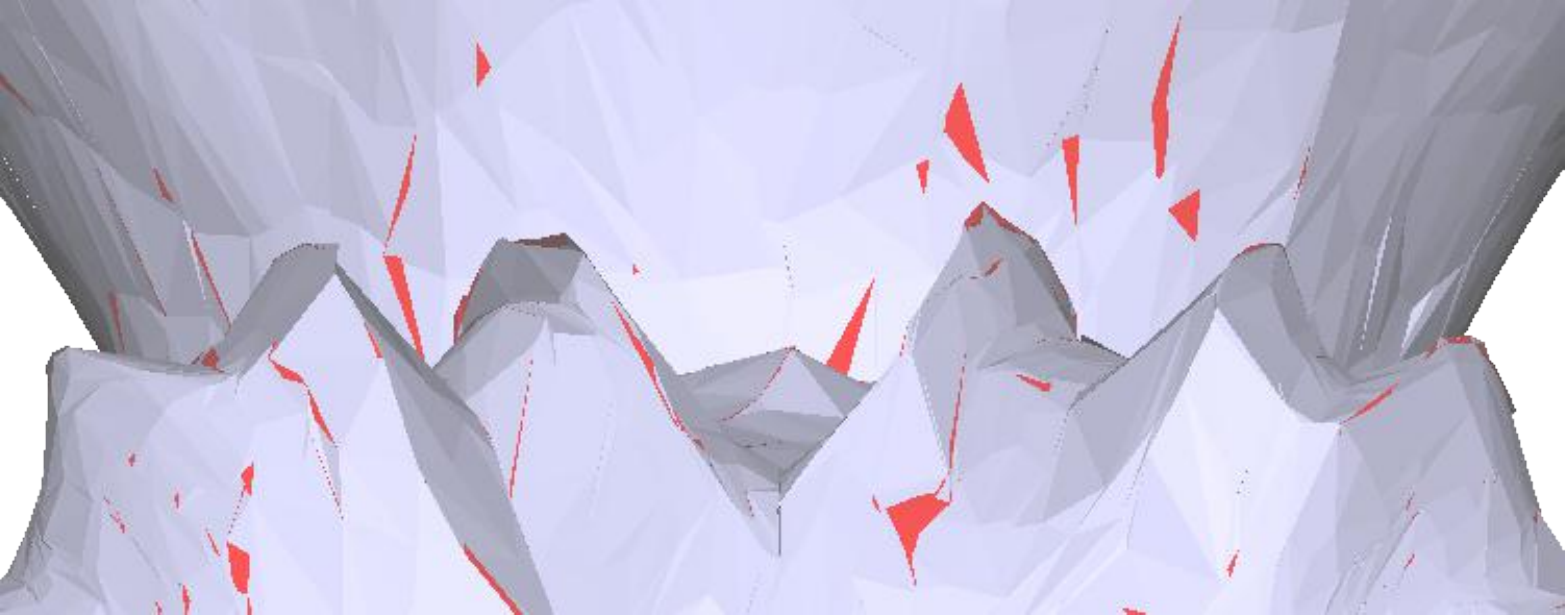}
	\includegraphics[width=2.1cm]{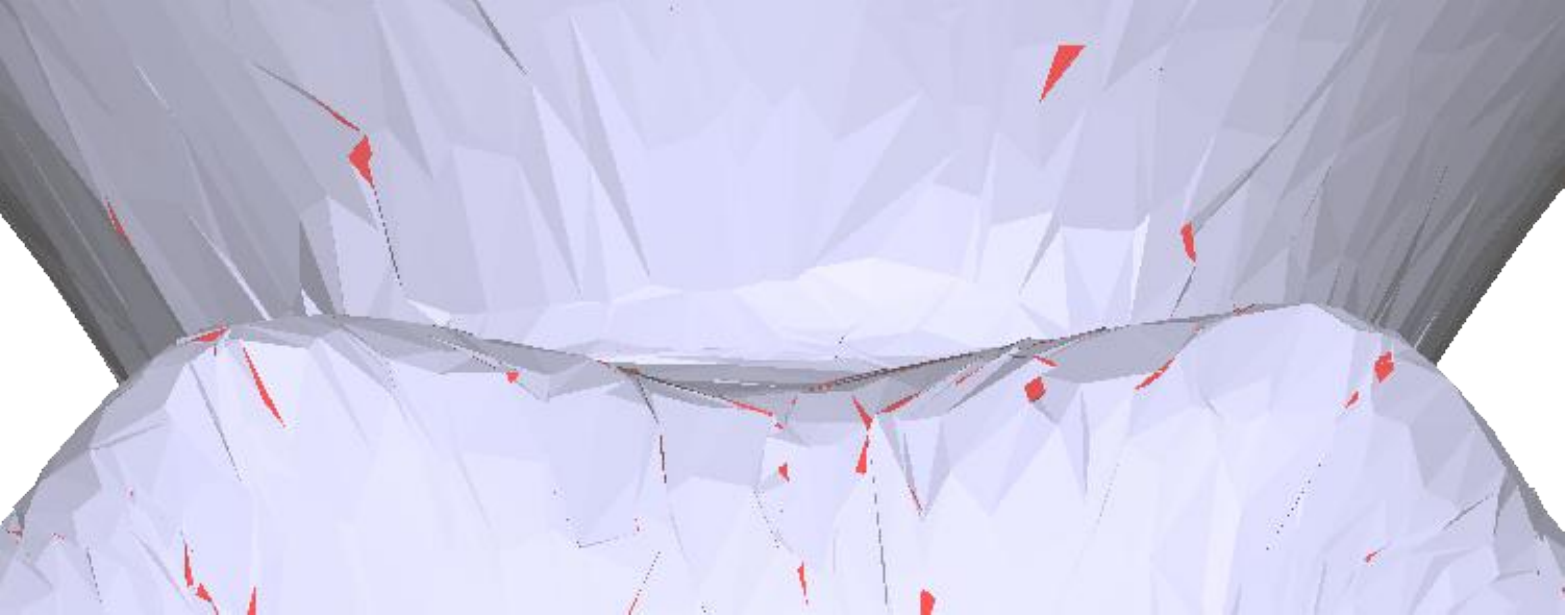}
	\includegraphics[width=2.1cm]{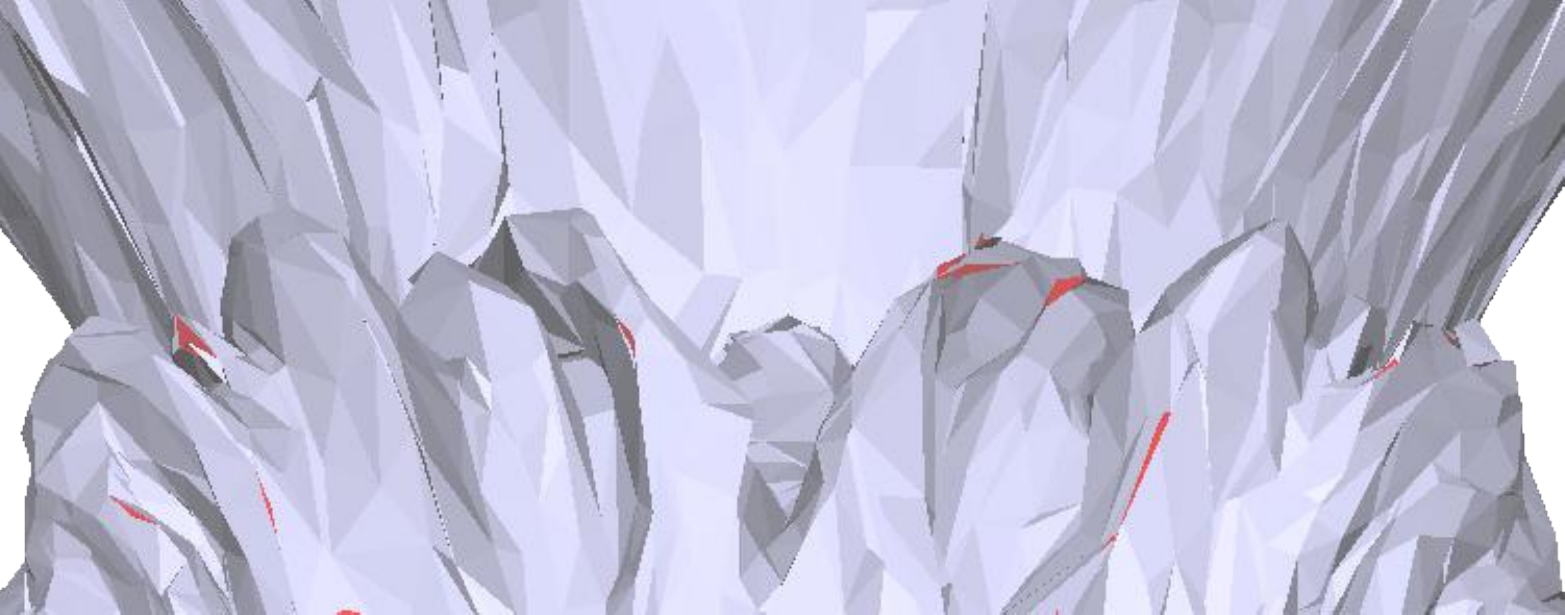}
		\includegraphics[width=2.1cm]{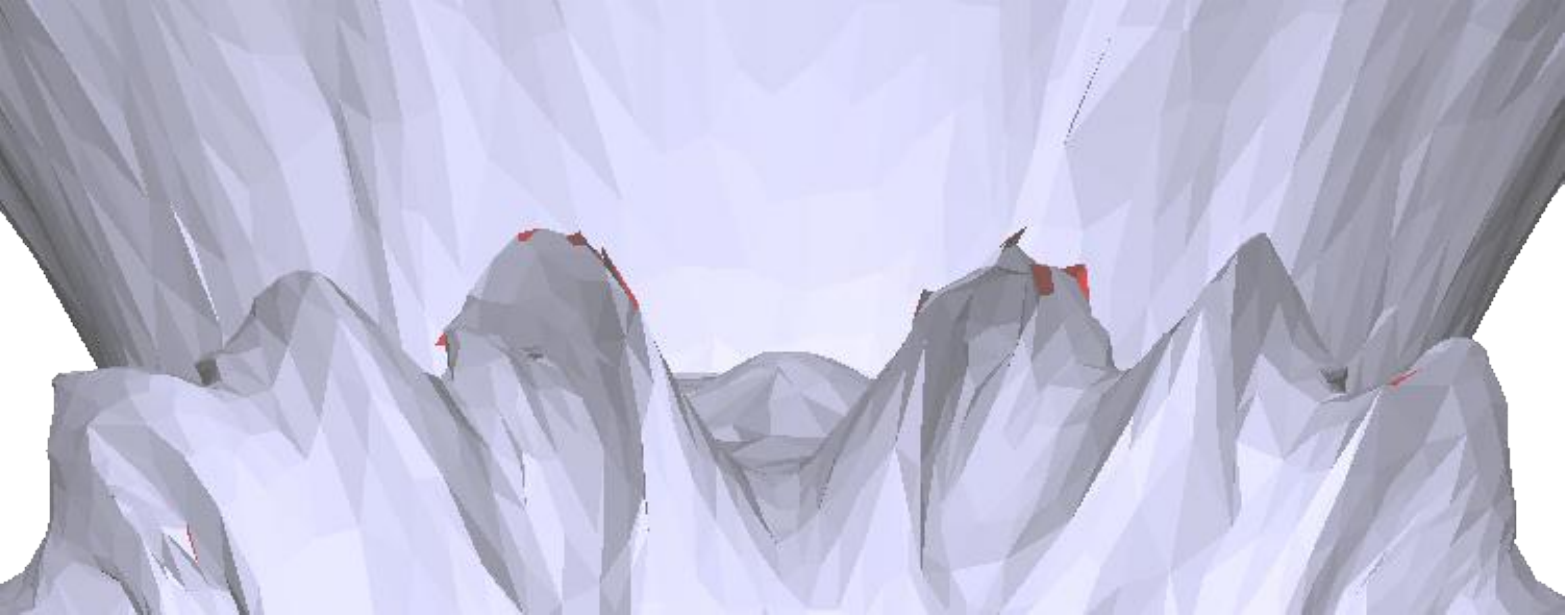}\\		
			\includegraphics[width=2.1cm]{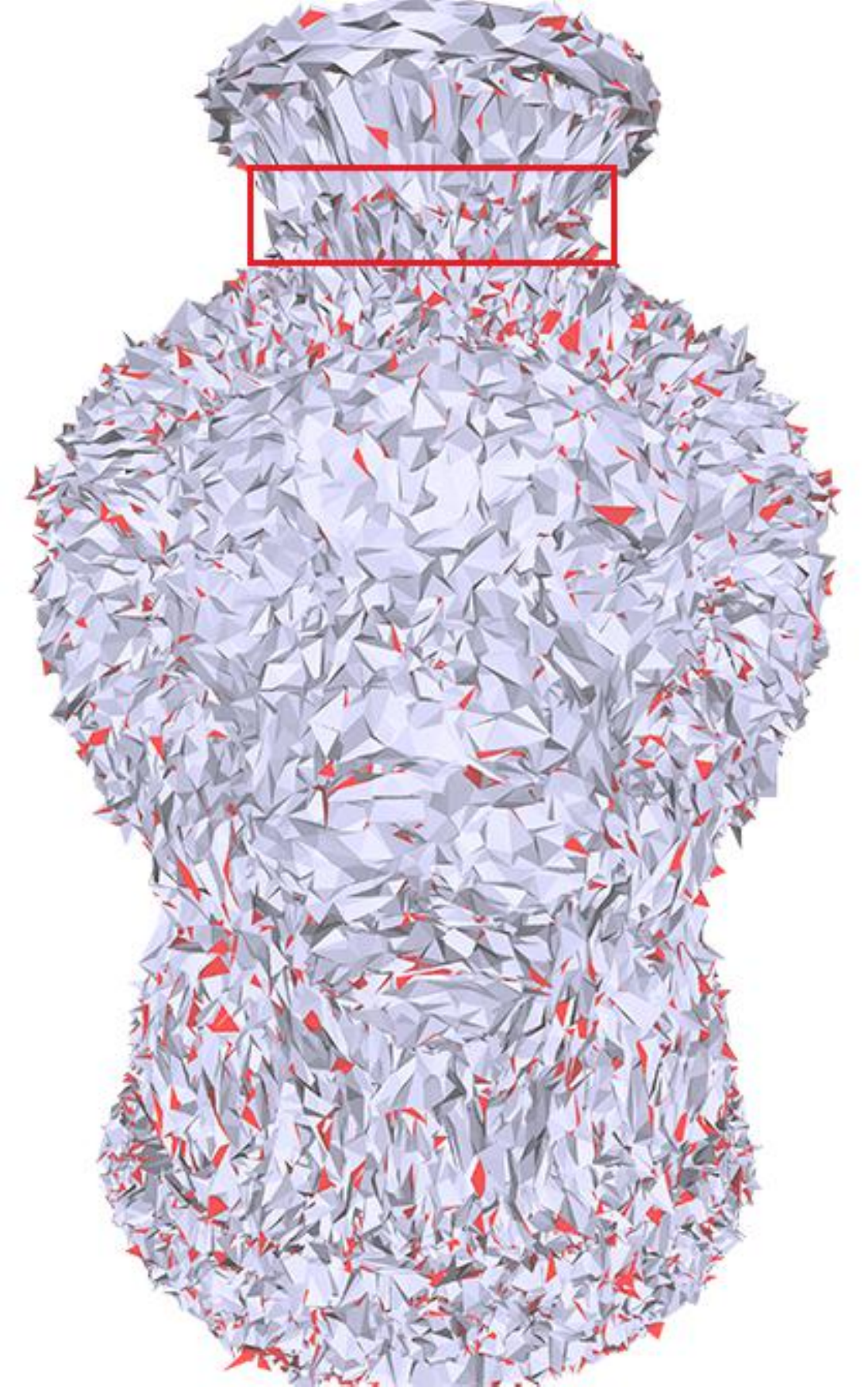}
			\includegraphics[width=2.1cm]{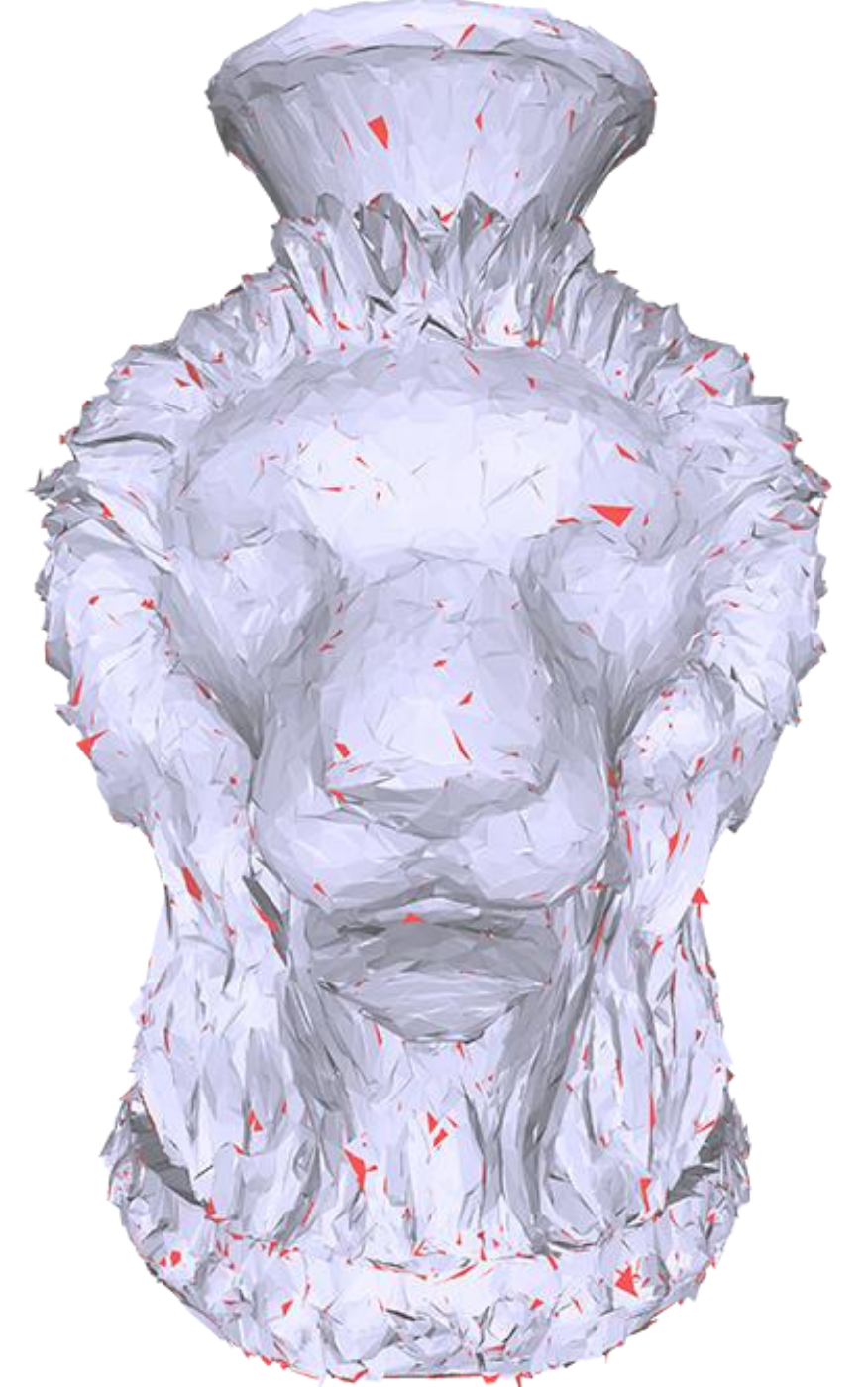}
			\includegraphics[width=2.1cm]{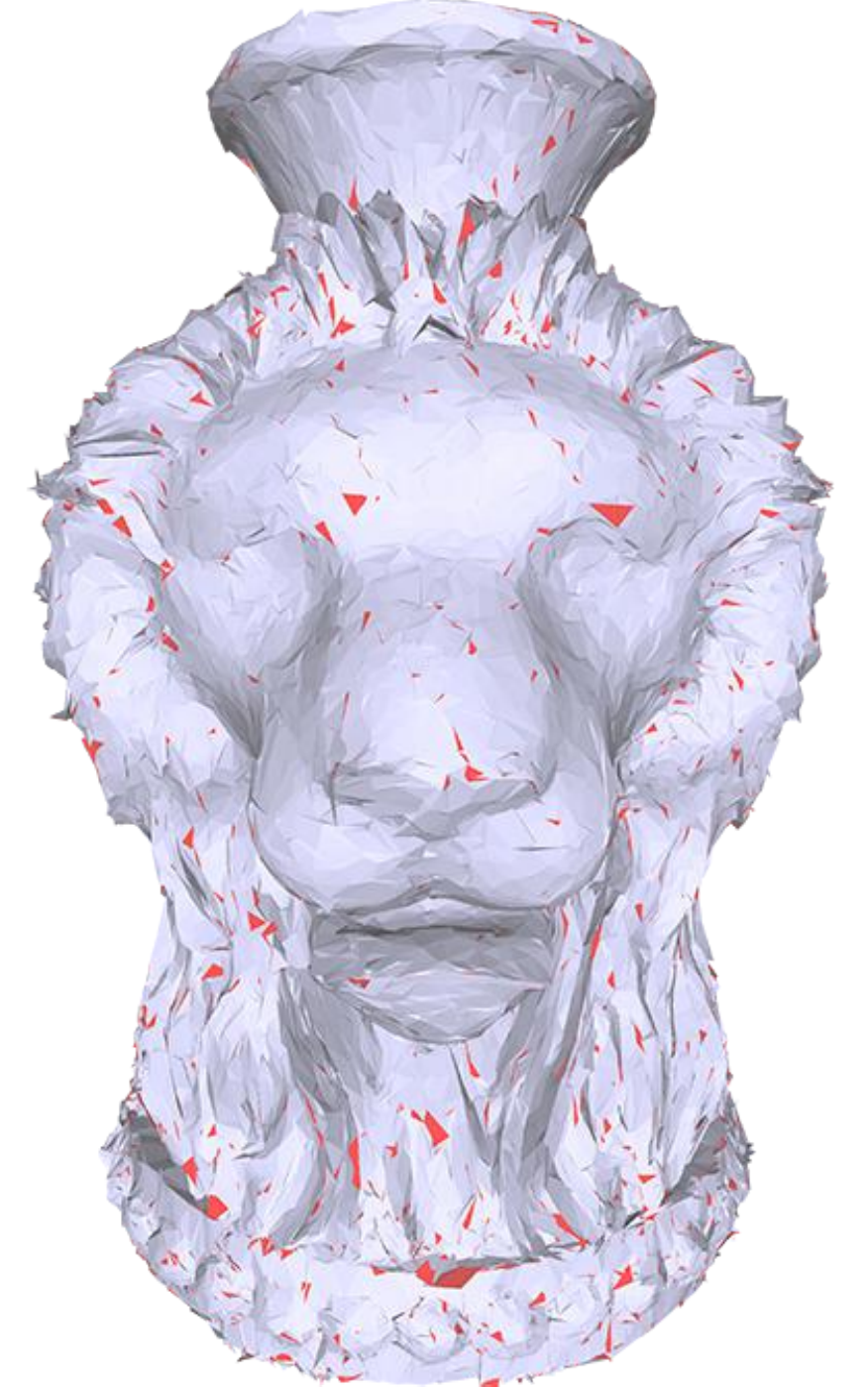}
			\includegraphics[width=2.1cm]{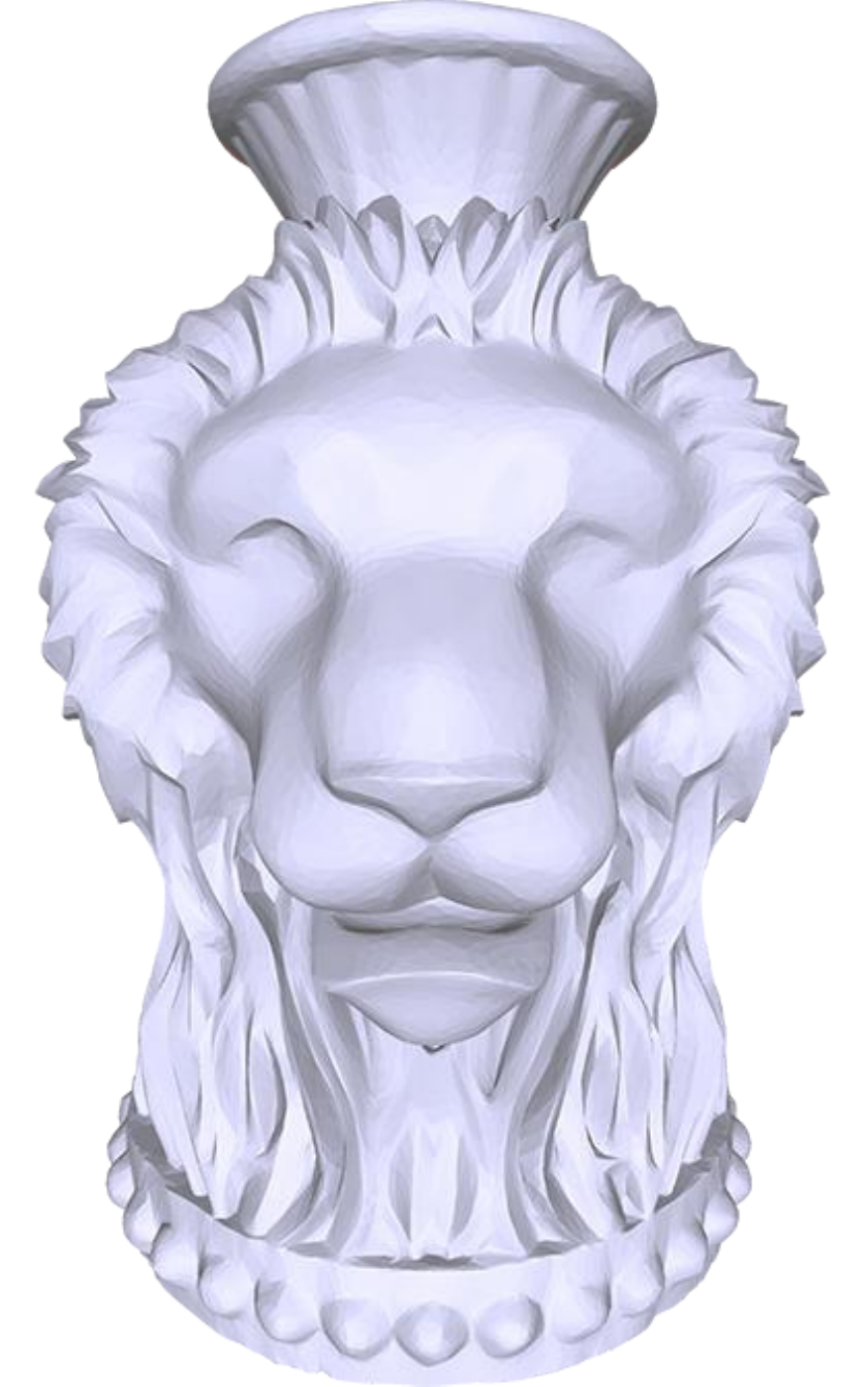}
			\includegraphics[width=2.1cm]{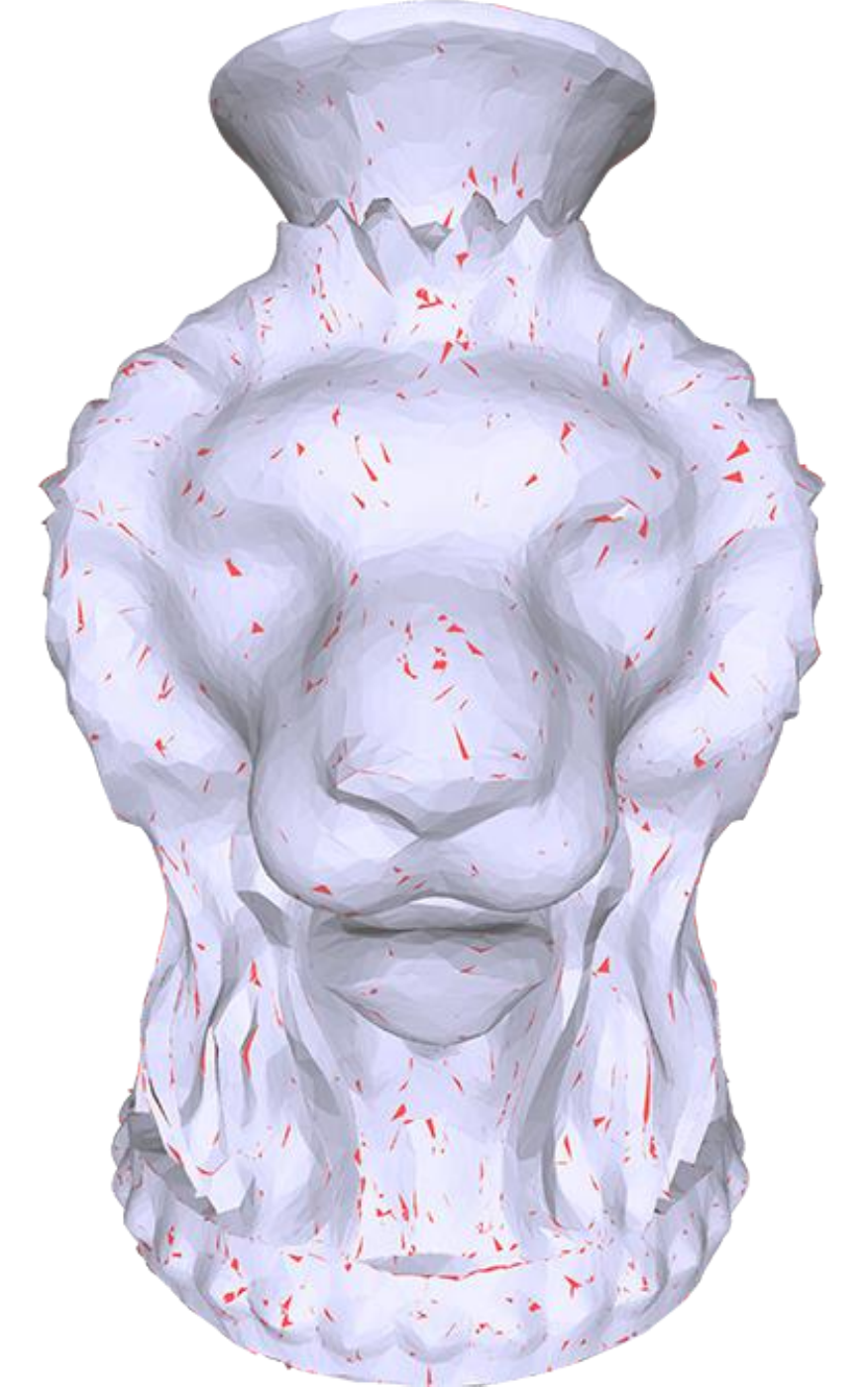}
			\includegraphics[width=2.1cm]{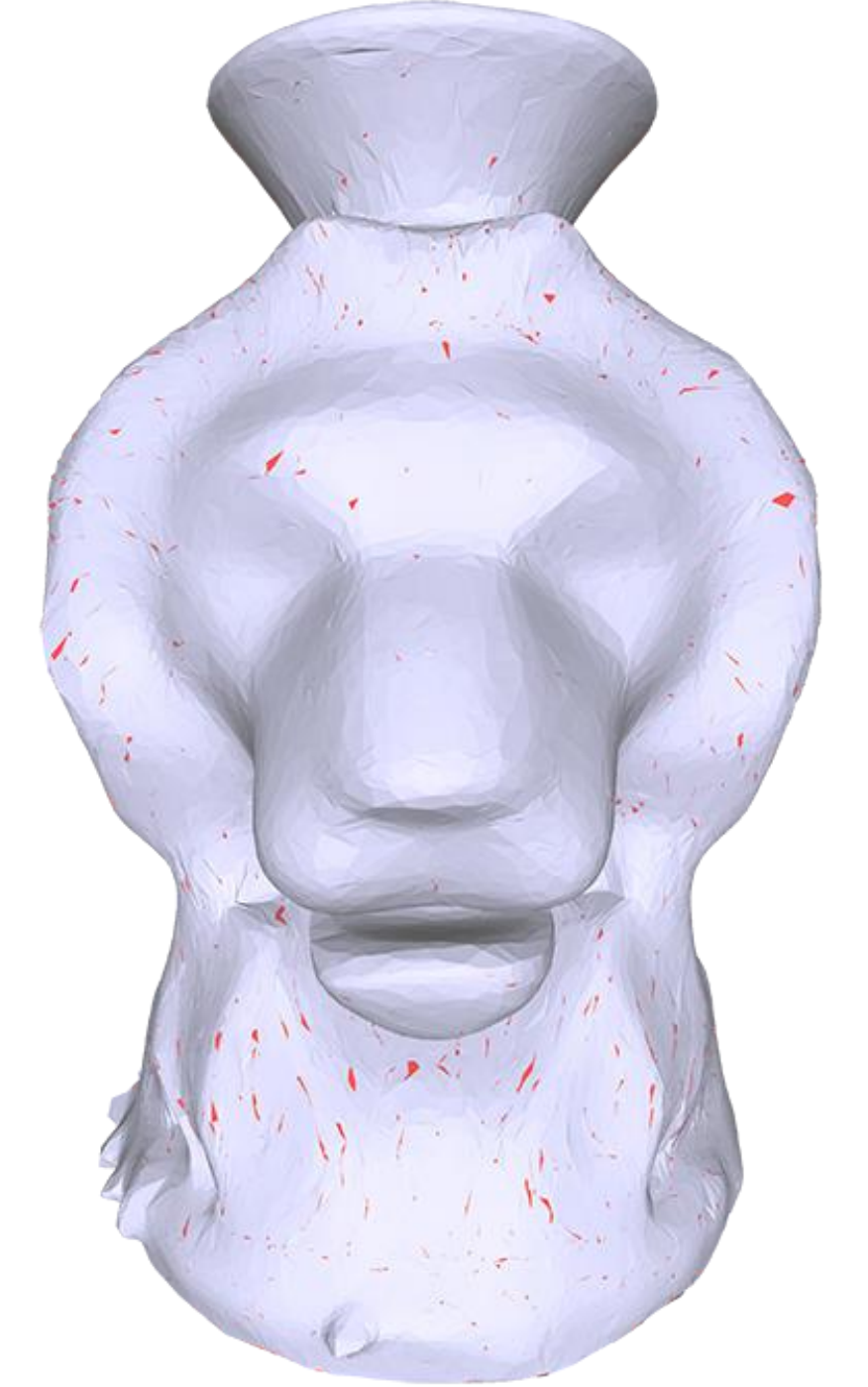}
			\includegraphics[width=2.1cm]{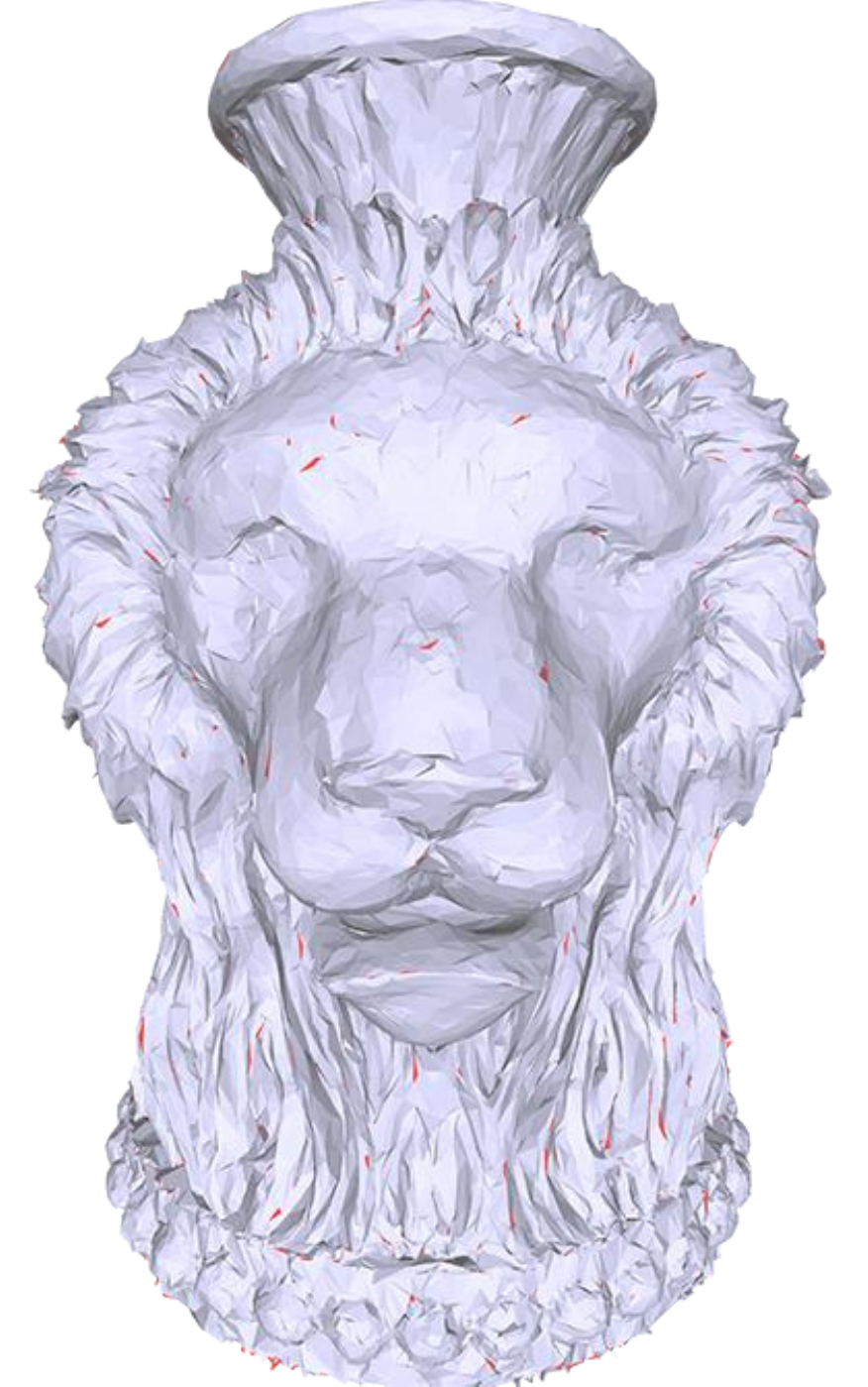}
			\includegraphics[width=2.1cm]{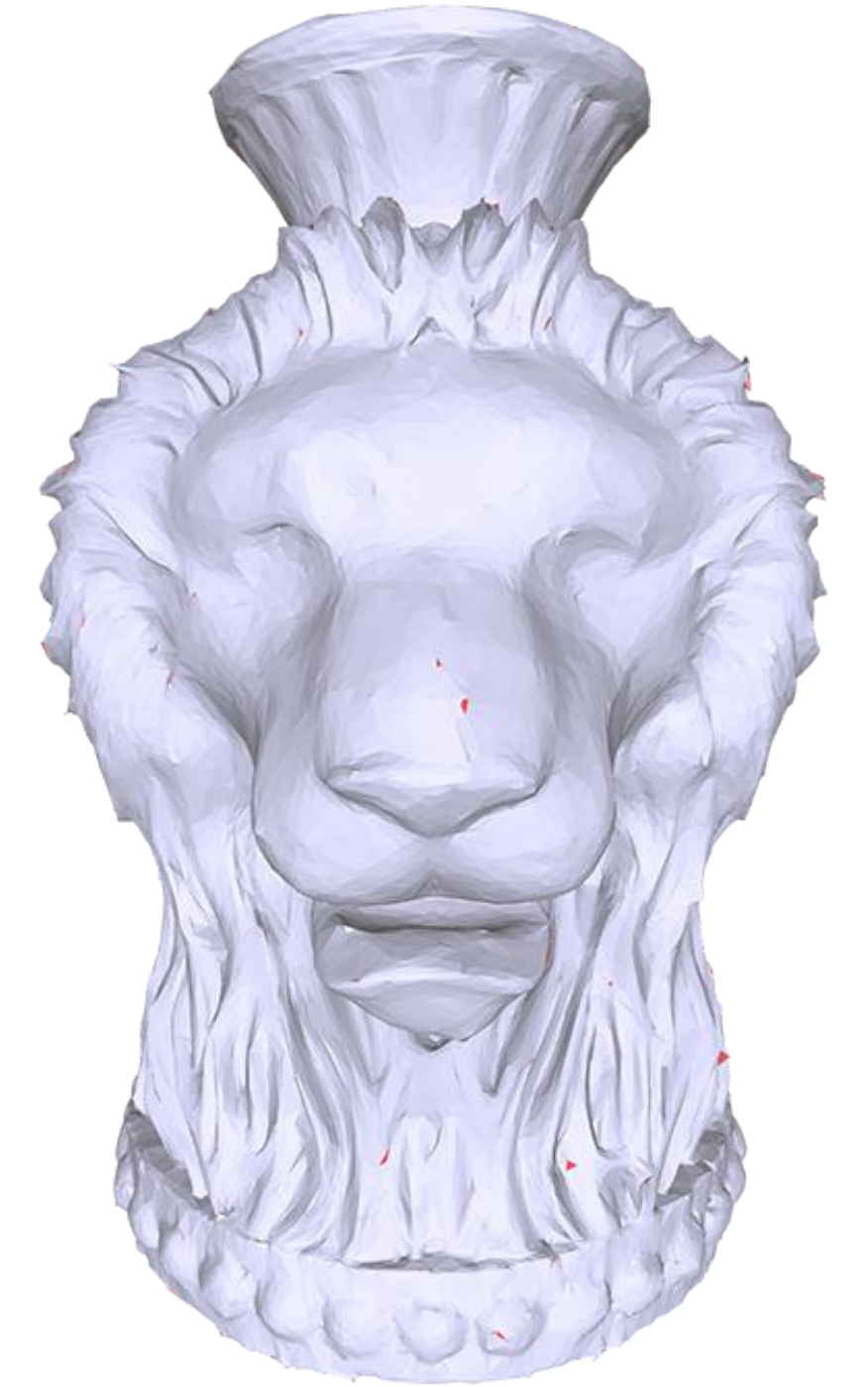}\\
			\makebox[2.1cm]{Noisy}
			\makebox[2.1cm]{UNF~\cite{Sun2007}}
	    \makebox[2.1cm]{BNF~\cite{Zheng2011}}
	    \makebox[2.1cm]{L0~\cite{He2013}}
	    \makebox[2.1cm]{GNF~\cite{Zhang2015}}
	\makebox[2.1cm]{CNR~\cite{Wang-2016-SA}}
	\makebox[2.1cm]{NLLR~\cite{Li2018}}
	\makebox[2.1cm]{Ours}	\\
\caption{Models corrupted with Gaussian noise: Bunny $\sigma_n=0.5l_e$, Nicolo $\sigma_n=0.2l_e$ and $\sigma_n=0.5l_e$, Vaselion $\sigma_n=0.2l_e$ and $\sigma_n=0.8l_e$. The flipped triangles are rendered in red.}
\label{fig:somemodels}
\end{figure*}

The uniform Laplacian operator can unfold the flipped triangles during surface smoothing. Thanks to this property, our approach is less sensitive to high-level noise than most of the compared methods, as shown in Figure~\ref{fig:com_amadillo0.5} and \ref{fig:somemodels} (Bunny: $\sigma_{n}=0.5l_{e}$, Nicolo: $\sigma_{n}=0.5l_{e}$, Vaselion: $\sigma_{n}=0.8l_{e}$ ). \textit{The flipped triangles are rendered in red. See the close-up views. }

\begin{figure*} [htbp]
	\begin{center}
		\begin{tabular}{c} 
			\includegraphics[width=2.5cm]{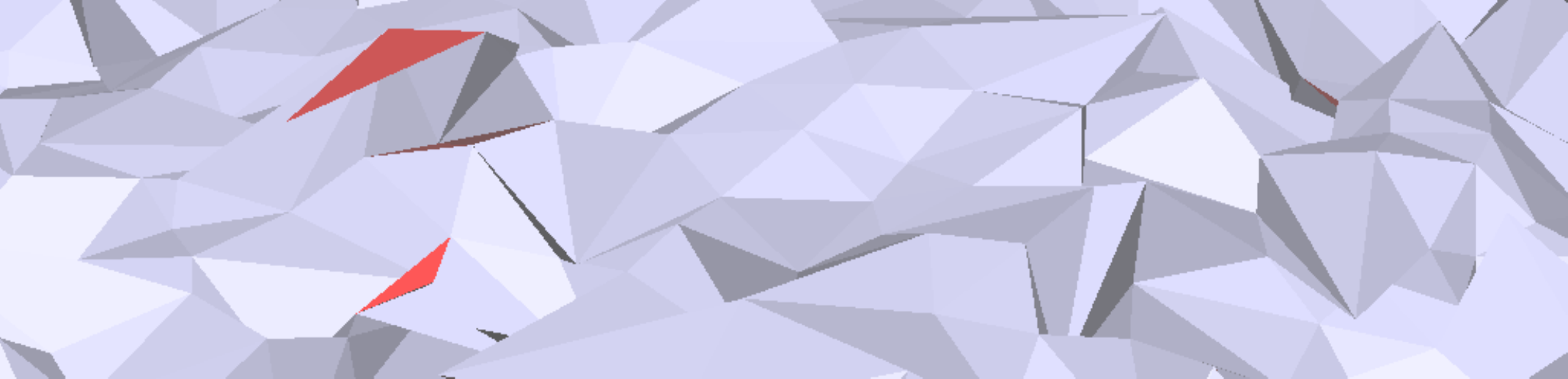}
			\includegraphics[width=2.5cm]{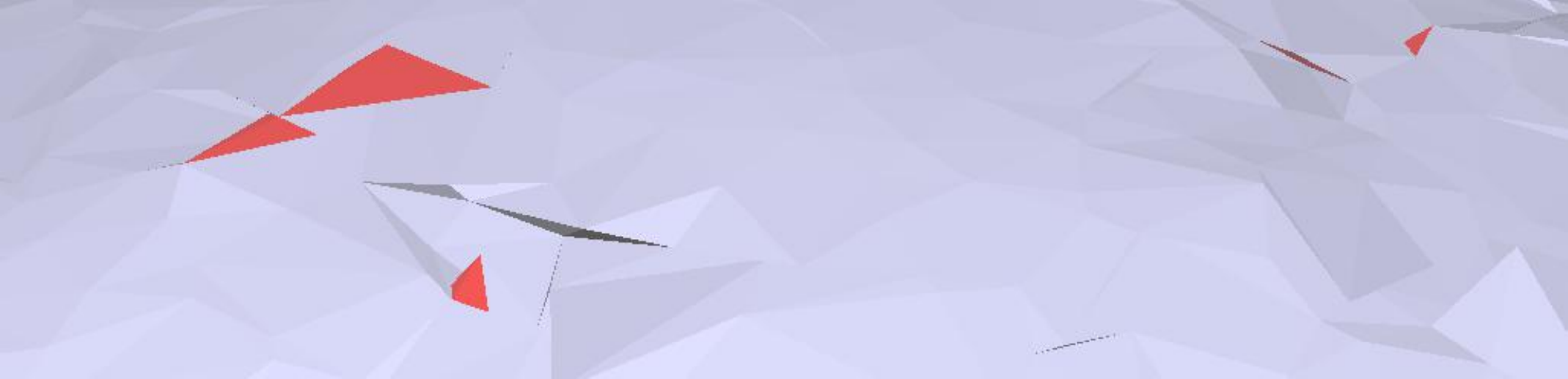}
			\includegraphics[width=2.5cm]{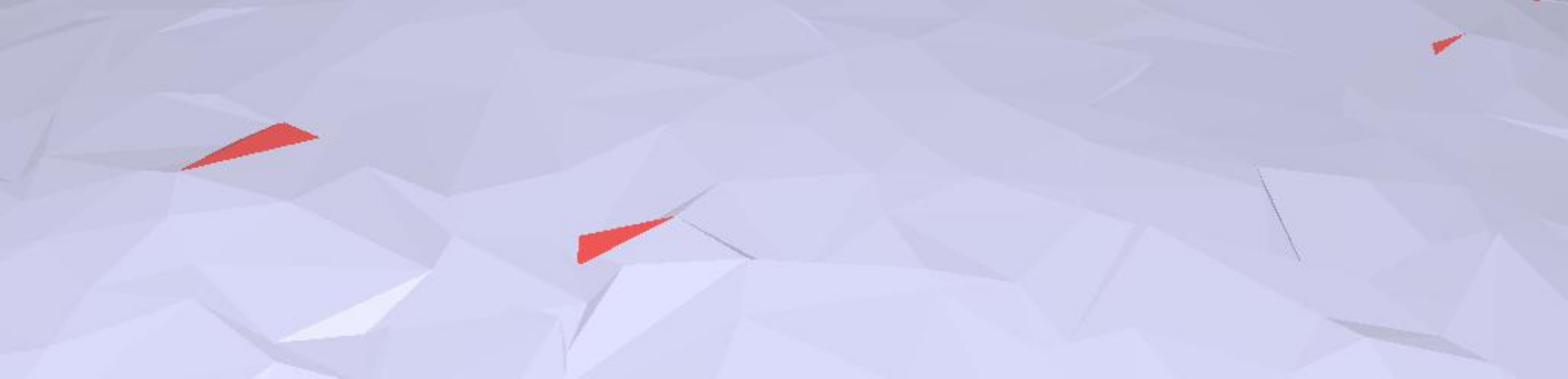}
			\includegraphics[width=2.5cm]{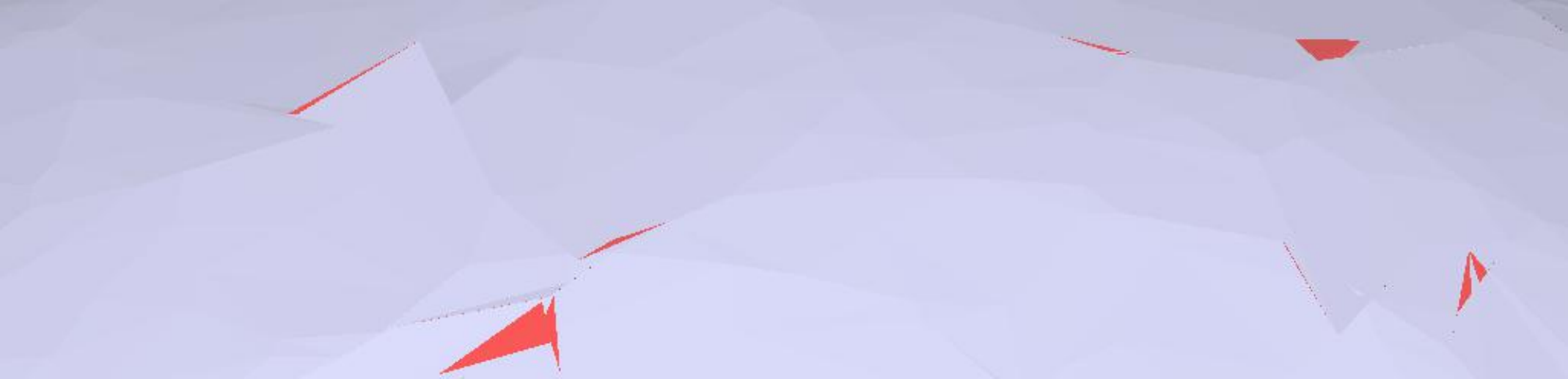}
			\includegraphics[width=2.5cm]{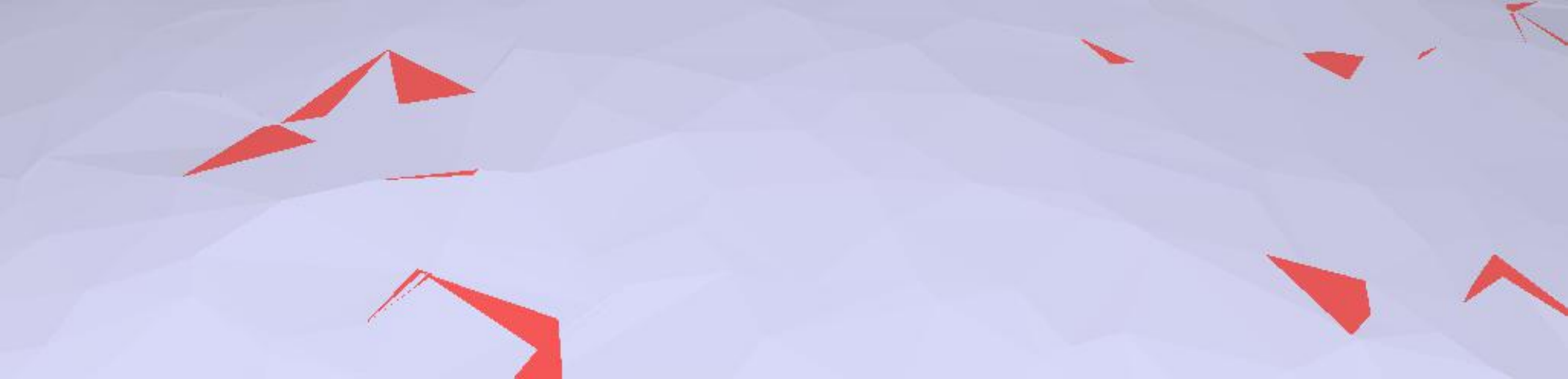}
			\includegraphics[width=2.5cm]{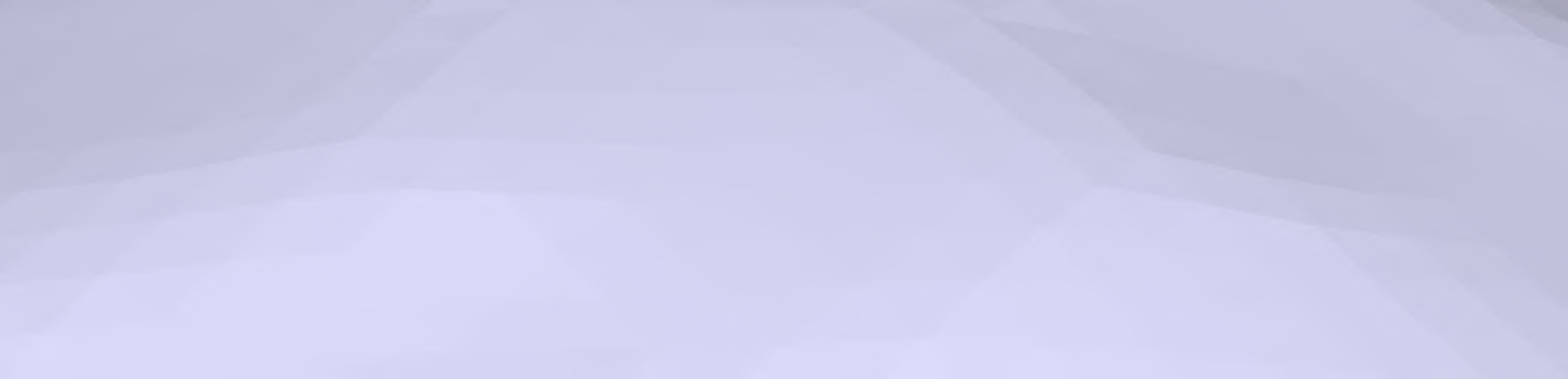}
		\includegraphics[width=2.5cm]{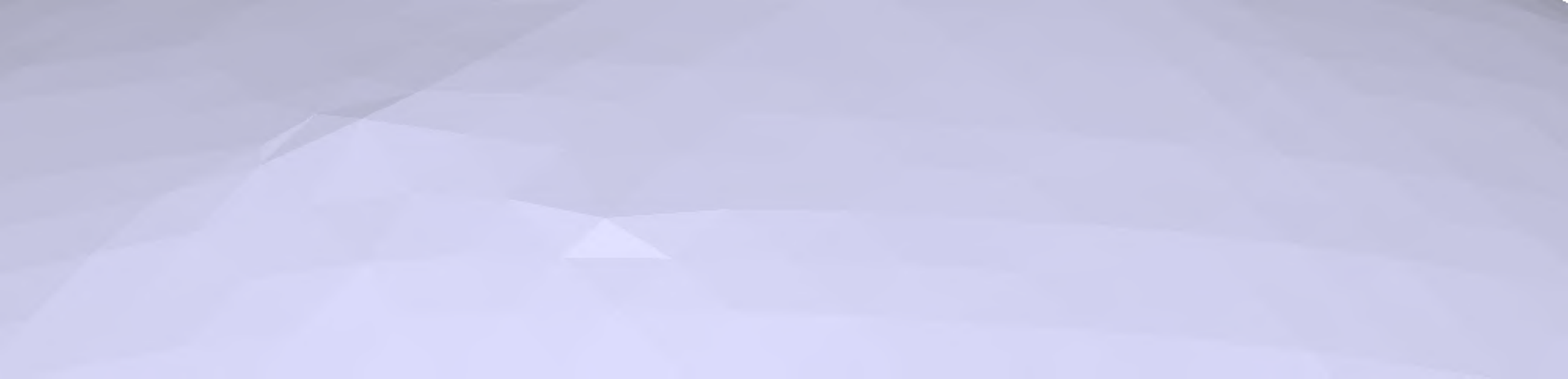}\\
	\subfigure[Noisy]{
   	\includegraphics[width=2.5cm]{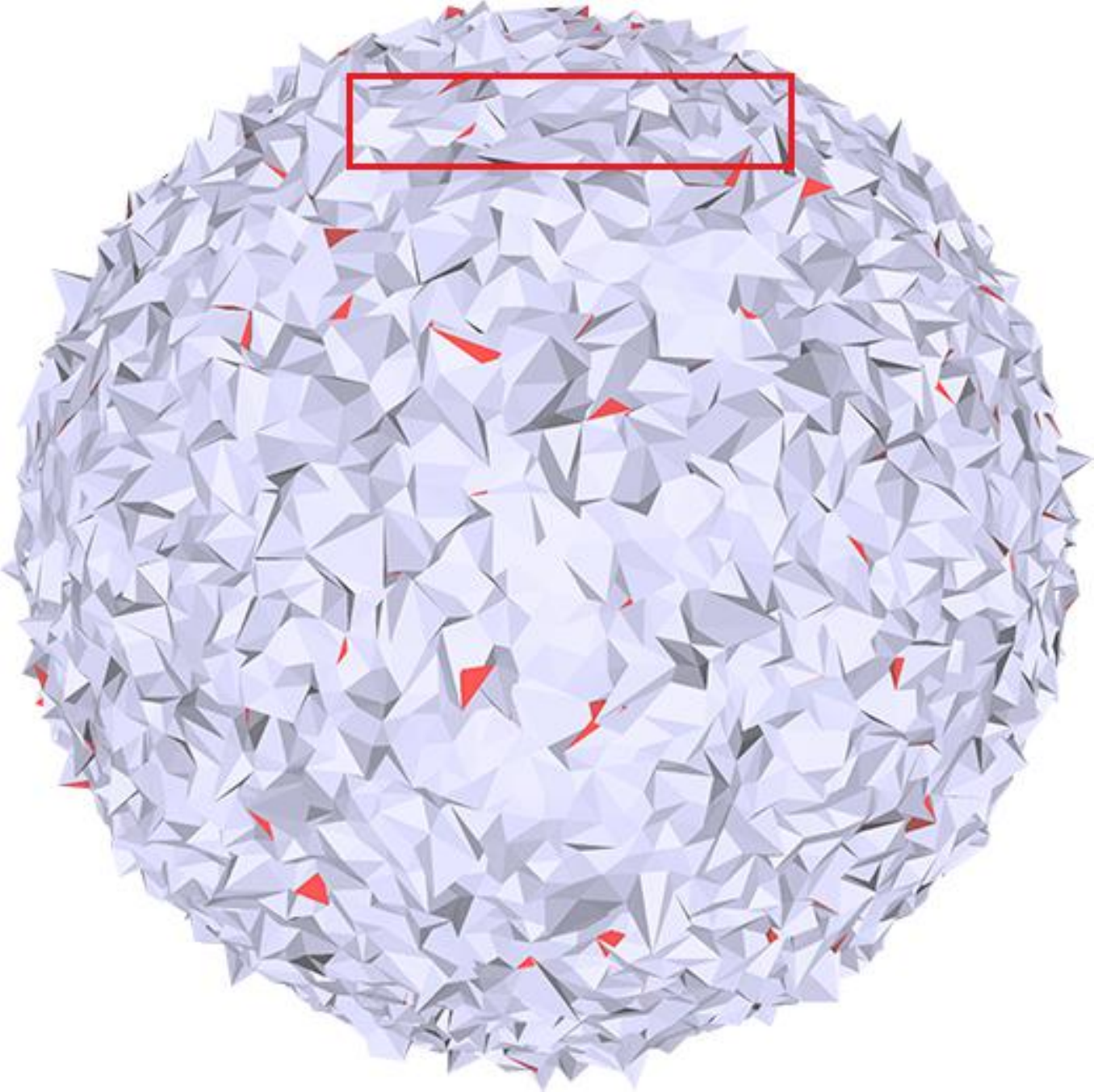}}
    \subfigure[UNF~\cite{Sun2007}]{
	\includegraphics[width=2.5cm]{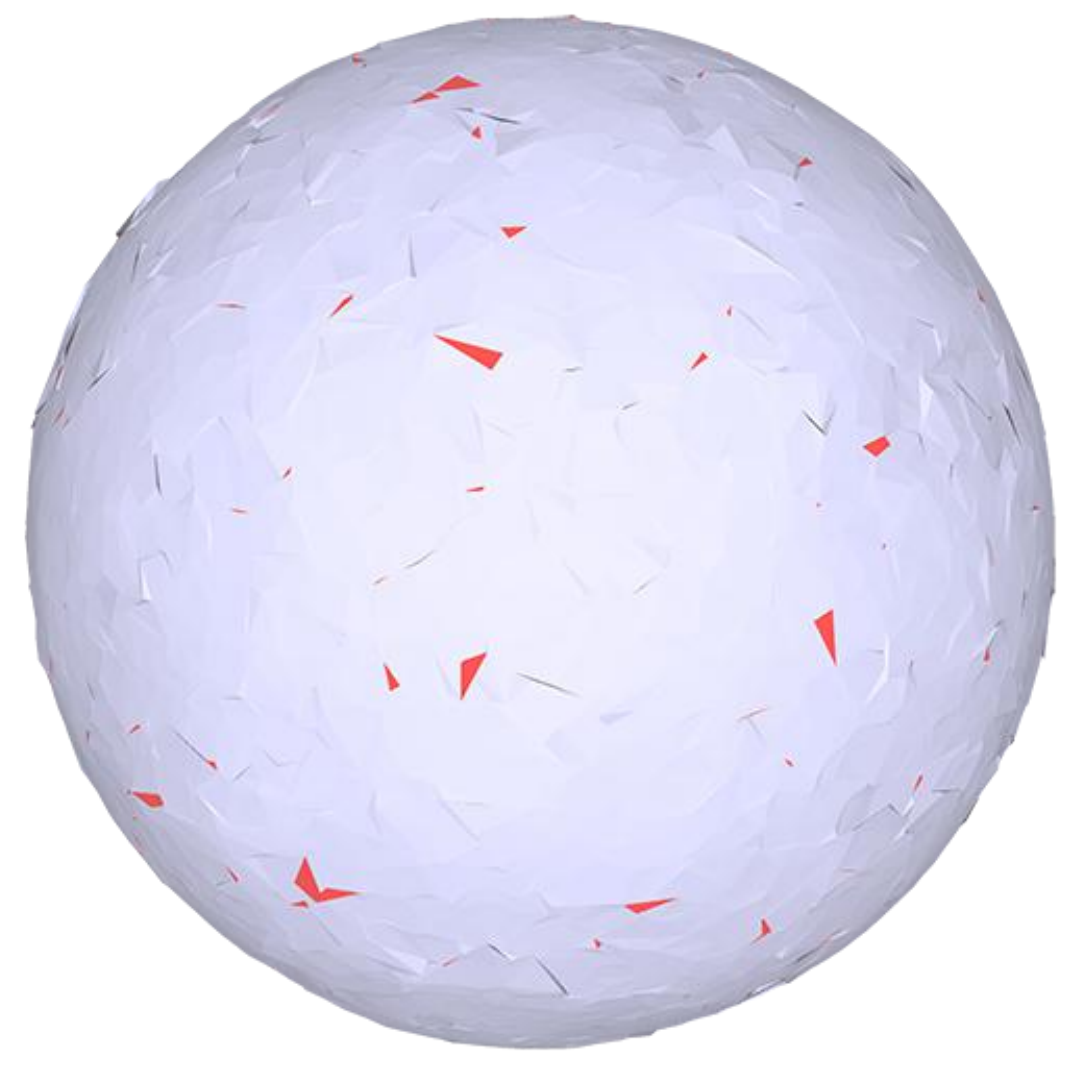}}
	\subfigure[BNF~\cite{Zheng2011}]{
	\includegraphics[width=2.5cm]{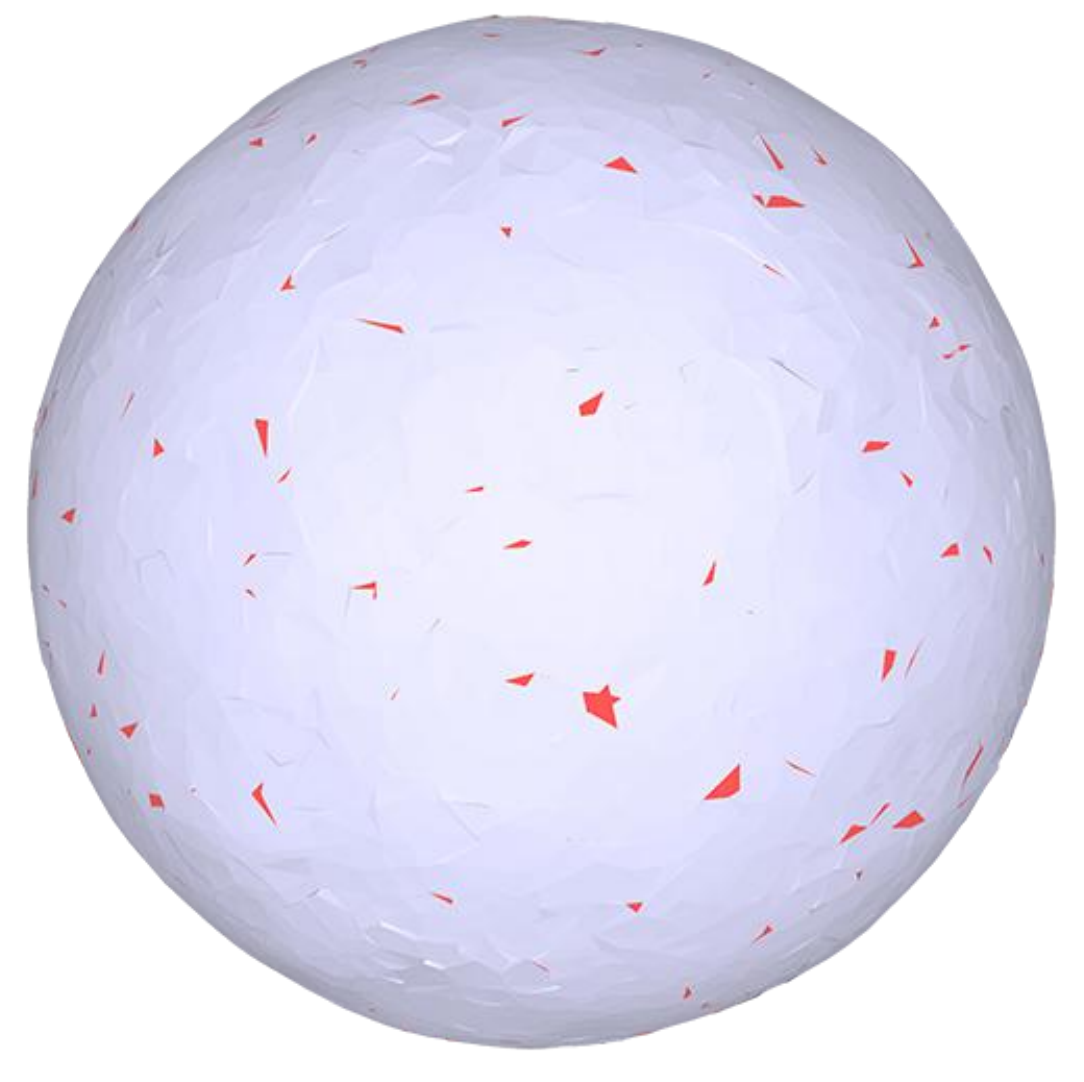}}
	\subfigure[L0~\cite{He2013}]{
	\includegraphics[width=2.5cm]{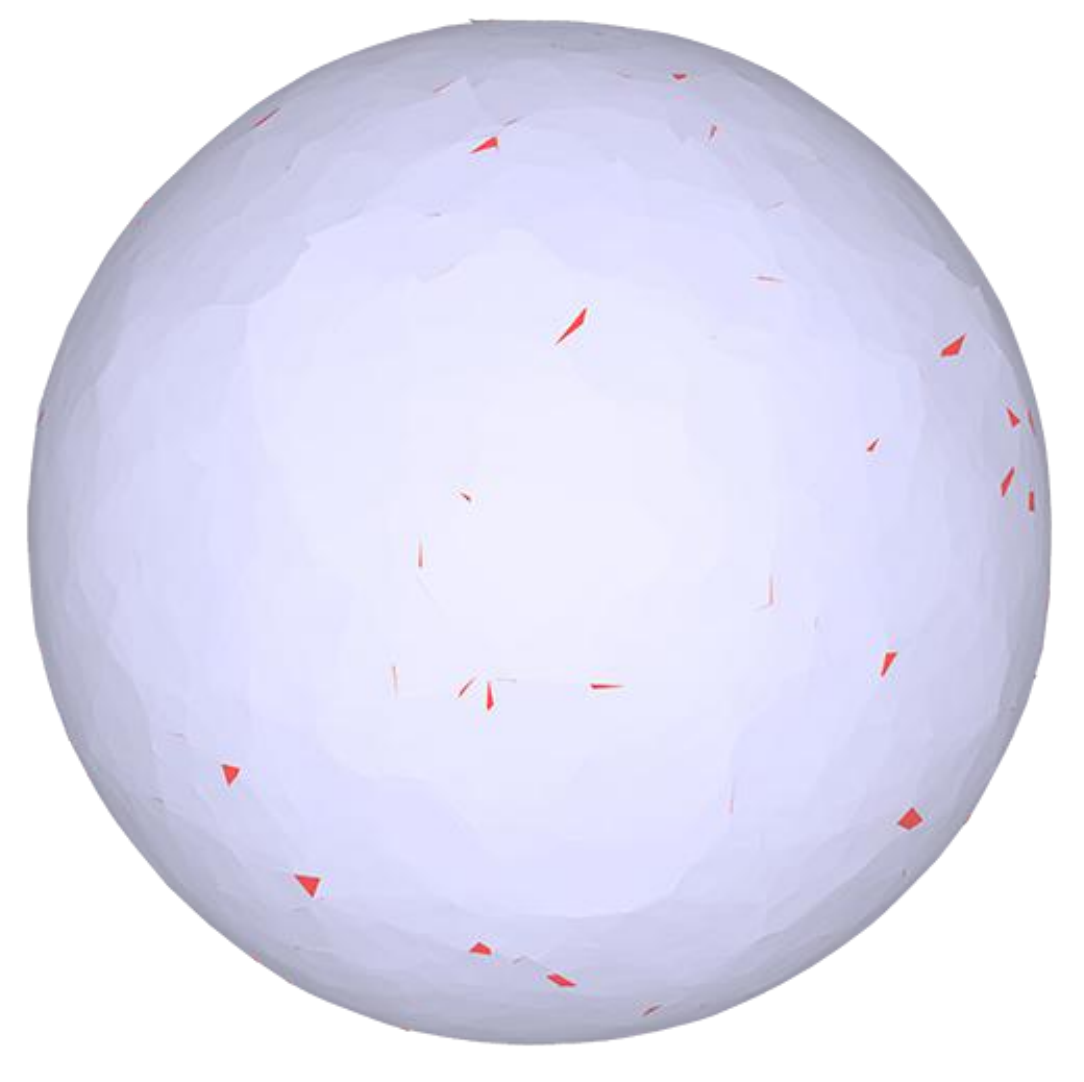}}
		\subfigure[GNF~\cite{Zhang2015}]{
	\includegraphics[width=2.5cm]{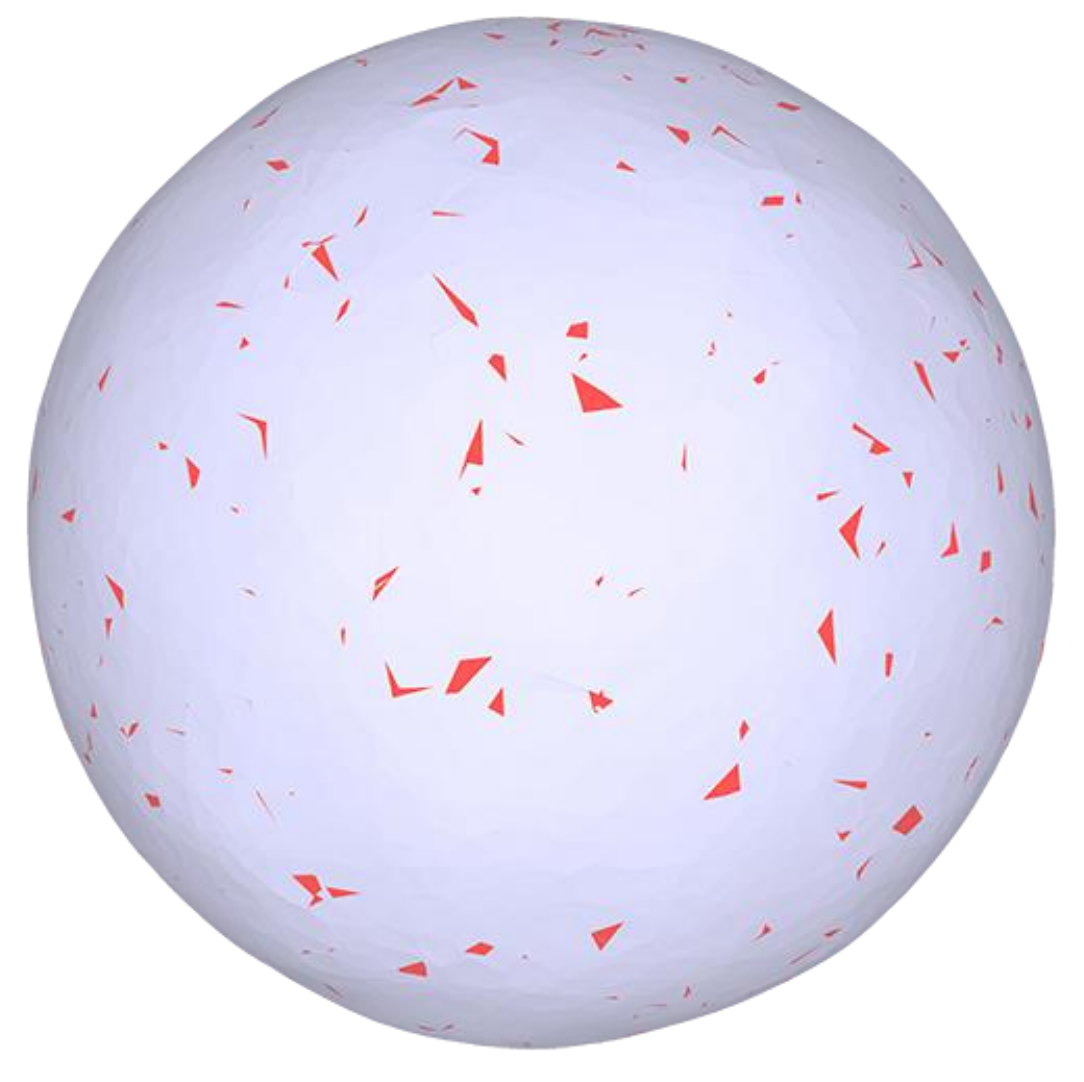}}
		\subfigure[LU~\cite{Lu2017-1}]{
	\includegraphics[width=2.5cm]{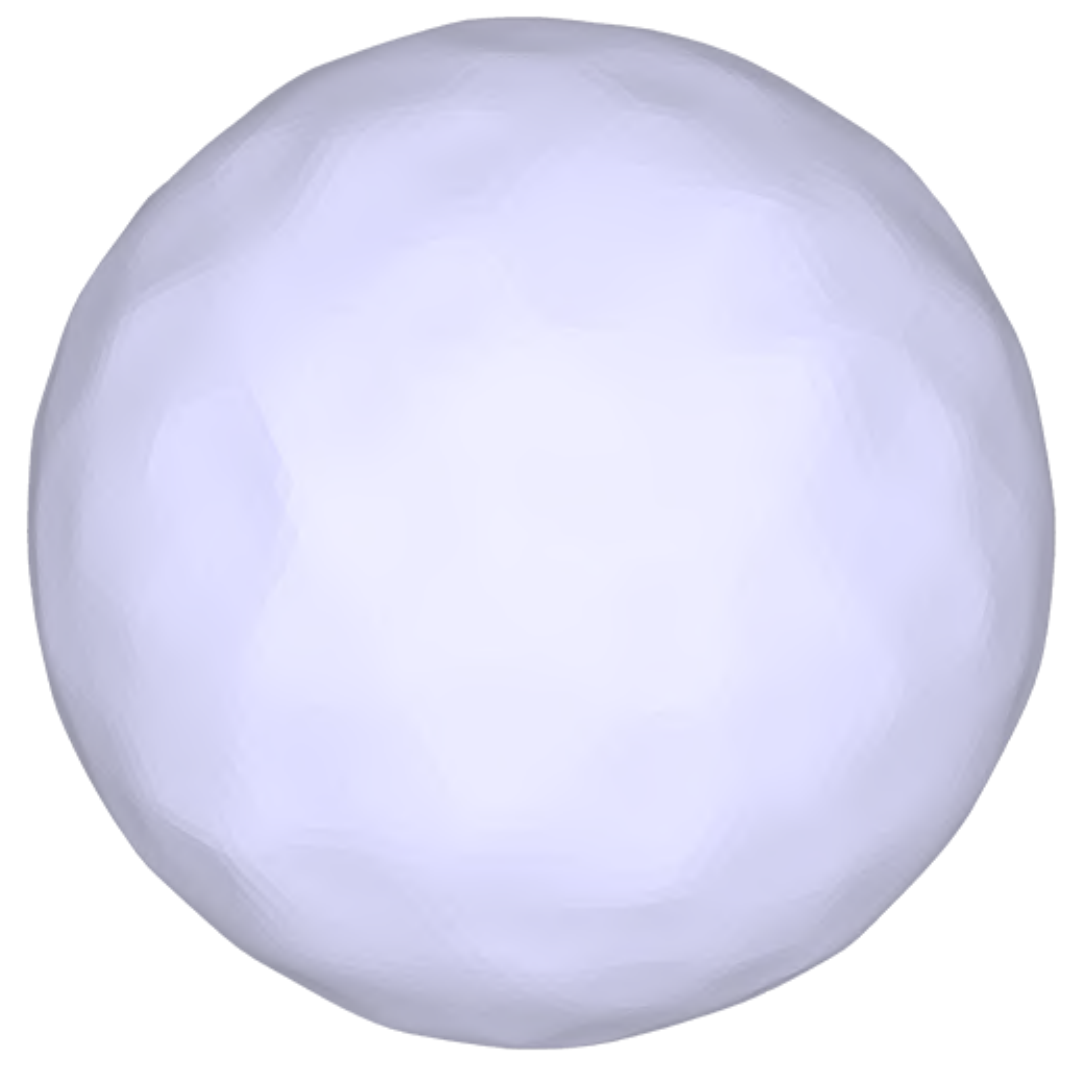}}
		\subfigure[Ours]{
\includegraphics[width=2.5cm]{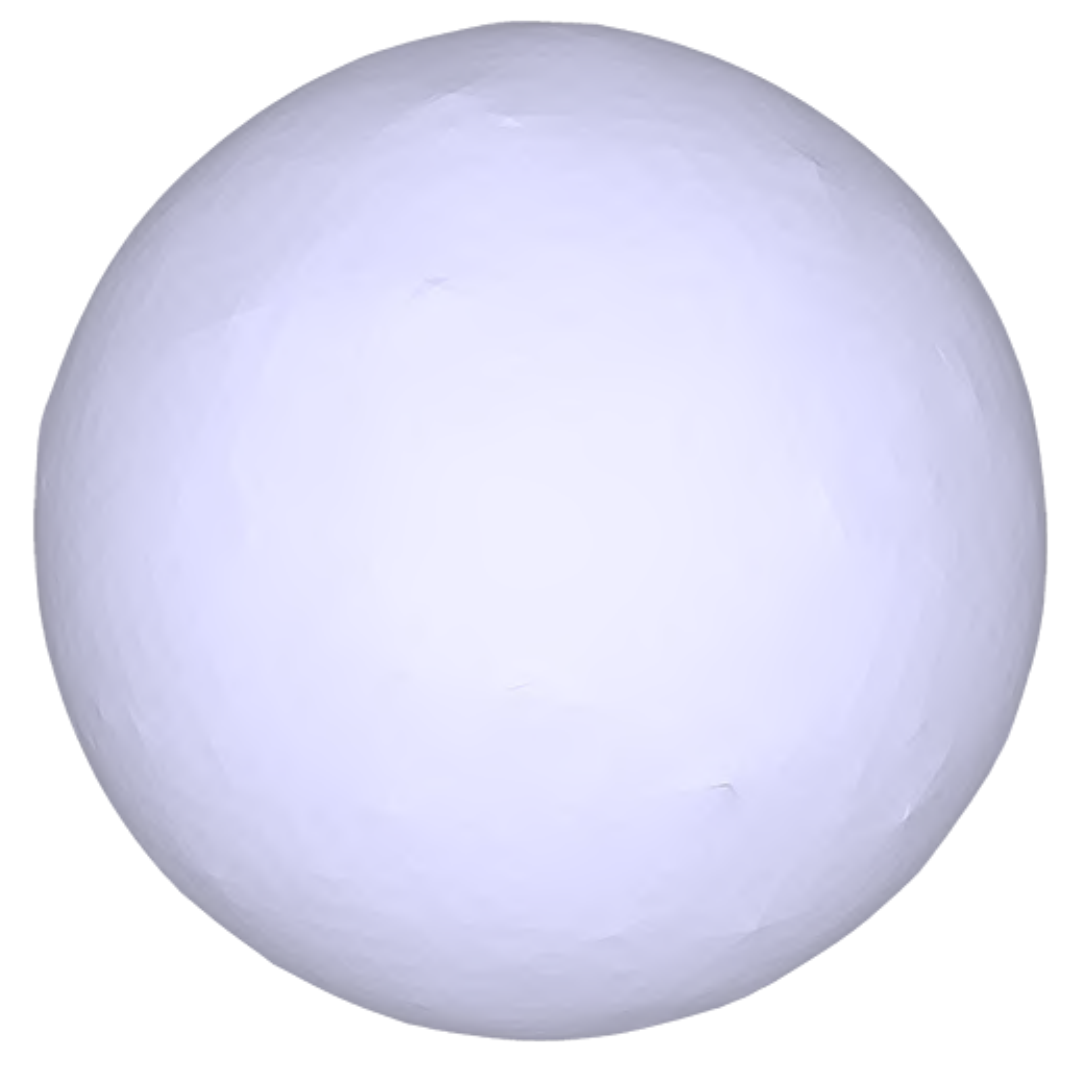}}
		\end{tabular}
	\end{center}
	\caption[res] 
	{ \label{fig:com_sphere0.7} 
		Sphere with heavy Gaussian noise ($\sigma_{n}=0.7l_{e}$). The flipped triangles are rendered in red.  See the close-up views for details. 
	}
\end{figure*}

Figure~\ref{fig:com_amadillo0.5} shows visual results over the Armadillo model contaminated by Gaussian noise with $\sigma_n=0.5e_{l}$. We can see from the blown-up windows that the state-of-the-art methods tend to oversmooth or oversharpen the fine details, or retain excessive noise in the model. We can observe from the mouth and eyes that our method well preserves the small-scale features which are usually lost to some extent in the results by the other methods. Figure \ref{fig:somemodels} (Bunny:$\sigma_{n}=0.5l_{e}$ and Nicolo:$\sigma_{n}=0.5l_{e}$) demonstrates that our approach preserves better surface details. Some methods may oversharpen the ear in Figure \ref{fig:somemodels} (Bunny:$\sigma_{n}=0.5l_{e}$) and nose in Figure \ref{fig:somemodels} (Nicolo:$\sigma_{n}=0.5l_{e}$). In Figure \ref{fig:somemodels} (Vaselion:$\sigma_{n}=0.8l_{e}$), there are many curved features which are particularly difficult to recover by most existing methods when removing high-level noise. The method \cite{He2013} can robustly remove the noise while it results in oversharpening in certain areas and loses some fine details. By contrast, our method outputs better results, in terms of features and details preservation. In Figure \ref{fig:com_sphere0.7}, our method well recovers the sphere shape which has fewer folded triangles than other methods. \textit{Notice that CNR \cite{Wang-2016-SA} may produce better results if the model is trained on surface meshes with large noise.}

Undoubtedly, our method can also well handle small or medium noise. Figure \ref{fig:somemodels} (Nicolo:$\sigma_{n}=0.2l_{e}$ and Vaselion:$\sigma_{n}=0.2l_{e}$) shows two such examples. We can observe from Figure \ref{fig:somemodels} (Nicolo:$\sigma_{n}=0.2l_{e}$) that our method and NLLR \cite{Li2018} generates similar results which are the best among all results. NLLR \cite{Li2018} uses non-local information for mesh smoothing while our method uses local information only. Figure \ref{fig:somemodels} (Vaselion:$\sigma_{n}=0.2l_{e}$) demonstrates that the result by our HLO is better than the results by other techniques, in terms of noise removal and features preservation.



%

\begin{figure*} [htbp]
	\begin{center}
		\begin{tabular}{c} 
			\includegraphics[width=2.2cm]{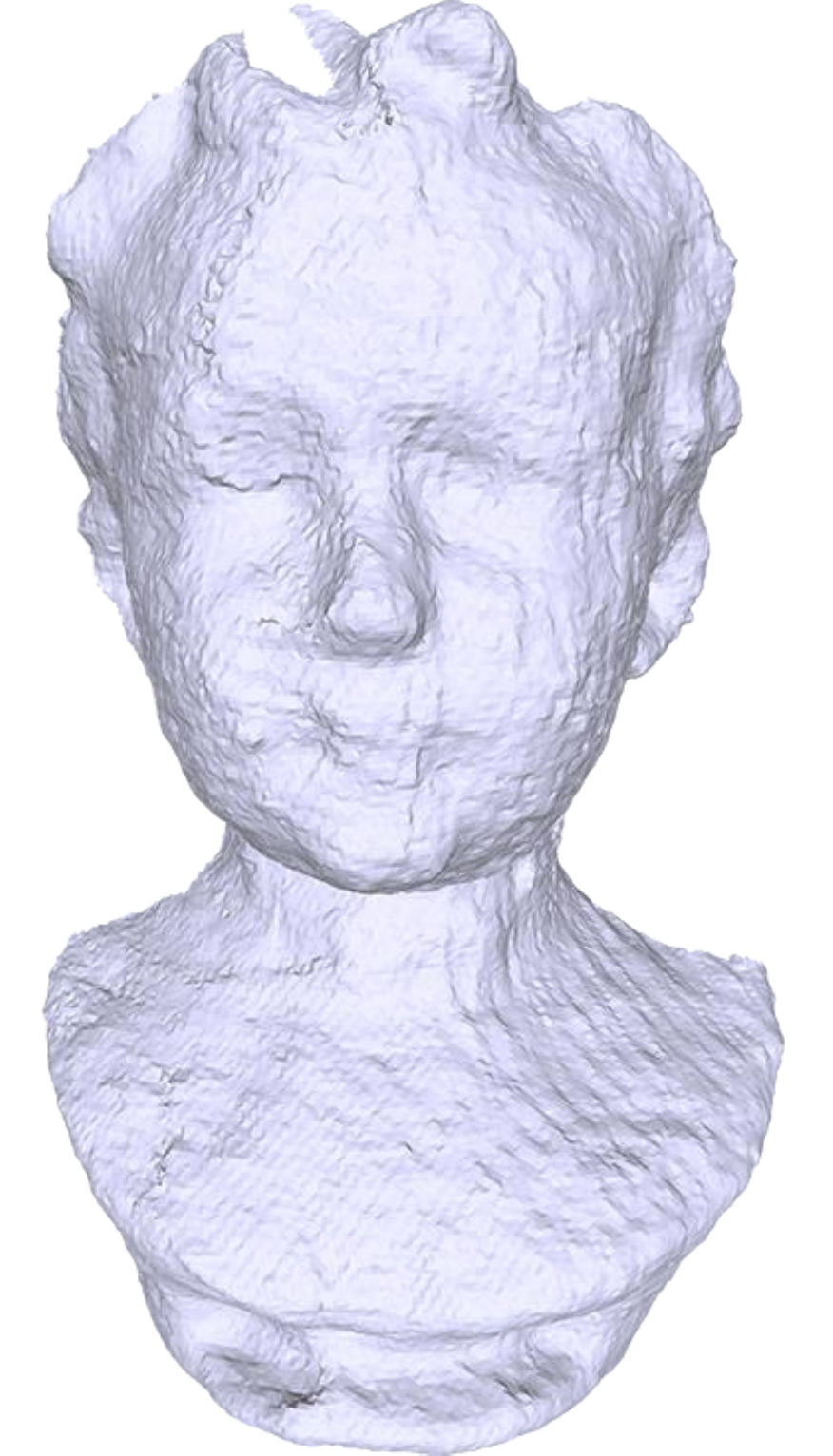}
			\includegraphics[width=2.2cm]{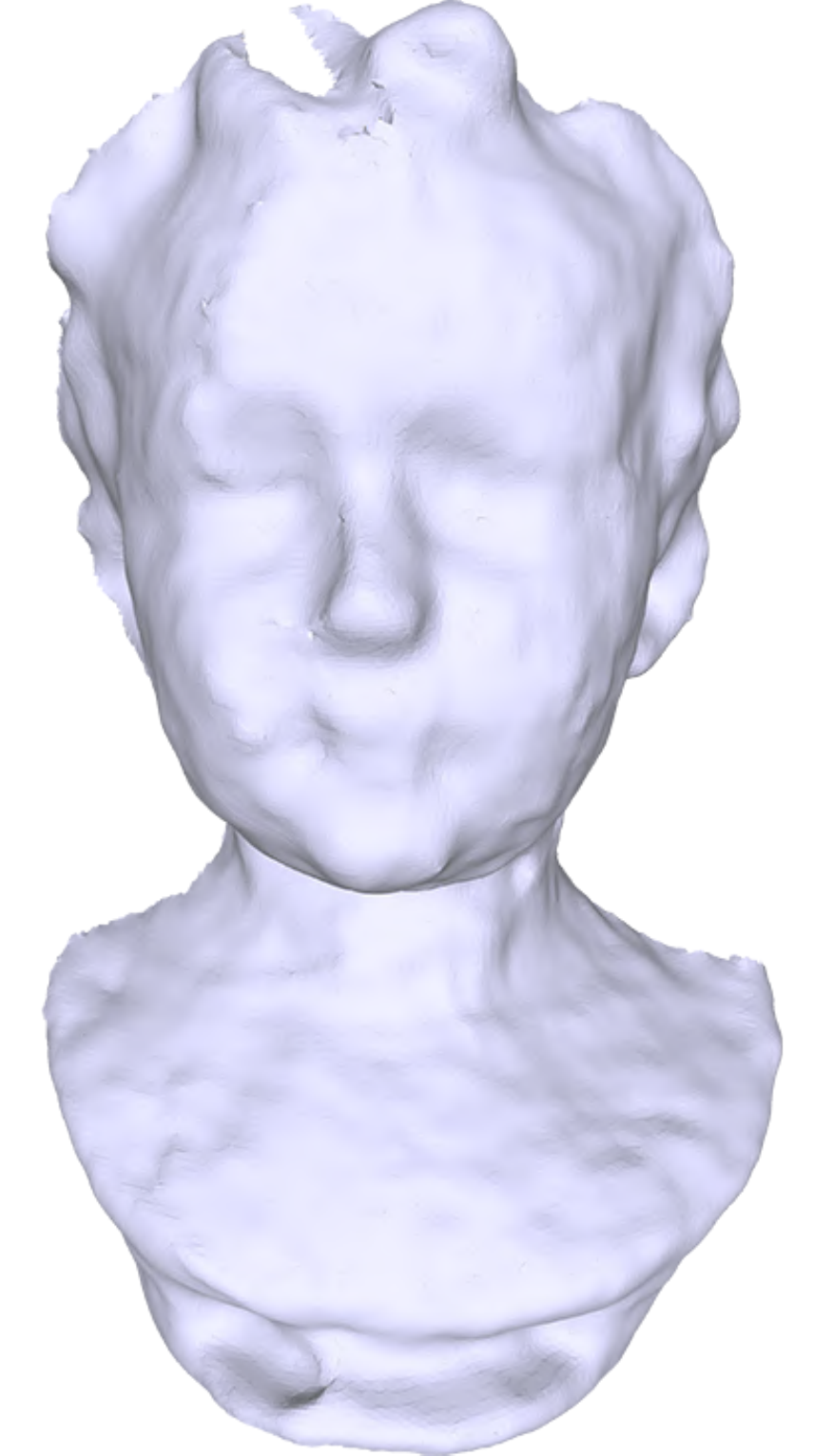}
			\includegraphics[width=2.2cm]{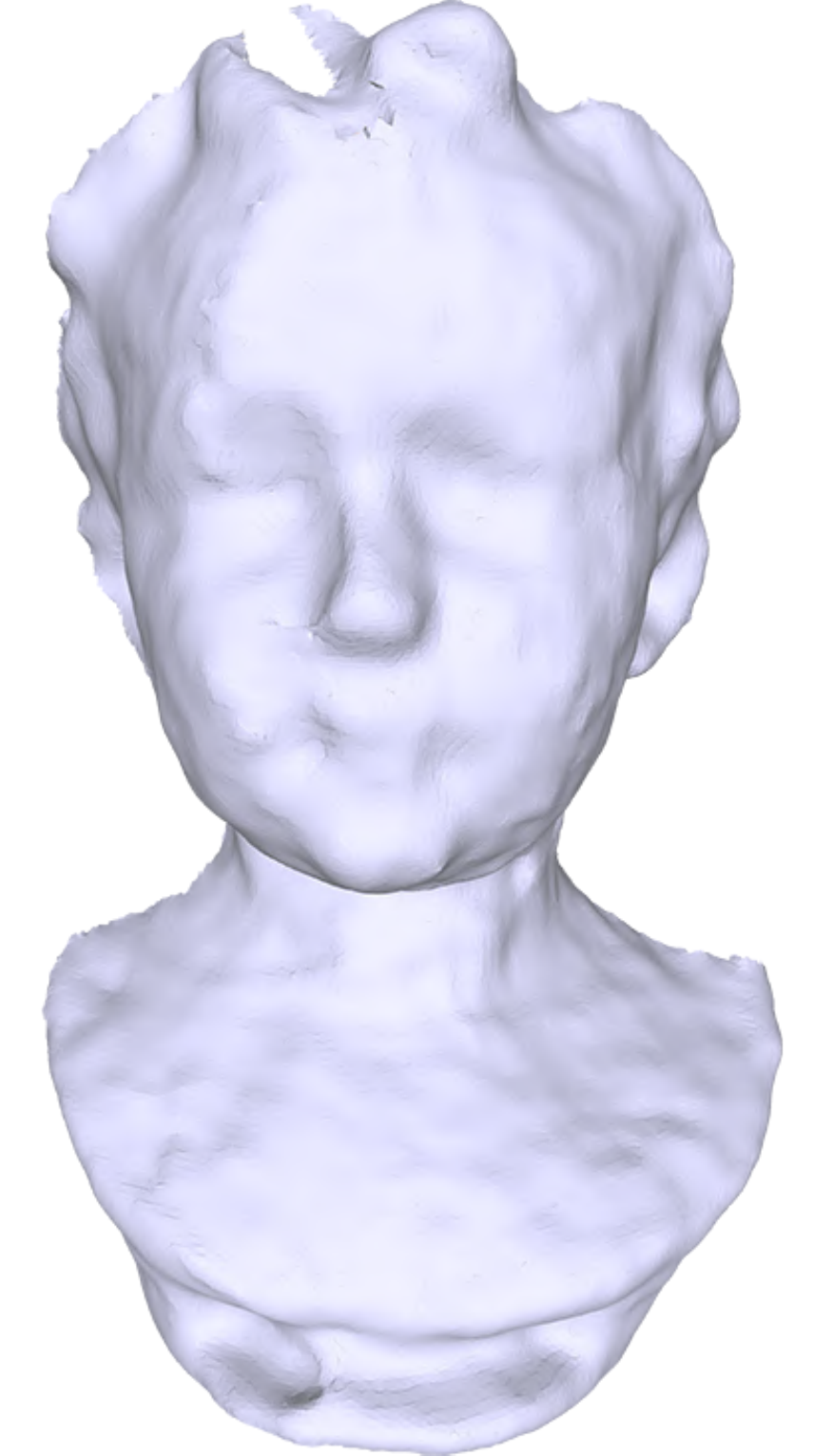}
			\includegraphics[width=2.2cm]{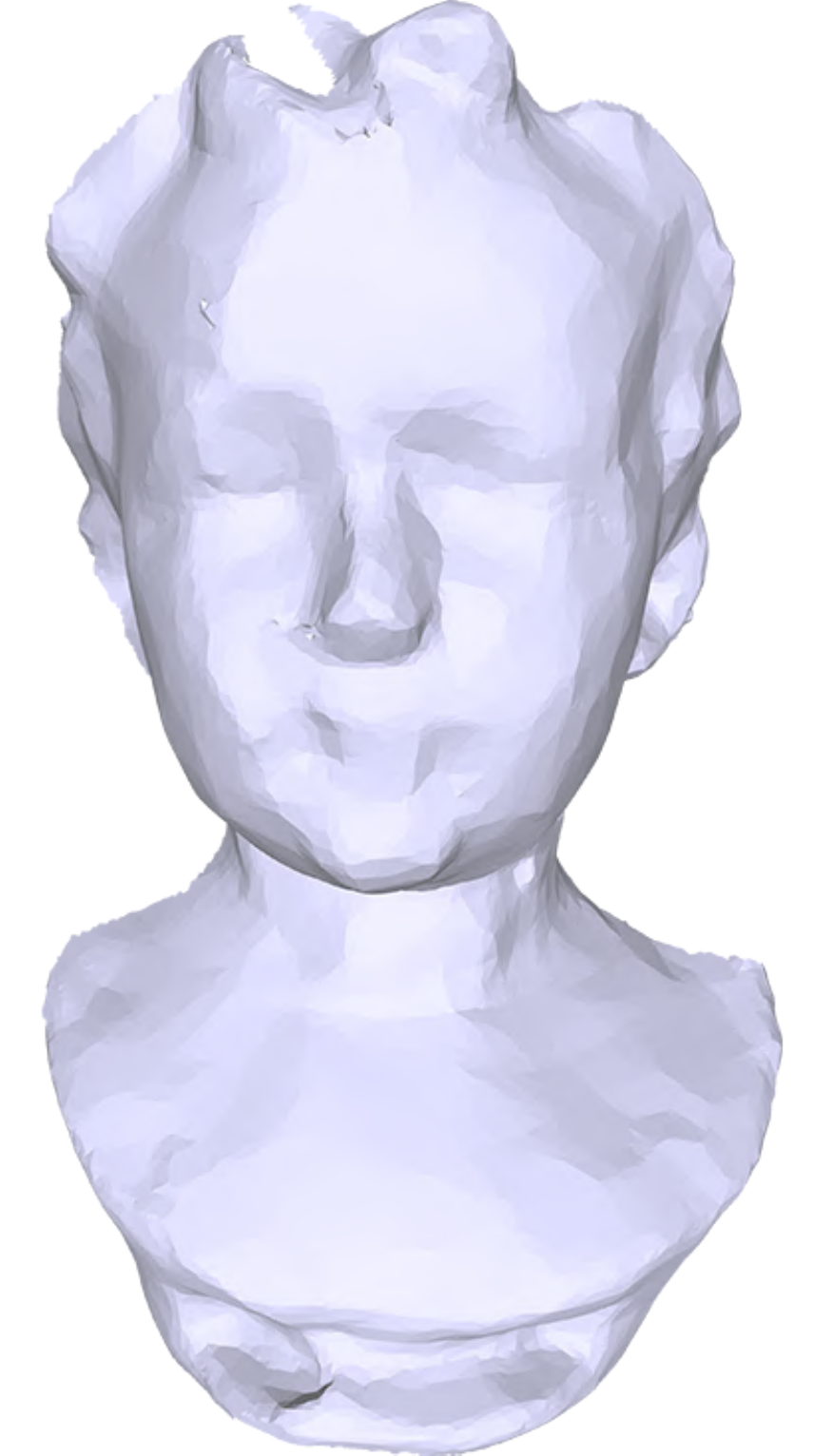}
			\includegraphics[width=2.2cm]{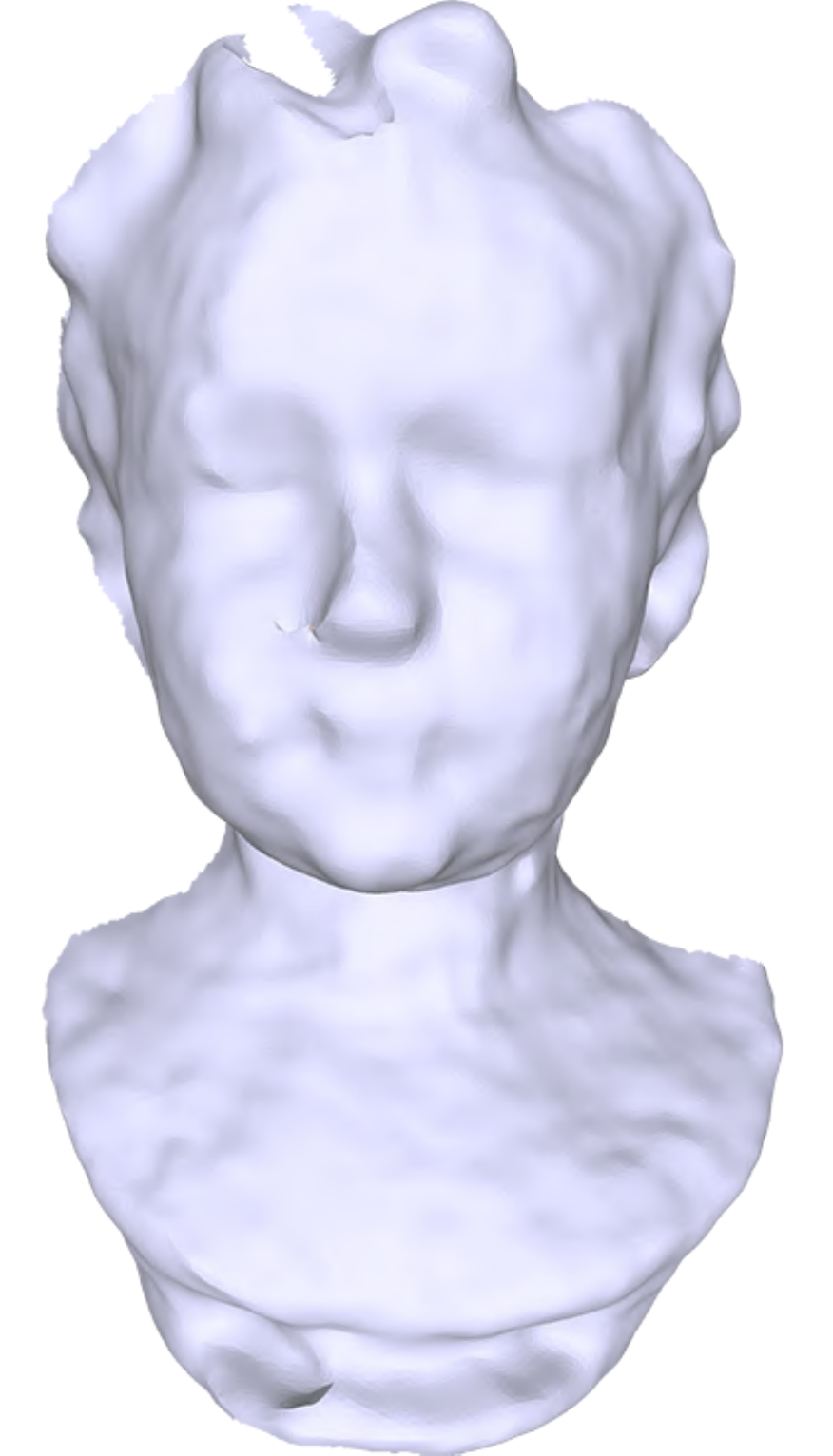}
			\includegraphics[width=2.2cm]{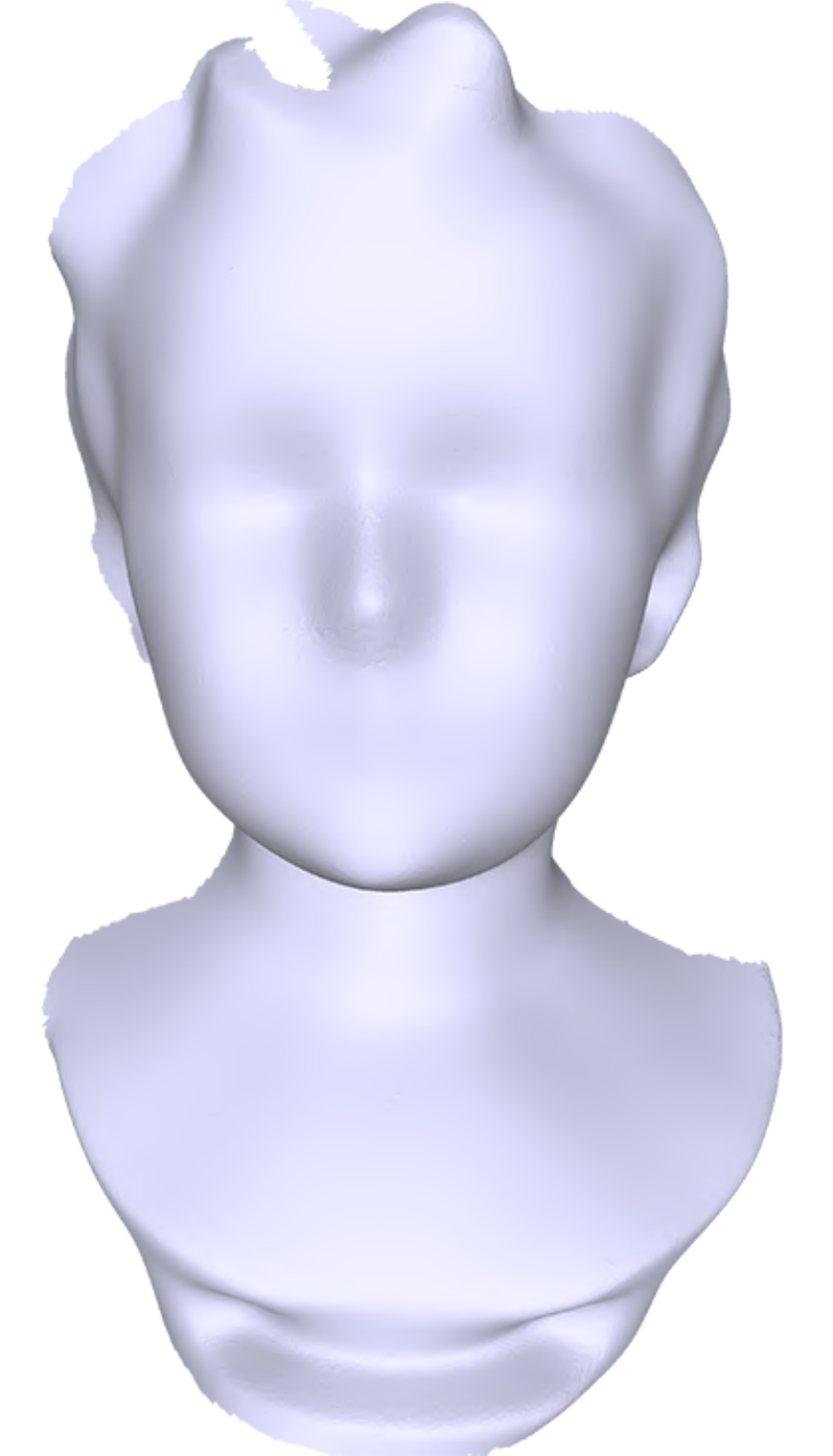}
			\includegraphics[width=2.2cm]{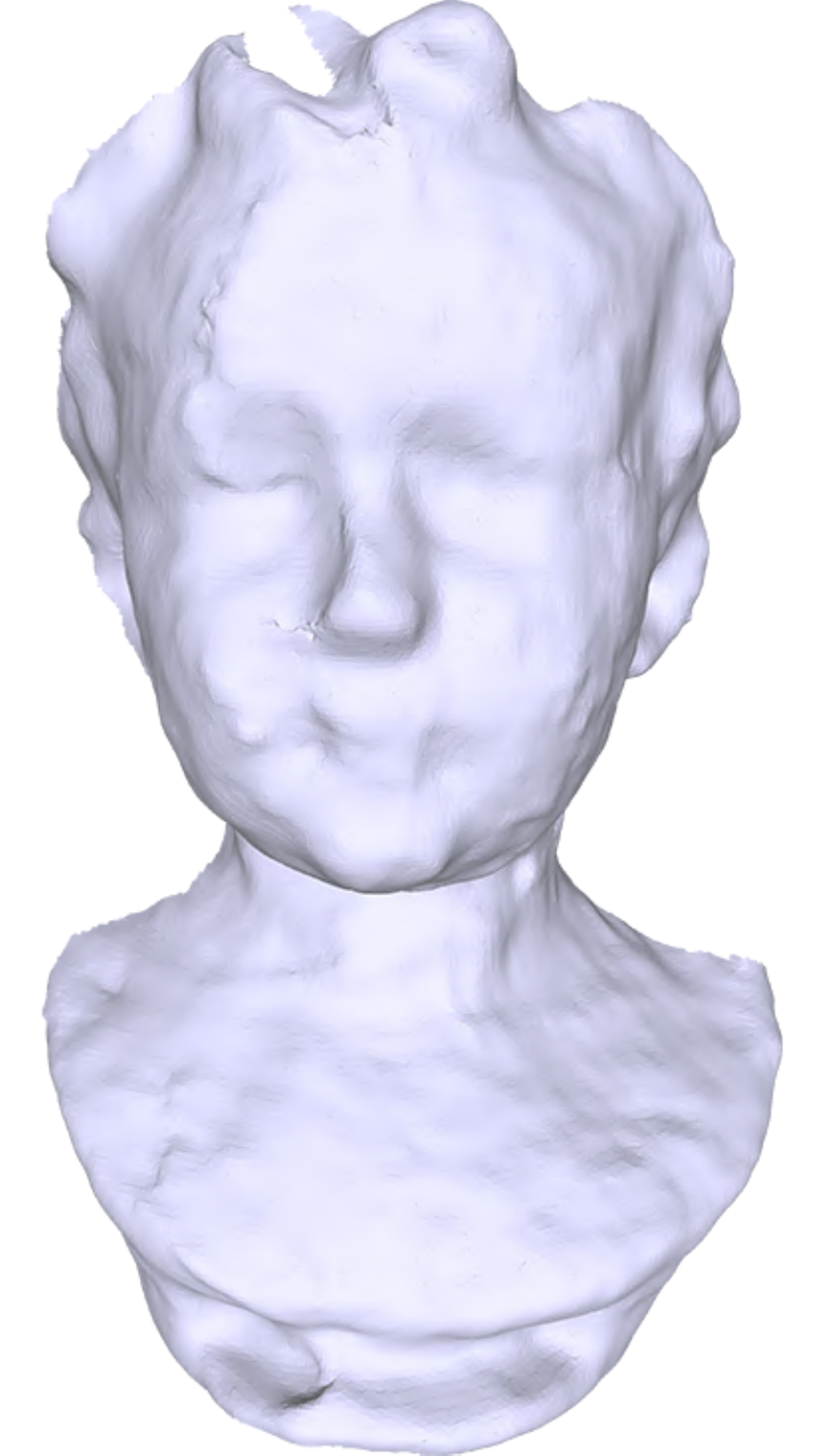}
			\includegraphics[width=2.2cm]{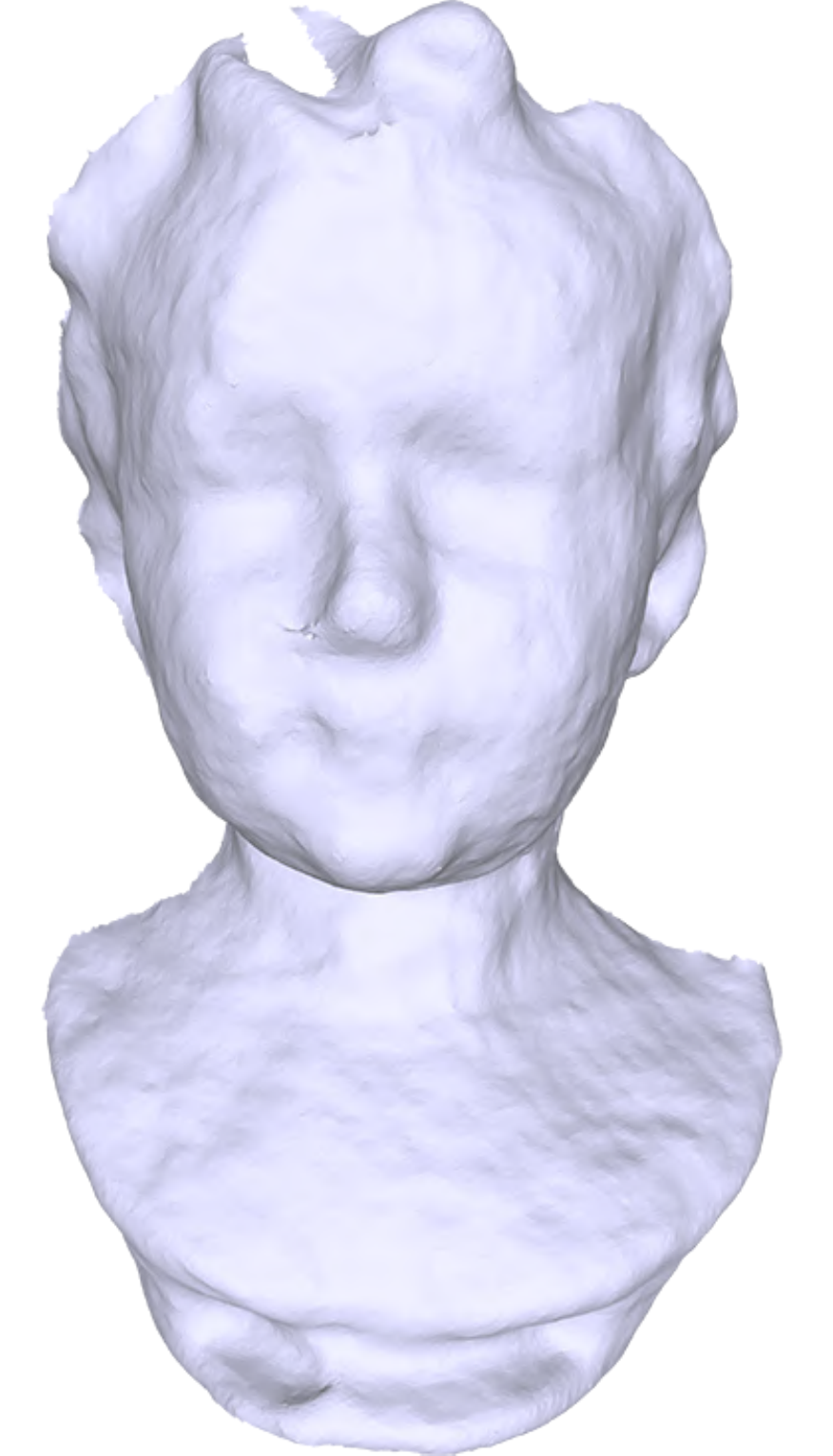}	\\
			\includegraphics[width=2.2cm]{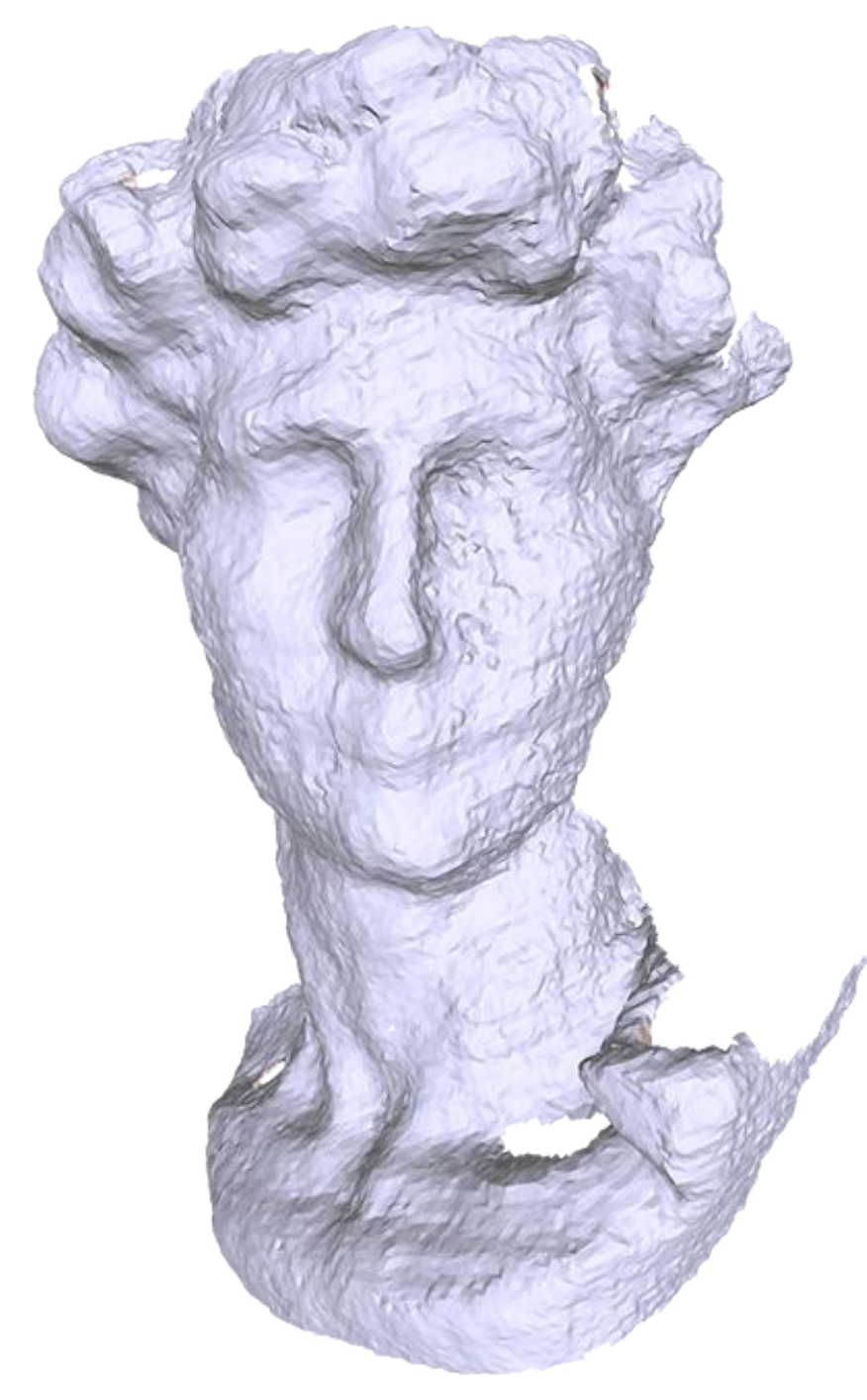}
			\includegraphics[width=2.2cm]{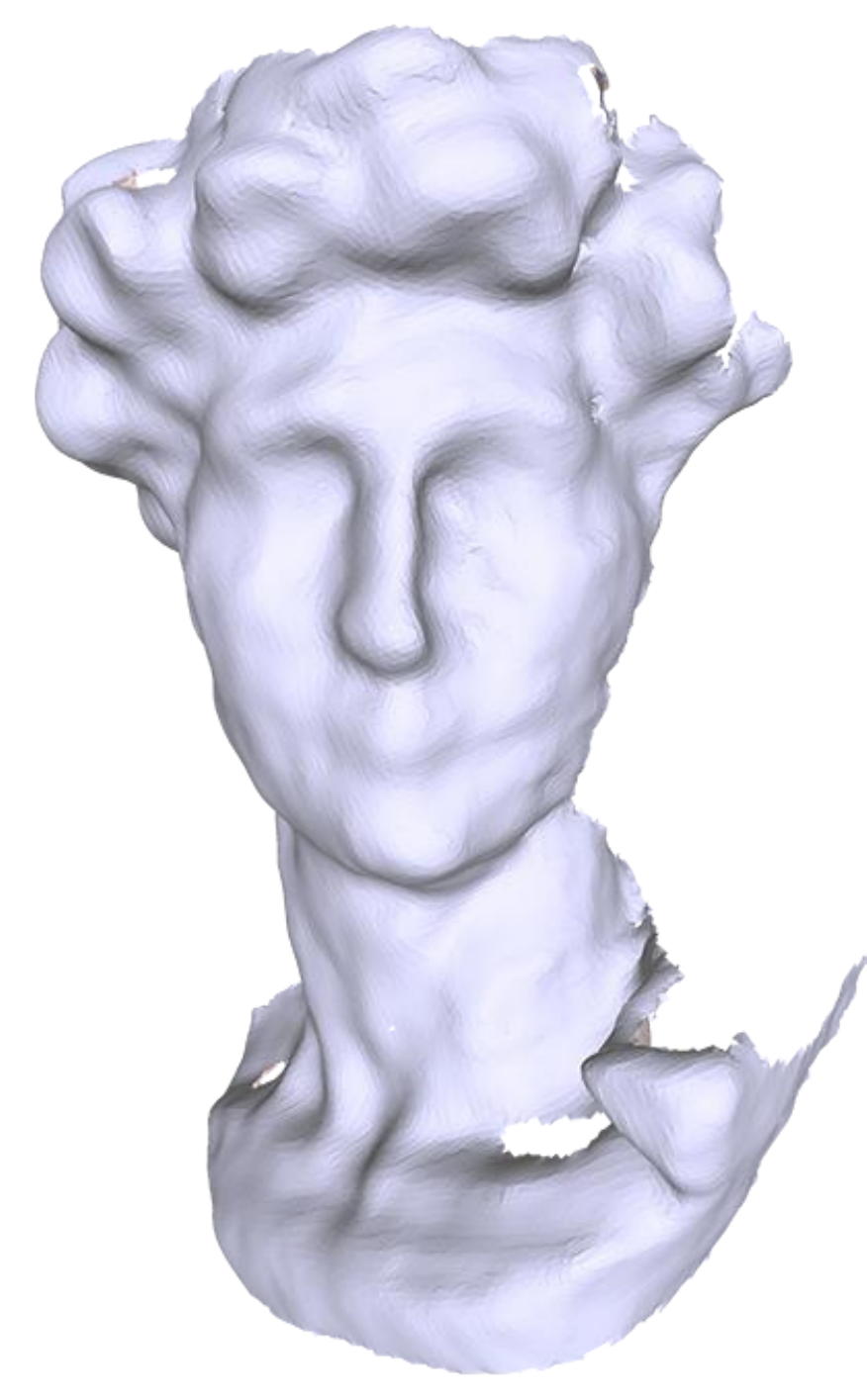}
			\includegraphics[width=2.2cm]{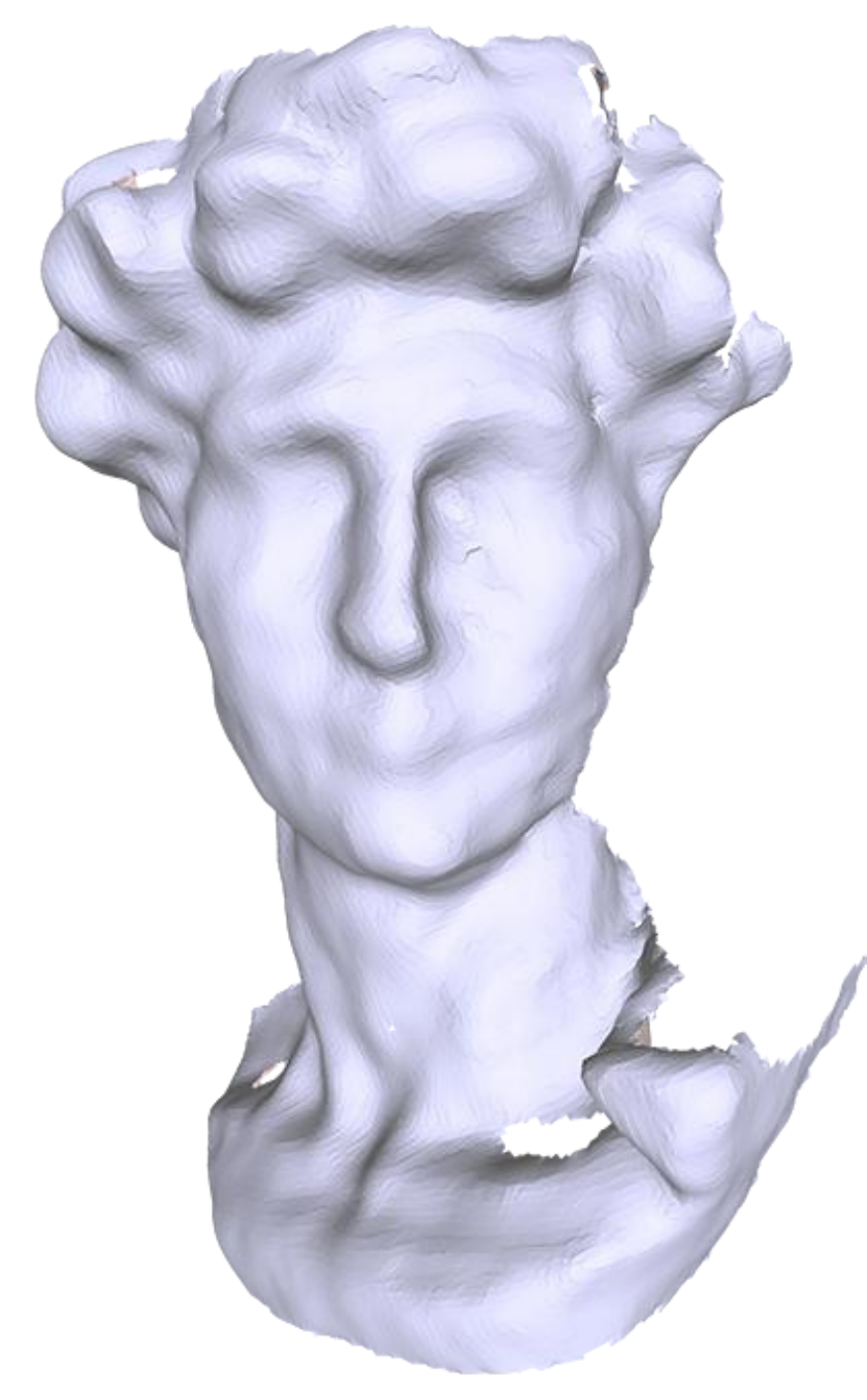}
			\includegraphics[width=2.2cm]{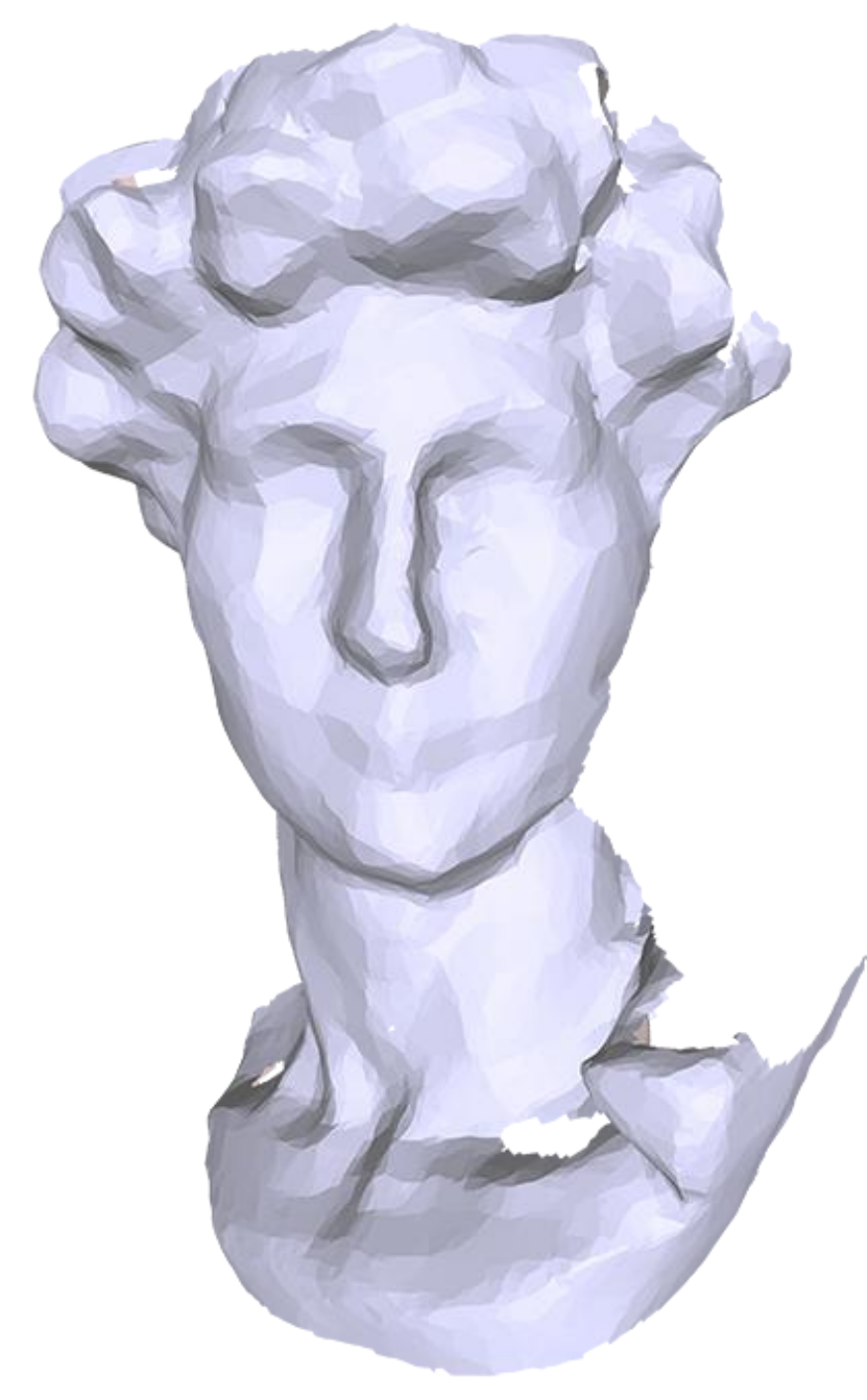}
			\includegraphics[width=2.2cm]{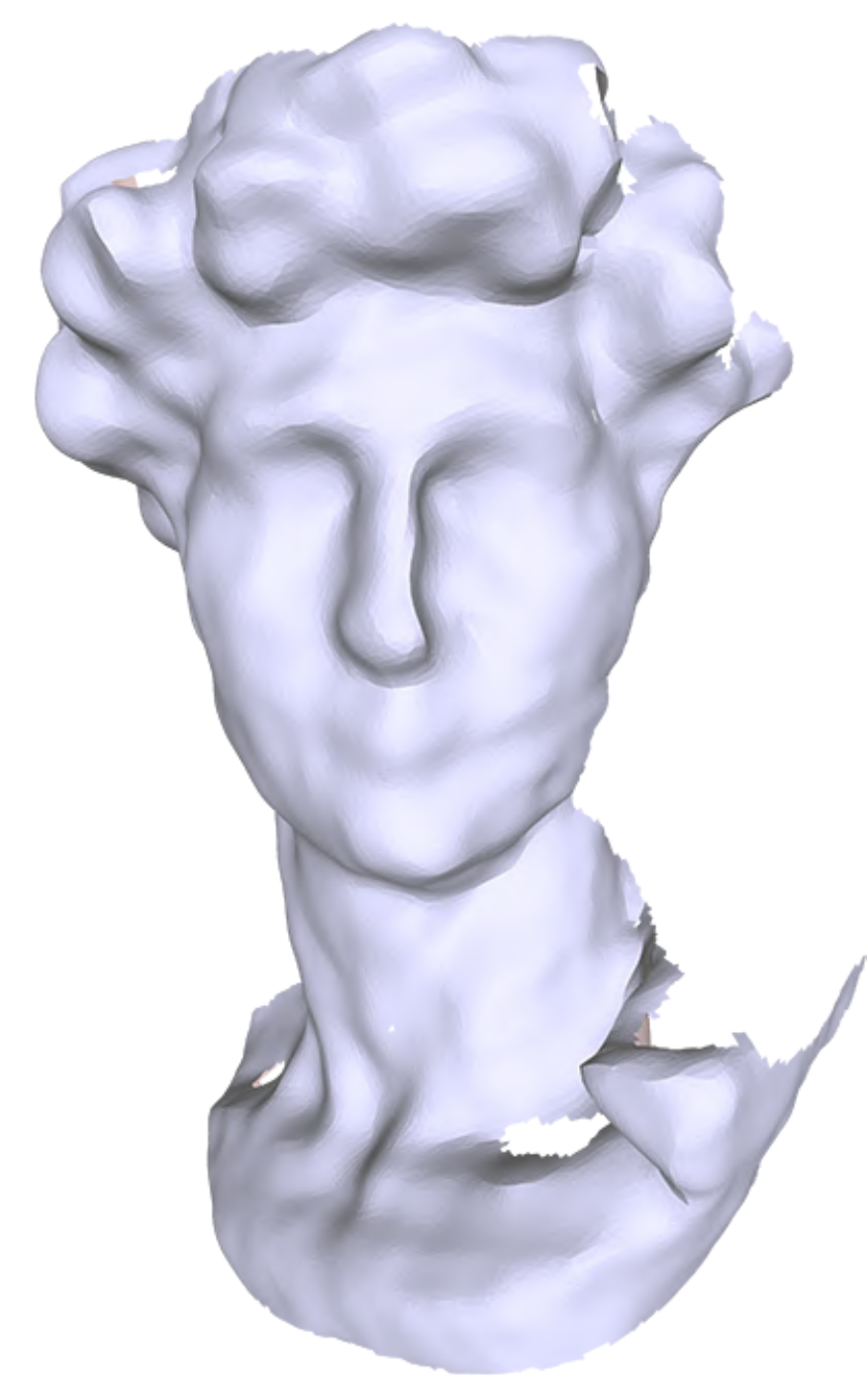}
			\includegraphics[width=2.2cm]{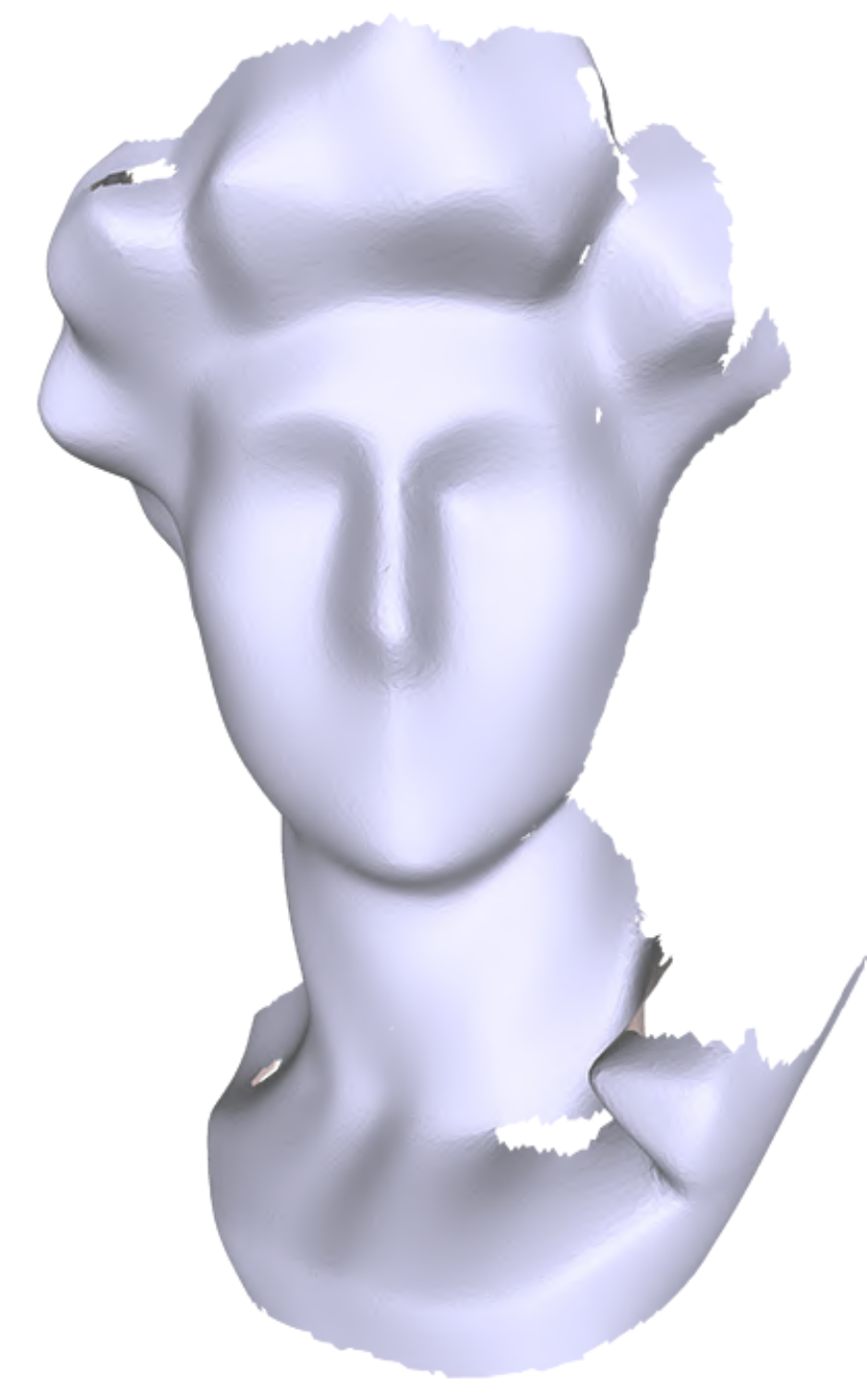}
			\includegraphics[width=2.2cm]{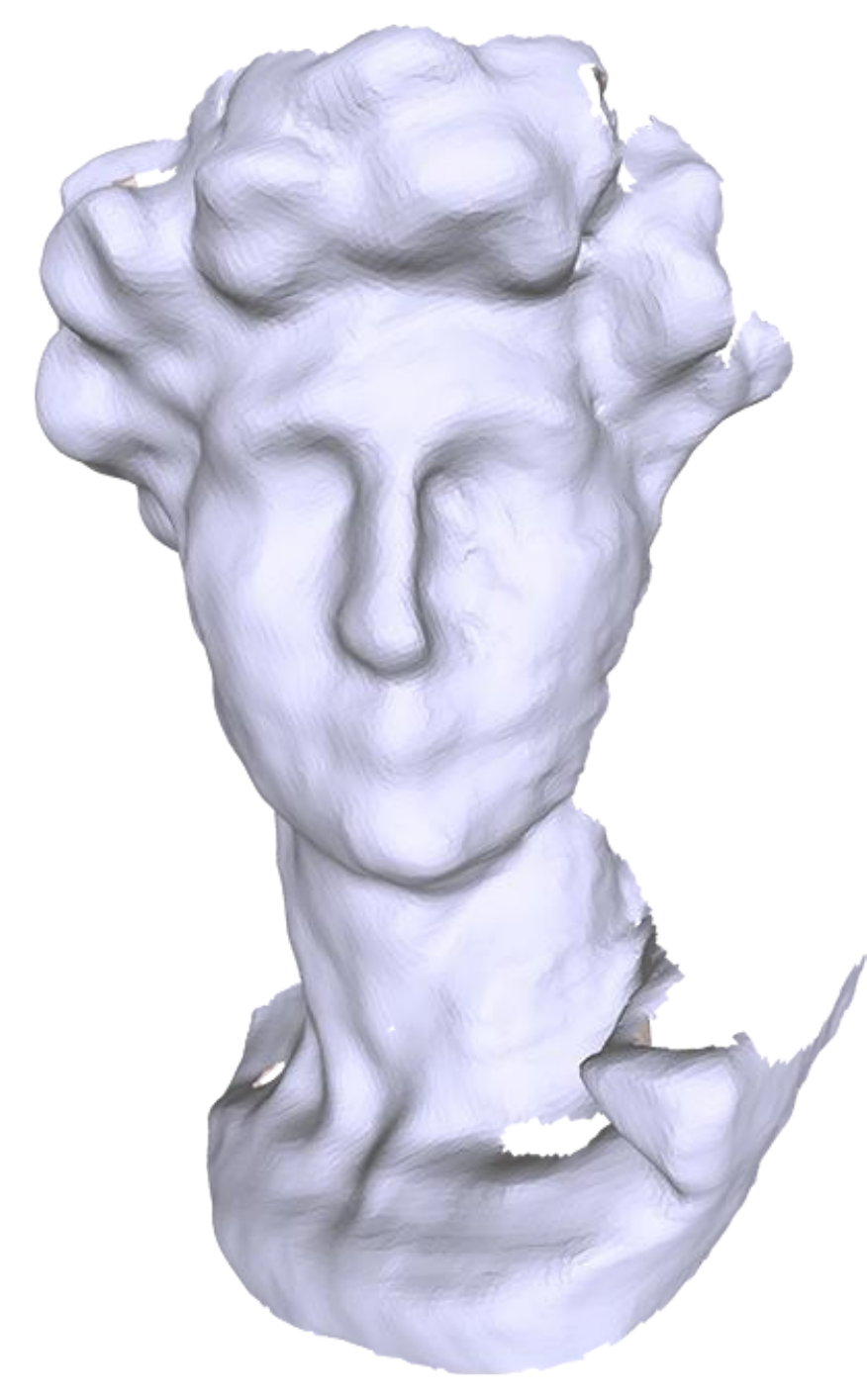}
			\includegraphics[width=2.2cm]{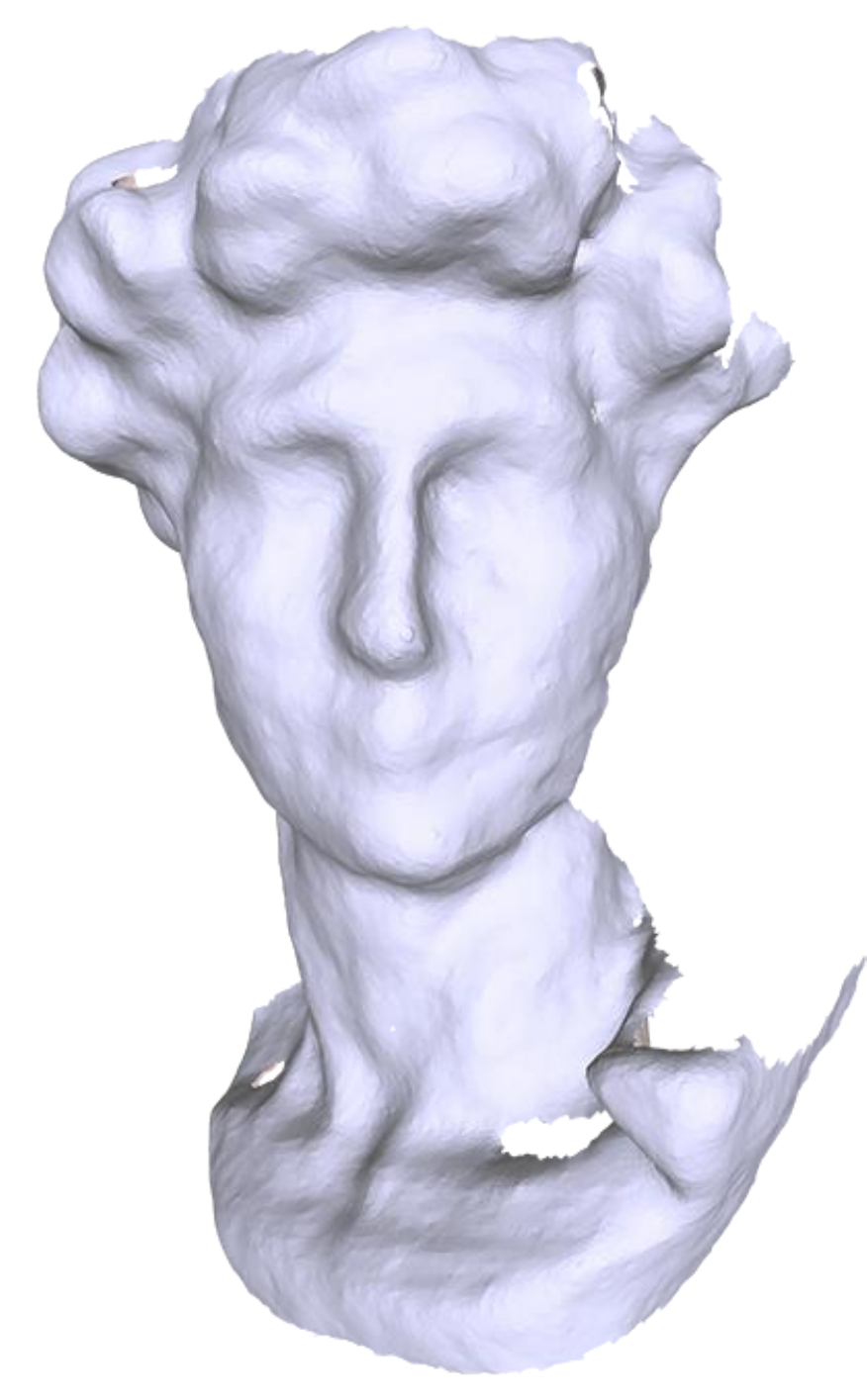} \\
		\subfigure[Raw scan]{
			\includegraphics[width=2.2cm]{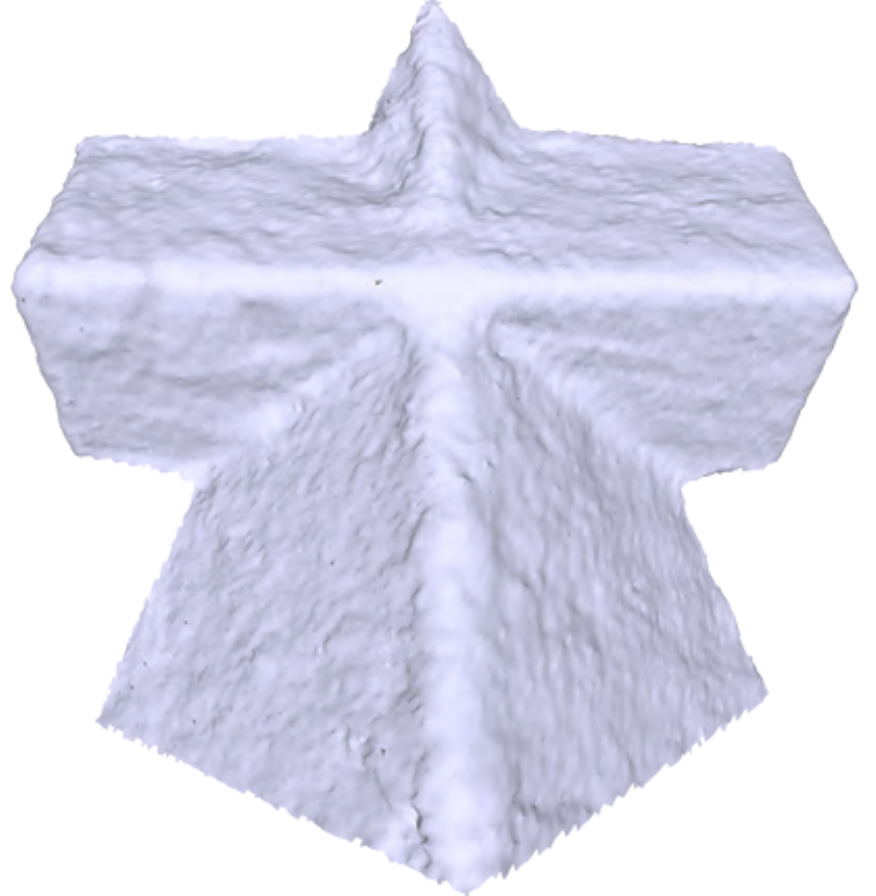}}
			\subfigure[UNF~\cite{Sun2007}]{
			\includegraphics[width=2.2cm]{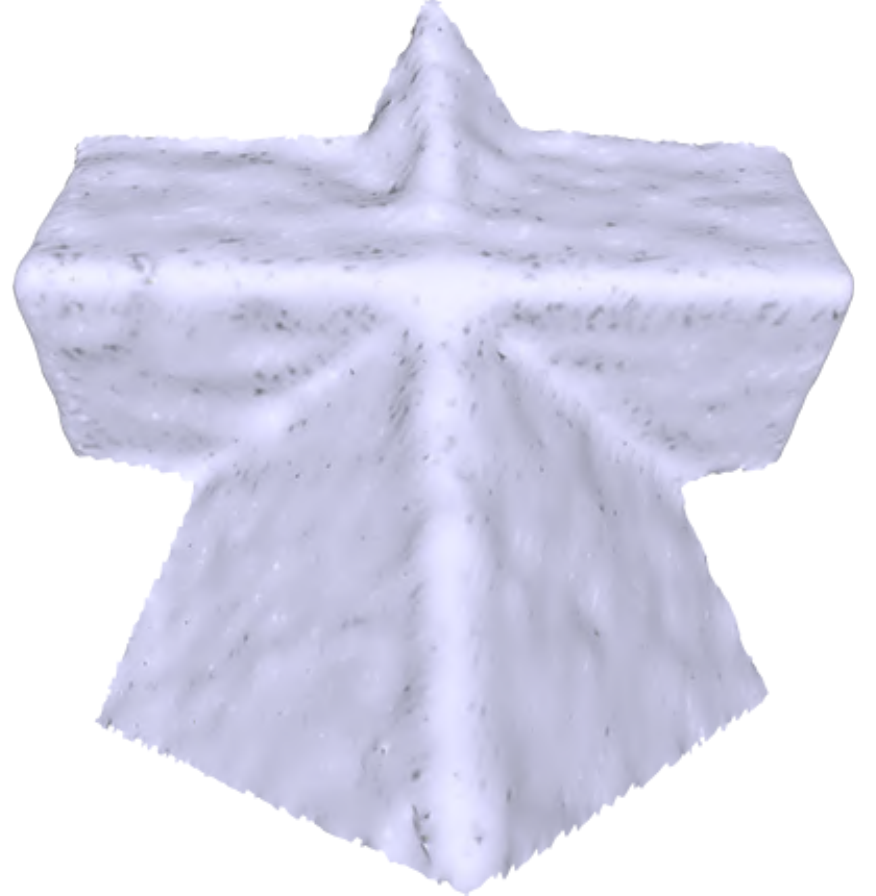}}
	\subfigure[BNF~\cite{Zheng2011}]{			
			\includegraphics[width=2.2cm]{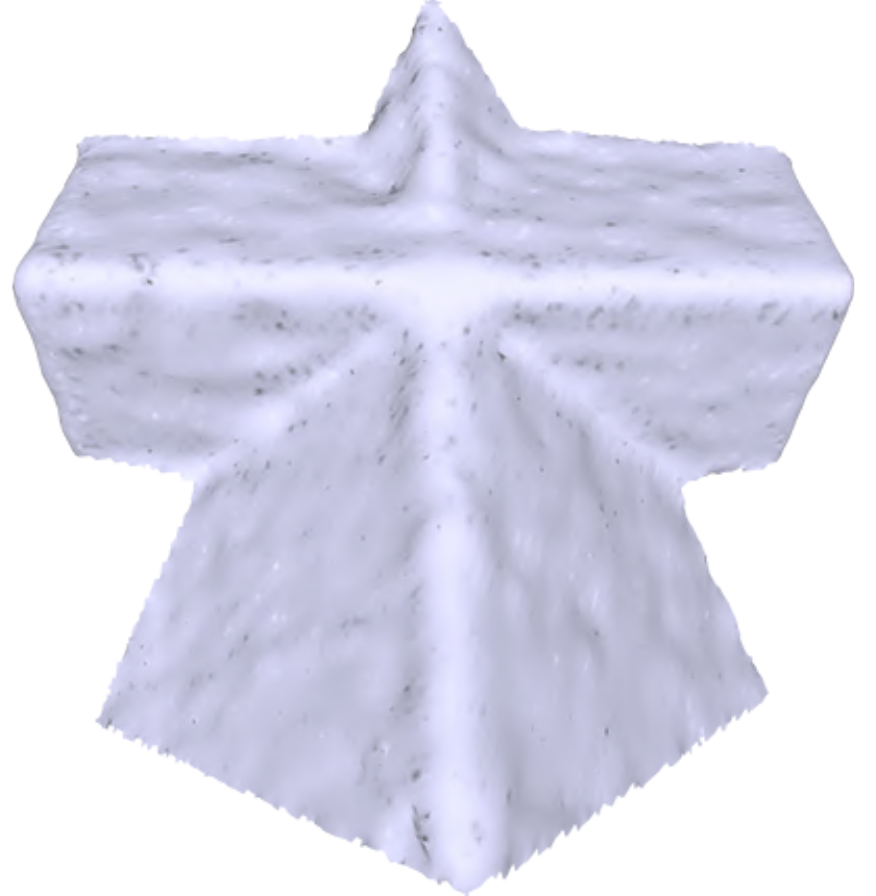}}
	\subfigure[L0~\cite{He2013}]{
			\includegraphics[width=2.2cm]{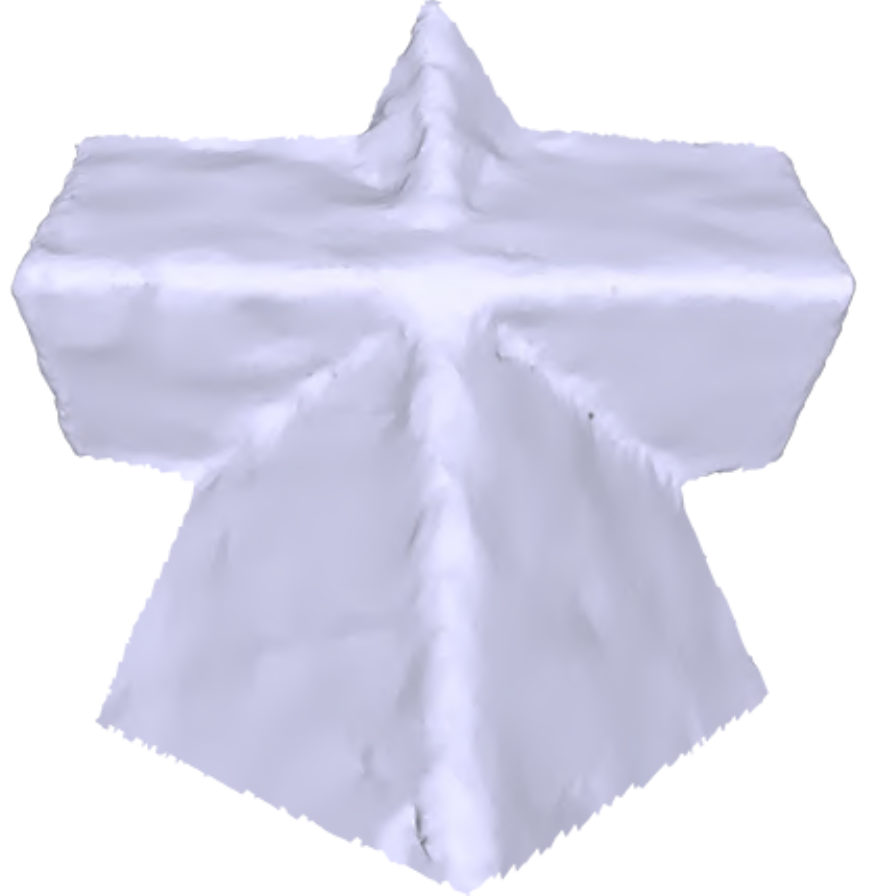}}
	\subfigure[GNF~\cite{Zhang2015}]{			
			\includegraphics[width=2.2cm]{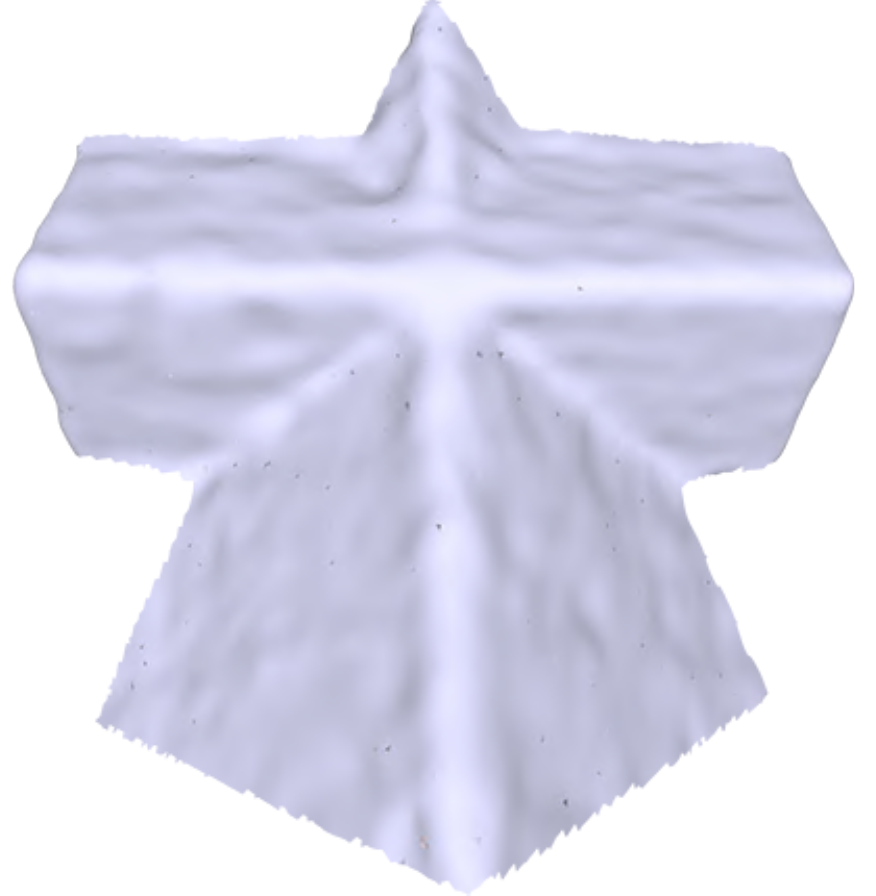}}
	\subfigure[CNR~\cite{Wang-2016-SA}]{			
			\includegraphics[width=2.2cm]{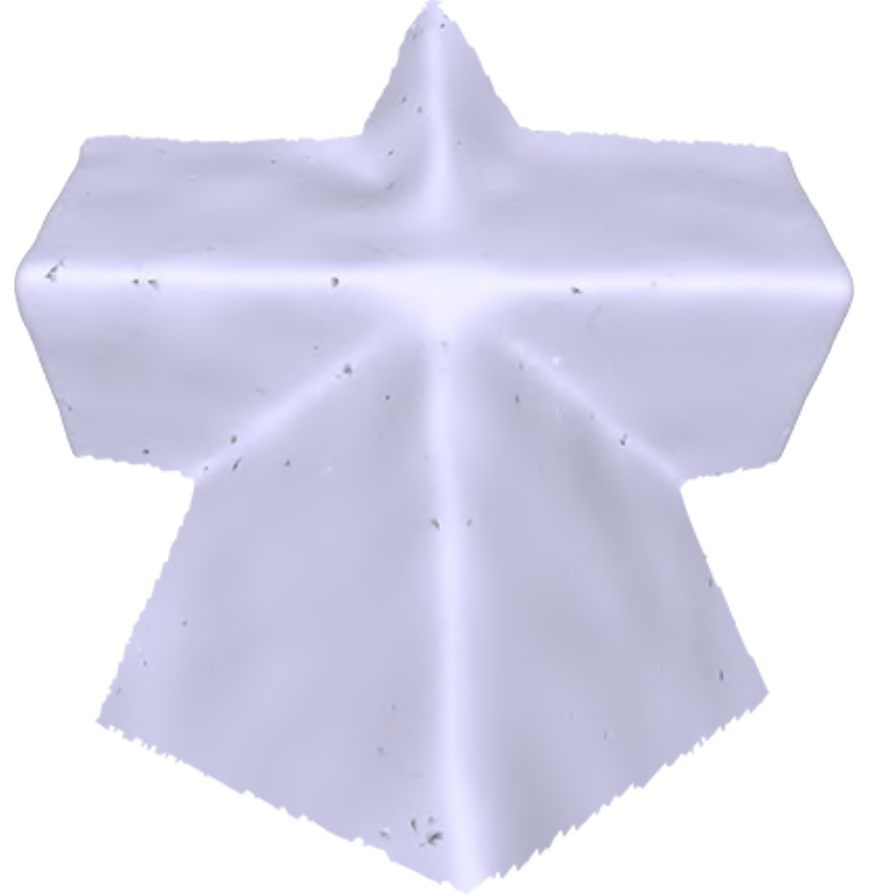}}
	\subfigure[NLLR~\cite{Li2018}]{					
			\includegraphics[width=2.2cm]{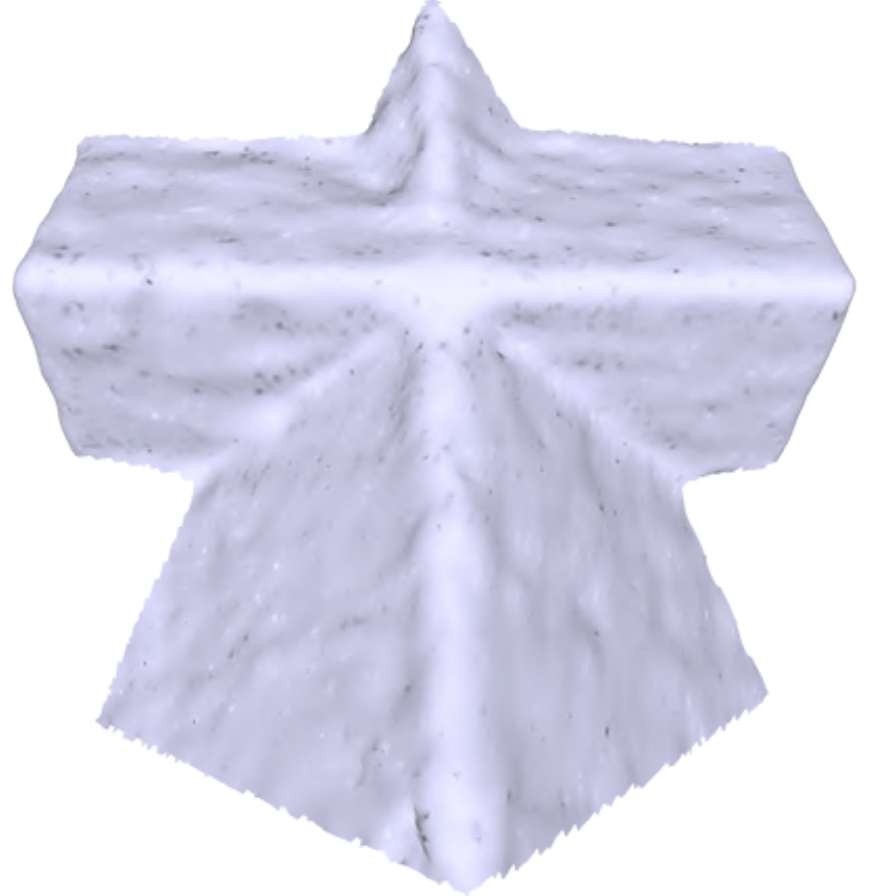}}
	\subfigure[Ours]{				
			\includegraphics[width=2.2cm]{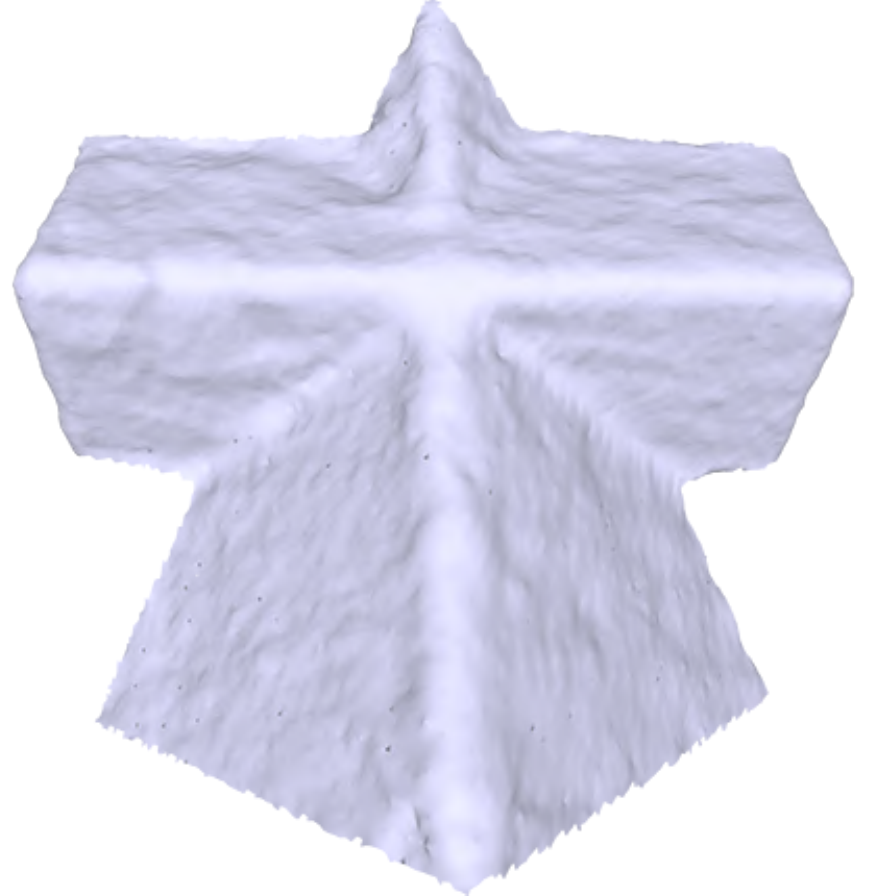}}
		\end{tabular}
	\end{center}
	\caption[res] 
	{ \label{fig:com_scan} 
		Results of models scanned by Microsoft Kinect. 
		}
\end{figure*}

\begin{figure} [htbp]
\centering
\includegraphics[width=1.6in]{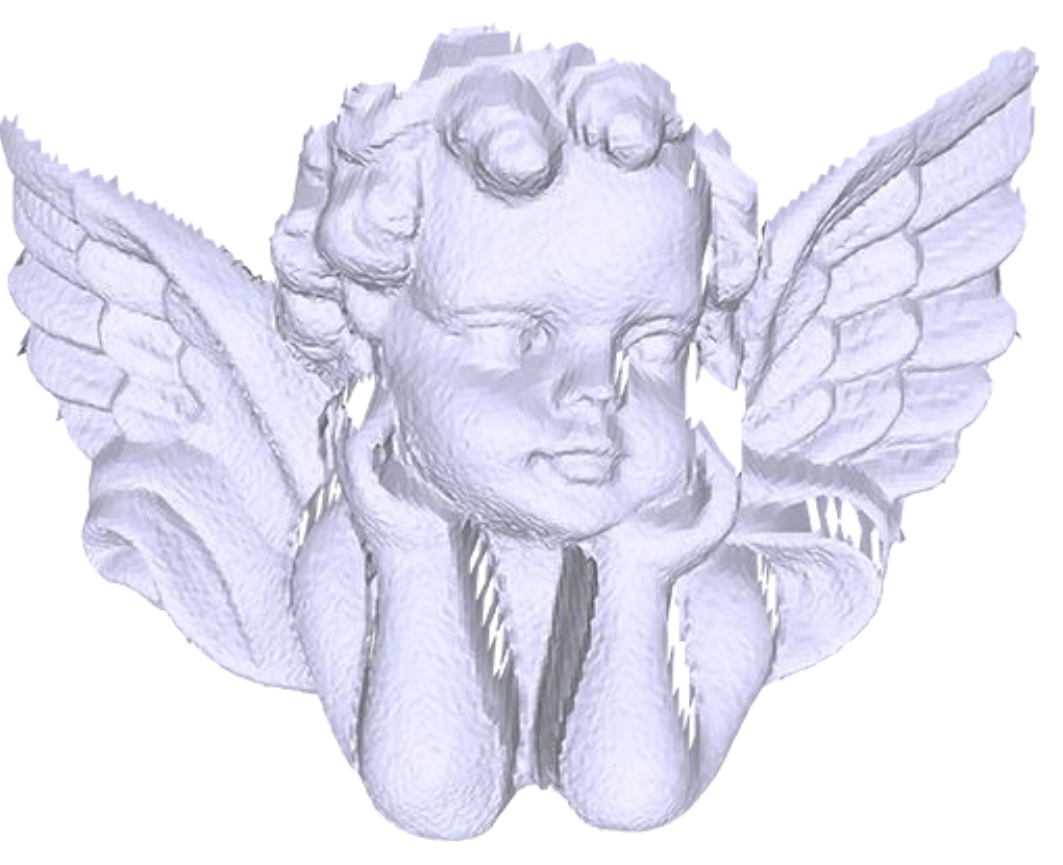}
\includegraphics[width=1.6in]{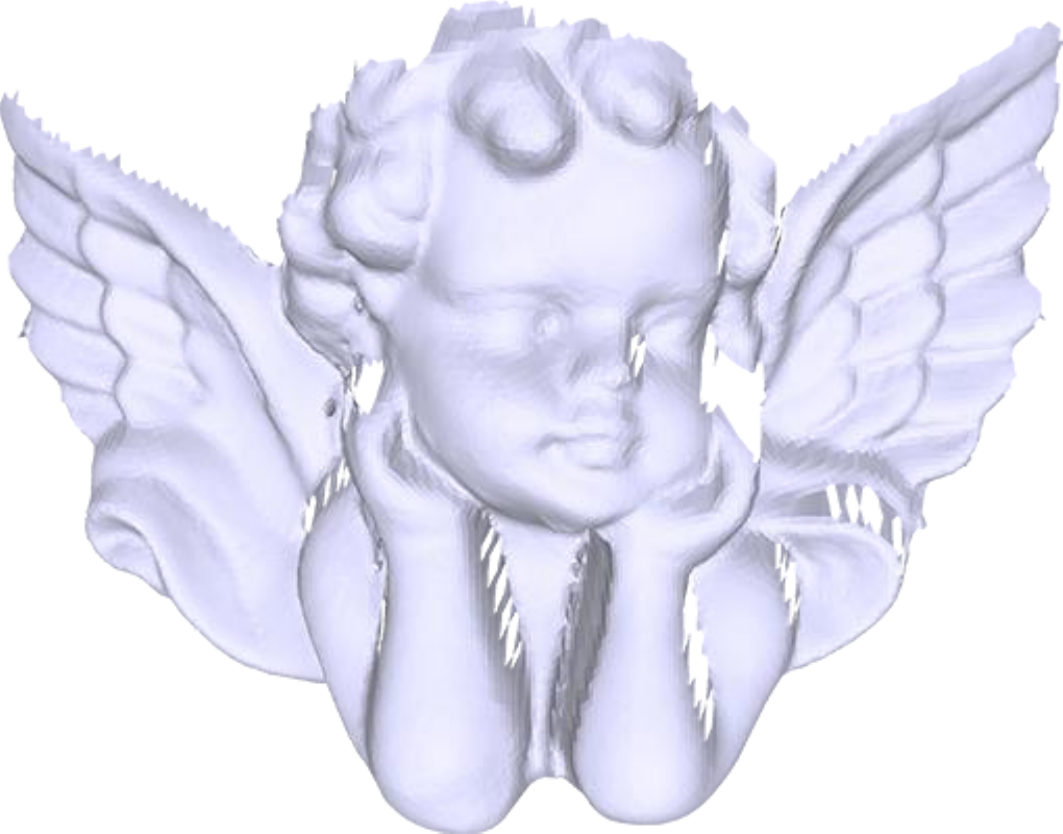}\\
\includegraphics[width=1.6in]{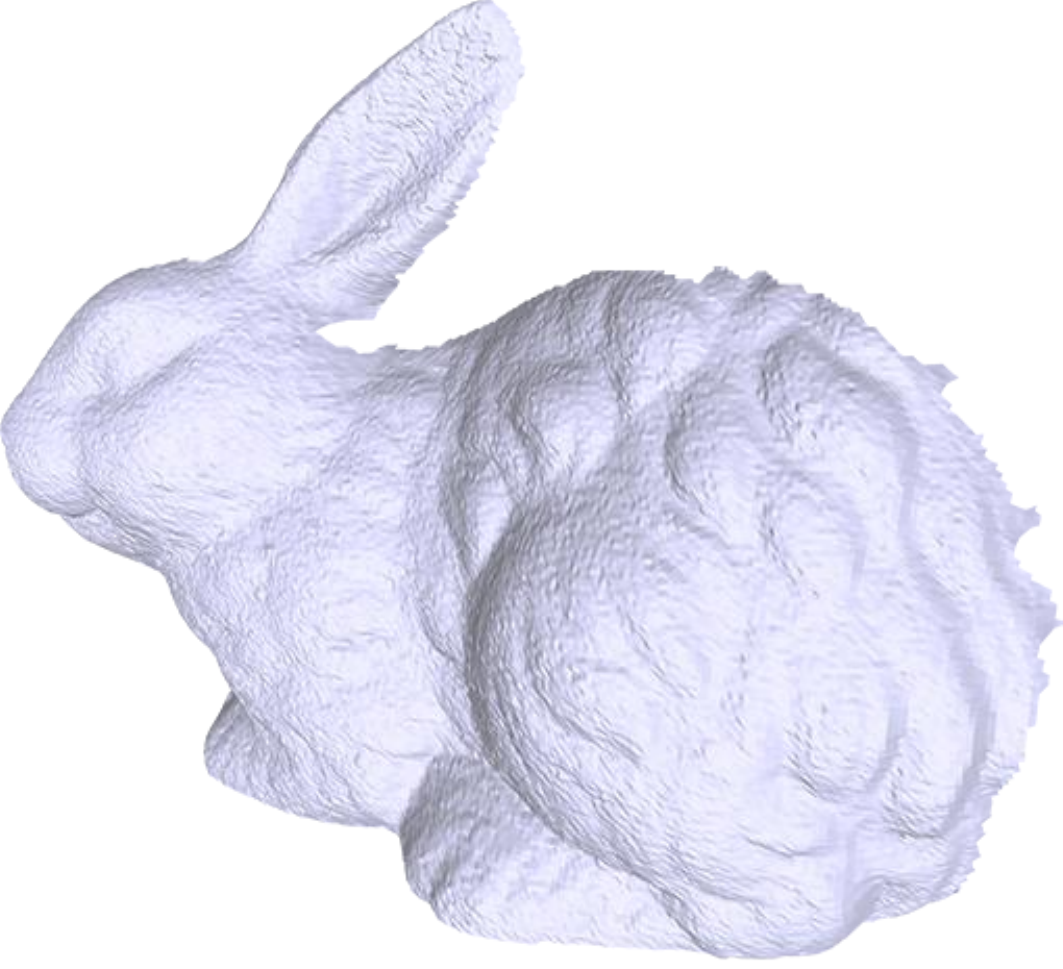}
\includegraphics[width=1.6in]{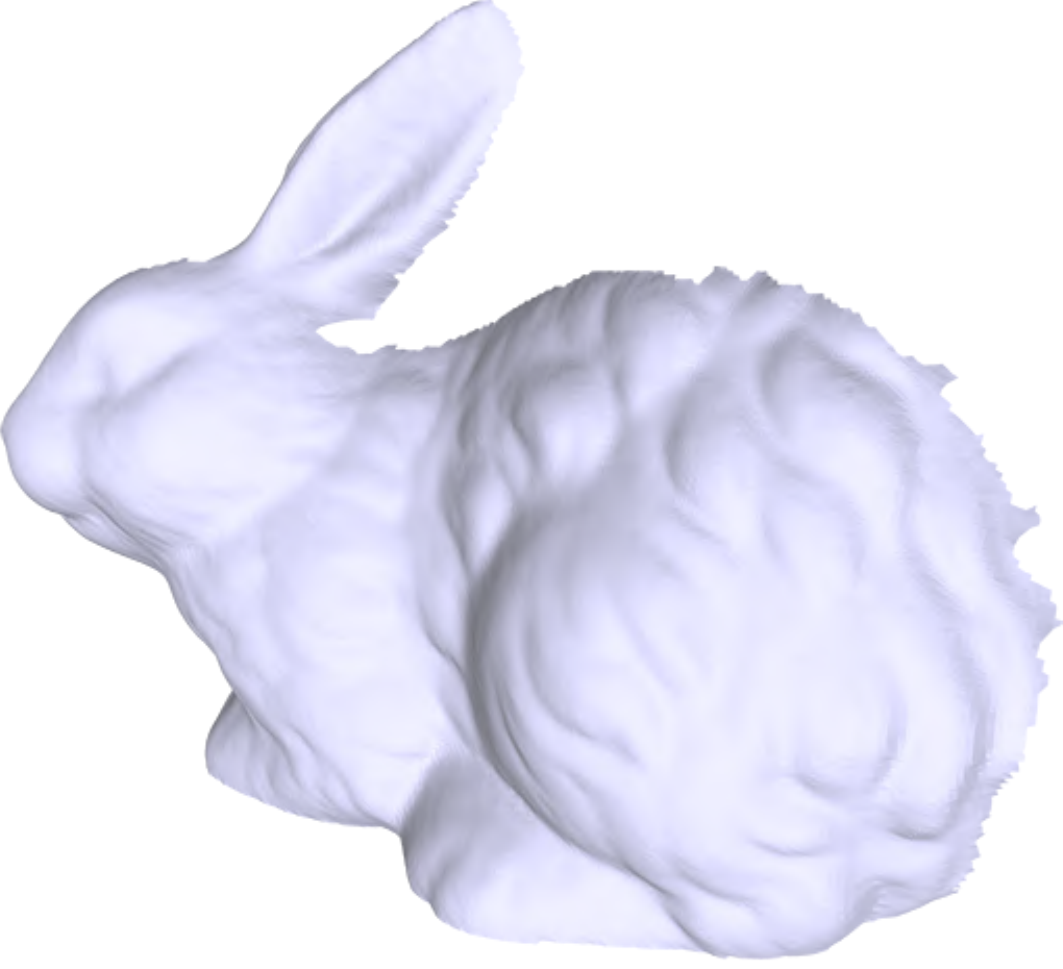}\\
\makebox[1.6in]{(a) Raw scan}
\makebox[1.6in]{(b) Our result}

\caption	{ \label{fig:com_scan2} 
		Results of scanned models: Angel and Rabbit. 
	}
\end{figure} 

\begin{table}[htbp]\scriptsize
    \centering
    \caption{Quantitative comparisons with the representative mesh smoothing methods.}\label{table1}
    \begin{tabular}{|l|l|l|l|l|}\hline
    Models & Methods & \tabincell{l}{MSAE\\$(\times10^{-2})$} & \tabincell{l}{$E_v$\\$(\times10^{-3})$} & Parameters \\
    \hline
    \tabincell{l}{Amadillo\\($ \sigma_{n}=0.5l_{e} $)\\(Figure~\ref{fig:com_amadillo0.5})\\$\left|V\right|$: 43243 \\$\left|F\right|$: 86482} &
    \tabincell{l}{ \uppercase\expandafter{BMF} \\
    \uppercase\expandafter{BNF} \\
    \uppercase\expandafter{CNR}\\
    \uppercase\expandafter{GNF}\\
    \uppercase\expandafter{UNF} \\
    \uppercase\expandafter{L0}\\
    \uppercase\expandafter{NLLR}\\
    \uppercase\expandafter{OURS}\\
    }
    & \tabincell{l}{47.54\\53.58\\34.28\\44.90\\49.20\\9.59\\32.18\\\textbf{9.33}}
    & \tabincell{l}{3.34\\3.11\\3.39\\3.10\\3.07\\3.52\\\textbf{3.03}\\3.57}
	& \tabincell{l}{($ 30 $)\\$ (10,0.35,10) $\\$ (synthetic,10) $\\$(10,0.35,10) $\\$ (0.5,10,10) $\\$ (Default) $\\$ (Default) $\\($5 $)}
    \\ \hline
   \tabincell{l}{iHbunny\\($ \sigma_{n}=0.5l_{e} $)\\(Figure~\ref{fig:somemodels} row 1)\\$\left|V\right|$: 34834 \\$\left|F\right|$: 69451} &
 \tabincell{l}{ \uppercase\expandafter{BMF} \\
    \uppercase\expandafter{BNF} \\
    \uppercase\expandafter{CNR}\\
    \uppercase\expandafter{GNF}\\
    \uppercase\expandafter{UNF} \\
    \uppercase\expandafter{L0}\\
    \uppercase\expandafter{NLLR}\\
    \uppercase\expandafter{OURS}\\
    }
    & \tabincell{l}{5.07\\3.57\\29.7\\ \textbf{3.39}\\3.87\\4.32\\25.24\\6.25}
    & \tabincell{l}{4.99\\4.39\\3.96\\4.25\\4.38\\4.70\\ \textbf{3.58}\\	4.32}
   	& \tabincell{l}{($ 30 $)\\$ (10,0.35,20) $\\$ (synthetic,10) $\\$(10,0.35,20) $\\$ (0.4,20,10) $\\$ (Default) $\\$ (Default) $\\($5 $)}
       \\ \hline
      \tabincell{l}{Nicolo\\($ \sigma_{n}=0.2l_{e} $)\\(Figure~\ref{fig:somemodels} row 2)\\$\left|V\right|$: 14846 \\$\left|F\right|$: 29437} &
    \tabincell{l}{ \uppercase\expandafter{BMF} \\
       \uppercase\expandafter{BNF} \\
       \uppercase\expandafter{CNR}\\
       \uppercase\expandafter{GNF}\\
       \uppercase\expandafter{UNF} \\
       \uppercase\expandafter{L0}\\
       \uppercase\expandafter{NLLR}\\
       \uppercase\expandafter{OURS}\\
       }
       & \tabincell{l}{16.20 \\8.97 \\8.56 \\9.42 \\9.77 \\4.44 \\7.13 \\ \textbf{3.35} }
       & \tabincell{l}{0.96 \\0.92 \\0.96 \\0.94  \\0.93 \\0.93 \\0.93 \\ \textbf{0.89}  }
      	& \tabincell{l}{($ 30 $)\\$ (10,0.35,20) $\\$ (synthetic,3) $\\$(10,0.35,20) $\\$ (0.5,20,10) $\\$ (Default) $\\$ (Default) $\\($3 $)}
          \\ \hline
        \tabincell{l}{Nicolo\\($ \sigma_{n}=0.5l_{e} $)\\(Figure~\ref{fig:somemodels} row 3)\\$\left|V\right|$: 14846 \\$\left|F\right|$: 29437} &
        \tabincell{l}{ \uppercase\expandafter{BMF} \\
                 \uppercase\expandafter{BNF} \\
                 \uppercase\expandafter{CNR}\\
                 \uppercase\expandafter{GNF}\\
                 \uppercase\expandafter{UNF} \\
                 \uppercase\expandafter{L0}\\
                 \uppercase\expandafter{NLLR}\\
                 \uppercase\expandafter{OURS}\\
                 }
                 & \tabincell{l}{56.33 \\65.33 \\41.35 \\58.15  \\61.66 \\\textbf{5.67} \\38.44 \\10.33                 }
                 & \tabincell{l}{1.13 \\1.00 \\1.22 \\0.93 \\0.94 \\1.02 \\\textbf{0.92} \\1.08 }
                	& \tabincell{l}{($ 50 $)\\$ (10,0.45,20) $\\$ (synthetic,10) $\\$(10,0.45,20) $\\$ (0.45,20,10) $\\$ (Default) $\\$ (Default) $\\($5 $)}
                    \\ \hline
        \tabincell{l}{Vaselion\\($ \sigma_{n}=0.2l_{e}$)\\(Figure~\ref{fig:somemodels} row 4)\\$\left|V\right|$: 38728 \\$\left|F\right|$: 77452} &
        \tabincell{l}{ \uppercase\expandafter{BMF} \\
           \uppercase\expandafter{BNF} \\
           \uppercase\expandafter{CNR}\\
           \uppercase\expandafter{GNF}\\
           \uppercase\expandafter{UNF} \\
           \uppercase\expandafter{L0}\\
           \uppercase\expandafter{NLLR}\\
           \uppercase\expandafter{OURS}\\
           }
           & \tabincell{l}{38.33 \\33.76 \\45.25 \\44.75  \\39.07 \\35.84 \\35.00 \\ \textbf{14.49}   }
           & \tabincell{l}{34.47 \\14.20 \\14.49 \\14.40 \\13.84 \\15.30 \\ \textbf{12.51} \\14.80    }
          	& \tabincell{l}{($ 30 $)\\$ (10,0.35,10) $\\$ (synthetic,1) $\\$(10,0.35,10) $\\$ (0.5,10,10) $\\$ (Default) $\\$ (Default) $\\($3 $)}
              \\ \hline
  		\tabincell{l}{Vaselion\\($ \sigma_{n}=0.8l_{e} $)\\(Figure~\ref{fig:somemodels} row 5)\\$\left|V\right|$: 38728 \\$\left|F\right|$: 77452} &
        \tabincell{l}{ \uppercase\expandafter{BMF} \\
           \uppercase\expandafter{BNF} \\
           \uppercase\expandafter{CNR}\\
           \uppercase\expandafter{GNF}\\
           \uppercase\expandafter{UNF} \\
           \uppercase\expandafter{L0}\\
           \uppercase\expandafter{NLLR}\\
           \uppercase\expandafter{OURS}\\
           }
           & \tabincell{l}{167.9 \\171.9 \\126.0 \\166.6  \\158.7 \\ \textbf{14.72} \\82.25 \\32.35 
           }
           & \tabincell{l}{30.06 \\12.58 \\14.82 \\13.98 \\13.26 \\16.49 \\ \textbf{11.29} \\15.87          }
          	& \tabincell{l}{($ 50 $)\\$ (10,0.45,30) $\\$ (synthetic,10) $\\$(10,0.45,30) $\\$ (0.5,20,10) $\\$ (Default) $\\$ (Default) $\\($10 $)}
              \\ \hline
  		\tabincell{l}{Boy\\(kinect fusion) \\(Figure~\ref{fig:com_scan} row 1)\\$\left|V\right|$: 76866 \\$\left|F\right|$: 152198} &
        \tabincell{l}{ \uppercase\expandafter{BMF} \\
           \uppercase\expandafter{BNF} \\
           \uppercase\expandafter{CNR}\\
           \uppercase\expandafter{GNF}\\
           \uppercase\expandafter{UNF} \\
           \uppercase\expandafter{L0}\\
           \uppercase\expandafter{NLLR}\\
           \uppercase\expandafter{OURS}\\
           }
           & \tabincell{l}{10.03 \\9.67 \\9.70 \\8.50 \\9.42 \\11.21 \\9.61 \\ \textbf{8.06}
           }
           & \tabincell{l}{5.00 \\4.89 \\4.95 \\5.06  \\4.90 \\4.90 \\4.91 \\ \textbf{4.86}
           }
          	& \tabincell{l}{($ 30 $)\\$ (10,0.35,20) $\\$ (kinect fusion,10) $\\$(10,0.35,20) $\\$ (0.5,10,10) $\\$ (Default) $\\$ (Default) $\\($10$)}
              \\ \hline
  		\tabincell{l}{David\\(kinect fusion) \\(Figure~\ref{fig:com_scan} row 2)\\$\left|V\right|$: 51789 \\$\left|F\right|$: 101937} &
        \tabincell{l}{ \uppercase\expandafter{BMF} \\
           \uppercase\expandafter{BNF} \\
           \uppercase\expandafter{CNR}\\
           \uppercase\expandafter{GNF}\\
           \uppercase\expandafter{UNF} \\
           \uppercase\expandafter{L0}\\
           \uppercase\expandafter{NLLR}\\
           \uppercase\expandafter{OURS}\\
           }
           & \tabincell{l}{10.74 \\10.70 \\15.85  \\ \textbf{9.14} \\10.28 \\12.59 \\10.39 \\9.32 
            }
           & \tabincell{l}{1.60 \\1.572 \\2.09 \\1.65  \\ \textbf{1.566} \\1.60 \\1.58 \\2.12 
                     }
          	& \tabincell{l}{($ 30 $)\\$ (10,0.45,20) $\\$ (kinect fusion,5) $\\$(10,0.45,20) $\\$ (0.45,10,10) $\\$ (Default) $\\$ (Default) $\\($10$)}
              \\ \hline
  		\tabincell{l}{Pyramid\\(kinect fusion)\\(Figure~\ref{fig:com_scan} row 3)\\$\left|V\right|$: 35261 \\$\left|F\right|$: 69611} &
        \tabincell{l}{ \uppercase\expandafter{BMF} \\
           \uppercase\expandafter{BNF} \\
           \uppercase\expandafter{CNR}\\
           \uppercase\expandafter{GNF}\\
           \uppercase\expandafter{UNF} \\
           \uppercase\expandafter{L0}\\
           \uppercase\expandafter{NLLR}\\
           \uppercase\expandafter{OURS}\\
           }
           & \tabincell{l}{6.70 \\6.98 \\5.41 \\\textbf{5.40}  \\7.43 \\5.59 \\6.79 \\6.62 
                    }
           & \tabincell{l}{3.68 \\3.45 \\3.45 \\\textbf{3.28}  \\3.43 \\3.29 \\3.29 \\3.58 
           }
          	& \tabincell{l}{($ 30 $)\\$ (10,0.45,20) $\\$ (kinect fusion,3) $\\$(10,0.45,20) $\\$ (0.6,20,10) $\\$ (\sqrt{2},0.1,0.1) $\\$ (Default) $\\($5$)}
              \\ \hline                            
    \end{tabular}
\end{table}

\begin{table}[htbp]\scriptsize
    \centering
    \caption{Runtime comparisons of several state of the art methods with our method (in seconds). The vertex iterations of all the methods are set to 10. }\label{table:time}
    \begin{tabular}{|l|l|l|l|l|l|l|}\hline
    Models & UNF & BNF & GNF & L0 & NLLR & OURS\\
    \hline
   \tabincell{l}{Amadillo\\(Figure~\ref{fig:com_amadillo0.5})\\$\left|V\right|$: 43243 \\$\left|F\right|$: 86482}  & 5.68 & 7.12 &117.5 & 55.8 & 75.9 & \textbf{1.81}
    \\ \hline
    \tabincell{l}{iHbunny\\(Figure~\ref{fig:somemodels}\\ Row 1)\\$\left|V\right|$: 34834 \\$\left|F\right|$: 69451} & 3.80 &3.15 &35.6 & 102.4 & 21.4  & \textbf{1.39}
    \\ \hline 
   \tabincell{l}{Nicolo\\(Figure~\ref{fig:somemodels}\\ Row 2)\\$\left|V\right|$: 14846 \\$\left|F\right|$: 29437}  & 0.89 &0.83 & 28.0 & 23.8 & 19.0& \textbf{0.59}
    \\ \hline
  \tabincell{l}{Vaselion\\(Figure~\ref{fig:somemodels}\\ Row 5)\\$\left|V\right|$: 38728 \\$\left|F\right|$: 77452} & 6.52 &7.23 &84.6  & 54.2 & 27.8  & \textbf{1.54}
    \\ \hline    
   \tabincell{l}{Sphere\\(Figure~\ref{fig:com_sphere0.7})\\$\left|V\right|$: 10242 \\$\left|F\right|$: 20480} & 0.39 &0.54&8.97 & 16.1 & 15.1 & \textbf{0.38}
        \\ \hline             
    \end{tabular}
\end{table}

\textit{Raw scanned models.} In addition to the synthetic models, we also tested the proposed HLO on scanned models corrupted with raw noise. Figure~\ref{fig:com_scan} show the smoothing results of all methods on three real scanned surface meshes. Our method produces very competitive results, in terms of preserving features. Taking the second row as example, our HLO and NLLR \cite{Li2018} produce the best results while other methods may oversharpen or oversmooth certain regions. Note that NLLR \cite{Li2018} uses non-local information while ours utilizes local information only.

There are several reasons for why our method outperforms or is comparable to state-of-the-art techniques. (1) Our method is position based which largely mitigates the issue of flipped triangles. By contrast, most normal-based approaches (e.g., \cite{Sun2007,Zheng2011,Li2018}) cannot effectively overcome this issue, without using any pre-processing strategy like \cite{Lu2017-1,Lu2016}. \cite{He2013} is also position based; however it may over-sharpen some regions unexpectedly, due to the edge-based operator on the whole shape. CNR \cite{Wang-2016-SA} is a normal regression method that relies on immediate position update in each iteration; that is, it cannot directly predict the final normals. It thus easily smooths out features without a good mapping. (2) Multiple steps and parameters make those methods more complicated to search the solution space, thus leading to less automation and less robustness in handling different levels of noise, especially heavy noise. On the contrary, our method involves a single parameter, and is more simpler, automatic and robust in dealing with various noise.

\subsection{Quantitative Evaluations}
Besides the above visual comparisons, we also compare the state-of-the-art techniques with our approach from a quantitative perspective. Specifically, we employ $E_v$ and MSAE (mean square angular error) to respectively evaluate the positional error and normal error, as suggested by previous works \cite{Sun2007,Zheng2011,Lu2016}. These two metrics are calculated between the smoothing results and their corresponding ground truth. 

According to ~\cite{Sun2007}, $ E_v $ is the $ L^2 $ vetex-based mesh-to-mesh error metric, and MSAE measures the mean square angular error between the face normals of the denoised mesh and those of the ground truth.

\begin{equation}
E_v = \sqrt{\frac{1}{3\sum_{k\in F}A_{k}}\sum_{i\in V}\sum_{j\in F_{V}(i)}A_{j}dist(x^{'}_{i},T)^2},
\end{equation}

where $ A_k $ is the area of face $ k $, and $ dist(x
^{'}_{i},T) $ is the $ L^2 $ distance between the updated vertex $ x^{'}_{i} $ and a triangle of the reference mesh $ T $ which is closest to $ x^{'}_{i} $.

\begin{equation}
\text{MSAE}=\frac{\sum_{k\in F}\theta_k^2}{N_{F}} 
\end{equation}

where $ \theta_k $ is the angle between the $k$-th face normal of the denoised model and its corresponding normal in the ground-truth model, and $ N_{F} $ is the number of faces in the 3D shape.

Table \ref{table1} summarizes the statistical numbers of $E_v$ and MSAE over most models for all the compared methods. As with previous works \cite{Sun2007,Lu2016}, we also found that the visual comparison results might not necessarily agree with the $E_v$. This is because our method smooths surfaces via unfolding the flipped triangles in the noisy input, which is similar to the behavior of \cite{He2013}. As a consequence, vertices may walk far from their ground-truth locations. Regarding MSAE, the $L_0$ minimization \cite{He2013} and our method are usually the best two among all techniques. We also compare the runtime of our method with those methods and list them in Table \ref{table:time}. Our approach outperforms those methods in terms of speed. It has excellent speed performance in handling models with large sizes (e.g., Amadillo and Vaselion in Figure~\ref{fig:com_sphere0.7}). 
It should be noted that our code has not been optimized or speeded up via parallelization.

\section{Conclusion}
\label{sec:conclusion}
In this work, we have introduced a novel method for feature-preserving surface smoothing. Motivated by the shrinking and feature-wiping issues of the uniform Laplacian operator, we analyzed the differential property at feature points. We developed a Half-kernel Laplacian Operator (HLO) which can preserve surface features and resist shrinkage. Various experiments show that our approach is better or comparable to state-of-the-art surface smoothing techniques, in terms of visual quality and quantitative evaluations. 
Our method is robust to high noise and it has a single parameter (the number of vertex update iterations) which is easy and intuitive to tune.

\begin{figure} [htbp]
	\begin{center}
		\begin{tabular}{c} 
			\includegraphics[width=0.9 in]{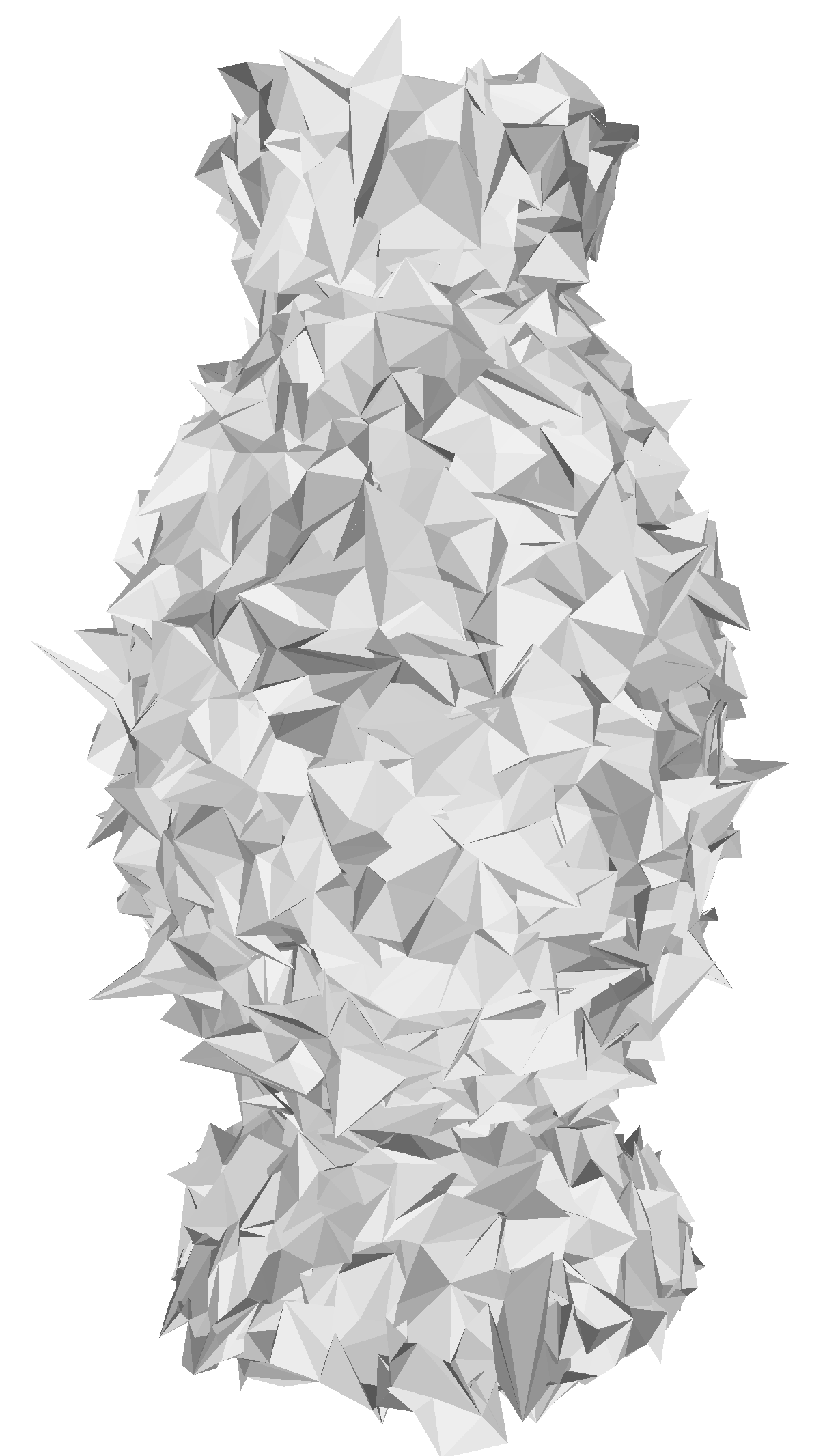}
			\includegraphics[width=0.9 in]{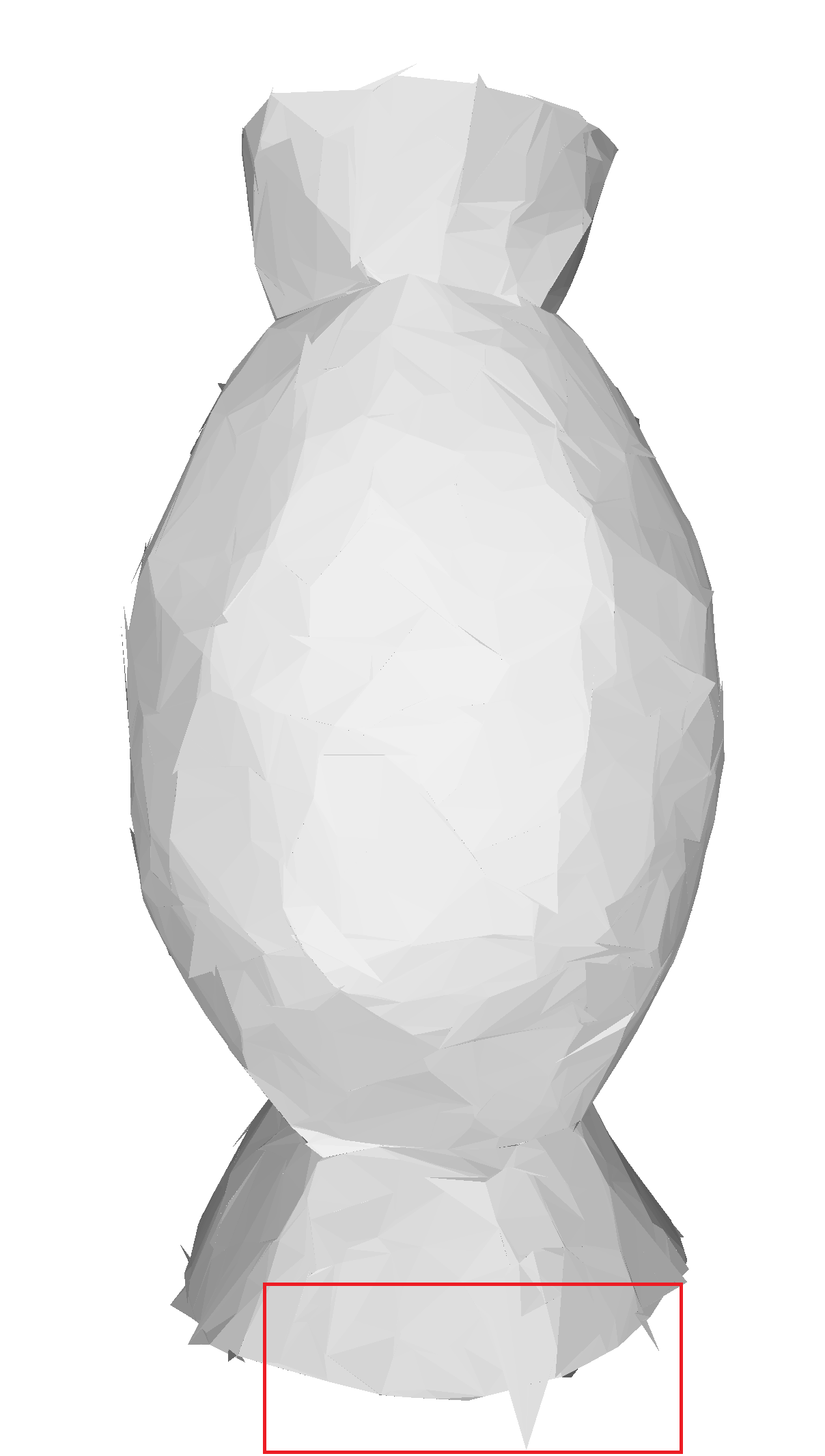}
			\includegraphics[width=0.9 in]{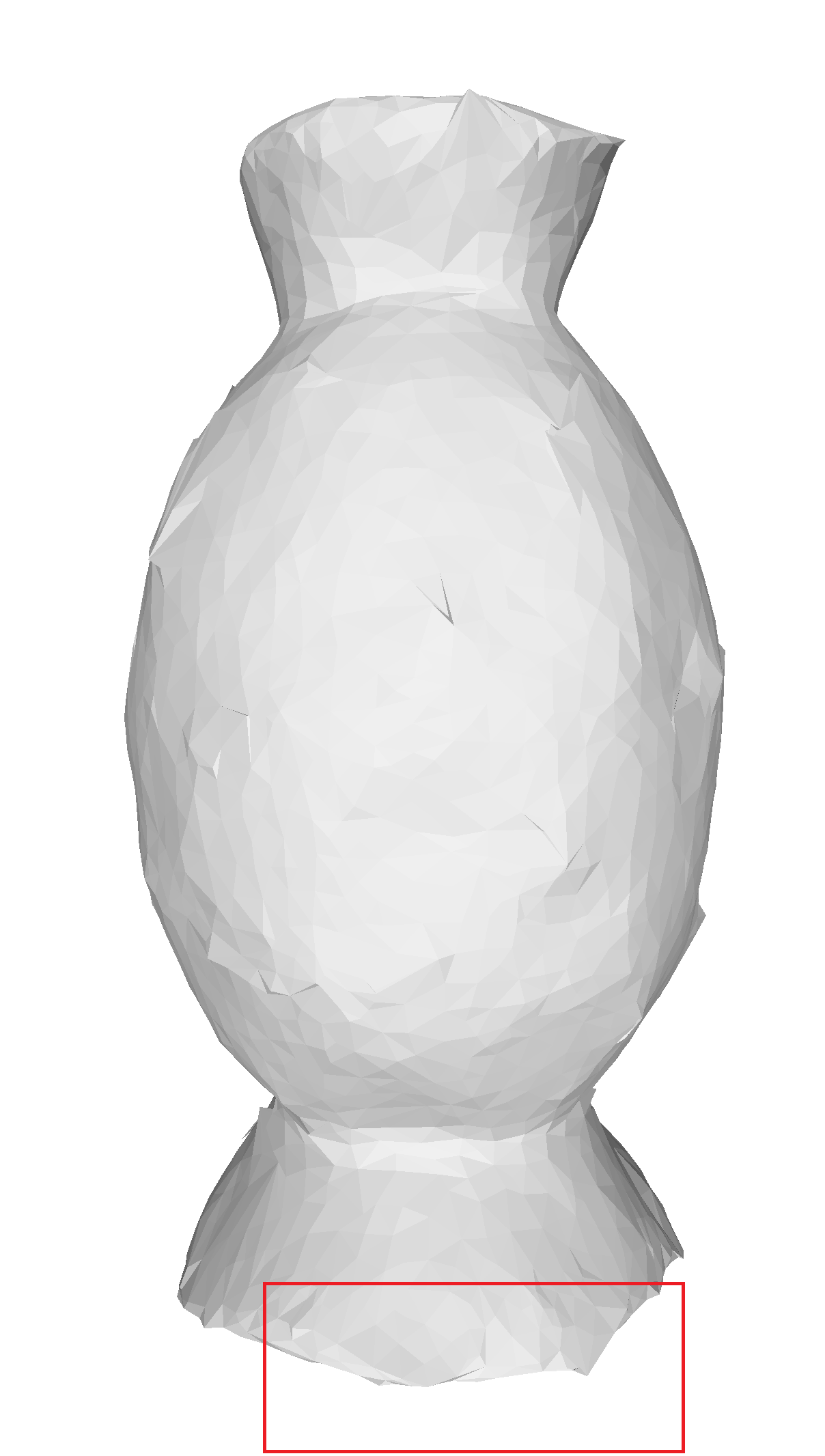}		\\
			\makebox[0.9 in]{(a) Noisy}
			\makebox[0.9 in]{(b) L0}	
			\makebox[0.9 in]{(c) Ours}	
		\end{tabular}
	\end{center}
	\caption[weights] 
	{ \label{fig:limit} 
	(a) The vase model with noise ($ \sigma_{n} = 1.5l_{e} $). (b) The smoothing result of L0 \cite{He2013}. (c) The smoothing result of our HLO after 10 iterations. }
\end{figure} 

Our method still suffers from a few limitations.  First, it has difficulty to recover smooth sharp edges for CAD-like models with a small number of vertices (Figure~\ref{fig:limit}). This is because HLO is not an edge-based operator \cite{He2013}. For real scanned CAD-like models with large number of vertices, our method can produce results with comparable quality to the state-of-the-art methods (Figure \ref{fig:com_scan} row 3). As with previous works \cite{Zheng2011,Lu2016}, we also fixed the open boundaries of the input noisy mesh, which might sometimes lead to undesired results.

As the future work, we would like to incorporate new schemes to overcome the issue of recovering smooth sharp edges. We would also like to improve the performance via parallelization. 



\section*{Acknowledgements}
This work was supported in part by the National Natural Science Foundation of China under Grant 61907031. Xuequan Lu is supported by Deakin CY01-251301-F003-PJ03906-PG00447. Ying He is supported by MOE RG26/17.

\section*{References}

\bibliography{arxiv}
 
\end{document}